# Modélisations et inversions des diagrammes de diffusion particulaire enregistrés par un photodétecteur organique conformable

par

*Matthias Sentis*

Thèse présentée à la faculté d'Aix-Marseille en vue de l'obtention du grade de

Docteur en Physique-Énergique

Thèse soutenue à Marseille, le 12 Décembre 2014

**Composition du jury :**

| | | | |
|---|---|---|---|
| Séverine BARBOSA | IUSTI | Maître de conférences | Examinatrice |
| Fabien CHAUCHARD-RIOS | INDATECH | Directeur général | Examinateur |
| Olivier DHEZ | ISORG/CEA | Directeur R&D | Examinateur |
| Laurent HESPEL | ONERA/DOTA | Directeur de recherche | Examinateur |
| Fabrice LEMOINE | LEMTA | Professeur | Rapporteur |
| Fabrice ONOFRI | IUSTI | Directeur de recherche | Directeur de thèse |
| Claude ROZE | CORIA | Professeur | Rapporteur |
| Maria-Rosaria VETRANO | VKI | Maître de conférences | Examinatrice |

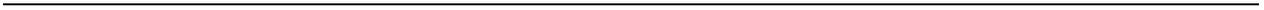

# Remerciements

Cette thèse de doctorat est financée par le fond unique interministériel, OSEO et les collectivités régionales que je tiens à remercier pour leur support durant ces trois ans.

Je remercie les rapporteurs de cette thèse, Fabrice Lemoine et Claude Rozé, ainsi que les autres membres du jury, Maria Rosaria Vetrano, Fabien Chauchard, Laurent Hespel, Olivier Dhez qui ont accepté de juger ces travaux

Je tiens particulièrement à remercier Fabrice Onofri qui m'a encadré pendant trois ans, pour ses nombreuses idées et conseils scientifiques très pertinents, pour sa patience et sa passion pour la recherche ainsi que pour son honnêteté et son franc-parler. Je remercie également Séverine Barbosa pour son encadrement et son aide pour la rédaction de cette thèse de doctorat.

J'exprime mes remerciements à Chantal Pelcé qui m'a aidé à corriger ce manuscrit et qui a toujours été à l'écoute durant ces trois ans.

Je remercie le laboratoire IUSTI qui m'a accueilli dans ses locaux et ses techniciens, ingénieurs et personnels administratifs qui m'ont beaucoup aidé :

- Joyce Bartolini pour son aide sur les missions et pour ses discutions
- Pascal Campion pour son aide administrative, ses conseils avisés ainsi que sa zénitude.
- Jean-Marie Felio pour les conceptions assistées par ordinateur (Catia) ainsi que pour la réalisation du masque de filtres optiques optimisé,
- Yann Jobic pour son aide sur les clusters de calcul
- Pierre Lantoine pour le contrôleur pas à pas et sa bonne humeur,
- Delphine Logos pour son aide sur les commandes de matériel
- Stéphane Martinez pour son énorme travail sur l'électronique d'acquisition développé et pour ses conseils,
- Sady Noél pour son aide dans la conception mécanique du banc goniométrique, sa motivation et son sourire de deuxième moitié de semaine,
- Julien Ouali-Gaffric (JOG) pour son aide en informatique, m'avoir initié au métal, et sa patience en colocation,
- Jeanne Pulino pour son aide administrative sur les différents contrats

Un grand merci aux thésards ou autre du laboratoire qui ont été pour moi, un vrai soutien pendant cette thèse : Anne-Marie l'allemande, Ariane la toulousaine, Benjamin le belge, Cedric la relève,

Claudia la collègue de bureau, Didier l'artiste amateur de bière, Florian l'américain, guillaume l'handicapé (puisses tu un jour marcher normalement), Jorge le chilien campeur, Lionel le randonneur, Quentin l'espion et certains que j'ai dû oublié (désolé).

Je remercie également ma famille: mes parents et mes trois frères Thomas, Arnaud et Raphaël qui ne m'ont pas beaucoup vu durant ces trois ans, pour leur soutien et leur patience.

Un grand merci aux amis qui m'ont soutenu et pris soins de moi alors que je ne le rendais pas forcément, les gars vous êtes les meilleurs : Benben, Charles, Edouard, Damien, Doudou, Guillaume, Joris, Laforeuse, Ludo, Lucho, Monique, Momo, Papy, Shorty, petit Vinz. J'espère sincèrement n'avoir oublié personne.

Pour finir, j'exprime toute ma gratitude à Eugénie qui m'a non seulement supporté malgré mes changements d'humeur mais qui m'a également soutenu et réconforté pendant ces trois ans. Sans toi, je n'aurai probablement pas fini, merci...

# Table des matières

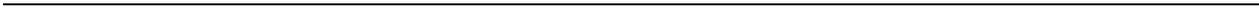

# Chapitre 1

## Introduction

Ces travaux de thèse de doctorat ont été réalisés au sein du laboratoire IUSTI, unité mixte de recherche n°7343 du CNRS et d'Aix-Marseille Université, et plus particulièrement dans le Groupe d'Optique des Systèmes Particulaires (GOSP). Ce dernier étudie les propriétés de diffusion de la lumière des systèmes particulaires et développe des méthodes optiques de caractérisation de ces milieux.

Cette thèse a également été conduite dans le cadre d'un projet "technologie Optique de rupture pour le *Process Analytical Technology* (Opti-PAT)" supporté par le fond unique interministériel, Bpifrance (OSEO) et les collectivités régionales. Il a rassemblé durant trois années un consortium de laboratoires industriels (INDATECH, ISORG, CEA/Liten, Ondalys, Sanofi) et académiques (IUSTI, Mines d'Alès) visant au développement de méthodes optiques non intrusives de contrôle des écoulements particulaires en milieu industriel. Ce sujet de recherche est directement issu du contexte du PAT, défini à l'origine par l'agence américaine des produits alimentaires et médicamenteux [**Balboni 2003**], et dont l'objectif est d'imposer à terme, dans les industries pharmaceutique, agroalimentaire et de traitement des effluents, l'emploi de capteurs permettant de contrôler en temps réel les procédés industriels, de même que les produits manufacturés ou traités, ceci, dans le but ultime d'améliorer la productivité, les performances et la compétitivité de ces industries. Des directives européennes vont également dans ce sens.

L'objectif scientifique de ce projet de recherche, et donc de cette thèse de doctorat, est de démontrer l'intérêt des photodétecteurs organiques conformables pour la caractérisation de milieux particulaires, en prenant en compte un cahier des charges lié aux applications PAT mais aussi aux limites technologiques actuelles de ces nouveaux détecteurs. Pour ce faire, nous avons développé différents outils de modélisation et d'optimisation de la réponse de ces capteurs du point de vue granulométrie optique uniquement. Des méthodes d'inversion très simples ont également été développées, de même qu'une expérience de laboratoire.

Dans la suite de cette introduction nous passons en revue, de manière succincte, les différents enjeux et points auxquels cette thèse de doctorat se réfère : les écoulements particulaires, les techniques de granulométrie optique, le PAT et les photodétecteurs organiques.

## 1.1 Les écoulements et suspensions particulaires

Les écoulements multiphasiques, les suspensions sont composés d'une phase continue (liquide ou gazeuse) et d'une phase dispersée (particules, gouttes ou bulles). Ils sont rencontrés, étudiés et employés dans bien des domaines, allant de la physique de l'atmosphère (formation des nuages, nucléation, agrégation et transport des aérosols,....) [**Mishchenko 2006**] à l'industrie lourde (formation des suies, traitement des effluents,...) ou l'industrie de haute technologie (pharmacie galénique, nanotechnologies,...), voir la Figure 1. Pour tous ces domaines, il est primordial de pouvoir caractériser et contrôler les propriétés intrinsèques de la phase dispersée et notamment la morphologie, la taille, la concentration, le matériau,...

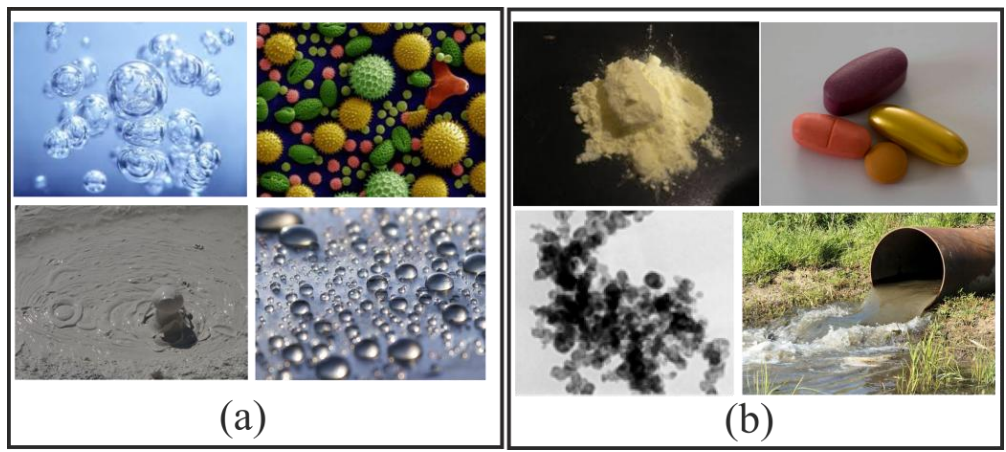

Figure 1 : Exemples de systèmes particulaires (a) naturels (bulles, pollens, boues et gouttes d'eau) et (b) industriels (poudres et comprimés pharmaceutiques, suies, eaux usées)

Les premières méthodes d'analyse granulométrique consistaient essentiellement à prélever un échantillon sur la ligne de production, puis à le tamiser [**Gulink 1943**]. D'autres méthodes se sont développées par la suite, comme les méthodes de sédimentation [**Gessner 1936**] ou les analyses par microscopie optique puis microscopie électronique (Microscope Électronique à Balayage,

MEB, Microscopie Électronique en Transmission, TEM). Ces techniques, robustes et relativement simples, sont cependant très intrusives. De plus, elles sont biaisées par la méthode de prélèvement et de manipulation de l'échantillon. Avec l'apparition du laser autour des années 70, les techniques optiques (ou "laser") d'analyse granulométrique se sont considérablement développées. Ces techniques permettent des analyses à distance et sont, de ce fait, considérées comme peu intrusives.

## 1.2  Les granulomètres optiques

Il existe un très grand nombre de techniques de granulométrie optique, leur classification devenant dès lors particulièrement difficile. On peut cependant différencier les techniques dynamiques (reposant, en pratique, sur une approximation quasi-statique) des techniques statiques, selon que la mesure est implicitement ou non liée au temps. On peut opérer un deuxième classement, selon la propriété véhiculant l'information : intensité, état de polarisation, fréquence ou phase de l'onde diffusée [**Onofri 2012**]. La forme des particules mesurables constitue également un critère de différentiation important, de même que la taille et la concentration minimale et/ou maximale de la phase dispersée. Le fait est que, à de rares exceptions près, les granulomètres optiques sont limités à l'analyse de milieux optiquement dilués et transparents, composés de particules uniquement (ou quasiment) sphériques et homogènes. Les milieux denses peuvent être analysés via des dilutions mais cette possibilité, parfois critique pour le système, n'est évidemment pas souhaitable. Ajoutons que pour un même appareil, la dynamique sur les tailles mesurables est souvent limitée. Le Tableau 1 dresse un panorama non exhaustif des principales techniques de granulométrie optique utilisées par les laboratoires et certaines industries pour caractériser les écoulements diphasiques. On constate sans surprise qu'un même système, et donc un même principe de mesure, ne peut pas couvrir l'intégralité de nos besoins en termes granulométriques (taille, forme, concentration, dynamique,...).

| Mesures / Techniques | Quantité | Concentration | Morphologie | Taille |
|---|---|---|---|---|
| *Diffractomètre* | Intensité angulaire | Faible | 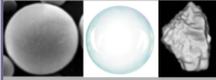 | 100nm-3mm |
| *Arc-en-ciel* | Intensité | Faible | 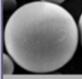 | 20µm-3mm |
| *Diffusion critique* | Intensité | Faible | 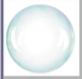 | 20µm-3mm |
| *Ellipsometrie* | Etat de polarisation | Faible | 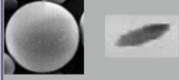 | 10nm-1µm |
| *Dynamic Light Scattering (DLS)* | Intensité | Faible à élevée | 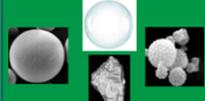 | 0.1nm-8µm |
| *Diffusive Wave Spectroscopy (DWS)* | Intensité | Très élevée | 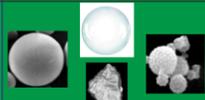 | 1mm-1µm |
| *Interférométrie Phase doppler* | Déphasage | Faible à intermédiaire | 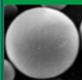 | 0.5µm-1mm |
| *Holographie* | Amplitude-Phase | Très faible | 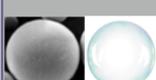 | 10µm-1mm |
| *Imagerie* | Taux de transmission | Très faible | 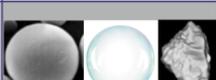 | >10µm |

Mesures dynamiques    Particules sphériques    Ellipsoïde    Forme quelconque

Mesures statiques    Bulles    Fractales

Tableau 1 : Principales techniques granulométriques optiques utilisées pour la caractérisation de systèmes particulaires en écoulement.

## 1.3 Le PAT et les problématiques de la mesure

Au fil des années, le monde industriel voit ses contraintes environnementales, économiques et technologiques s'alourdir. Ces changements nécessitent que les industriels s'adaptent et anticipent, en développant la maîtrise et la qualité des procédés de fabrication tout en réduisant leurs coûts et leurs impacts environnementaux.

Les autorités et instances normatives encouragent et obligent de plus en plus les fabricants à adopter des nouvelles techniques de mesure en temps réel de leurs produits [**El-Hagrasy 2006**]. Le *process analytical technology* cherche à répondre à ce besoin [**Moes 2008**]. Les fabricants d'instruments de mesure ont rapidement proposé des solutions s'appuyant sur des méthodes d'analyse en laboratoire ou avec des appareils de laboratoire. Ces premières solutions "déportées" nécessitent le prélèvement d'un échantillon sur la ligne de production. Elles sont donc inadaptées au contrôle en temps réel du procédé.

Le but du PAT est donc de développer une solution *in-line* pour la mesure des paramètres du produit. Il est important de distinguer ce type de mesure [**De Beer 2008**] de celles qui sont effectuées (les termes anglais font référence dans le domaine,...) :

- *"Off-line"* : l'échantillon est prélevé sur la chaîne de production (manuellement) pour être ensuite analysé par un laboratoire (physiquement très distant du procédé).

- *"At-Line"* : l'échantillon est prélevé sur la chaîne de production pour être analysé par un équipement situé près de (voire sur) la zone de production.

- *"On-line"* : l'échantillon est prélevé et analysé de manière automatique directement sur la ligne de production.

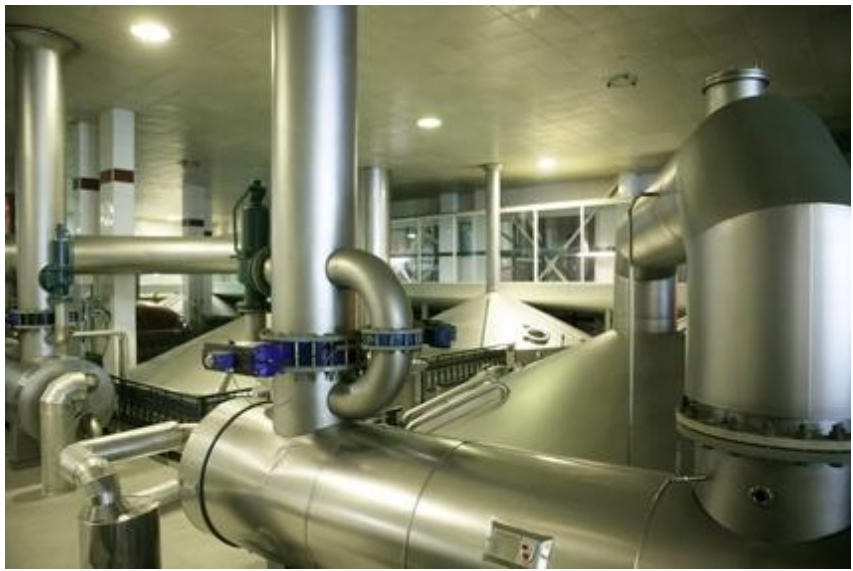

Figure 2 : Configuration type ou "idéale", selon le PAT, d'une instrumentation optique (en bleu) *in-line* implantée sur une ligne de production (pharmaceutique dans cet exemple) - image INDATECH

L'approche à adopter pour la caractérisation *in-line* est clairement d'utiliser sur la ligne de production une mesure sans contact et donc bien souvent optique. Les mesures optiques dans un tel environnement sont cependant très difficiles à mettre en œuvre. En effet, la morphologie, l'encombrement, les vibrations, les sources de contamination, etc… de la ligne de production peuvent être très complexes et variés (voir par exemple la Figure 2). Les contraintes de forme et d'encombrement sont particulièrement gênantes puisque la plupart des granulomètres actuels (diffractomètres, DLS,...) utilisent des cuves d'analyse calibrées et internes (généralement à section rectangulaire). Ils pèsent par ailleurs quelques dizaines de kilogrammes, doivent être positionnés sur une paillasse de laboratoire (voire une table optique) et ils sont suffisamment coûteux pour qu'il soit impensable de multiplier leur nombre sur une ligne de production.

On comprend ainsi mieux l'ambition du projet OPTIPAT qui est, en résumé, de permettre des mesures optiques *in-line* et à bas coût, ceci notamment par le biais de nouveaux photodétecteurs développés par le CEA-LITEN et la société ISORG, deux partenaires du projet.

Le premier "cahier des charges" de ce travail de thèse peut-être résumé ainsi :

- proposer des solutions pour permettre la mesure de la granulométrie de milieux particulaires dilués et si possible denses,

- particules de composition connue (indice) et de forme simple (sphère),

- distribution granulométrique pouvant être raisonnablement approximée par une distribution analytique monomodale à deux paramètres (diamètre moyen et écart-type),

- gamme granulométrique se situant entre 0.1 et 10µm pour le traitement des eaux et 10µm et 1mm pour les procédés pharmaceutiques,

- mesure effectuée au travers d'un cylindre de verre répondant aux normes du PAT (diamètre de 22 mm),

- l'instrumentation doit être aussi compacte que possible et peu coûteuse.

De plus, comme nous l'avons déjà évoqué, ces contraintes répondent également aux besoins de l'instrumentation scientifique et donc de la recherche académique.

## 1.4 Les photodétecteurs organiques

D'une manière générale, un photodétecteur (aussi appelé détecteur optique ou détecteur photosensible) est un composant qui produit un signal électrique (une tension ou un courant) dépendant du rayonnement électromagnétique incident sur sa surface. Il existe déjà une très grande variété de détecteurs optiques. La conversion photon/électron est généralement réalisée par un semi-conducteur et plus spécifiquement, dans le visible, par le silicium : photodiode, photodiode avalanche, phototransistor, photomultiplicateur, Charged-Coupled Device (CCD), Complementarity Metal-Oxide-Semiconductor (CMOS),.... On classe fréquemment ceux-ci en trois catégories, selon l'effet physique et le mode de fonctionnement mis en jeu [**Desvignes 1992**] :

> - *effet photoconductif* : les photons incidents produisent des électrons libres qui se déplacent de la bande de valence vers la bande de conduction. La conductivité électrique du détecteur varie alors en fonction du flux incident,

> - *effet photovoltaïque* : ce type de détecteur possède une jonction en matériau semi-conducteur. Les paires électrons-trous créées par l'impact des photons sur la surface sont séparées par le champ électrique au sein de la jonction P-N. Quand le flux de photons est suffisamment énergétique, une tension est ensuite générée,

> - *effet photoémissif* : les photons incidents libèrent des électrons de la surface du matériau de détection (en couche fine). Ces électrons sont par la suite collectés dans un circuit externe.

Certains détecteurs reposent sur d'autres semi-conducteurs, ou bien utilisent l'effet thermique pour les applications infrarouge par exemple, mais nous n'en dirons pas plus ici (pour plus de détails voir [**Desvignes 1992**]). Comme nous l'avons déjà évoqué, les photodétecteurs les plus utilisés pour la métrologie des écoulements utilisent le silicium en mode photoélectrique (CCD, CMOS, photodiodes) et dans certains cas, en mode photoémissif ou photoconductif (photodiodes). Les grandeurs les plus importantes pour juger de la qualité d'un photodétecteur sont les suivantes [**Desvignes 1992**] :

> - *sensibilité spectrale* : rapport de la puissance du signal de sortie sur la puissance du signal en entrée du détecteur en fonction de la longueur d'onde,

> - *efficacité quantique* : rapport du nombre d'électrons produits sur le nombre de photons incidents durant une unité de temps,

> - *constante de temps* : durée nécessaire pour que la réponse du détecteur atteigne 63% de sa valeur maximale,

> - *domaine de linéarité* : domaine pour lequel la réponse du détecteur est proportionnelle au flux lumineux incident,

- *ratio signal sur bruit* : rapport entre la puissance du signal et la puissance du bruit (notamment thermique).

Malgré le succès remarquable des technologies des semi-conducteurs "classiques", la production industrielle de photodétecteurs de forme complexe et de très grande taille reste problématique. De plus, même si les "wafers" présentent une certaine flexibilité, on ne peut pas parler de conformabilité. C'est pour cela notamment que, durant ces dernières années, de nombreux développements autour des technologies organiques ont vu le jour afin de dépasser les limitations évoquées précédemment, tout en diminuant le coût financier du développement de la mesure [**Nomura 2004, Ng 2008, Arca 2013**].

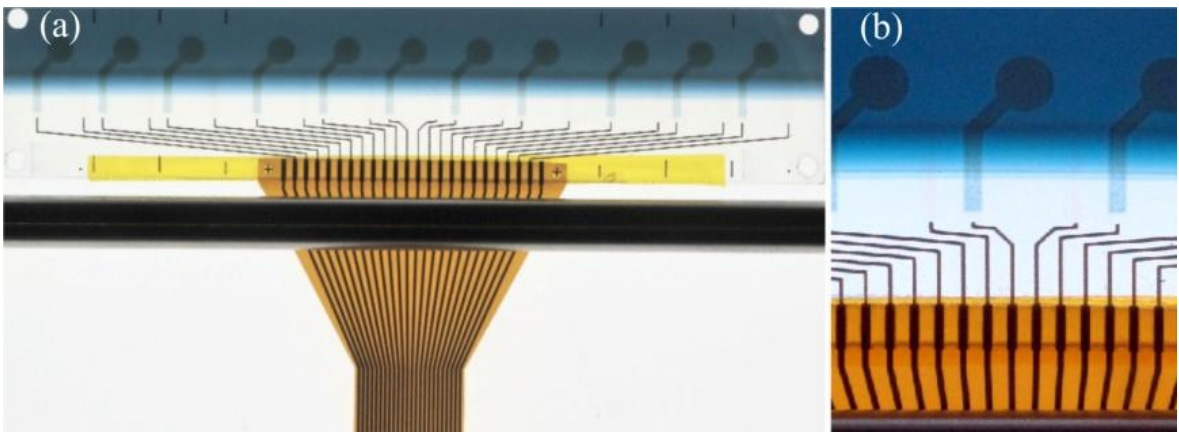

Figure 3 : Photographies (a) d'une feuille de photodétecteurs non optimisés et (b) zoom sur quelques zones photosensibles et leur connectique.

Avec des capteurs organiques photosensibles (OPS, voir la Figure 3), on peut imaginer produire, sur demande, et pour un prix raisonnable, des zones photosensibles de taille et de forme arbitraire. La conformabilité de ces films (de type plastique - Rhodoïd), permet d'envisager leur implantation dans des zones très complexes et confinées. En effet, ces surfaces peuvent être pliées jusqu'à atteindre des rayons de courbure de quelques centimètres, sans perte notable de rendement.

A ce stade, dans la mesure où ces détecteurs sont encore en développement, on ne peut pas en dire beaucoup sur les caractéristiques photosensibles et électroniques des OPS. Cependant, on peut comparer quelques grandeurs caractéristiques de cette technologie par rapport aux technologies de détection classiques. La Figure 4 compare ainsi les efficacités quantiques.

On remarque que l'efficacité quantique (QE) des photodétecteurs organiques est du même ordre de grandeur que celle des détecteurs conventionnels, quoique plus faible. Elle est maximale dans le jaune-rouge (580-620nm) avec QE#70%. On remarque également que leur efficacité s'écroule dans le proche infrarouge. On sait par ailleurs que le rayonnement UV endommage ces films organiques.

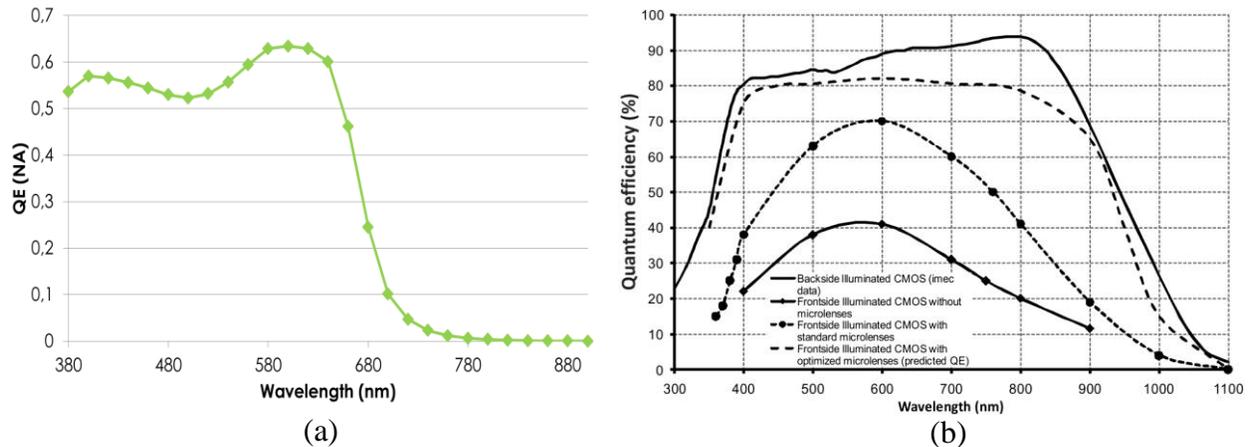

(a)            (b)

Figure 4 : Comparaison des efficacités quantiques (a) d'un photodétecteur organique (CEA-ISORG) et (b) différents photodétecteurs à semi-conducteurs [**Nikolic 2011**]

Cette technologie progressant à grands pas, on peut s'attendre dans les années à venir à une augmentation des performances de ces capteurs. Dans le cadre du projet OPTIPAT, les autres caractéristiques et limites principales des films produits par le CEA-LITEN et ISORG sont les suivantes :

- le film plastique est limité à une taille de 320x380mm avec une surface utile de 280x340mm,

- sur un même connecteur, le nombre de zones photosensibles est limité à 12

- la technologie d'impression multicouche (de type *Offset*) utilisée pour produire ces barrettes a un rendement de 90%. Cela signifie que, statistiquement, sur 10 zones photosensibles imprimées sur un même film, la probabilité qu'une ne fonctionne pas est non négligeable,

- la taille et la surface minimales d'une zone photosensible sont de 1mm - 1mm$^2$,

- la taille maximale n'est pas véritablement connue, mais disons qu'elle est non limitante pour le projet OPTIPAT (le taux de bruit et le temps de réponse sont probablement les deux points critiques),

- l'espacement minimal entre deux zones photosensibles est de 1mm,

- le temps de réponse est beaucoup plus élevé qu'avec la technologie silicium, de l'ordre de plusieurs dizaines de millisecondes pour une zone de 20mm$^2$. Dans l'état actuel, ceux-ci ne peuvent qu'être utilisés pour caractériser des écoulements moyens.

Ces atouts et limitations ont constitué, en quelque sorte, le second "cahier des charges" de ce travail de thèse, même si au début de celui-ci nous espérions des performances accrues sur, notamment, le nombre de détecteurs par barrette, la taille minimale des zones photosensibles et la possibilité de multiplier les tests.

## 1.5 Organisation du manuscrit de thèse

Pour répondre aux problématiques introduites précédemment, et de façon fondamentale sur l'intérêt des photodétecteurs organiques conformables actuels pour la granulométrie optique, nous avons procédé en plusieurs étapes. Ces dernières sont détaillées au travers des six chapitres principaux de ce manuscrit de thèse.

- Le **chapitre 2** synthétise, ou détaille, les principaux modèles et théories de diffusion de la lumière par une particule isolée : théories électromagnétiques, modèles d'optique physique et géométrique. Sur la base des travaux de Van de Hulst [**Hulst 1957**], un modèle "hybride" y est également développé pour introduire différentes notions fondamentales mais aussi comme cas test pour le modèle détaillé dans le chapitre suivant.

- Le **chapitre 3** présente un modèle de type Monte-Carlo permettant de simuler la diffusion de la lumière dans un environnement complexe. Ce dernier, intégralement développé durant cette thèse, permet de prendre en compte les nombreux paramètres influents dans un montage optique: forme du faisceau incident, forme des cuves et détecteurs, caractéristiques physiques et propriétés optiques du système particulaire,... Ce modèle a été implémenté puis parallélisé en Fortran/MPI. Il permet de modéliser, sous certaines hypothèses qui seront explicitées, la réponse attendue des OPS dans les conditions du PAT.

- Le **chapitre 4** propose quelques solutions de mesure pour les milieux particulaires dilués. Les différentes phases de développement et d'optimisation d'un prototype néphélométrique (multi-angulaire) sont présentées, ainsi que certaines méthodes inverses très simples. En complément, un prototype de mesure en configuration diffractométrique est également abordé.

- Le **chapitre 5** offre quelques pistes de mesure pour les milieux fortement concentrés. La réponse du prototype néphélométrique développé dans le **chapitre 4** y est simulée et quelques cas complémentaires sont présentés.

- Le **chapitre 6** présente le banc goniométrique expérimental développé au laboratoire pour tester les configurations optimisées, ainsi que des résultats expérimentaux préliminaires.

- Le **chapitre 7** est une conclusion générale avec mise en avant des perspectives attendues de ce travail de thèse de doctorat.

Ce manuscrit compte également 4 annexes qui sont placées, comme il est d'usage, en fin de texte.

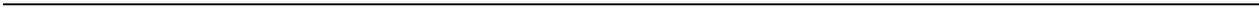

# Chapitre 2

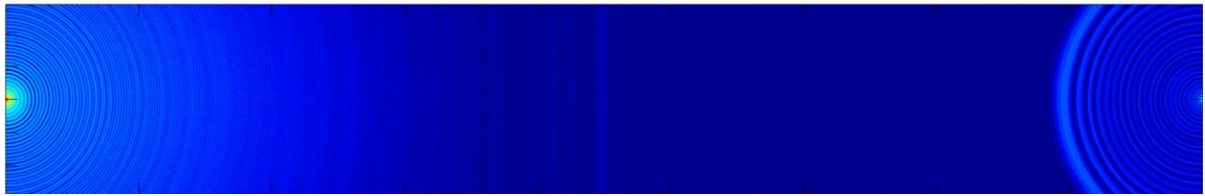

## Modèles de diffusion de la lumière par des particules et surfaces spéculaires

Ce chapitre a pour objectif d'introduire les principaux modèles et théories décrivant la diffusion et l'absorption de la lumière par des particules (sphériques ou de formes quelconques) et certaines surfaces spéculaires (décrites par des fonctions continues dérivables). Ces outils vont des modèles électromagnétiques (théorie de Lorenz-Mie, approximation en dipôles discrets ,...) aux modèles asymptotiques (optique géométrique, approximations d'optique physique,...), voir la Figure 5.

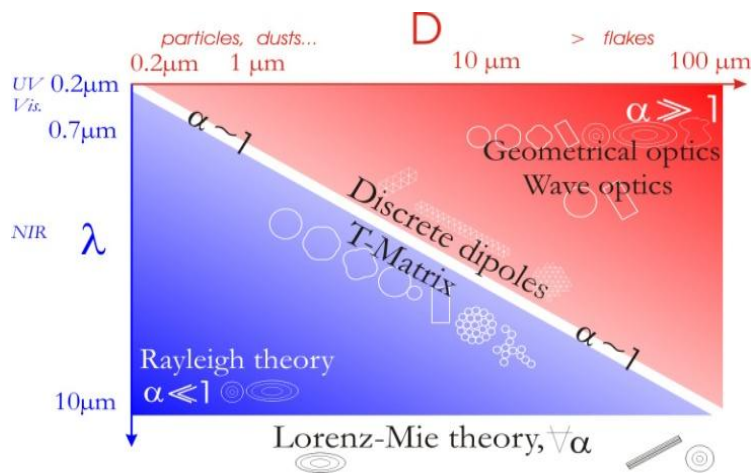

Figure 5 : Illustration de certaines des conditions de validité des approches les plus couramment utilisées pour modéliser les propriétés d'absorption et de diffusion de la lumière par des "particules"

Sans chercher à être exhaustif, compte tenu de l'ampleur du domaine, on introduira également un modèle "hybride" qui combine modèles géométriques et physiques. Ce modèle, implémenté dans un code Fortran, permet de simuler la diffusion d'une onde plane par une particule sphérique de grand paramètre de taille (rapport de la circonférence de la particule et de la longueur d'onde incidente). Les notions introduites dans ce chapitre, comme le modèle hybride mentionné ci-dessus, constituent des bases importantes pour la compréhension et l'élaboration du modèle de type Monte-Carlo détaillé par la suite.

## 2.1 Modèles électromagnétiques

### 2.1.1 Théorie de Lorenz-Mie

Depuis que cette dernière a été introduite en 1908 par Gustav Mie [**Mie 1908**], la "théorie de Lorenz-Mie" (LMT), est la plus largement utilisée pour la description des propriétés de diffusion et d'absorption de la lumière par des petites particules (cf. [**Bohren 1998**], [**Xu 2002**])) La LMT décrit de manière exacte l'interaction entre une onde électromagnétique plane monochromatique et une particule sphérique, homogène, isotrope à matériau linéaire et non magnétique, placée dans un milieu non absorbant. Cette théorie permet de résoudre les équations de Maxwell en utilisant une méthode de séparation des variables en coordonnées sphériques avec conditions aux limites. Dans ce qui suit, en suivant Bohren et Huffman [**Bohren 1998**], nous en décrivons les principales étapes et résultats.

Pour être solution des équations de Maxwell, l'onde électromagnétique doit satisfaire simultanément les équations d'onde pour les champs Électriques ($\mathbf{E}$) et Magnétiques ($\mathbf{H}$) :

$$\begin{cases} \nabla^2\mathbf{E} + k^2\mathbf{E} = 0 \\ \nabla^2\mathbf{H} + k^2\mathbf{H} = 0 \end{cases} \tag{1}$$

où $k^2 = \omega^2\varepsilon\mu$ est le vecteur d'onde, $\omega$ la pulsation d'onde, $\varepsilon$ et $\mu$ la permittivité électrique et la perméabilité magnétique respectivement. La résolution du système d'équations (1) revient dans un premier temps à résoudre une unique équation d'onde scalaire :

$$\nabla^2\psi + k^2\psi = 0 \tag{2}$$

où $\psi$ est une fonction liée aux harmoniques sphériques $\mathbf{M} = \nabla\times(\mathbf{r}\psi)$ et $\mathbf{N} = (\nabla\times\mathbf{M})/k$. L'équation d'onde scalaire peut être exprimée dans le système de coordonnées sphériques par :

$$\frac{1}{r}\frac{\partial}{\partial r}\left(r^2\frac{\partial\psi}{\partial r}\right)+\frac{1}{r^2\sin\theta}\frac{\partial}{\partial\theta}\left(\sin\theta\frac{\partial\psi}{\partial\theta}\right)+\frac{1}{r^2\sin\theta}\frac{\partial^2\psi}{\partial\phi^2}+k^2\psi=0 \qquad (3)$$

Une des idées clés de la théorie de Lorenz-Mie est d'utiliser une méthode de séparation variable (SVM) pour trouver des solutions particulières de l'équation (3) :

$$\psi\left(r,\theta,\phi\right)=R(r)\Theta(\theta)\Phi(\phi) \qquad (4)$$

où les trois fonctions $R$, $\Theta$ et $\Phi$ sont indépendantes. On se ramène alors à un système à trois équations, avec $m$ et $n$ des constantes de séparation, de la forme :

$$\begin{cases} \dfrac{d^2\Phi}{d\phi^2}+m^2\Phi=0 & (a) \\[2mm] \dfrac{1}{\sin\theta}\dfrac{d}{d\theta}\left(\sin\theta\dfrac{d\Theta}{d\theta}\right)+\left[n\left(n+1\right)-\dfrac{m^2}{\sin^2\theta}\right]=0 & (b) \\[2mm] \dfrac{d}{dr}\left(r^2\dfrac{dR}{dr}\right)+\left[k^2r^2-n\left(n+1\right)\right]R=0 & (c) \end{cases} \qquad (5)$$

Les solutions linéairement indépendantes de l'équation (5)-(a) sont du type suivant:

$$\begin{cases} \Phi_e=\cos\left(m\phi\right) \\ \Phi_o=\sin\left(m\phi\right) \end{cases} \qquad (6)$$

où les indices $e$ et $o$ désignent les ordres pair et impair respectivement.

Les solutions de l'équation (5)-(b) (qui sont finies en $\theta=0$ et $\theta=\pi$) sont les polynômes de Legendre $P_n^m(\cos\theta)$. La résolution de l'équation (5)-(c) nécessite l'introduction de la variable adimensionnelle $\rho=kr$ (pour un changement de variable) et de la fonction $Z=R\sqrt{\rho}$ de sorte que cette dernière puisse être réécrite :

$$\rho\frac{d}{d\rho}\left(\rho\frac{dZ}{d\rho}\right)+\left[\rho^2-\left(n-\frac{1}{2}\right)^2\right]Z=0 \qquad (7)$$

Les solutions linéairement indépendantes de l'équation (7) sont des combinaisons de fonctions de Bessel sphériques de première et deuxième espèces $j_n$, $y_n$, $k_n^{(1)}$, $k_n^{(2)}$. En coordonnées sphériques, les solutions qui satisfont l'équation d'onde scalaire (2) sont de la forme suivante [**Bohren 1998**] :

$$\begin{aligned} \psi_{emn}&=\cos(m\phi)P_n^m\left(\cos\theta\right)z_n\left(kr\right)\\ \psi_{omn}&=\sin(m\phi)P_n^m\left(\cos\theta\right)z_n\left(kr\right) \end{aligned} \qquad (8)$$

Dans le cadre de l'optique linéaire, les champs diffusés ($\mathbf{E}_S,\mathbf{H}_S$) et les champs internes à la particule ($\mathbf{E}_p,\mathbf{H}_p$) sont simplement proportionnels au champ incident et peuvent être exprimés sous la forme de séries infinies d'harmoniques sphériques pondérées par des coefficients complexes $a_n, b_n, c_n, d_n$ :

$$\begin{cases} \mathbf{E}_p = \sum_{n=1}^{\infty} E_n \left( c_n \mathbf{M}_{o1n}^{(1)} - i d_n \mathbf{N}_{e1n}^{(1)} \right) \\ \mathbf{H}_p = \frac{-k_p}{\omega \mu_p} \sum_{n=1}^{\infty} E_n \left( d_n \mathbf{M}_{e1n}^{(1)} - i c_n \mathbf{N}_{o1n}^{(1)} \right) \end{cases} \begin{cases} \mathbf{E}_s = \sum_{n=1}^{\infty} E_n \left( i a_n \mathbf{N}_{e1n}^{(3)} - b_n \mathbf{M}_{o1n}^{(3)} \right) \\ \mathbf{H}_s = \frac{k}{\omega \mu} \sum_{n=1}^{\infty} E_n \left( i b_n \mathbf{N}_{o1n}^{(3)} - a_n \mathbf{M}_{e1n}^{(3)} \right) \end{cases} \quad (9)$$

où $E_n = i^n E_0 (2n+1)/n(n+1)$. Dans l'équation (9), les coefficients $a_n, b_n$ sont appelés les "coefficients de diffusion externe" ou "coefficients de Mie", les coefficients $c_n, d_n$ sont les "coefficients de diffusion interne".

L'écriture des conditions limites au centre de la particule et à l'infini, permet de rejeter (ou retenir) certaines des fonctions de Bessel utilisées pour les combinaisons linéaires introduites par l'équation (7). Pour les ondes électromagnétiques, les composantes tangentielles des champs électromagnétiques doivent être continues sur la surface des particules, d'où pour la seconde série de conditions limites :

$$\begin{cases} E_{i\theta} + E_{s\theta} = E_{p\theta} \\ H_{i\theta} + H_{s\theta} = H_{p\theta} \end{cases} \begin{cases} E_{i\phi} + E_{s\phi} = E_{p\phi} \\ H_{i\phi} + H_{s\phi} = H_{p\phi} \end{cases} \quad (10)$$

En introduisant les fonctions de Ricatti-Bessel définies par :

$$\psi_n(\rho) = \rho j_n(x), \quad \xi_n(x) = \rho h_n^{(1)}(x), \quad (11)$$

avec $x = \pi D/\lambda$ pour le paramètre de taille, $m$ pour l'indice relatif de la particule (différent de la variable de séparation) de perméabilité relative $\mu = 1$, les coefficients de diffusion externe s'expriment alors de la manière suivante :

$$a_n = \frac{m \psi_n(mx) \psi'_n(x) - \psi_n(x) \psi'_n(mx)}{m \psi_n(mx) \xi'_n(x) - \xi_n(x) \psi'_n(mx)}$$

$$b_n = \frac{\psi_n(mx) \psi'_n(x) - m \psi_n(x) \psi'_n(mx)}{\psi_n(mx) \xi'_n(x) - m \xi_n(x) \psi'_n(mx)} \quad (12)$$

Pour des raisons numériques, ces coefficients $a_n, b_n$ sont généralement reformulés à l'aide des dérivées logarithmiques des fonctions de Riccati-Bessel :

$$a_n = D_n^{(3)} \frac{m D_n^{(1)}(x) - D_n^{(1)}(mx)}{m D_n^{(2)}(x) - D_n^{(1)}(mx)}$$

$$b_n = D_n^{(3)} \frac{D_n^{(1)}(x) - m D_n^{(1)}(mx)}{D_n^{(2)}(x) - m D_n^{(1)}(mx)} \quad (13)$$

où :

$$D_n^{(1)}(z) = \frac{\psi_n^{'}(z)}{\psi_n(z)}, \quad D_n^{(2)}(z) = \frac{\xi_n^{'}(z)}{\xi_n(z)}, \quad D_n^{(3)}(z) = \frac{\psi_n(z)}{\xi_n(z)} \tag{14}$$

Dans l'hypothèse de champ lointain ($kr \gg \lambda$), la relation entre les champs électriques incident et diffusé peut être exprimée pour les composants parallèles $\parallel$ et perpendiculaires $\perp$ au plan de diffusion comme suit :

$$\begin{pmatrix} E_{s\parallel} \\ E_{s\perp} \end{pmatrix} = \frac{e^{ik(r-z)}}{-ikr} \begin{pmatrix} S_2 & S_3 \\ S_4 & S_1 \end{pmatrix} \begin{pmatrix} E_{i\parallel} \\ E_{i\perp} \end{pmatrix} \tag{15}$$

Pour une particule sphérique nous avons $S_3 = S_4 = 0$ et $S_1, S_2$ sont donnés par :

$$S_1 = \sum_{n=1}^{\infty} \frac{(2n+1)}{n(n+1)} \left( a_n \pi_n + b_n \tau_n \right)$$
$$S_2 = \sum_{n=1}^{\infty} \frac{(2n+1)}{n(n+1)} \left( a_n \tau_n + b_n \pi_n \right) \tag{16}$$

avec pour les fonctions angulaires :

$$\pi_n = \frac{P_n^1}{\sin\theta}, \quad \tau_n = \frac{dP_n^1}{\sin\theta} \tag{17}$$

En utilisant le vecteur de Poynting il est possible de calculer les relations pour les intensités de diffusion $i_\parallel \sim |S_2|^2$ et $i_\perp \sim |S_1|^2$ [**Bohren 1998**].

Les sections efficaces d'extinction, de diffusion et d'absorption se calculent à l'aide des séries :

$$C_s = \frac{2\pi}{k^2} \sum_{n=1}^{\infty} (2n+1) \left( |a_n|^2 + |b_n|^2 \right) \qquad \text{(a)}$$
$$C_{ext} = \frac{2\pi}{k^2} \sum_{n=1}^{\infty} (2n+1) \operatorname{Re}\{a_n + b_n\} \qquad \text{(b)} \tag{18}$$
$$C_a = C_{p,ext} - C_{p,sca} \qquad \text{(c)}$$

Numériquement, les séries d'expansion infinie sont tronquées à partir de $n > n_{stop}$ :

$$n_{stop} = x + 4x^{1/3} + 2, \tag{19}$$

On note que, selon le principe de l'approximation localisée [**Hulst 1957**], les termes d'expansion $n$ peuvent être associés à des rayons (au sens de l'optique géométrique) qui impactent sur la surface de la particule à la distance $R_n$ de son centre :

$$R_n = \left( n + \frac{1}{2} \right) \frac{\lambda}{2\pi}. \tag{20}$$

Depuis que celle-ci a été introduite, la théorie de Lorenz-Mie a été largement étendue aux cas d'une sphère homogène éclairée par un faisceau de forme arbitraire [**Barton 1988, Gouesbet 1988**], d'une sphère multicouche ou hétérogène [**Onofri 1995**], d'un sphéroïde [**Ren 1997**]. De nombreux codes de calcul et de nombreuses applications utilisant la LMT ont été développées (cf. [**Wriedt 2009**]).

### 2.1.2 Séries de Debye

En 1909, Debye a formulé [**Debye 1909**] d'une façon un peu différente le problème de la diffusion dite de « Mie ». Il découle de ses travaux que l'on peut réécrire les expressions obtenues avec la LMT sous la forme de contributions imputables à des ondes partiellement réfléchies et partiellement transmises par la particule. Du fait de la géométrie du problème, ces ondes sont nécessairement sphériques pour une sphère et cylindriques pour un cylindre éclairé sous incidence normale. On distingue les ondes diffusées par la particule, de celles qui se propagent à l'intérieur de cette dernière. Ceci amène à l'introduction de coefficients de «réflexion» et de «transmission » pour ces ondes partielles.

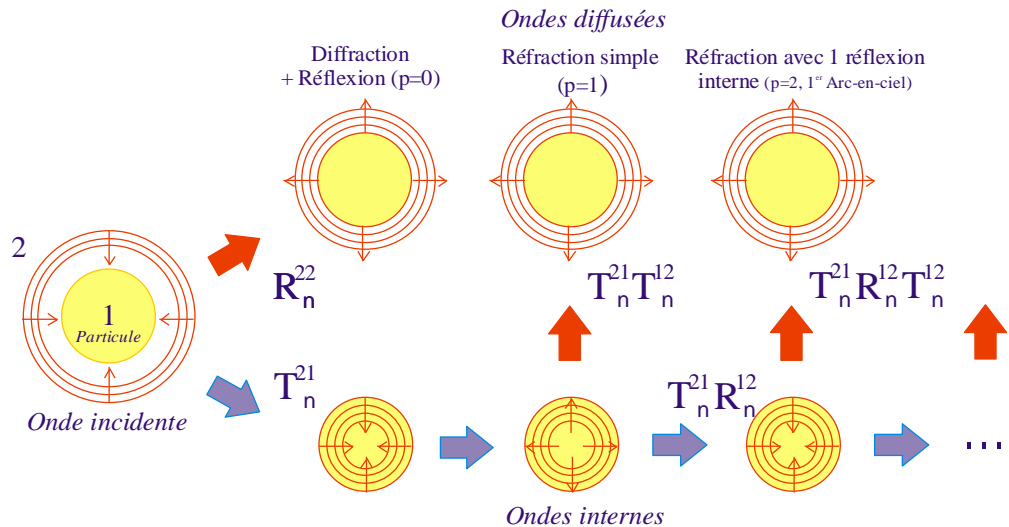

Figure 6 : Décomposition de Debye : le champ incident et le champ diffusé sont décomposés en ondes partielles ayant subi $p$ réflexions sur l'interface particule/milieu extérieur [**Hovenac 1992**]

La Figure 6 présente un schéma d'interprétation de la méthode de décomposition de Debye, telle qu'elle a été reformulée par Hovenac et Lock [**Hovenac 1992**] (voir également les références [**Albrecht 2003, Gouesbet 2004**]). La particule est associée au milieu 1 et le milieu extérieur au milieu 2, avec :

- $R_n^{(22)}$ : coefficient de "réflexion" de l'onde incidente partiellement "réfléchie" (réflexion spéculaire et diffraction) vers le milieu extérieur,

- $T_n^{(21)}$ : coefficient de "transmission" des ondes externes partielles vers l'intérieur de la particule,

- $R_n^{(11)}$ : coefficient de "réflexion" des ondes partielles internes sur la surface interne de la particule,

- $T_n^{(12)}$ : coefficient de "transmission" des ondes partielles internes vers le milieu extérieur.

La méthode de décomposition de Debye permet d'obtenir les coefficients de transmission et de réflexion des ondes partielles qui, dans le cas d'une particule sphérique, s'écrivent comme suit [**Hovenac 1992**] (voir également la Figure 6) :

$$R_n^{(22)} = \frac{\alpha H_n'^{(1)}(x) H_n^{(2)}(y) - \beta H_n^{(1)}(x) H_n'^{(2)}(y)}{D_n(x,y)} \tag{21}$$

$$R_n^{(11)} = \frac{\alpha H_n'^{(1)}(x) H_n^{(1)}(y) - \beta H_n^{(1)}(x) H_n'^{(1)}(y)}{D_n(x,y)} \tag{22}$$

$$T_n^{(21)} = -\frac{2i(m_1/m_2)}{D_n(x,y)} \tag{23}$$

$$T_n^{(12)} = -\frac{2i}{D_n(x,y)} = \frac{T_n^{(21)}}{(m_1/m_2)} \tag{24}$$

avec

$$D_n(x,y) = -\alpha H_n'^{(1)}(x) H_n^{(2)}(y) + \beta H_n^{(1)}(x) H_n'^{(2)}(y) \tag{25}$$

$$x = \frac{\pi D}{\lambda} = kr \qquad y = \left(\frac{m_1}{m_2}\right) x = mkr \tag{26}$$

avec $m_1$ l'indice de réfraction de la particule et $m_2$ l'indice du milieu environnant.

Les fonctions de Hankel, qui sont souvent symbolisées par les lettres grecques "Xi" et "Dzêta": $H_n^{(1)}(z) = \xi_n(z)$ et $H_n^{(2)}(z) = \zeta_n(z)$ [**Abramowitz 1964**] avec $z = x$ ou $z = y$, peuvent être explicitées comme suit :

$$H_n^{(1)}(z) = \psi_n(z) + i\chi_n(z) = z\left[j_n(z) + iy_n(z)\right]$$
$$H_n^{(2)}(z) = \psi_n(z) - i\chi_n(z) = z\left[j_n(z) - iy_n(z)\right] \tag{27}$$

où $j_n(z), y_n(z)$ sont les fonctions de Bessel sphériques du premier et second ordre (ou fonctions sphériques de Bessel et de Neumann) avec $z \in \mathbb{C}$, $n \in \mathbb{N}$. Elles sont reliées aux fonctions de Ricatti-Bessel par :

$$\psi_n(z) = z j_n(z) \qquad \chi_n(z) = z y_n(z) \tag{28}$$

En utilisant pour celles-ci la loi de récurrence suivante :

$$B'_n(z) = B_{n-1}(z) + \frac{n}{z} B_n(z) \tag{29}$$

les dérivées des fonctions de Hankel s'expriment comme suit :

$$H'^{(1)}_n(z) = H^{(1)}_{n-1}(z) + \frac{n}{z} H^{(1)}_n(z)$$
$$H'^{(2)}_n(z) = H^{(2)}_{n-1}(z) + \frac{n}{z} H^{(2)}_n(z) \tag{30}$$

Au final, les coefficients de diffusion externe des différents ordres de la décomposition de Debye sont donnés par l'équation (32).

Dans l'équivalence donnée par l'équation (33), entre les coefficients des séries de Debye et ceux de la théorie de la Lorenz-Mie, ces derniers sont de la forme :

$$a_n = \frac{m \psi_n(y) \psi'_n(x) - \psi_n(x) \psi'_n(y)}{m \psi_n(y) H'^{(1)}_n - H^{(1)}_n \psi'_n(y)}$$
$$b_n = \frac{\psi_n(y) \psi'_n(x) - m \psi_n(x) \psi'_n(y)}{\psi_n(y) H'^{(1)}_n - m H^{(1)}_n \psi'_n(y)} \tag{31}$$

Pour l'onde partielle $p$ et l'ordre d'expansion $n$, les coefficients de diffusion externe s'écrivent :

$$\left. \begin{array}{c} a_n(p) \\ b_n(p) \end{array} \right\} = \frac{1}{2} \left\{ \begin{array}{l} 1 - R_n^{(22)} \\ -T_n^{(21)} \left( R_n^{(11)} \right)^{p-1} T_n^{(12)} \end{array} \right. \quad \text{pour} \quad \left\{ \begin{array}{l} \text{p=0} \\ \text{p} \geq 1 \end{array} \right. \tag{32}$$

où $p = 0$ correspond à la diffraction et la réflexion spéculaire (indissociables ici), $p = 1$ à la réfraction simple, $p = 2$ à l'onde partielle qui a subi une réflexion interne, $p = 3$ avec deux réflexions internes, etc…

Il est important de noter qu'il existe une stricte équivalence entre les coefficients externes de diffusion de la théorie de Lorenz-Mie et ceux de la théorie de Debye (à condition que l'ordre de la décomposition $p \to \infty$, bien qu'en pratique $p \geq 100$ s'avère souvent suffisant) :

$$\left\{ \begin{array}{c} a_n \\ b_n \end{array} \right\}_{LMT} = \frac{1}{2} \left[ 1 - R_n^{(22)} - \sum_{p=1}^{\infty} T_n^{(21)} \left( R_n^{(11)} \right)^{p-1} T_n^{(12)} \right] \text{ pour } \alpha = \left\{ \begin{array}{c} m \\ 1 \end{array} \right. \text{ et } \beta = \left\{ \begin{array}{c} 1 \\ m \end{array} \right. \text{ et p} \to \infty \tag{33}$$

Le calcul numérique direct de ces séries est relativement stable et si l'on compare les diagrammes de diffusion obtenus avec les séries de Debye et la LMT, on trouve un très bon accord. De petites différences apparaissent néanmoins à certains angles. En fait, les séries de Debye nécessitent la sommation complexe d'un plus grand nombre de fonctions que la LMT. Le calcul complet de ces séries est donc nécessairement plus sensible au développement du bruit numérique.

Pour plus de détails, on se reportera à l'habilitation à diriger des recherches de F. Onofri [**Onofri 2005**].

## 2.1.3 Méthodes numériques

Comme nous l'avons vu dans les précédents paragraphes, les théories "classiques" permettent d'étudier la diffusion de la lumière par des particules de forme canonique : sphère, ellipsoïde, cylindre,.... Le calcul des propriétés de diffusion de particules de forme complexe (comme des suies, des poussières, des aérosols...) requiert l'utilisation de méthodes "numériques" appropriées. Cette section a pour but de présenter brièvement les principales méthodes numériques utilisées pour calculer les propriétés de diffusion de la lumière par des particules de forme quelconque.

### 2.1.3.1 Approximation en dipôles discrets (DDA)

L'idée de base de cette méthode a été introduite en 1964 par Howard DeVoe [**DeVoe 1964**] qui l'a utilisée pour étudier les propriétés optiques d'agrégats moléculaires. Cette méthode n'était alors valide que pour des agrégats de très petite taille par rapport à la longueur d'onde. Purcell et Pennypacker ont généralisé la DDA à des agrégats de plus grande taille pour étudier les contributions des grains de poussières interstellaires [**Purcell 1973**].

De manière très succincte, cette méthode repose sur une discrétisation de la particule diffusante en N dipôles élémentaires de diamètre $D \ll \lambda$ et de polarisabilité $\alpha$ [**Onofri 2012**]. Le champ total est obtenu en calculant le champ rayonné par chacun des N dipôles ainsi que tous les contre-champs induits. Pour obtenir des résultats satisfaisants, il faut un grand nombre de dipôles. La DDA nécessite de ce fait des ressources informatiques conséquentes (CPU et surtout, mémoire). En plus de certaines instabilités numériques, tout ceci fait que cette méthode ne s'applique qu'aux particules dont le paramètre de taille $x$ et l'indice de réfraction $m_p$ sont limités ( $\left| m_p k x \right| \le 0.25$ , $\left| m_p - 1 \right| < 2$ ) [**Draine 1994**]). Le principal avantage de cette méthode réside dans le fait que cette dernière permet de calculer les propriétés de diffusion de particules de forme quelconque, simplement en choisissant un maillage adapté aux particules étudiées. Le second avantage de la DDA est son domaine d'application qui s'étend avec le développement des capacités de calcul et de stockage des ordinateurs.

### 2.1.3.2 La méthode de la T-Matrice

La méthode de la T-Matrice est une méthode numérique très utilisée pour calculer les propriétés de diffusion de la lumière par des particules de forme quelconque. Cette méthode introduite par Waterman [**Waterman 1965**] en 1965, qui s'appuie sur une méthode aux conditions aux limites étendues, a été depuis largement développée et améliorée.

La méthode de la T-matrice repose sur le principe d'équivalence de Schelkunoff [**Schelkunoff 1992**] selon lequel le champ électromagnétique à l'extérieur d'une surface régulière S est équivalent à celui produit par une distribution de courants superficiels électriques et magnétiques répartis sur cette surface S. A l'intérieur de la surface S, les sources produisent un champ électrique et magnétique nul (la T-Matrice est également connue comme la méthode des champs nuls). Par conséquent, le champ total (incident et diffusé) à l'extérieur de la surface S peut être exprimé sous forme d'intégrales de surface. Cette dernière hypothèse n'est pas valable sur la surface S elle-même. Elle est valable uniquement à l'intérieur de la sphère inscrite ou à l'extérieur de la sphère circonscrite à la surface S. Les courants de surface équivalents apparaissent comme étant la somme de n - harmoniques sphériques pondérée par des coefficients inconnus. Au sein de la sphère inscrite, l'équation $\mathbf{E} = \mathbf{0}$ est transformée en un système d'équations linéaires qui relie les coefficients inconnus du champ diffusé avec les coefficients connus des champs incidents. La résolution du système matriciel résultant (d'où le nom de T-matrice), et dont les coefficients sont des combinaisons d'intégrales harmoniques sphériques sur S, donne alors accès au champ diffusé. Le nombre d'harmoniques n dépend de la forme, de la taille et de l'indice de réfraction de la particule diffusante.

De nombreux codes de calcul implémentant la méthode de la T-Matrice ont été développés (par exemple [**Barber 1990, Mishchenko 1996, Auger 2007, Nieminen 2007**]). Elle a rendu possible la caractérisation optique de systèmes particulaires (par exemple [**Doicu 2006, Martin 2006, Binek 2007, Onofri 2013**]).

Pour plus de détails, on se référera à la thèse de M. Wozniak [**Wozniak 2012**].

## 2.2 Modèles (ou "approximations") d'optique physique

L'optique physique ou "ondulatoire", est apparue au XVII$^{ème}$ afin d'expliquer certaines observations (irisations, anneaux de diffraction,...) incompatibles avec une vision purement géométrique des phénomènes de diffusion. Nous n'évoquerons ici que les points clés et principaux résultats des modèles utilisés dans la suite de ce travail. Le lecteur pourra approfondir ces différents points en se rapportant aux références bibliographiques fournies.

### 2.2.1 La théorie de la diffraction de Fraunhofer

Le principe d'Huygens-Fresnel est à la base de l'interprétation de la diffraction. Ce principe est exclusivement ondulatoire et se décompose en deux contributions que l'on peut tenter de résumer ainsi :

- la contribution d'Huygens [**Huygens 1690**] : la lumière se propage de proche en proche. Chaque élément de surface atteint par celle-ci se comporte comme une source secondaire qui émet des ondelettes sphériques dont l'amplitude est proportionnelle à cet élément.

- la contribution de Fresnel [**Fresnel 1816**] : l'amplitude complexe de la vibration lumineuse en un point est la somme des amplitudes complexes des vibrations produites par toutes les sources secondaires. Toutes ces sources interfèrent pour reformer à "l'identique" la vibration au point considéré.

L'approximation de Fraunhofer traite le cas particulier de la diffraction à l'infini d'une onde plane par une ouverture circulaire de dimension caractéristique D. Cette théorie scalaire de la diffraction utilise différentes approximations pour obtenir une forme simplifiée du front d'onde dont on étudie la propagation à l'infini au moyen de l'intégrale d'Huygens-Fresnel [**Born 1980**]. Parmi celles-ci, on compte l'approximation des petits angles et la grande distance $d$ entre l'objet et l'observateur :

$$d \geq \frac{2D^2}{\lambda} \qquad (34)$$

La théorie de Fraunhofer permet ainsi d'obtenir l'amplitude du champ diffracté à l'infini par une ouverture circulaire :

$$S_F(\theta) = x^2 \frac{J_1(x\sin(\theta))}{x\sin(\theta)} \qquad (35)$$

où $x$ est le paramètre de taille de l'ouverture circulaire, $\theta$ l'angle de diffusion et $J_1$ la fonction de Bessel de premier ordre et de première espèce.

Cette formule est également valide pour un objet circulaire. En effet, le principe de Babinet [**Poicelot 1957**] stipule que la figure de diffraction est la même pour une ouverture et son conjugué (c'est-à-dire corps opaque). C'est-à-dire qu'une ouverture circulaire a la même figure de diffraction qu'une particule sphérique de diamètre équivalent à l'ouverture. La figure 3 montre le diagramme de diffusion aux petits angles d'une goutte d'eau de diamètre D=10µm dans l'air. L'approximation de Fraunhofer a été longtemps utilisée pour inverser les données obtenues par les diffractomètres laser [**Xu 2002**] de par sa simplicité de mise en œuvre ainsi que sa rapidité de calculs. On sait cependant qu'elle devient très imprécise lorsque le diamètre des particules n'est pas très grand devant la longueur d'onde, ou lorsque l'indice relatif de ces dernières est faible. De plus, son caractère scalaire (pas d'effet de la polarisation) peut poser problème dans certaines situations.

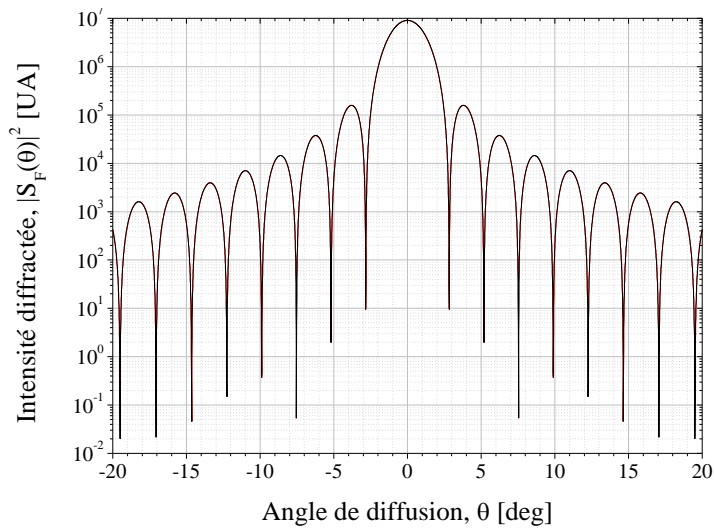

Figure 7 : Figure de diffraction obtenue pour une goutte d'eau de diamètre $D = 10\mu m$ dans l'air, une longueur d'onde de $\lambda = 405nm$ et en polarisation perpendiculaire.

## 2.2.2   La théorie d'Airy de l'arc-en-ciel

Selon l'optique géométrique, le phénomène d'arc-en-ciel est lié à l'existence d'un angle de déviation limite des rayons lumineux subissant une réflexion interne (p=2) dans une particule sphérique et transparente. Depuis les travaux de Descartes, on sait que le premier arc-en-ciel produit par une particule individuelle, sphérique, est localisé précisément à l'angle de diffusion $\theta_{rg}$ :

$$\theta_{rg} = 2\tau_{rg} - \cos^{-1}\left(\cos\tau_{rg}\,/\,m\right) \quad avec \quad \sin\tau_{rg} = \sqrt{\left(m^2 - 1\right)/3} \tag{36}$$

Cette équation indique que $\theta_{rg}$ ne dépend que de l'indice relatif $m>1$ de la particule[1] (voir Figure 14).

Young, Fresnel et Airy ont développé une théorie ondulatoire du phénomène d'arc-en-ciel. Cette dernière, désormais connue sous le nom de "théorie d'Airy", décrit l'arc-en-ciel comme un phénomène de diffraction lié à la caustique observée en $\theta=\theta_{rg}$. De fait, dans sa théorie de l'arc-en-ciel, Airy utilise l'intégrale d'Huygens-Fresnel pour étudier la propagation à l'infini du front d'onde des rayons p=2 au voisinage de $\theta=\theta_{rg}$. La forme initiale du front d'onde, approximée en série de Taylor, est déduite de calculs purement géométriques qui négligent notamment les coefficients de Fresnel. Les bases physiques et mathématiques de cette théorie ayant fait l'objet de nombreux travaux et publications, nous n'en présentons ici que les principaux résultats. Ainsi, selon la théorie d'Airy, l'intensité lumineuse diffusée au voisinage de $\theta_{rg}$ par une particule sphérique isolée [**Watson 2004**] est de la forme :

$$I\left(\theta\right) = \frac{cE_0^2}{8\pi\lambda^2R^2}\left(\frac{3\lambda D^2}{16h}\right)^{2/3}\left(\int_0^\infty \cos\left[\frac{\pi}{2}\left(z\eta - \eta^3\right)\right]d\eta\right)^2 \tag{37}$$

où, de manière équivalente à la POA (voir plus loin), $\eta$ est une variable d'intégration :

$$\eta = v\left(4h\,/\,\lambda D^2\right)^{1/3} \tag{38}$$

$v = D\left(\cos\tau - \cos\tau_{rg}\right)/2$, $h^{-1} = \tan\tau_{rg}\sin^2\tau_{rg}$, R est la distance au point d'observation et $z(\theta)$ un angle de déviation par rapport à la position de l'arc-en-ciel géométrique :

$$z = \left(\theta - \theta_{rg}\right)\left(\frac{16D^2}{h\lambda^2}\right)^{1/3} \tag{39}$$

L'équation (37) laisse apparaitre la fonction d'Airy[2] dont la représentation intégrale [**Abramowitz 1964, Chang 1996**], est :

$$\int_0^\infty \cos\left[\frac{\pi}{2}\left(z\eta - \eta^3\right)\right]d\eta = \int_0^\infty \cos\left[\frac{\pi}{2}\eta^3 - \frac{\pi z}{2}\eta\right]d\eta = \left(\frac{2\pi^2}{3}\right)^{1/3}Ai\left[-\left(\frac{\pi^2}{12}\right)^{1/3}z\right] \tag{40}$$

---

[1] Pour une goutte d'eau, m=1.332 pour $\lambda$=0.5 μm et T=20°C. Les couleurs de l'arc-en-ciel viennent de la dépendance de l'indice avec la longueur d'onde (dispersion)

[2] Cette intégrale a été nommée par Airy : "Rainbow Integral"

La distribution d'intensité au voisinage de l'angle d'arc-en-ciel $\theta_{rg}$ est donc proportionnelle à :

$$I(\theta, m, D, \lambda) \propto \frac{E_0^2}{h^{1/3}(m)} \left(\frac{\pi D}{\lambda}\right)^{4/3} Ai^2 \left[-\left(\frac{\pi^2}{12}\right)^{1/3} z\right] \quad (41)$$

La théorie d'Airy permet d'améliorer significativement la description de l'arc-en-ciel introduite par Descartes. L'équation (41) montre, par exemple, de manière explicite, que les caractéristiques de l'arc-en-ciel ne dépendent pas seulement de l'indice des particules, mais également de leur diamètre selon une loi de puissance en $D^{4/3}$. De fait, dans l'équation (41), la dépendance avec le diamètre est également implicite *via* l'argument de la fonction d'Airy qui joue notamment sur la fréquence angulaire des "franges d'Airy".

Elle permet également d'effectuer des calculs rapides (comparé à la LMT), ce qui est un facteur non négligeable pour l'inversion de données expérimentales. Cependant, la qualité des prédictions de la théorie d'Airy décroît assez rapidement à mesure que l'on s'éloigne de l'angle d'arc-en-ciel prédit par l'optique géométrique ou lorsque la taille de la particule n'est plus très grande devant la longueur d'onde. En fait, la théorie d'Airy tend à minimiser la décroissance de l'intensité et la fréquence angulaire des arcs d'ordre élevé (lobes) ; et ceci, d'autant plus que la particule est petite. Elle néglige également les effets de polarisation. Il est également important de noter que, numériquement, les séries de Debye prévoient une dépendance de l'intensité de l'arc-en-ciel en $I \propto D^\gamma$ avec $\gamma = 2.34013 \approx 7/3$ [**Ouedraogo 2005**]. Les références [**Hulst 1957, Nussenzweig 1992, Lemaitre 2004, Watson 2004**] ne donnent pas d'explication concernant cette différence notable avec la théorie d'Airy. Selon nous, cette différence provient du fait que la théorie d'Airy néglige les effets de courbure de la surface de la particule perpendiculaire au plan de diffusion (voir la section sur notre généralisation de la POA). Aussi, dans ce qui suit, de façon pragmatique, nous avons choisi de reformuler l'équation (41) sous la forme :

$$I(\theta, m, D, \lambda) \propto \frac{E_0^2}{h^{1/3}(m)} \left(\frac{\pi D}{\lambda}\right)^{7/3} Ai^2 \left[-\left(\frac{\pi^2}{12}\right)^{1/3} z\right] \quad (42)$$

Roth et al. [**Roth 1991, Roth 1991, Roth 1994**] semblent avoir été les premiers à utiliser le phénomène d'arc-en-ciel et la théorie d'Airy pour caractériser expérimentalement des particules individuelles en écoulement. Van Beeck [**Beeck 1994, Beeck 1997**] et Sankar et al. [**Sankar 1993**] ont également été des précurseurs dans ce domaine. A noter que si l'analyse de l'arc-en-ciel produit par une particule individuelle permet, en théorie, d'obtenir la corrélation entre sa taille et son indice (matériau, composition, température), cet arc-en-ciel est bien souvent fortement perturbé par le voisinage d'autres particules ou le profil du faisceau.

### 2.2.3 Approximation d'optique physique : diffusion critique par une sphère

Cette approximation s'applique à décrire la diffusion par des bulles au voisinage de l'angle critique. Bien moins connue que la théorie d'Airy de l'arc-en-ciel, elle en reprend le cheminement [**Marston 1979, Langley 1984**]. Nous détaillerons et généraliserons cette approche dans la section, § 2.2.5.

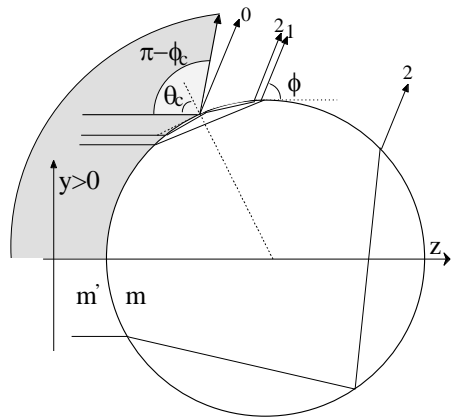

Figure 8 : Géométrie du modèle de diffusion critique (dans cette figure, pour conserver les notations de Marston, nous avons permuté $\theta$ et $\phi$ ) [**Onofri 1999**]

Optiquement, une bulle est définie comme une particule dont l'indice de réfraction relatif par rapport au milieu environnant est inférieur à l'unité, $m < 1$, voir la Figure 8. Il peut donc aussi bien s'agir d'une "bulle" d'air dans de l'eau $(m \approx 0.751)$ que d'une "goutte" d'eau dans une huile $(m \approx 0.919)$. Les lois de Snell-Descartes indiquent qu'il existe un angle de réfraction limite pour les rayons lumineux qui se propagent d'un milieu d'indice élevé vers un milieu d'indice faible. Cet angle " critique " conduit à une brusque transition vers la réflexion totale pour $\theta > \theta_c(m)$ avec $\theta_c(m) = 2\pi - 2\arcsin(m^{-1})$. Au-delà de cet angle (ou point) critique, la surface de la bulle se comporte comme un miroir parfait.

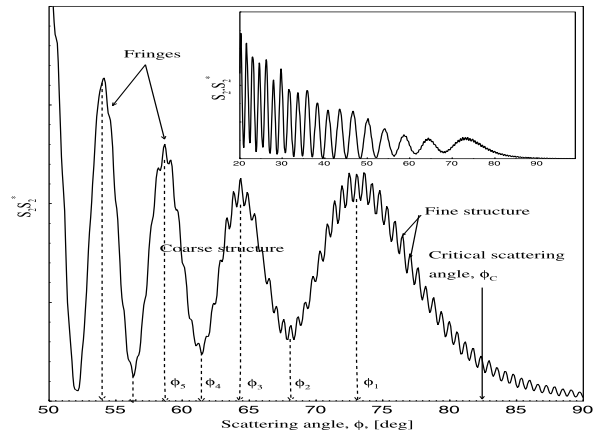

Figure 9 : Diagramme de diffusion dans la région de l'angle de diffusion critique d'une bulle ( $D = 100\mu m$ , air/eau avec m=0.751, LMT)

La Figure 9 montre une simulation d'une partie du diagramme de diffusion d'une bulle d'air dans l'eau, dans la région de l'angle de diffusion critique. On observe près de l'angle prédit par l'optique géométrique $\theta_c(m) \approx 82.7°$, une structure périodique, ou "franges de diffusion critique", présentant certaines similarités avec celles de l'arc-en-ciel. Si l'optique géométrique pure fait apparaitre une dépendance explicite entre $\theta_c(m)$ et l'indice de réfraction de la bulle étudiée, elle n'explique pas la dépendance du diagramme de diffusion au voisinage de l'angle critique avec le diamètre des bulles.

Au tournant des années 70 à 80, Marston et ses collègues ont développé une Approximation d'Optique Physique (POA) ([**Marston 1979**],[**Marston 1981**], [**Arnott 1991**]) modélisant la diffusion de la lumière par des bulles sphériques au voisinage de l'angle critique de diffusion. Comme nous l'avons déjà évoqué, la POA présente de grandes similarités avec la théorie de l'arc-en-ciel développée par Airy [**Airy 1838**] et la théorie de la diffraction de Fraunhofer. Ces trois théories décrivent au premier ordre la diffusion dans le champ lointain d'un front d'onde virtuel généré par une singularité. Dans le cas de la diffusion critique, cette singularité est qualifiée de caustique faible [**Fiedler-Ferrari 1991**] [**Lock 2003**]. Elle est induite par la non dérivabilité des coefficients de Fresnel (voir plus loin) des rayons réfléchis $(p = 0)$ au voisinage du point critique [**Davis 1955, Hulst 1981, Fiedler-Ferrari 1991**]. Nous ne détaillerons pas les étapes des calculs de Marston dans la mesure où nous généraliserons sa théorie dans la partie 2.2.5. Retenons simplement que pour une sphère de rayon[3] $a \gg \lambda$ , la

---

[3] Pour conserver les notations de Marston, $a$ représente ici le rayon de la particule.

distribution angulaire de l'intensité diffractée dans la région de l'angle de diffusion est de la forme :

$$I(\theta_d) = I_0 \left(\frac{a}{R}\right)^2 \frac{g(\theta_d)\overline{g}(\theta_d)}{8}$$  (43)

où $\theta_d$ est l'écart angulaire par rapport à l'angle de diffusion critique prédit par l'optique géométrique $\theta_c$, voir la Figure 8. Ce paramètre dépend lui-même des différents paramètres du problème :

$$\omega(\theta, m, a, \lambda) = \sin(\theta_c(m) - \theta)\sqrt{(a/\lambda)\cos(\theta_c(m))}$$  (44)

où $\theta$ est l'angle de diffusion dans le plan de diffusion *(OXZ)*, $I_0$ l'intensité de l'onde incidente et $R$ est la distance de la particule au point d'observation. La fonction $g\overline{g}(\theta_d) = \left[C(\theta_d) + 1/2\right]^2 + \left[S(\theta_d) + 1/2\right]^2$ est similaire à l'intégrale de Huygens-Fresnel obtenue pour la diffraction en champ proche d'un front d'onde par un coin [**Nussenzweig 1992**] avec, pour les intégrales de Fresnel en cosinus $C(\omega)$ et sinus $S(\omega)$ :

$$F(\omega) = C(\omega) + iS(\omega) = \int_0^\omega \cos\left(\frac{\pi z^2}{2}\right)dz + i\int_0^\omega \sin\left(\frac{\pi z^2}{2}\right)dz$$  (45)

Comparée à l'optique géométrique pure, la POA, avec les équations (43)-(45), améliore significativement la description de la diffusion au voisinage de l'angle critique. Elle reproduit certaines des caractéristiques fondamentales du diagramme de diffusion au voisinage de l'angle critique : anneaux (ou "franges") de type diffraction, basses et hautes fréquences, se détachant après une zone de relative faible diffusion. Des écarts sont observés avec les prédictions de la théorie de Lorenz-Mie. Ceux-ci sont attribués le plus souvent au fait que les rayons d'ordre p>0 sont négligés par la POA classique, de même que le phénomène de Goos-Hänchen [**Lötsch 1971**]. Néanmoins, différents travaux permettent de prendre en compte ces effets [**Langley 1984, Onofri 2009, Wozniak 2012**] rendant les prédictions de la POA bien plus quantitatives.

L'étude de ces diagrammes au voisinage de l'angle critique est utilisée à des fins de caractérisation expérimentale d'écoulements à bulles, en traitant les bulles individuellement [**Onofri 1999**] ou globalement [**Onofri 2007, Onofri 2009, Onofri 2011**]. D'un point de vue expérimental, pour la caractérisation des bulles individuelles notamment, il est très intéressant de disposer d'expressions analytiques reliant la position des extremas des lobes de diffusion aux caractéristiques des bulles : chose impossible avec la LMT. Pour obtenir ces relations à partir de la POA, il suffit de rechercher les extrema locaux de l'intégrale :

$$H(\theta_d) = \left(C(\theta_d) + 1/2\right)^2 + \left(S(\theta_d) + 1/2\right)^2 \tag{46}$$

et donc, les zéros de sa dérivée :

$$\left(C(\theta_d) + 1/2\right)\cos\left(\frac{\pi\,\theta_d^{\,2}}{2}\right) + \left(S(\theta_d) + 1/2\right)\sin\left(\frac{\pi\,\theta_d^{\,2}}{2}\right) = 0 \tag{47}$$

L'évolution de $H$ de même que les valeurs tabulées de ses premiers zéros sont données dans l'article [**Onofri 1999**]. Les zéros sont indexés par l'indice $j$. Les maxima et minima de $H(\alpha)$ correspondent respectivement aux valeurs impaires et paires de $j$. $H(\alpha)$ est équivalent à $I(\alpha)$ avec le changement de variable: $\alpha = \omega(\theta, m_r, a, \lambda)$.

La position des extrema ($\theta_j$, c'est-à-dire: franges "brillantes" et "sombres") est obtenue par la résolution de l'équation suivante :

$$\alpha_j = \sin(\theta_c(m_r) - \theta_j)\sqrt{(a/\lambda)\theta_c(m_r)} \tag{48}$$

La dépendance angulaire de la frange $j$, en fonction du diamètre et de l'indice de la particule, est alors donnée par :

$$\theta_j = \frac{180}{\pi}\left[\pi - \arcsin\left(\alpha_j\sqrt{\frac{m_r\lambda/a}{m_r^2 - 1}}\right) - 2\arcsin\left(m_r^{-1}\right)\right] \tag{49}$$

Cette dernière expression indique que pour un indice donné, la position angulaire des franges de diffusion critique est liée au diamètre de la particule par une loi en $(\theta_c - \theta) \propto 1/\sqrt{a}$. Cette dépendance est plus importante que celle prédite par la théorie d'Airy de l'arc-en-ciel. Pour cette dernière la dépendance est en $(\theta_r - \theta) \propto 1/a^{2/3}$, où $\theta_r$ est la position angulaire de l'arc-en-ciel prédit par l'optique géométrique. Après quelques manipulations mathématiques, on peut montrer que de la mesure de la position angulaire de deux extrema du diagramme de diffusion, $j \equiv p$ et $j \equiv q$, avec $p < q$, on peut déterminer le diamètre et l'indice de réfraction relatif de la particule détectée [**Onofri 2009**] :

$$a_{pq} = \frac{\alpha_p^2 + \alpha_q^2 - 2\cos\left(\theta_q - \theta_p\right)\alpha_p\alpha_q}{\sin^2\left(\theta_q - \theta_p\right)} \times \frac{\lambda}{\sqrt{1 - \cos\left[\dfrac{\theta_p}{2} - \dfrac{1}{2}\arctan\left(\sin\left(\theta_q - \theta_p\right)\dfrac{\alpha_p}{\alpha_q - \alpha_p\cos\left(\theta_q - \theta_p\right)}\right)\right]^2}} \tag{50}$$

$$m_{pq} = \left\{ \sin\left[ \frac{1}{2}(\pi - \theta_p) - \frac{1}{2}\arctan\left( \sin(\theta_q - \theta_p) \middle/ \left[ \cos(\theta_q - \theta_p) - \frac{\alpha_q}{\alpha_p} \right] \right) \right] \right\}^{-1} \tag{51}$$

## 2.2.4 Les coefficients de Fresnel

Ces coefficients ont été introduits par Augustin Jean Fresnel au XIX$^{\text{ème}}$ siècle, afin de décrire les relations existantes entre les ondes électromagnétiques incidente, réfléchie et transmise à l'interface entre deux milieux d'indice de réfraction différent, comme les milieux 1 et 2 représentés en Figure 10.

L'onde plane harmonique incidente sur l'interface (indice i) est décrite par le système d'équations suivant :

$$\begin{aligned} \mathbf{E}_i &= \mathbf{E}_{i0} e^{i(\mathbf{k}_i \cdot \mathbf{r} - \omega t)} \\ \mathbf{H}_i &= \mathbf{H}_{i0} e^{i(\mathbf{k}_i \cdot \mathbf{r} - \omega t)} \end{aligned} \tag{52}$$

où $\mathbf{E}_i$ et $\mathbf{H}_i$ représentent respectivement les champs électrique et magnétique incidents, $\mathbf{k}_i$ le vecteur d'onde dans le milieu considéré, $\mathbf{r}$ le vecteur position, $\omega$ la pulsation de l'onde et $t$ le temps. La forme générale des ondes réfléchie et transmise est formellement identique à l'indice de réfraction près. On différencie ces ondes par les indices $r$ pour réfléchie et $t$ pour transmise, voir Figure 10.

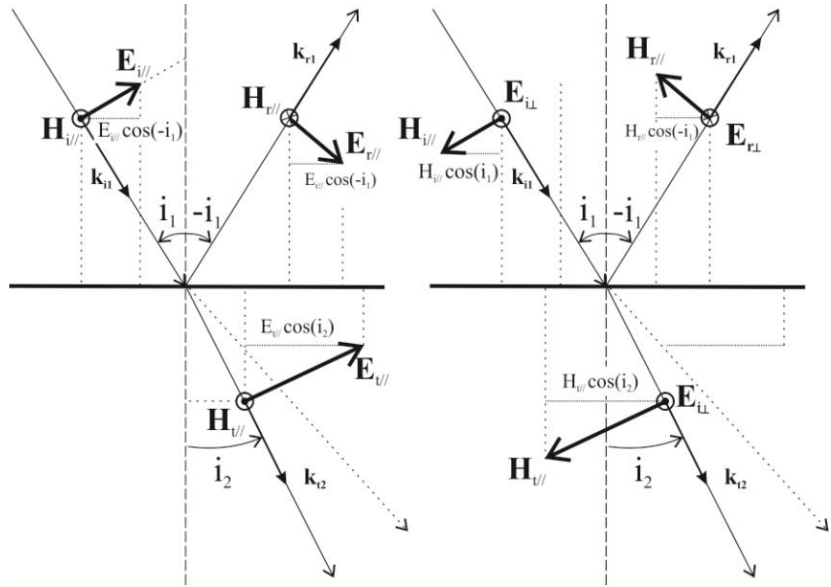

Figure 10 : Mise en évidence des différentes composantes de l'onde par rapport au plan d'indice et la surface

L'onde, solution des équations de Maxwell, satisfait notamment ces deux conditions :

$$\mathbf{k} \times \mathbf{E} = \omega \mu \mathbf{H} \tag{53}$$

$$\mathbf{k}.\mathbf{k} = \omega\varepsilon\mu \tag{54}$$

avec pour l'indice complexe $m$ du milieu :

$$m = n_m + ik_m = c\sqrt{\varepsilon\mu} = \frac{\omega\lambda_0}{2\pi}\sqrt{\varepsilon\mu} \tag{55}$$

où $n_m$ et $k_m$ désignent la partie réelle et imaginaire du milieu pour la longueur d'onde $\lambda_0$ (définie dans le vide) et $\varepsilon$ et $\mu$ la permittivité électrique et la perméabilité magnétique du milieu pour cette longueur d'onde.

Pour dériver les coefficients de Fresnel, on part de l'écriture de la continuité des composantes tangentielles des champs sur la surface. Soit pour un champ incident de polarisation parallèle (champ électrique appartenant au plan d'incidence et satisfaisant à l'équation (53)), et d'angle d'incidence $i_1$ :

$$E_{i\parallel}\cos i_1 + E_{r\parallel}\cos(-i_1) = E_{t\parallel}\cos(i_2) \tag{56}$$

$$H_{i\parallel} + H_{r\parallel} = H_{t\parallel} \tag{57}$$

En utilisant la relation (53), l'équation (57) devient :

$$\frac{\mathbf{k}_{i1} \times \mathbf{E}_{i\parallel}}{\omega\mu_1} + \frac{\mathbf{k}_{r1} \times \mathbf{E}_{r\parallel}}{\omega\mu_1} = \frac{\mathbf{k}_{t2} \times \mathbf{E}_{t\parallel}}{\omega\mu_2} \tag{58}$$

Le produit scalaire est appliqué en remarquant le changement de signe (déphasage de $\pi$) pour l'onde réfléchie :

$$\frac{k_{i1}E_{i\parallel}}{\omega\mu_1} - \frac{k_{r1}E_{r\parallel}}{\omega\mu_1} = \frac{k_{t2}E_{t\parallel}}{\omega\mu_2} \tag{59}$$

En utilisant cette équation il vient :

$$E_{i\parallel} - E_{r\parallel} = \frac{\mu_1}{\mu_2}\frac{m_2}{m_1}E_{t\parallel} \tag{60}$$

Avec les équations (56) et (60), le coefficient de réflexion devient :

$$\left(E_{i\parallel} - E_{r\parallel}\right) = AE_{t\parallel} \text{ avec } A = \frac{\mu_1}{\mu_2}\frac{m_2}{m_1}$$

$$AE_{i\parallel}\cos(i_1) + AE_{r\parallel}\cos(i_1) - E_{i\parallel}\cos(i_2) + E_{r\parallel}\cos(i_2) = 0 \tag{61}$$

$$r_\parallel = \frac{E_{r\parallel}}{E_{i\parallel}} = \frac{\cos(i_2) - \dfrac{\mu_1}{\mu_2}\dfrac{m_2}{m_1}\cos(i_1)}{\cos(i_2) + \dfrac{\mu_1}{\mu_2}\dfrac{m_2}{m_1}\cos(i_1)}$$

et le coefficient de transmission :

$$E_{i\parallel} - AE_{r\parallel} = E_{r\parallel}, \text{ avec } A = \frac{\mu_1}{\mu_2}\frac{m_2}{m_1}$$

$$\cos(i_1)\left[E_{i\parallel} + E_{i\parallel} - AE_{r\parallel}\right] = E_{t\parallel}\cos(i_2)$$

$$2E_{i\parallel}\cos(i_1) = AE_{t\parallel}\cos(i_1) + E_{t\parallel}\cos(i_2) \qquad (62)$$

$$t_{\parallel} = \frac{E_{t\parallel}}{E_{i\parallel}} = \frac{2\cos(i_1)}{\cos(i_2) + \dfrac{\mu_1}{\mu_2}\dfrac{m_2}{m_1}\cos(i_1)}$$

L'énergie et la direction de propagation d'une onde électromagnétique se calculent à l'aide du vecteur de Poynting dont la forme générale (moyennée sur une période $T = 1/\nu$) est :

$$\langle S \rangle_\nu = \frac{1}{2}\operatorname{Re}\left\{\mathbf{E} \times \mathbf{H}^*\right\} \qquad (63)$$

avec l'équation (53), il vient :

$$\langle \mathbf{S} \rangle_\mathbf{v} = \frac{1}{2\omega\mu}\operatorname{Re}\left\{\mathbf{E} \times (\mathbf{k} \times \mathbf{E})^*\right\} = \frac{1}{2\omega\mu}\operatorname{Re}\left\{\left(\mathbf{k}.(\mathbf{E}.\mathbf{E}) - \mathbf{E}(\mathbf{k}.\mathbf{E})\right)^*\right\}$$

$$= \frac{E^2}{2\omega\mu}\mathbf{k} = \frac{\pi m E^2}{\omega\mu\lambda_0}\frac{\mathbf{k}}{\|\mathbf{k}\|} \qquad (64)$$

Écrivons la conservation de l'énergie entre les trois ondes et dans les deux milieux :

$$\left\|\langle \mathbf{S}_i \rangle_\mathbf{v}\right\| + \left\|\langle \mathbf{S}_r \rangle_\mathbf{v}\right\| = \left\|\langle \mathbf{S}_t \rangle_\mathbf{v}\right\| \qquad (65)$$

Ce qui donne :

$$\frac{\pi m_1 E_{i\parallel}^2}{\omega\mu_1\lambda_0} + \frac{\pi m_1 E_{r\parallel}^2}{\omega\mu_1\lambda_0} = \frac{\pi m_2 E_{t\parallel}^2}{\omega\mu_2\lambda_0} \qquad (66)$$

$$1 + \frac{E_{r\parallel}^2}{E_{i\parallel}^2} = \frac{m_2}{m_1}\frac{\mu_1}{\mu_2}\frac{E_{t\parallel}^2}{E_{i\parallel}^2} \qquad (67)$$

On trouve alors :

$$1 + r_{\parallel}^2 = \frac{m_2}{m_1}\frac{\mu_1}{\mu_2}t_{\parallel}^2 \qquad (68)$$

Dans ce qui suit, nous préférons utiliser ces coefficients sous la forme introduite par Van de Hulst [**Hulst 1981**], en fonction des angles complémentaires $\tau$ et $\tau'$, et en posant $\mu_1 = \mu_2$ :

$$r_{\perp} = \frac{m_2\sin(\tau) - m_1\sin(\tau')}{m_2\sin(\tau) + m_1\sin(\tau')}$$

$$r_{\parallel} = \frac{m_1\sin(\tau) - m_2\sin(\tau')}{m_1\sin(\tau) + m_2\sin(\tau')} \qquad (69)$$

Les coefficients de transmission en amplitude se déduisent directement de l'équation (68). Ces coefficients complexes permettent de déterminer les ratios d'amplitude, la phase ou l'énergie des rayons réfractés et réfléchis par une surface localement plane.

## 2.2.5 Approximation d'optique physique : diffusion critique par un sphéroïde

### 2.2.5.1 Introduction

La POA introduite par Marston et al. ([**Marston 1979**],[**Marston 1981**], [**Arnott 1991**]) permet de modéliser de façon très correcte (et efficace) la diffusion de la lumière par des bulles sphériques au voisinage de l'angle critique de diffusion. Cependant, il est manifeste que les bulles millimétriques, ou fortement cisaillées, sont généralement sphéroïdales (voire de formes beaucoup plus complexes) [**Grace 1976, Arnott 1991**]. Nous détaillons, dans ce qui suit, le travail que nous avons conduit pour généraliser la POA au cas de bulles sphéroïdales [**Onofri 2012**]. Ce travail original est présenté après la section sur les coefficients de Fresnel car ces derniers sont utilisés dans notre modèle.

### 2.2.5.2 Géométrie du problème et description du front d'onde

Le problème traité est schématisé dans la Figure 11. Dans le système de coordonnées cartésiennes $(Oxyz)$, l'équation canonique d'un ellipsoïde axisymétrique (c'est-à-dire un sphéroïde) par rapport à l'axe x et de semi-axes $a_x, a_y = a_z$, s'écrit $x^2 / a_x^2 + \left( y^2 + z^2 \right) / a_y^2 = 1$. Suivant la valeur du rapport d'aspect $\gamma = a_x / a_y$, le sphéroïde est dit oblate $(\gamma \leq 1)$ ou prolate $(\gamma \geq 1)$. Pour une onde plane incidente de longueur d'onde $\lambda_0$ et vecteur d'onde $k = 2\pi m_2 / \lambda_0$, l'angle critique de diffusion[4] vaut $\theta_c = \pi - 2\phi_c$ avec pour l'angle critique $\sin \phi_c = m^{-1}$ où, comme dans ce qui suit, $m(\lambda_0) = m_1 / m_2 < 1$ est l'indice de réfraction relatif de la bulle. L'indice "$c$" indique que le paramètre ou la variable est défini pour l'angle critique, voir la Figure 11.

---

[4] Nous utilisons ici des notations standards. Pour retrouver celles de Marston, il faut notamment permuter $\theta$ avec $\phi$.

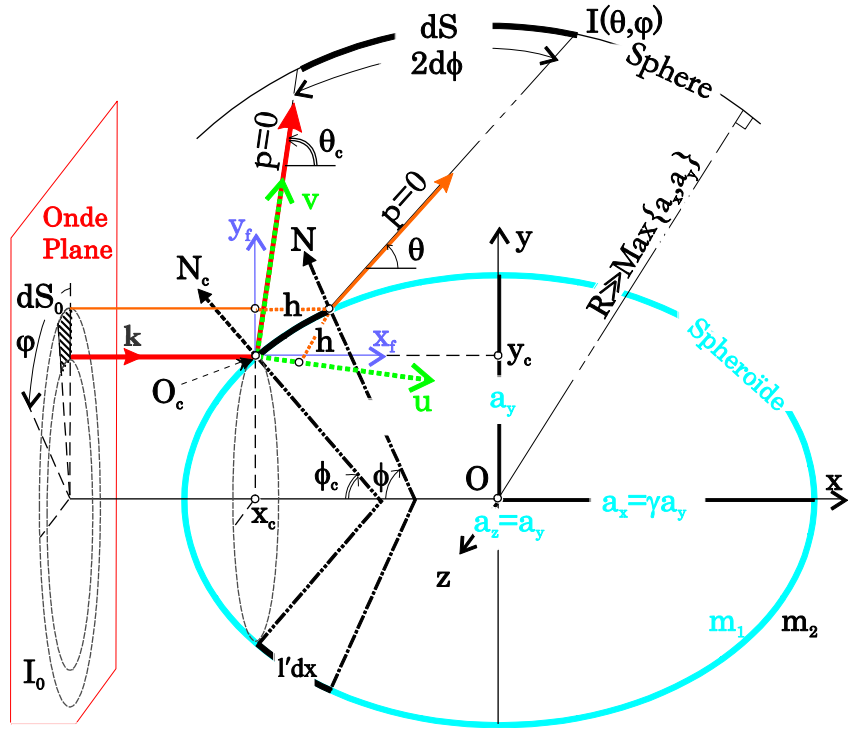

Figure 11 : Approximation d'optique physique : géométrie du modèle de sphéroïde

En raison de la symétrie du problème traité, on peut dans un premier temps se limiter à l'étude de la diffusion d'une surface elliptique d'équation : $x^2/a_x^2 + y^2/a_y^2 = 1$. Dans la région telle que $(x \leq 0, y \geq 0)$, la normale sortante en tout point de la surface de l'ellipse a pour module au carré : $N^2 = x^2/a_x^4 + y^2/a_y^4$. Le rayon de courbure local s'écrit $R = a_x^2 a_y^2 N^3$. Les composantes du vecteur directeur de la normale s'écrivent $\cos(\phi) = -x/(a_x^2 N)$ et $\sin(\phi) = y/(a_y^2 N)$, avec :

$$\cos 2\phi = 2x^2/(a_x^4 N^2) - 1,$$
$$\sin 2\phi = -2xy/(N^2 a_x^2 a_y^2). \tag{70}$$

Pour décrire le front d'onde virtuel associé aux rayons réfléchis $(p = 0)$ et au voisinage de l'angle critique, nous introduisons un second repère cartésien $(O_c x_f y_f)$ centré sur le point critique $O_c(x_c, y_c)$ :

$$x_c = -\sqrt{a_x^4/(a_x^2 + a_y^2 \tan^2 \phi_c)},$$
$$y_c = \sqrt{a_y^4/(a_y^2 + a_x^2/\tan^2 \phi_c)}. \tag{71}$$

Les axes $x_f$ et $x$ sont parallèles entre eux avec $h = x - x_c > 0$ et $y_f < 0$, où $x_f = h(1 + \cos 2\phi)$ et $y_f = (y - y_c) - h \sin 2\phi$. L'équation du front d'onde peut être paramétrée en fonction de $\phi$, comme dans le travail de Marston. Néanmoins, pour les bulles de section elliptique, nous avons préféré choisir un paramétrage en $x$ [**Onofri 2012**]. Pour décrire ce front d'onde, nous introduisons un troisième système de coordonnées cartésiennes $(O_c uv)$, centré sur le point critique, où l'axe $v$ est aligné suivant la direction du rayon réfléchi correspondant à l'angle critique. La matrice de passage entre les systèmes de coordonnées $(O_c x_f y_f)$ et $(O_c uv)$ s'écrit:

$$\begin{pmatrix} u \\ v \end{pmatrix} = \begin{pmatrix} \sin 2\phi_c & \cos 2\phi_c \\ -\cos 2\phi_c & \sin 2\phi_c \end{pmatrix} \begin{pmatrix} x_f \\ y_f \end{pmatrix}. \tag{72}$$

Nous recherchons une relation décrivant le profil du front d'onde, du type $v = v(u)$ et qui soit valide au voisinage de la condition de l'angle critique, c'est-à-dire pour $\phi \approx \phi_c$ et $u \approx v \approx 0$. D'un point de vue formel, en omettant les termes cubiques et d'ordre supérieur, le développement en série de Taylor du front d'onde au voisinage de $x = x_c$ s'écrit :

$$v \approx 0 + (dv/du)u + (d^2v/du^2)u^2/2! + O(u)^3. \tag{73}$$

Pour simplifier les notations, on pose que $q \equiv u'$ et $p \equiv v'$, et $(\ )'$ et $(\ )''$ désignent la dérivée première $(d/dx)$ et la dérivée seconde $(d^2/dx^2)$ par rapport à $x$. Toujours de manière générale, on a $dv/du = p/q$ et $d^2v/du^2 = (qp' - pq')/q^3$. Avec ces notations, la dérivation de l'équation (72) au point critique nous donne que :

$$\begin{aligned} p_c &= -\cos 2\phi_c x_f' + \sin 2\phi_c y_f', \\ q_c &= \sin 2\phi_c x_f' + \cos 2\phi_c y_f'. \end{aligned} \tag{74}$$

Pour déterminer les termes, à priori, non nuls, de l'équation (73), il faut tout d'abord évaluer les dérivées premières en $x = x_c$ :

$$\begin{aligned} x_f' &= 2x_c^2/\left(a_x^4 N_c^2\right), \\ y_f' &= \frac{a_y^2 x_c}{a_x^2 y_c N_c^2} \left[ \frac{1}{a_y^2} - \frac{x_c^2}{a_x^2} \left( \frac{1}{a_x^2} + \frac{1}{a_y^2} \right) \right]. \end{aligned} \tag{75}$$

Il est aisé de démontrer que $p_c = 0$ et $q_c \neq 0$. Ce qui fait que dans l'équation (73) la première dérivée s'annule, alors que la seconde se réduit à $d^2v/du^2 = q^{-2}p'$ où en $x = x_c$ :

$$p' = -\cos 2\phi_c x_f'' + \sin 2\phi_c y_f''. \tag{76}$$

Pour les dérivées secondes en $x = x_c$, on a :

$$x_f'' = 8x_c / \left(a_x^4 a_y^2 N_c^4\right),$$

$$y_f'' = \frac{4}{a_x^2 a_y^3}\left[\frac{2x_c^2 y_c}{a_x^2 N_c^4}\left(\frac{1}{a_y^2}-\frac{1}{a_x^2}\right)+\frac{a_y^2}{y_c N_c^2}\left(\frac{y_c^2}{a_y^2}-\frac{x_c^2}{a_x^2}\right)\right]-\frac{a_y^4}{a_x^2 y_c^3}. \tag{77}$$

Au final, on obtient pour l'expression du front d'onde virtuel [**Onofri 2012**] :

$$v = \alpha_e u^2 \text{ with } \alpha_e = p'\big/\left(2q_c^2\right). \tag{78}$$

Deux remarques importantes peuvent être faites concernant l' équation (78) :

- premièrement, le front d'onde virtuel est quadratique, comme pour la sphère. A noter que dans la théorie d'Airy, et pour une particule sphérique, le front d'onde associé à l'arc-en-ciel est cubique. La valeur du coefficient $\alpha_e$ du terme quadratique peut être déterminée à partir des équations (74)-(77). Ces équations vérifient qu'en posant $a_x = a_y = a$, on obtient comme cas particulier les résultats établis par Marston dans le cas de la sphère, $\alpha_e = \alpha_s = -\left(a\cos\phi_c\right)^{-1}$ [**Marston 1979**].

- deuxièmement, le coefficient $\alpha_e$ que nous venons d'obtenir pour un ellipsoïde correspond à celui d'une sphère dont le rayon égale le rayon de courbure de l'ellipse dans le plan $(xy)$, au point critique, c.-à-d. que $\alpha_e \equiv \alpha_s\left(a = a_{c\parallel} = a_x^2 a_y^2 N_c^3\right)$.

### 2.2.5.3 Propagation et forme du front d'onde dans le champ lointain

Calculons maintenant le champ résultant de la propagation dans le champ lointain de ce front d'onde. Pour simplifier les notations, on omet la dépendance temporelle du champ (en $e^{i\omega t}$). En conservant pour cette partie les notations de Marston [**Marston 1979**], l'amplitude du champ diffusé est approximée comme étant proportionnelle à la transformée du front d'onde $r_j(u)\exp\left(-ik\alpha u^2\right)$ :

$$f_j(\eta) = \int_{-\infty}^{\infty} r_j(u) e^{ik\left(-\alpha u^2 - \eta u\right)} du \tag{79}$$

où $\eta = \theta - \theta_c$ représente une petite déviation angulaire à l'angle de diffusion critique, $r_j$ une approximation au premier ordre du coefficient de Fresnel (voir la section 2.2.4) du rayon réfléchi et où l'indice $j = 1,2$ différencie la polarisation perpendiculaire (1) de parallèle (2). Pour $y \leq y_c$ et $\varepsilon_> = \phi_c - \phi \geq 0$, ces coefficients de réflectivité se réduisent à $r_1 \simeq 1 - \sqrt{8\varepsilon_>/\left(m\cos\phi_c\right)}$ et $r_2 \approx -1 + m^2\sqrt{8\varepsilon_>/\left(m\cos\phi_c\right)}$. Pour $y \geq y_c$ et $\varepsilon_< = \phi_c - \phi \leq 0$, $r_1$ and $r_2$ sont tous les deux des nombres complexes de module unité que l'on peut noter $r_j = \exp\left(-i\delta_j\right)$ avec $\delta_1 \simeq \sqrt{8\varepsilon_< / \left(m\cos\phi_c\right)}$ et $\delta_2 \simeq m^2 \sqrt{8\varepsilon_< / \left(m\cos\phi_c\right)}$. On remarquera que les expressions que nous avons obtenues pour ces coefficients de réflectivité sont en tout point identiques à celles obtenues par Marston dans le cas de la sphère. Pour exprimer les relations précédentes dans le repère $\left(O_c uv\right)$, nous faisons l'approximation selon laquelle :

$$u \approx 0 + \left(x_f - x_c\right)u' = q_c\left(\phi - \phi_c\right)\phi' \tag{80}$$

avec $\phi' = 1 / \left(N_c^3 a_x^2 a_y^2 \sin\phi_c\right)$ pour $x = x_c$, il vient que pour une bulle de section elliptique nous avons $\delta_1 \simeq \beta_1 \sqrt{u}$ et $\delta_2 \simeq \beta_2 \sqrt{u}$, où $\beta_1 = B^{-1}\sqrt{8 / \left(m\cos\phi_c\right)}$, $\beta_2 = m^2 \beta_1$ et $B^2 = u / \varepsilon_< = q_c / \left(N_c^3 a_x^2 a_y^2 \sin\phi_c\right)$.

L' équation (79) prend ainsi la forme suivante :

$$f_j\left(\eta\right) = \pm \int_0^\infty e^{-i\left[k\left(\alpha u^2 + \eta u\right) + \beta_j u^{1/2}\right]} du. \tag{81}$$

Comme pour la sphère, l'intégration précédente peut être facilitée en complétant le carré de l'intégrante :

$$f_j\left(\eta\right) = \pm e^{ik\eta^2/(4\alpha)} \int_0^\infty e^{-i\left(k\alpha\left[u+\eta/(2\alpha)\right]^2 + \beta_j u^{1/2}\right)} du. \tag{82}$$

Il est intéressant d'opérer le changement de variable $u \rightarrow z$ avec $z = Au - w$, $dz = Adu$, $w = -A\eta / \left(2\alpha\right)$ où $A$ et $w$ sont des constantes que nous expliciterons plus tard. Pour les conditions limites $\varepsilon_< \leq 0$, on a $z = -w$ pour $u = 0$ et $z \rightarrow \infty$ quand $u \rightarrow \infty$. L' équation (82) devient ainsi :

$$f_j\left(\eta\right) = \pm \frac{e^{ik\eta^2/(4\alpha)}}{A} \int_{-w}^\infty e^{-i\left(\frac{k\alpha}{A^2}z^2 + \frac{\beta_j}{A^{1/2}}[z+w]^{1/2}\right)} dz. \tag{83}$$

Pour retomber sur une intégrale du type intégrale de Fresnel, nous posons $A = \sqrt{-2k\alpha / \pi}$, ce qui donne $w = -\eta\sqrt{k / \left(2\pi\alpha\right)}$ avec $\alpha \leq 0$. Comme dans le cas de la sphère, nous introduisons la fonction de phase $\psi_j\left(z,w\right) = \beta_j A^{-1/2}\left[\left(z+w\right)^{1/2} - w^{1/2}\right]$. $\psi_j\left(z,w\right)$ varie lentement avec $w$ quand $z \rightarrow w$ et $w \rightarrow 0$. De ce fait, comme pour la sphère, en utilisant la méthode des phases stationnaires, on peut approximer la fonction de phase par $\psi_j\left(z \rightarrow w, w \rightarrow 0\right) \approx \psi_j\left(w, 0\right)$ $= \beta_j A^{-1/2} w^{1/2}$. En réarrangeant l'équation (83), on peut mettre celle-ci sous la forme $f_j(\eta) \approx \pm g/A$, où la nouvelle fonction $g(w)$ dépend explicitement et implicitement de $\eta$ (avec $\eta \leq 0$) [**Onofri 2012**] :

$$g(w) \approx e^{\left[ \frac{k\eta^2}{4\alpha} - \psi_j(w,0) \right]} \int_{-\infty}^{w} e^{i(\pi z^2/2)} dz. \tag{84}$$

L'intégrale de l'équation (84) peut être évaluée à l'aide de l'intégrale de Fresnel (et ses termes en sinus et cosinus, voir par exemple [**Marston 1979, Onofri 2009**]). Le produit $g\bar{g}$ (où $\bar{g}$ est le complexe conjugué de $g$) est une fonction oscillante dont la modulation décroît à mesure que la déviation angulaire $\theta_c - \theta$ croît positivement (c'est-à-dire $g\bar{g} \rightarrow 2$). La modulation et l'intensité du produit $g\bar{g}$ s'atténuent lorsque $\theta - \theta_c$ devient de plus en plus négatif (c'est-à-dire $g\bar{g} \rightarrow 0$) mais, dans ce cas, notre modèle est formellement incorrect du fait de l'approximation faite sur $\psi_j(z,w)$.

Pour généraliser les résultats précédents au cas du sphéroïde et ainsi obtenir l'intensité absolue diffusée à très grande distance $R \gg Max\{a_x, a_y\}$ par celui-ci, on utilise la conservation de l'énergie entre les rayons incidents et diffusés: $I_0 dS_0(\theta, \varphi) = I(\theta, \varphi) dS(\theta, \varphi)$ [**Davis 1955**]. La surface différentielle associée à un rayon incident s'écrit $dS_0 \approx |(l'dx \times y \times d\varphi)\cos\phi_c|$ avec $\varphi = [0, 2\pi]$. Au voisinage du point critique, l'élément différentiel de longueur de l'ellipse vaut $l'dx = \sqrt{1 + (y')^2} dx$. Il peut être réécrit sous la forme $dl_c = a_{c,\parallel} d\theta$, où $a_{c,\parallel}$ correspond bien évidement au rayon du cercle tangent à l'ellipse au point critique. En rappelant que $\theta = \pi - 2\phi$, nous obtenons pour les rayons diffusés $dS = |-2Rd\phi \times R\sin2\phi_c \times d\varphi|$. Il vient *in fine* que pour de grandes valeurs de $w(\theta)$ et pour $\theta_c - \theta \geq 0$, et dans la limite où $g\bar{g} \rightarrow 2$, la diffusion au voisinage de l'angle critique d'un sphéroïde éclairé selon son axe principal s'écrit :

$$I(\theta, \varphi) = I_0 \left( \frac{a_{c,\parallel} a_{c,\perp}}{R^2} \right) \frac{g(\theta)\bar{g}(\theta)}{8}, \tag{85}$$

où nous avons introduit $a_{c,\perp} = y_c / \sin\phi_c$, le second rayon de courbure principal du sphéroïde.

### 2.2.5.4 Résultats numériques préliminaires et perspectives

La Figure 12 compare les profils des fronts d'onde de sphères et sphéroïdes, selon que l'on utilise la POA de Marston ou notre généralisation de cette théorie au cas de sphéroïdes. Force est de constater que l'accord est parfait lorsque l'on prend pour la sphère celle dont le rayon de courbure est égal à celui du sphéroïde au point critique. Cela conforte l'idée que la diffusion critique est sensible au rayon de courbure local du diffuseur.

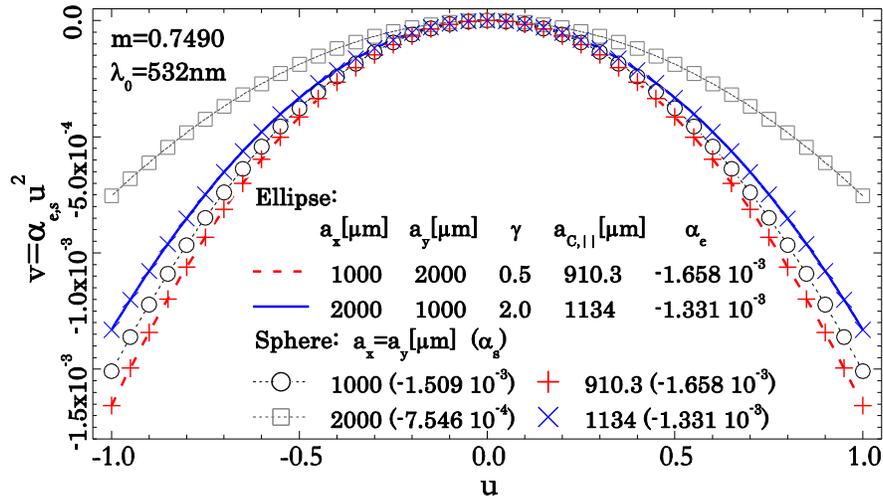

Figure 12 : Approximation d'optique physique : comparaison des fronts d'onde

Dans la mesure où, à l'heure actuelle, aucun modèle électromagnétique n'est en mesure de calculer les propriétés de diffusion de sphéroïdes de grand paramètre de taille, la Figure 13 compare simplement les prédictions de notre modèle d'optique physique avec celles de la théorie de Lorenz-Mie. La comparaison est effectuée pour des bulles d'air dans l'eau et des gouttes d'eau dans de l'huile silicone. Dans le cas de la LMT, le rayon retenu est celui de la section elliptique au point critique (avec $a = a_{c,||}; \varphi = 0°$). Au-delà du relatif bon accord entre les deux approches, les différences sont principalement dues au fait que, contrairement au phénomène d'arc-en-ciel, les rayons d'ordre élevé $(p \geq 1)$ contribuent fortement au champ diffusé au voisinage de l'angle critique [**Marston 1981, Fiedler-Ferrari 1991, Lock 2003, Wu 2007, Onofri 2009**]. Des raisons similaires ont été évoquées et quantifiées dans le cas de la sphère (voir par exemple [**Marston 1981, Onofri 2009**]).

De fait, ces rayons peuvent être pris en compte dans le cadre de l'optique géométrique [**Wu 2007, Ren 2011**] puisque pour ces ordres de diffusion les coefficients de réflectivité et l'angle de diffusion sont continus et dérivables par rapport au paramètre d'impact. La combinaison de l'approche géométrique et de notre modèle d'optique physique est clairement une perspective à ce travail, de même que la généralisation de celui-ci à des ellipsoïdes d'orientation particulière, puis quelconque.

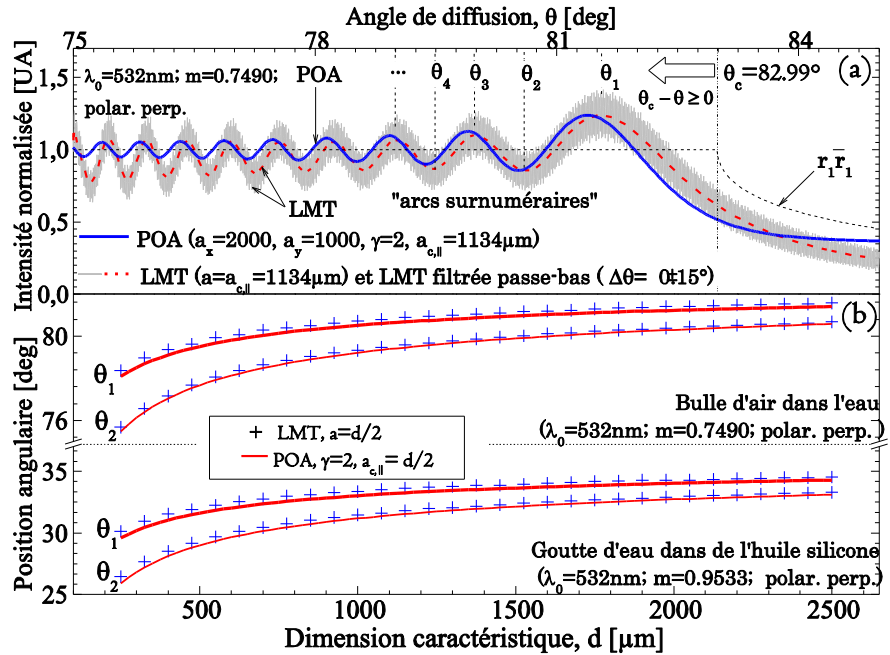

Figure 13 : Approximation d'optique physique : comparaison des prédictions de notre modèle avec ceux de la théorie de Lorenz-Mie (pour un rayon de courbure égal à celui du sphéroïde) (a) diagrammes de diffusion pour $a = a_{c,\parallel}$, (b) position angulaire des deux premières franges

## 2.3  Modèle d'optique géométrique

L'optique géométrique est le plus ancien modèle décrivant l'interaction lumière-matière. Il repose sur la notion de rayons lumineux (Euclide IV^ème siècle avant J.C.) qui sont traités de manière mécanistique. Une onde est alors décomposée en une multitude de rayons se propageant en ligne droite dans un milieu homogène (conséquence du principe de Fermat) qui, en interagissant avec une particule ou un dioptre, sont réfléchis, réfractés ou absorbés. Ce modèle est qualifié d'asymptotique à cause des particules très grandes devant la longueur d'onde ($D \gg \lambda$) et dont la surface est continue-dérivable. Il est également largement utilisé pour traiter les systèmes dioptriques et catadioptriques (lentilles, miroir...) employés pour former les faisceaux ou imager les systèmes particuliers. Dans ce qui suit, nous n'aborderons que le cas d'une particule sphérique.

La particule sphérique est au centre du repère $Oxyz$ du laboratoire, voir la Figure 14. Le faisceau incident se propage selon l'axe $Oy$. Ce dernier est décomposé en une multitude de rayons équiphases et parallèles entre eux.

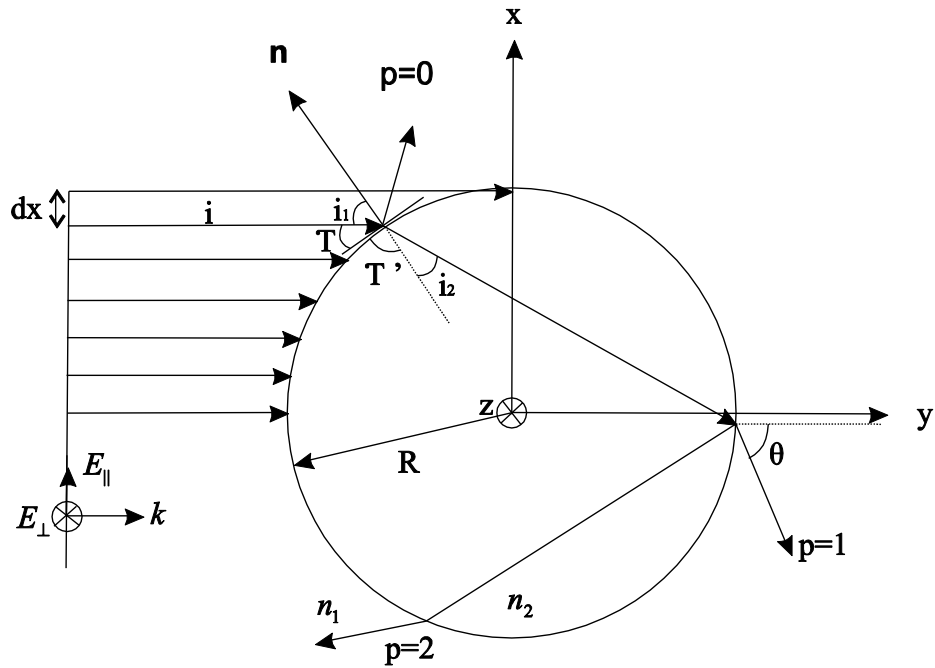

Figure 14 : Schéma pour la diffusion de la lumière par une sphère dans le cadre de l'optique géométrique

Nous utilisons les angles complémentaires aux angles " classiques " de l'optique géométrique (angles d'incidence), de même que :

- $m_1 = n_1 + ik_1$ l'indice de réfraction du milieu environnant,

- $m_2 = n_2 + ik_2$ l'indice de réfraction de la sphère,

- $\tau$ l'angle complémentaire à l'angle d'incidence $i_1$ ($\tau = \pi/2 - i_1$),

- $\tau'$ l'angle complémentaire à l'angle de réfraction $i_2$,

- $\theta$ l'angle de diffusion (déviation totale par rapport à $\tau$),

- $\lambda_0$ la longueur d'onde de l'onde incidente dans l'air,

- $R$ le rayon de la sphère,

- $\mathbf{n}$ le vecteur normal à la surface de la sphère au point d'impact du rayon incident,

- $\mathbf{i}$ le vecteur directeur du rayon incident.

Comme pour la théorie de Debye, l'ordre des différents rayons est identifié par le paramètre $p$ avec $p = 0$ pour le rayon réfléchi directement, $p = 1$ pour le rayon réfracté, $p = 2$ pour le rayon réfracté après une réflexion interne, $p = 3$ après deux réflexions internes, etc.

### 2.3.1 Les lois de Snell-Descartes vectorielles

L'optique géométrique repose notamment sur les deux lois de la réflexion et réfraction de Snell-Descartes :

$$\cos(i_1) = -\mathbf{i}.\mathbf{n} \tag{86}$$

$$\cos(i_2) = \sqrt{\left(1 - \left(\frac{m_1}{m_2}\right)^2 \left(1 - \cos(i_1)^2\right)\right)} \tag{87}$$

avec $\mathbf{i}$ pour le vecteur directeur du rayon incident et $\mathbf{n}$ la normale locale de la surface du dioptre, le symbole "." symbolisant le produit scalaire. Les angles complémentaires se ramènent aux angles classiques par :

$$\tau = \frac{\pi}{2} - \arccos(-\mathbf{i}.\mathbf{n})$$
$$\tau' = \frac{\pi}{2} - \arccos\left(\sqrt{\left(1 - \left(\frac{m_1}{m_2}\right)^2 \left(1 - \cos(i_1)^2\right)\right)}\right) \tag{88}$$

### 2.3.2 Cas particuliers des lois de Snell-Descartes

#### 2.3.2.1 Cas d'un milieu absorbant : $m_2$ complexe

On peut se demander ce qu'il advient des équations (86), (87) (et in fine (88)) lorsque l'indice de réfraction $m_2$ est complexe. En effet, une telle hypothèse nous amène à considérer des angles complexes qui, conceptuellement, posent problème. Cette question a été traitée pour le cas d'une interface transparente/absorbante ($m_1$ réel et $m_2$ complexe), pour les métaux notamment. De nombreux auteurs ([**Born 1980**], [**Kovalenko 2001**], [**Zhou 2007**] ) ont donné une nouvelle expression des Lois de Snell-Descartes dans ce cas particulier. Nous ne développerons pas ici les démarches permettant d'obtenir ces expressions qui sont détaillées dans les références précédemment citées.

Born et Wolf expriment la loi de la réfraction pour des matériaux absorbants en utilisant les angles classiques de l'optique géométrique :

$$\sin(i_2) = \frac{\sin(i_1)}{\sqrt{\sin^2(i_1) + n_2^2 q^2 \left(\cos(\gamma) - k_2 \sin(\gamma)\right)^2}} \tag{89}$$

avec $\gamma$ et q deux coefficients à déterminer. Pour exprimer ces coefficients, on utilise le système d'équations suivant :

$$\begin{cases} q^2 \cos(2\gamma) = 1 - \dfrac{1-k_2^2}{m_2^2}\sin^2(i_1) \\ q^2 \sin(2\gamma) = \dfrac{2k_2}{m_2^2}\sin^2(i_1) \end{cases} \tag{90}$$

qui après résolution, nous donne les expressions suivantes :

$$\begin{cases} \gamma = \dfrac{1}{2}\arctan\left(\dfrac{2k_2\sin^2(i_1)}{m_2^2 - \left(1-k_2^2\right)\sin^2(i_1)}\right) \\ q = \sqrt{\dfrac{2k_2}{m_2^2\sin(2\gamma)}\sin^2(i_1)} \end{cases} \tag{91}$$

Kovalenko donne une autre expression pour la loi de réfraction :

$$\sin(i_2) = \frac{\sqrt{2}\sin(i_1)}{\sqrt{n_2^2 - k_2^2 + \sin^2(i_1) + \sqrt{\left(n_2^2 - k_2^2 - \sin^2(i_1)\right)^2 + 4n_2^2 k_2^2}}} \tag{92}$$

Quant au modèle de Smith-Zhou, il introduit une expression différente pour l'indice de réfraction réel du milieu absorbant :

$$n_2' = \left(\frac{n_2^2 - k_2^2 + \sin^2(i_1) + \sqrt{\left(n_2^2 - k_2^2 + \sin^2(i_1)\right)^2 - 4\left(\left(n_2^2 - k_2^2\right)\sin^2(i_1) - n_2^2 k_2^2\right)}}{2}\right)^{\frac{1}{2}} \tag{93}$$

La Figure 15 compare les prédictions de ces modèles avec les lois classiques de Snell-Descartes, dans le cas d'un angle d'incidence fixe de $i_1 = 60\deg$ et pour différentes absorptions. On voit que pour des indices complexes du milieu absorbant $k_2$ élevés (supérieur à 0.1), la déviation des rayons par rapport à ce que prédit l'optique géométrique approche les 10 degrés. Les trois modèles présentés concordent de par leur tendance (diminution de l'angle de réfraction) et les ordres de grandeur (l'écart n'excède pas 1 degré), alors que les lois de Snell-Descartes avec l'utilisation du module pour le milieu absorbant surévaluent fortement la déviation des rayons (jusqu'à 35 degrés!). Le modèle de Born-Wolf prédit la plus grande déviation et le modèle de Smith-Zhou la plus petite déviation, voir la Figure 16.

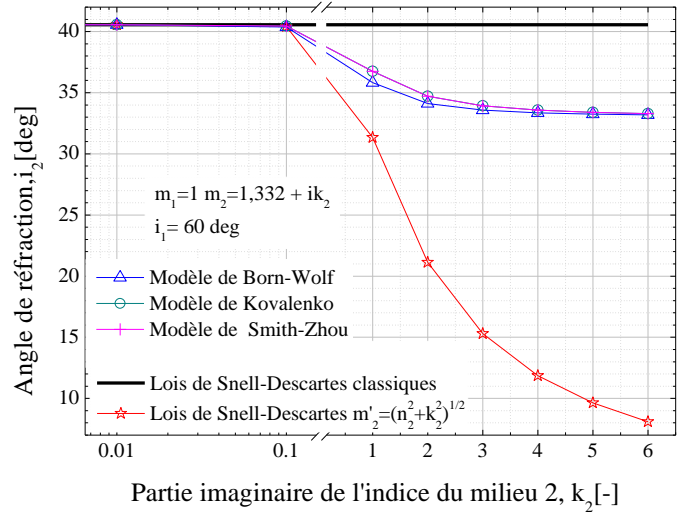

Figure 15 : Différents modèles pour la loi de réfraction entre un milieu transparent et un milieu absorbant d'absorption croissante.

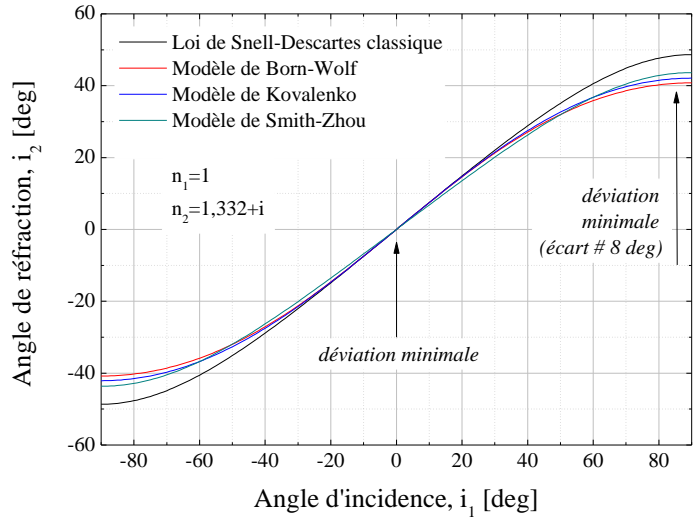

Figure 16 : Influence de l'angle d'incidence $i_1$ sur la déviation des rayons réfractés dans un milieu absorbant

L'absorption des rayons pour des valeurs de $k_2$ élevées se fait cependant sur des distances extrêmement courtes (de l'ordre du micromètre, on parle "d'effet de peau"). Ces rayons qui dévient de leur trajectoire "classique" n'influent donc que très peu les grandeurs considérées dans des milieux macroscopiques. D'ailleurs, dans le projet OPTIPAT, les milieux absorbants se résument aux masques métalliques et aux détecteurs. Nous reviendrons sur ce point dans le **chapitre 3**. Les lois de Snell-Descartes seront appliquées dans leur formulation classique (en prenant la partie réelle de $m_1$ et $m_2$).

**2.3.2.2  Cas $m_1 > m_2$ : Angle de diffusion critique**

Dans ce cas particulier du passage d'un milieu plus réfringent $m_1$ à un milieu moins réfringent $m_2$, il y a apparition d'un angle particulier appelé angle de diffusion critique (voir la section 2.2.3). L'angle complémentaire $\tau_c$ est associé à l'angle d'incidence critique. En-dessous de cette valeur particulière la réfraction n'existe plus, on parle alors de réflexion totale. Cet angle de diffusion critique se calcule mathématiquement de manière simple (angle limite pour lequel le cosinus dépasse 1) et s'exprime par :

$$\tau_c = \arccos(m_2/m_1) \tag{94}$$

**2.3.2.3  Cas $m_1 < m_2$ : Angle de réfraction limite**

Dans ce cas, lors du passage dans le milieu plus réfringent, le rayon incident se rapproche de la normale au point d'incidence. Lorsque l'incidence est rasante, c'est-à-dire $\tau = 0$, l'angle complémentaire $\tau'$ à l'angle de réfraction $i_2$ atteint sa valeur maximale, on parle d'angle de réfraction limite $\tau_{réfraction}$ avec :

$$\tau_{réfraction} = \arccos(m_1/m_2) \tag{95}$$

## 2.3.3  Calcul de l'angle de diffusion dans le cas d'une sphère

Pour une sphère, on peut définir l'angle de diffusion (déviation totale) $\theta_p$ pour les différents rayons d'ordre $p$ suivant la loi de récurrence :

$$\begin{aligned}
\theta_0 &= 2\tau \\
\theta_1 &= \theta_0 - 2\tau' \\
\theta_2 &= \theta_1 - 2\tau' \\
&... \\
\theta_p &= \theta_{p-1} - 2\tau'
\end{aligned} \tag{96}$$

On peut ainsi directement exprimer $\theta_p$ en fonction de $p$ par [**Hulst 1957**] :

$$\theta_p = 2\tau - 2p\tau' \tag{97}$$

Du fait de la symétrie du problème, seuls les angles complémentaires $\tau \in \left[0, \pi/2\right]$ sont considérés, avec un angle de diffusion ramené dans l'intervalle $\theta \in \left[0, \pi\right]$. Pour cela Van de Hulst introduit deux coefficients $q$ et $k$ tels que :

$$\theta' = 2k\pi + q\theta \tag{98}$$

où $k$ est un entier qui traduit le fait que pour les rayons d'ordre élevé, le nombre de rotations peut être important. $q = \pm 1$ est un entier qui permet de ramener les rayons dans le plan supérieur (correspondant au balayage de $\tau \in [0, \pi/2]$ ).

De fait, la détermination de $k$ et $q$ s'avère très délicate. Dans un premier temps, il nous est paru plus aisé d'utiliser la relation :

$$\theta' = \arccos\left(\cos(\theta)\right) \tag{99}$$

En effet, la parité de la fonction cosinus avec $\cos(\theta) = \cos(\theta + 2k\pi)$ permet d'obtenir une unique solution et la fonction arc cosinus est à valeur dans $[0, \pi]$. Cette approche a été implémentée numériquement avec succès.

Cependant, les coefficients $k$ et $q$ étant également utilisés pour le calcul des phases (voir section 2.4.3.2), il nous faut les déterminer de manière unique. Pour une valeur de $\theta$ plusieurs valeurs de $k$ et $q$ peuvent convenir pour obtenir un angle $\theta'$ dans $[0, \pi]$. Il faut s'assurer de garder le bon couple $k$ et $q$. Pour cela, l'angle trouvé grâce à l'équation (99) sera utilisé :

$$\text{Si } q = 1 \text{ alors } k = E\left(\frac{(\theta' - \theta)}{2\pi} + \frac{1}{2}\right)$$
$$\text{Si } q = -1 \text{ alors } k = E\left(\frac{(\theta' + \theta)}{2\pi} + \frac{1}{2}\right) \tag{100}$$

où E est la fonction partie entière. Cette méthode permet d'obtenir les deux couples de solutions de l'équation mais seul le couple tel que $2k\pi + q\theta = \arccos(\cos(\theta))$ est retenu.

## 2.4 Modèle hybride pour la diffusion par une particule

### 2.4.1 Introduction

Dans ce travail de thèse, l'implémentation numérique de l'optique géométrique a pour but premier de valider la modélisation de la diffusion de la lumière par la méthode de Monte-Carlo, où une discrétisation selon la coordonnée de diffusion dans le plan cartésien est utilisée. Nous utilisons donc la même discrétisation que Van de Hulst [**Hulst 1957**] et non une discrétisation selon l'angle d'incidence. Pour rendre plus quantitatif ce modèle "hybride", qui combine optique géométrique, optique électromagnétique et physique, nous prenons en compte les termes de phase, de divergence et certains effets ondulatoires (diffraction de Fraunhofer).

### 2.4.2 Modèle d'optique géométrique pure

#### 2.4.2.1 Direction de diffusion et intensité

Pour calculer la direction de diffusion, on utilise les résultats énoncés dans la section 2.3 et plus spécifiquement les équations (97) et (98).

L'intensité des rayons d'ordre $p$ est calculée grâce aux coefficients de Fresnel (de réflexion en amplitude perpendiculaire et parallèle) qui s'expriment par :

$$r_\perp = \frac{m_2 \sin(\tau) - m_1 \sin(\tau')}{m_2 \sin(\tau) + m_1 \sin(\tau')}$$
$$r_\parallel = \frac{m_1 \sin(\tau) - m_2 \sin(\tau')}{m_1 \sin(\tau) + m_2 \sin(\tau')}$$

(101)

A partir de l'équation (101), on peut déduire directement les coefficients de transmission en amplitude. Ces coefficients complexes permettent de déterminer les ratios d'amplitude (et *in fine* d'énergie) des rayons réfractés et réfléchis par une surface localement plane. On s'intéresse ici au cas où l'onde est polarisée perpendiculairement, le raisonnement étant identique pour le cas d'une polarisation parallèle. Pour $p = 0$ l'amplitude du champ associé au rayon réfléchi est noté $r_\perp$, on peut en déduire que celle du rayon réfracté $(p = 1)$ est égale à $(1 - r_\perp^2)$, puis on a $-r_\perp(1 - r_\perp^2)$ pour $p = 2$, etc…

On introduit la quantité $\varepsilon_\perp$ pour décrire ces amplitudes (respectivement $\varepsilon_\parallel$ pour une onde polarisée parallèlement) avec :

$$\varepsilon_\perp = r_\perp \qquad \text{pour } p = 0$$
$$\varepsilon_\perp = (1 - r_\perp^2)(-r_\perp)^{p-1} \qquad \text{pour } p \geq 1$$

(102)

Finalement l'intensité des rayons d'ordre p s'écrit [**Hulst 1957**] :

$$I_1(p) = \varepsilon_\perp^2(p) I_0$$

(103)

A titre indicatif, la Figure 17 compare les diagrammes de diffusion obtenus avec l'équation (103) et les séries de Debye. On note un bon accord global, même si certaines oscillations sont totalement minorées par le modèle géométrique.

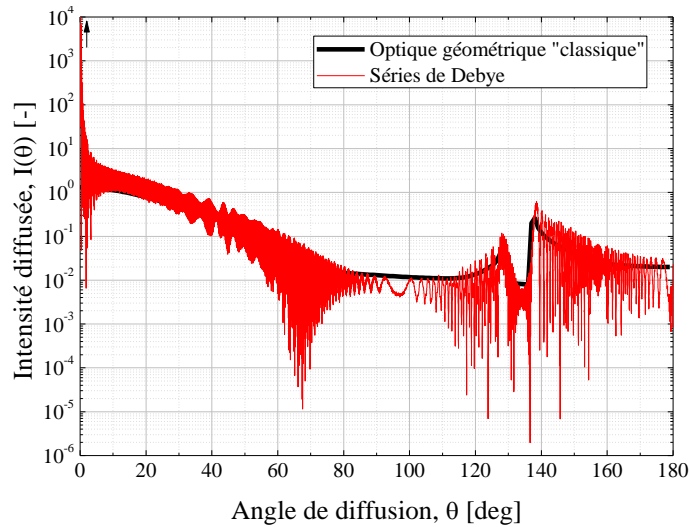

Figure 17 : Comparaison des diagrammes de diffusion obtenus par l'équation (103) et les séries de Debye (tronquées à partir de p=20)

### 2.4.3 Prise en compte des phénomènes ondulatoires

L'équation (103) ne prend pas en compte des effets comme la divergence, de même que certains effets ondulatoires tels que la phase, la diffraction… On peut rajouter les contributions individuelles de certains de ces phénomènes grâce au modèle de H.C. Van de Hulst [**Hulst 1981**].

#### 2.4.3.1 Divergence

La divergence rend compte de la courbure locale de la sphère qui, même à l'échelle infinitésimale, induit un étalement angulaire du faisceau incident lors de l'interaction avec la sphère.

Soit un rayon lumineux d'intensité $\mathbf{I_0}$ qui impacte une sphère de rayon $a$ avec un angle complémentaire à l'angle d'incidence, $\tau$. Ce rayon ressort de la sphère à l'angle de diffusion $\theta$.

On note $\mathbf{n}$ le vecteur normal à la surface élémentaire $dS$ éclairée par le rayon incident, $\varphi$ l'angle azimutal et $r$ la coordonnée radiale du repère dans lequel est centrée la particule, voir la Figure 18.

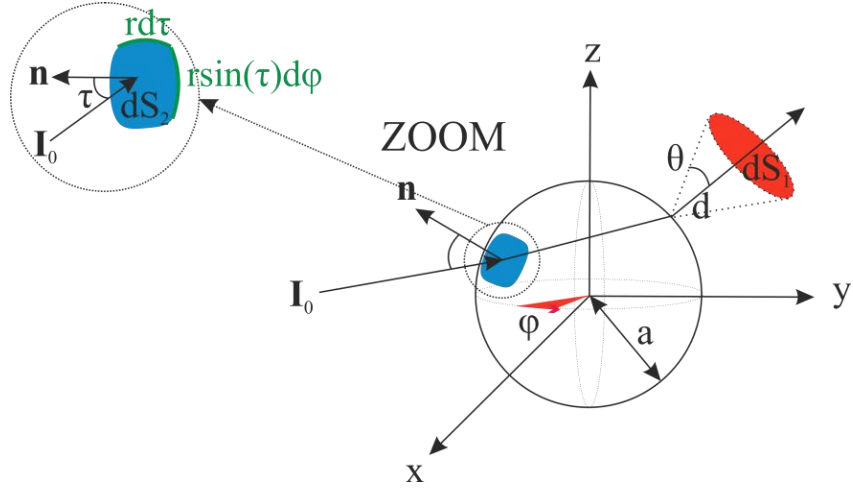

Figure 18 : Schéma illustrant la géométrie du problème pour le calcul de la divergence

Le flux d'énergie incident à travers la surface $dS_2$ s'écrit :

$$\phi_{entrée} = \mathbf{I}_0.\mathbf{dS}_2 = \mathbf{I}_0.\mathbf{n}dS_2 = I_0\cos(\tau)dS_2 \qquad (104)$$

Cette surface élémentaire est un rectangle de côté $rd\tau$ et $r\sin(\tau)d\varphi$, ce qui donne pour une sphère de rayon $a$ : $dS_2 = a^2\sin(\tau)d\tau d\varphi$ et donc pour le flux :

$$\phi_{entrée} = I_0 a^2\cos(\tau)\sin(\tau)d\tau d\varphi \qquad (105)$$

En sortie, à une grande distance $d \gg r$ de la sphère, le flux lumineux est contenu dans l'angle solide différentiel (cône de rayon 1, dans la Figure 18), il s'écrit :

$$\phi_{sortie} = I_1 dS_1 \qquad (106)$$

avec $dS_1 = d^2 d\Omega$, on a également $d\Omega = \sin(\theta)d\theta d\varphi$ d'où :

$$\phi_{sortie} = I_1 d^2\sin(\theta)d\theta d\varphi \qquad (107)$$

On peut supprimer, pour les énergies incidente et sortante, l'élément angulaire $d\varphi$ dans la direction perpendiculaire au plan de diffusion, car le problème est symétrique par rapport à ce plan. La conservation de l'énergie permet d'écrire que :

$$\varepsilon_p^{\,2}\phi_{entrée} = \phi_{sortie} \qquad (108)$$

Le facteur $\varepsilon_p^{\,2}$ rend compte de l'atténuation de l'énergie du rayon en entrée après $p$ réflexions internes, avec $\varepsilon_p \equiv \varepsilon_{p,\parallel}$ pour la polarisation parallèle et $\varepsilon_p \equiv \varepsilon_{p,\perp}$ pour la polarisation perpendiculaire, puisqu'une sphère conserve la polarisation de l'onde incidente. On obtient ainsi :

$$\varepsilon_p^{\,2} I_0 a^2\cos(\tau)\sin(\tau)d\tau = I_1 d^2\sin(\theta)d\theta \qquad (109)$$

In fine, l'intensité du rayon $p$ diffusée dans la direction $\theta$ est donnée par :

$$I = I_0 \varepsilon_p^{\;2} \frac{a^2}{d^2} \Theta \tag{110}$$

avec pour le facteur de divergence $\Theta$ de ce rayon :

$$\Theta = \frac{\cos(\tau)\sin(\tau)}{\sin(\theta)\dfrac{d\theta}{d\tau}} = \frac{\cos(\tau)\sin(\tau)}{\sin(\theta)\left|\dfrac{d\theta'}{d\tau}\right|} \tag{111}$$

Dans l'équation (110), l'état de polarisation est choisi *via* celui du coefficient de Fresnel.

### 2.4.3.2 Phase

Dans ce modèle on peut intégrer une partie de la nature ondulatoire de la lumière en associant une phase aux rayons. Pour ce faire, on commence par définir un rayon de référence qui passe par le centre de la sphère et qui ressort avec le même angle de diffusion que le rayon simulé. Van de Hulst [**Hulst 1981**] distingue alors trois types de déphasage par rapport au rayon de référence :

- ceux imputés aux réflexions et réfractions. Ils sont déjà pris en compte dans les coefficients de Fresnel (grandeurs complexes),

- ceux associés à la différence de chemin optique et dont l'expression générale est :

$$\delta_p = \frac{2\pi a n_1}{\lambda}(\sin(\tau) - p n_2 \sin(\tau')) \tag{112}$$

- ceux liés au passage du rayon par les points et lignes focales (qui traduisent la courbure du front d'onde et l'inversion de la direction du champ électrique). Van de Hulst propose l'expression suivante pour ce déphasage :

$$\frac{\pi}{2}(p - 2k + \frac{1}{2}s - \frac{1}{2}q) \tag{113}$$

Le déphasage total $\sigma_p$ du rayon $p$, dans la direction $\theta$, est la somme de tous les déphasages précédemment explicités :

$$\sigma_p = \delta + \frac{\pi}{2}(p - 2k + \frac{1}{2}s - \frac{1}{2}q) + phase(\varepsilon_p) + \frac{\pi}{2} \tag{114}$$

avec k, q les coefficients déterminés dans 2.4.2.1 et $s = \pm 1$.

A noter l'ajout pragmatique de $\pi/2$ à la phase pour mieux rendre compte du diagramme de diffusion obtenu avec la théorie de Lorenz-Mie (voir [**Hulst 1981**]).

La Figure 19 montre l'évolution de la phase pour les 4 premiers ordres, pour une goutte d'eau.

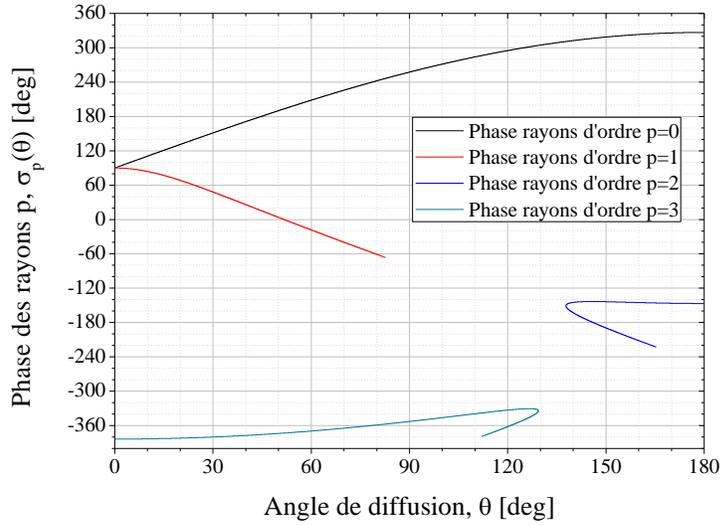

Figure 19 : Évolution, pour une goutte d'eau dans l'air de diamètre 200µm, de la phase des premiers ordres de diffusion : réflexion, réfraction simple, secondaire,...

### 2.4.3.3 Amplitude complexe et diffraction

Pour une polarisation donnée ( $\perp$ ou $\parallel$ ), l'amplitude complexe des rayons $p$ diffusés dans la direction $\theta$ est de la forme :

$$\mathbf{S}_p = \left| S_p \right| e^{-j\sigma_p} \tag{115}$$

avec $\left| \mathbf{S}_p \right| = \sqrt{\Theta} \left| \boldsymbol{\varepsilon}_p \right|$.

La diffraction peut être prise en compte en ajoutant la contribution modélisée par la théorie de Fraunhofer. L'intensité totale diffusée par la particule dans la direction $\theta$ est alors de la forme :

$$I(\theta) = \frac{1}{2} \operatorname{Re} \left\{ \left( \sum_{p=0}^{\infty} \mathbf{S}_p \right) \left( \sum_{p=0}^{\infty} \mathbf{S}_p \right)^* \right\} + \left| S_F(\theta) \right|^2 \tag{116}$$

où $S_F(\theta) = x^2 \left\{ J_1(x\sin(\theta))/x\sin(\theta) \right\}$ représente, dans l'approximation de Fraunhofer, l'amplitude de l'onde diffractée pour la particule. A noter que si dans l'équation (116) le premier terme dépend implicitement de l'état de polarisation du faisceau incident, ce n'est pas le cas du second terme.

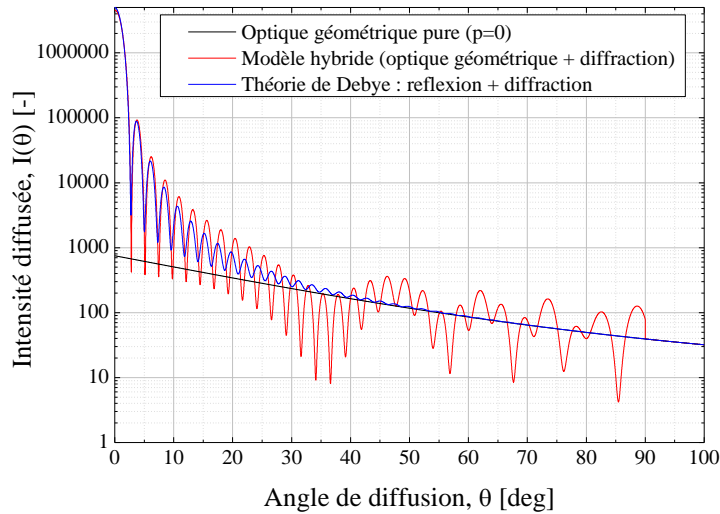

Figure 20 : Comparaison des diagrammes de diffusion pour les rayons d'ordre $p = 0$ : optique géométrique, optique géométrique avec diffraction, théorie de Debye. *D=10µm, λ=405nm*, polarisation perpendiculaire

On sait, de manière générale, que la diffraction prédomine aux petits angles pour les particules telles que $D \gg \lambda$. Cependant, c'est avec étonnement que, dans la littérature, nous n'avons trouvé aucuns critères sérieux sur la limite supérieure du domaine angulaire de validité de l'approximation de Fraunhofer. Seul un critère à $\pi/2$ ($\theta = 90°$) est évoqué du fait de la symétrie de la fonction $S_F(\theta)$. La Figure 20 en illustre les conséquences. Les prédictions de l'équation (116) avec et sans diffraction y sont comparées à celles de la théorie de Debye. On remarque que si l'on applique la diffraction jusqu'à 90°, son influence sur les rayons d'ordre 0 est très importante. La Figure 20 montre cependant que selon les séries de Debye, la diffraction devient négligeable pour des angles supérieurs à 35°. Pourtant, dans la littérature [**Keller 1961**],[**Xu 2006**], la diffraction semble appliquée comme sur la Figure 20. On peut alors se demander comment est-il possible qu'au final les diagrammes de diffusion concordent si bien ? A notre avis, l'amplitude des interférences entre les ordres p=0 et p=1 masque les valeurs excessives de la diffraction. La façon dont la diffraction est prise en compte dans ce modèle hybride n'est donc pas totalement satisfaisante.

### 2.4.4 Résultats et comparaisons aux séries de Debye

Un code Fortran 90 a été développé pour implémenter le modèle hybride précédemment décrit. Les différentes simulations ont été réalisées pour une goutte d'eau, d'indice $m_2 = 1.332$ et de diamètre $D = 200 \mu m$, placée dans l'air ($m_1 = 1.0$). La longueur d'onde de l'onde incidente est prise dans la plage du visible $\lambda = 405nm$. Les détecteurs sont quasi-ponctuels avec une ouverture angulaire de 0.01° ou 0.1°.

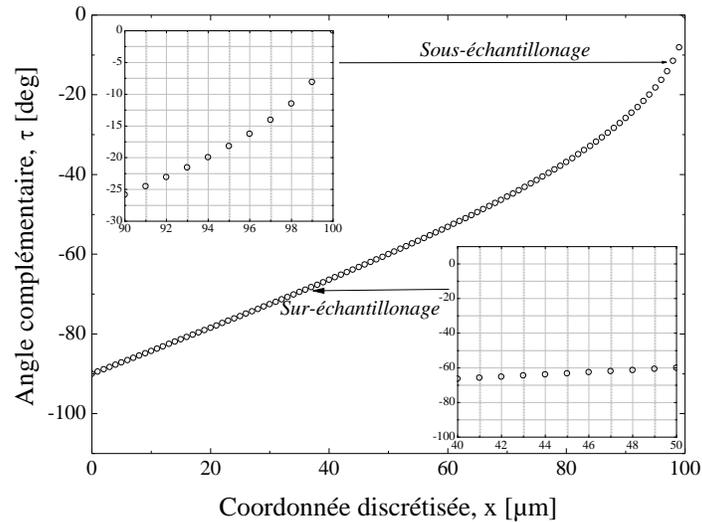

Figure 21 : Illustration du problème de discrétisation, par rapport au modèle de Van de Hulst, lorsqu'on utilise une discrétisation "cartésienne"

Dans le modèle de Van de Hulst, l'onde incidente est discrétisée *via* des variations infinitésimales $d\tau$ de l'angle complémentaire d'incidence. Dans la perspective d'élaborer un modèle de type Monte-Carlo, nous avons préféré une discrétisation en incréments $dx$ et $dz$ du faisceau se propageant suivant l'axe $y$. Si l'on n'y prend pas garde, notre approche, plus versatile et générale que celle de Van de Hulst, peut induire des problèmes de sous-échantillonnage. La Figure 21 illustre ce problème pour la coordonnée $x$, les rayons d'ordre $p = 0$, dans le cas d'une sphère de diamètre $D=100\mu m$ et un front d'onde discrétisé en 100 pas $dx = 1\mu m$. On remarque que dans la plage angulaire $[-10,0]$ degrés on a seulement 2 incréments, alors que dans la plage $[-90,-80]$ degrés on a énormément d'incréments $d\tau$. Pour résoudre ce problème nous avons implémenté une interpolation linéaire de l'amplitude et de la phase des rayons diffusés.

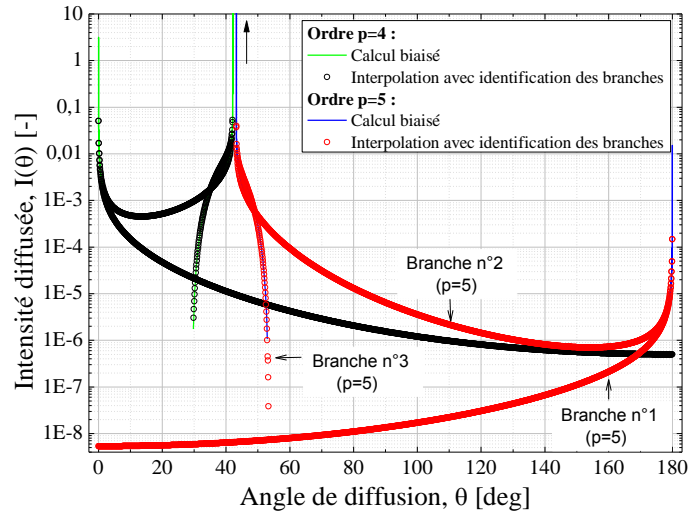

Figure 22 : Diagrammes de diffusion des rayons d'ordre $p = 4$ et $p = 5$ avant et après interpolation et identifications des branches pour une goutte d'eau

L'interpolation linéaire ne pose pas de problème technique pour les ordres 0 et 1 et elle s'applique de manière classique. Pour les ordres supérieurs, plusieurs branches peuvent être observées d'où la présence de plusieurs solutions pour un même angle de diffusion, voir la Figure 22. Pour pallier à cette difficulté, il faut interpoler chaque branche. En pratique, cette étape requiert d'étudier la monotonie de la fonction $\theta_p(x)$, d'isoler chaque branche et en rechercher les brusques variations associées aux singularités : arcs-en-ciel, diffusion critique,... Dans cet exemple, l'ouverture des détecteurs est de 0.01°, ce qui correspond à la maille du calcul d'interpolation.

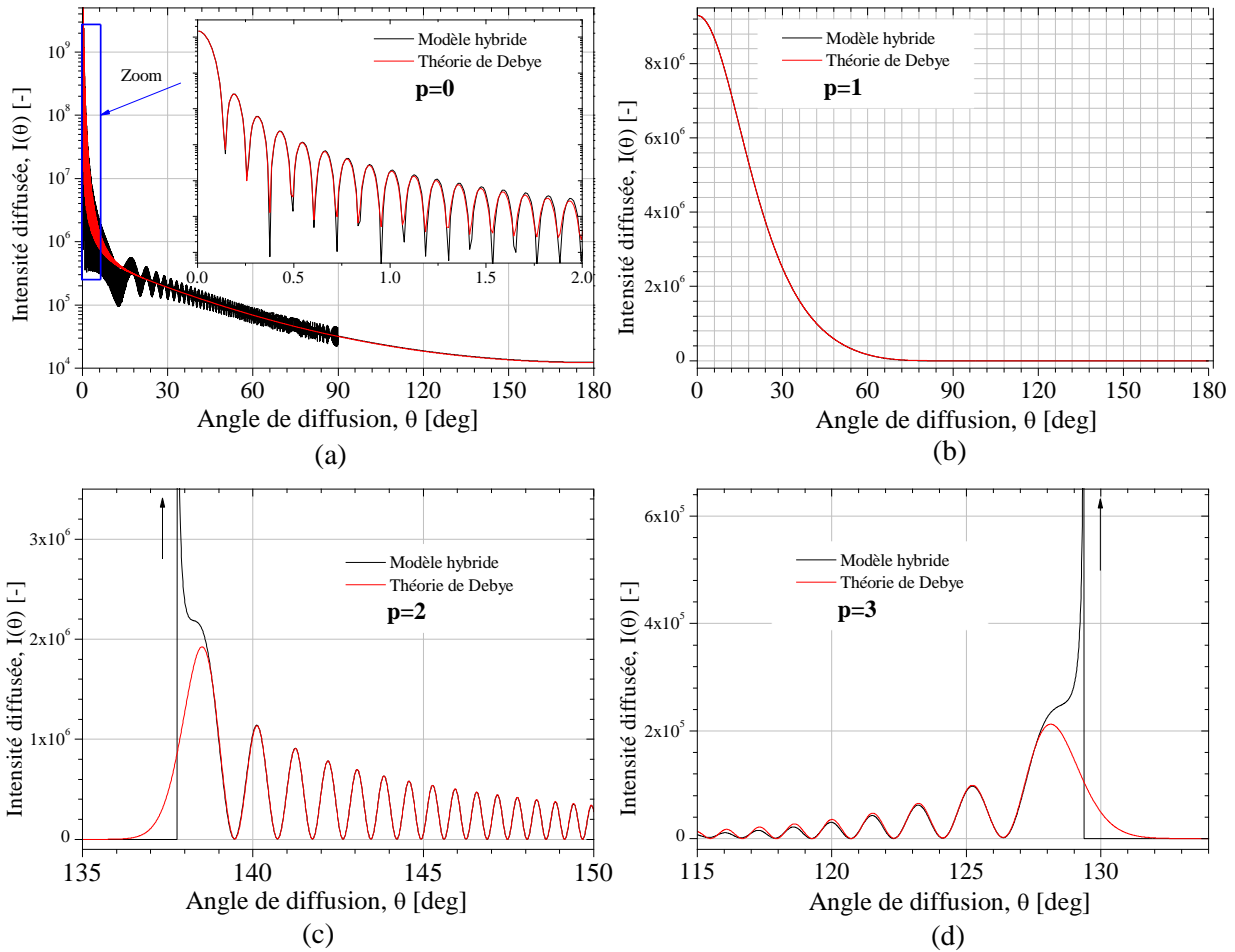

Figure 23 : Comparaison des prédictions du modèle hybride et de la théorie de Debye pour les quatre premiers ordre : (a) $p = 0$, (b) $p = 1$, (c) $p = 2$ et (d) $p = 3$. La particule, une gouttelette d'eau de 200µm de diamètre et d'indice de réfraction 1.332 est éclairée par une onde plane de polarisation perpendiculaire et de longueur d'onde λ=405nm

La Figure 23 confronte, au voisinage des principales singularités, les résultats obtenus avec le modèle hybride et la théorie de Debye. Pour permettre cette comparaison, il faut déterminer le coefficient de normalisation entre le modèle hybride et la théorie de Debye. Ici c'est l'ordre $p = 0$ au-delà de la zone de diffraction qui a été utilisé mais, de manière générale, tout ordre ne possédant localement qu'une branche et non oscillant peut convenir. Dans la Figure 23 (a), on observe des différences importantes entre 10° et 90°, alors que l'accord est très bon en dehors de cette zone. Comme nous l'avons déjà évoqué, cet écart est lié au raccordement trop brutal de la réflexion spéculaire (optique géométrique) et de la diffraction (optique physique). Pour l'ordre 1, dépourvu de singularité, l'accord est parfait sur tout le domaine angulaire. Il en est de même pour les ordres 2 et 3, à l'exception des zones angulaires associées aux caustiques correspondant à une divergence nulle, voir l'équation (111). Pour ces angles particuliers, l'intensité diffusée tend vers l'infini. Ceci n'a évidemment pas de sens et montre les limites de l'optique géométrique. Seule l'intégration d'un modèle d'optique physique, basé sur la théorie d'Airy [**Airy 1838**], pourrait résoudre ce problème (travail en cours). En dehors des régions singulières, cf. Figure 23 (c) et (d), les oscillations de haute et basse fréquence sont dues à l'existence simultanée de deux, voire trois branches, dont les rayons associés interfèrent entre eux. Ce phénomène est illustré par la Figure 24.

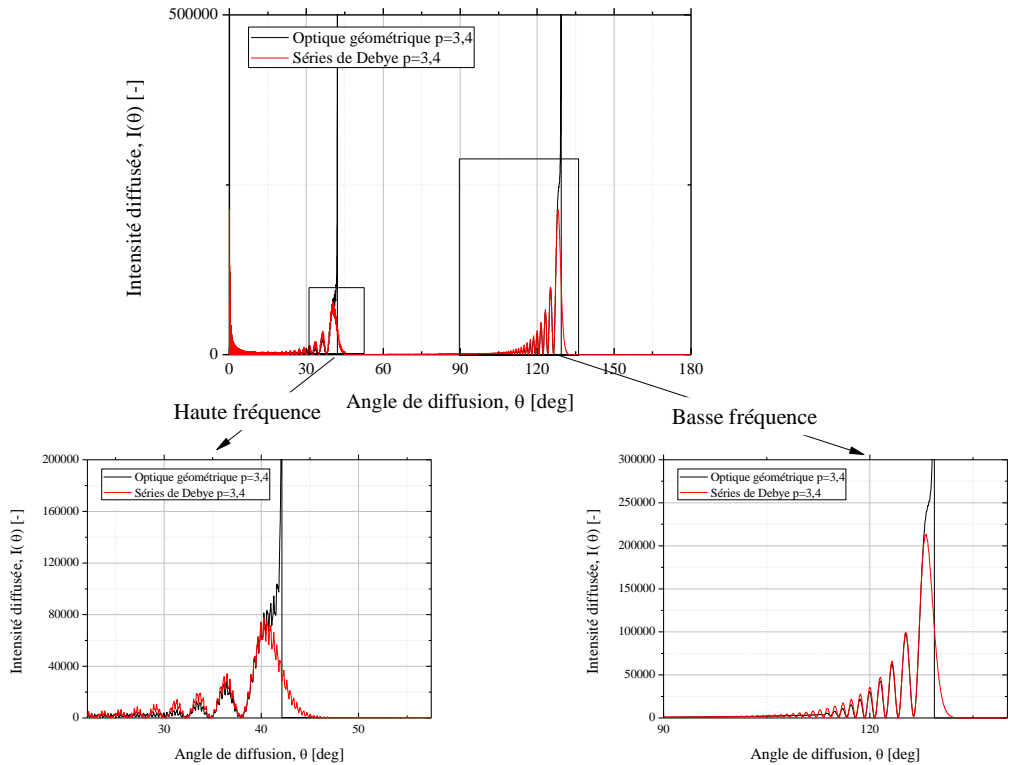

Figure 24 : Illustration du phénomène d'interférence pour des rayons d'ordre 3 et 4. La comparaison avec Debye est également réalisée. A gauche, les oscillations haute fréquence correspondent à une interférence à 3 branches. A droite, on observe les interférences à deux branches seulement qui entrainent des oscillations basse fréquence.

Au vue de ces résultats encourageants, on peut passer à l'observation d'un diagramme de diffusion complet. La Figure 25 montre celui calculé pour une goutte d'eau lorsque l'on prend en compte les 21 premiers ordres de diffusion, c'est-à-dire que $p_{max} = 20$. Pour plus de précision, on pourrait chercher à prendre en compte des ordres plus élevés. Cependant, les 20 premiers correspondent approximativement à près de 99,7% de l'intensité totale diffusée et le calcul des ordres supérieurs alourdirait considérablement les calculs.

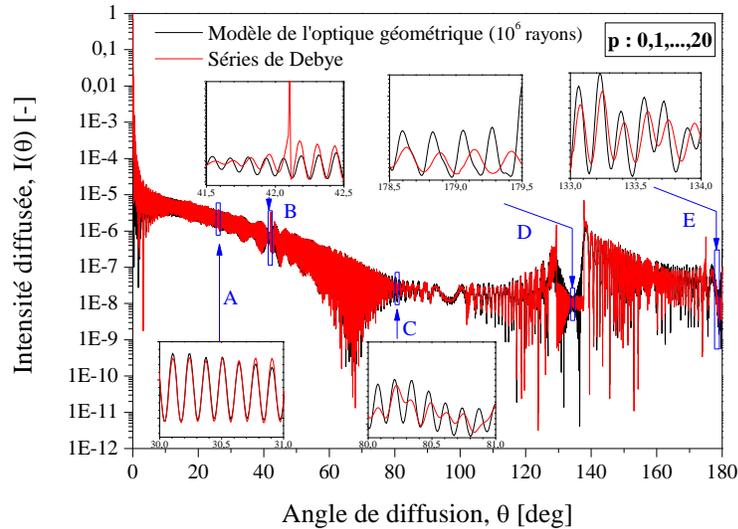

Figure 25 : Comparaison des diagrammes de diffusion prédits par l'optique géométrique et par les séries de Debye pour une gouttelette d'eau de diamètre $D = 200 \mu m$ avec zooms sur différentes régions (A, B, C, D et E)

L'accord global est bon et ceci tout particulièrement dans les zones éloignées des singularités ou bien dans celles où la réflexion ou la réfraction simple dominent (zone A par exemple). Des régions comme la bande sombre d'Alexander (zone D) ou la rétrodiffusion (zone E) sont moins bien décrites, mais celles-ci sont connues pour être très sensibles aux ondes de surface. La zone B est perturbée par la présence des arcs-en-ciel $p = 4$ et $p = 5$, comme nous l'avons vu avec la Figure 22. Les écarts observés autour de la zone C, qui correspond approximativement à l'angle de réfraction limite pour l'eau, sont plus difficiles à interpréter.

Quoiqu'il en soit, nous concluons de cette partie que le modèle hybride avec discrétisation cartésienne donne de très bons résultats. Au delà de son intérêt propre, il se révèle utile pour le développement et la validation du modèle de Monte-Carlo.

# Chapitre 3

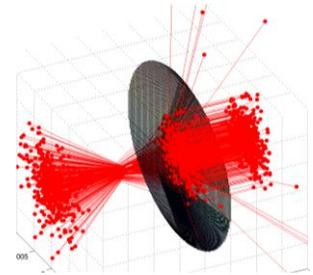

## Modélisation de la diffusion de la lumière par méthode de Monte-Carlo

La diffusion de la lumière par un système complexe, composé notamment de particules ou d'objets de tailles et de formes très différentes, nécessite de pouvoir traiter l'interaction de la lumière avec tout type d'élément. Il existe de nombreux modèles et théories traitant de la diffusion de la lumière (cf. **chapitre 2**) par des particules isolées. Parmi ceux-ci on citera les théories électromagnétiques utilisant une méthode de séparation des variables : théories de Lorenz-Mie [**Lorenz 1890, Mie 1908**], de Debye [**Debye 1909**], du Moment Complexe Angulaire [**Nussenzveig 1979**], des modèles plus ou moins élaborés qui reposent sur l'optique géométrique [**Hulst 1957, Bohren 1998, Xu 2006**] ou l'optique physique (au sens d'ondulatoire) : théories de Fraunhofer pour la diffraction vers l'avant [**Xu 2002**], d'Airy [**Airy 1838**] pour l'arc-en-ciel, de Marston [**Marston 1979**] pour la diffraction au voisinage de l'angle critique, etc…

Chaque approche a des avantages et des limites concernant la taille et la forme de l'objet, la forme de l'onde incidente, les temps de calcul, etc… De fait, modéliser la diffusion de la lumière dans un environnement complexe en utilisant une seule approche est impossible. Pour s'affranchir des limites de chaque approche, une modélisation de type Monte-Carlo (voir par exemple [**Buslenko 1966**]) est toute indiquée.

## 3.1  La méthode de Monte-Carlo

Cette méthode, qui fait référence aux jeux de hasard, consiste à calculer des valeurs numériques (comme des intégrales par exemple) en utilisant des procédés aléatoires. Les premières utilisations de cette méthode remontent à 1733 quand Buffon [**Buffon 1733**] tenta d'approximer la valeur de $\pi$ en lançant aléatoirement des aiguilles sur un plancher (d'où l'appellation expérience de l'aiguille de Buffon). On citera également Enrico Fermi qui, dans les années 30, a étudié la diffusion par des neutrons avec des méthodes probabilistes mais n'a jamais publié ses travaux [**Metropolis 1987**].

Le principe de cette méthode, telle qu'elle est utilisée maintenant, a été publié par Nicholas Metropolis, Stanislaw Ulam et Von Neumann [**Metropolis 1949**], peu après la seconde guerre mondiale, dans le cadre de recherches sur la bombe atomique. La dénomination "Monte-Carlo" a été inspiré par l'oncle de Stanislaw Ulam qui pariait très souvent des grosses sommes au casino de Monte-Carlo. Von Neumann conscient du potentiel de cette méthode participa à l'élaboration du premier ordinateur entièrement électronique, l'ENIAC, sur lequel il implémenta le premier générateur de nombres pseudo-aléatoires (méthode middle-square [**Neumann 1951**]). Ce modèle nécessite en effet de tirer une grande quantité de nombres aléatoires. Dans les années 40, ces nombres étaient lus dans des tables de nombres aléatoires rendant l'utilisation du modèle de Monte-Carlo extrêmement coûteuse en temps. C'est à cette époque que le développement de générateurs de nombres pseudo-aléatoires a fait l'objet d'efforts soutenus. Par la suite, la méthode de Monte-Carlo a été utilisée dans de nombreux domaines scientifiques (physique, biologie, microélectronique, mathématiques...).

L'application de cette technique pour la diffusion de la lumière a été développée notamment dans les années 1960 pour, entre autre, caractériser la diffusion multiple de la lumière. Chandrasekhar [**Chandrasekhar 1960**] avait développé une méthode d'ordonnée discrète pour le transfert radiatif dans des milieux denses. Cette approche, monodimensionnelle, traite le cas d'une onde plane incidente dans un milieu limité par deux plans parallèles. Cette approche est équivalente à la modélisation du transport de photons par l'Équation de Transfert Radiative (ETR).

Dans le cas de la modélisation de la diffusion de la lumière par des milieux particulaires, la méthode consiste à décomposer le faisceau incident en une multitude de photons (pas dans le sens quantique du terme mais plutôt de "rayon infinitésimal" [**Hulst 1957**]) et à suivre ces derniers de la source jusqu'au détecteur ou leur disparition du domaine de calcul (absorption par exemple). Lors de leur propagation, les photons interagissent avec des objets macroscopiques (supports optiques, cuves, lentilles...) et micrométriques (particules, rugosités des surfaces,...), transportant ainsi vers nous une information sur le milieu traversé. Chaque interaction induit un phénomène de diffusion ou d'absorption modélisé en une densité de probabilité à l'aide d'un modèle *ad hoc*. C'est sur ce point là que la méthode de Monte-Carlo s'avère particulièrement puissante. En effet, l'optique géométrique est encore le meilleur outil pour étudier la propagation et l'interaction de la lumière dans un environnement complexe composé d'objets macroscopiques avec des surfaces spéculaires (c'est-à-dire continues et dérivables). Quand ce n'est plus le cas, et notamment quand les objets ont des dimensions de l'ordre de la longueur d'onde ou bien qu'ils génèrent des caustiques, on utilise au choix l'optique physique (ondulatoire) ou bien des modèles électromagnétiques. Ainsi, dans le cas de nano et micro-diffuseurs (ou "particules"), on utilisera des modèles physiques et des théories électromagnétiques. La méthode de Monte-Carlo permet ainsi de découpler et de décrire ces phénomènes à différentes échelles, *via* une description successive et stochastique de tous les événements de diffusion et d'absorption. Néanmoins, il est important de noter qu'une limite bien connue de cette approche est que, les phénomènes les plus faibles étant les moins probables, cette dernière nécessite des ressources informatiques conséquentes. Certains auteurs utilisent pour cette approche la dénomination "Monte-Carlo de Tracé de Rayons" (MCTR [**Yuhan 2012**]). De nombreux codes de calcul ont été développés par différents auteurs suivant ce principe général (voir par exemple [**Collins 1972**], [**Bruscaglioni 1968**], [**Maheu 1988, Briton 1989**], [**Wang 1997**], etc.).

Dans cette thèse de doctorat, un code utilisant la méthode de Monte-Carlo a été intégralement pensé et implémenté pour répondre aux exigences du projet OPTIPAT (cf. **chapitre 1**). Ce code en Fortran 90, parallélisé, permet également au laboratoire de disposer d'un contrôle total sur cet outil de modélisation évolutif. L'algorithme général de traitement d'un photon depuis la source jusqu'au détecteur est présenté sur la Figure 26. Nous y reviendrons régulièrement dans ce chapitre qui en détaille le fonctionnement. Dans la section 2, nous présentons le générateur de nombres aléatoires de Mersenne-Twister. La section 3 traite de la modélisation de la source laser. La section 4 détaille l'interaction entre le photon et les différents éléments optiques (plan, cylindre, sphère...) du montage. La section 5 décrit l'interaction des photons avec le milieu particulaire. Pour finir, la section 6 permet de valider notre code Monte-Carlo sur des cas concrets. Parmi nos hypothèses, voici les principales posées:

- *Les particules sont distribuées* de manière homogène au sein du milieu particulaire. Cette hypothèse forte n'est pas valable pour certaines conditions d'écoulement.

- *Le régime de diffusion* peut être simple ou multiple. *La diffusion multiple* (indépendante) est vue comme une succession de diffusions simples. On suppose donc que chaque particule diffuse ou absorbe la lumière indépendamment de la présence des autres. Cette hypothèse est valide tant que la concentration n'est pas trop importante.

- *La diffusion de la lumière* est traitée comme un processus purement élastique. Les particules sont donc considérées comme immobiles.

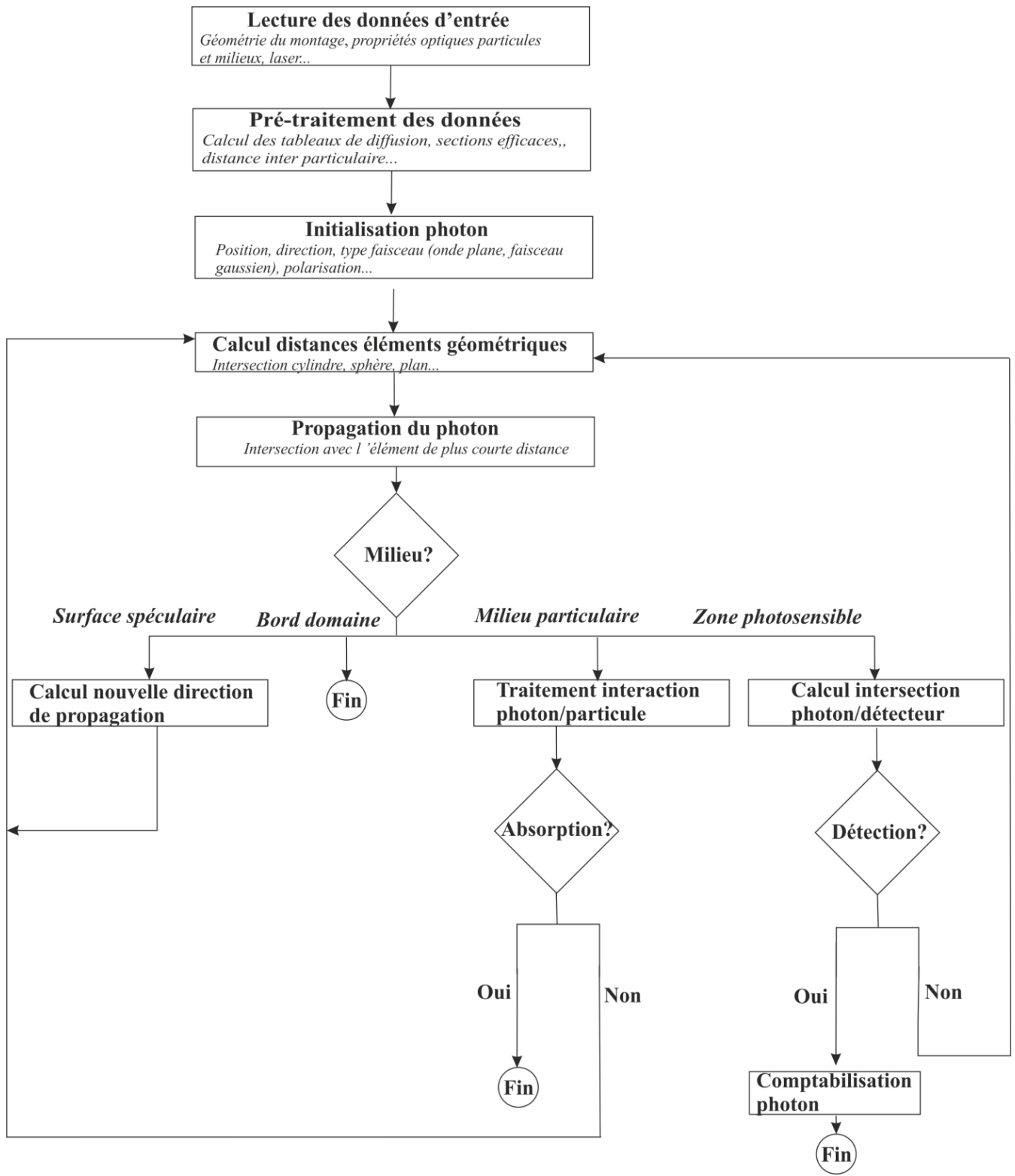

Figure 26 : Algorithme de traitement d'un photon depuis la source laser jusqu'à sa disparition (détection, absorption particulaire, sortie du domaine de calcul...)

## 3.2   Le générateur de nombres aléatoires de Mersenne-Twister

La méthode de Monte-Carlo est une méthode stochastique. Pour chaque probabilité calculée, un nombre aléatoire doit être tiré. Étant donné le nombre extrêmement élevé d'événements probabilistes à déterminer, le générateur aléatoire doit être choisi avec minutie. Celui-ci doit être uniforme et posséder une période très importante, pour assurer l'indépendance des photons et modéliser les phénomènes les plus faibles. Les fonctions natives en Fortran permettant de générer des nombres aléatoires ne possèdent pas une période suffisamment longue pour notre étude.

Après une étude bibliographique, nous avons retenu le générateur aléatoire de Mersenne-Twister [**Matsumoto 1998**] qui, de fait, est très utilisé par les codes implémentant une méthode de Monte-Carlo. Avec un choix approprié de paramètres (voir plus loin), la période de ce générateur va jusqu'à $T = 2^{19937} - 1$ avec une équi-distribution de dimension 623 et une exactitude de 32 ou 64 bits selon les machines. Ce générateur satisfait les tests de "DIEHARD" particulièrement exigeants en termes de qualité [**Marsaglia 1995**].

L'autre particularité de ce dernier tient au fait qu'il réalise des opérations sur des booléens et non sur des entiers ou nombres réels le rendant particulièrement rapide. Il génère des nombres aléatoires distribués de manière uniforme dans l'intervalle $\left[0, 2^{w-1}\right]$ avec $w$ le nombre de bits machine (32 bits ou 64 bits en général). L'algorithme du générateur de Mersenne-Twister passe par trois étapes principales : à savoir l'initialisation de la graine du générateur aléatoire, une opération de récurrence et une opération de *tempering*.

### 3.2.1   Initialisation de la graine

L'initialisation de la graine est particulièrement importante. En effet, de cette dernière, dépend toute la suite des nombres aléatoires. De plus, à chaque simulation, la graine doit être réinitialisée au risque d'obtenir constamment la même séquence de nombres aléatoires. Dans notre cas, le code étant parallélisé, la graine est générée avec une valeur différente sur chaque processeur. Pour cela, nous avons mis en place une procédure d'initialisation de la graine selon le numéro de processeur :

$$RandSeed = E\left[\left|64979t(i_{num}-83)\bmod\left(104729\right)\right|\right] \tag{117}$$

avec E la fonction partie entière, $|\ |$ la fonction valeur absolue, mod la fonction modulo, $i_{num}$ le numéro du processeur et t une variable temporelle définie comme la somme du temps en millisecondes, du temps en secondes et du temps en minutes.

### 3.2.2  Pseudo-algorithme

La graine du générateur aléatoire s'écrit en entrée du programme comme une combinaison linéaire de bits (représentation machine) :

$$RandSeed = \sum_{i=0}^{w-1}x_i 2^i \tag{118}$$

On détermine la suite des $x_i$ selon la relation de récurrence suivante :

$$x_{k+n} = x_{k+m} XOR\left(x_k^u \mid x_{k+1}^l\right)A \text{ pour } k = 0,1,... \tag{119}$$

avec XOR l'opérateur OU exclusif, | la fonction de concaténation des bits, A une matrice $(w,w)$ à valeurs dans $F_2 = \{0;1\}$, $X_k^u$ les $w-r$ bits les plus grands dans $X_k$ et $X_k^l$ les r bits les plus petits de $X_k$. La matrice A s'écrit :

$$\begin{pmatrix} 0 & I_{w-1} \\ a_{w-1} & \left(a_{w-2}...a_0\right) \end{pmatrix} \tag{120}$$

avec $I_{w-1}$ la matrice identité de taille $(w-1,w-1)$ et $a = \left(a_0...a_{w-1}\right)$ un coefficient *ad hoc*.

On multiplie alors les $X_{k+n}$ bits générés par une matrice de *tempering* T afin de mélanger les bits pour augmenter l'imprédictibilité des valeurs générées. Cette étape constitue ce que l'on appelle une opération de *tempering* [**Matsumoto 1994**]. On applique alors différentes opérations pour ce mélange [**Matsumoto 1998**] :

$$\begin{aligned} y &\leftarrow x_{k+n}...x_0 \\ y &\leftarrow yXOR\left(y \gg u\right) \\ y &\leftarrow yXOR\left(\left(y \ll s\right)ANDb\right) \\ y &\leftarrow yXOR\left(\left(y \ll t\right)ANDc\right) \\ y &\leftarrow yXOR\left(y \gg l\right) \end{aligned} \tag{121}$$

avec $\gg$ et $\ll$ les opérations de décalage de bits à droite et à gauche respectivement, $u,s,t,l,b,c$ des paramètres *ad hoc*.

### 3.2.3  Définition des paramètres et sorties du générateur

Tout au long de l'algorithme, plusieurs coefficients ont été utilisés sans être définis, que ce soit dans l'étape de récurrence ou de *tempering*. Matsumoto [**Matsumoto 1998**] propose des paramètres maximisant la période et l'efficacité du générateur :

$$
\begin{aligned}
&\text{Paramètres de récurrence :} \\
&w = 32, r = 31, a = 2567483615, n = 624, m = 397 \\
&\text{Paramètres de } tempering : \\
&u = 11, t = 15, b = 2636928640, s = 7, l = 18, c = 4022730752
\end{aligned}
\tag{122}
$$

Une fois ces paramètres définis et l'algorithme mis en place, le générateur aléatoire produit des nombres tels que $x \in \left[0, 2^{w-1}\right]$. Pour obtenir des variables à valeurs dans $[0,1]$, il suffit d'écrire notre nombre aléatoire comme suit :

$$
\xi = \frac{x}{2^{w-1}}
\tag{123}
$$

Les différents tests de "DIEHARD", ou du $\chi^2$, n'ont pas été réalisés étant donné que ce générateur est très largement utilisé depuis son invention. Seules la période, la moyenne, la médiane et la variance (uniformité dans $[0,1]$) ont été vérifiées.

## 3.3  Description du faisceau laser

Le faisceau laser incident est décomposé en une multitude de rayons. Il faut donc trouver une discrétisation respectant le profil du faisceau laser.

Dans la plupart des modèles et théories de la diffusion (**cf. Chapitre 2**), l'onde plane est un cas de référence (pour la théorie de Lorenz-Mie notamment). Une onde plane est une onde dont l'amplitude et la phase sont les mêmes sur le plan normal à la direction de propagation de l'onde. Ce type d'onde est facilement décomposable en rayons distribués selon deux lois uniformes indépendantes.

On considère une onde localement plane se propageant selon l'axe $y$ et centrée sur ce dernier. La largeur de ce "faisceau" selon $x$ vaut $w_x$ et $w_z$ selon $z$. On peut alors discrétiser les coordonnées $x$ et $z$ selon deux lois uniformes :

$$
x \sim U\left(\left[\frac{-w_x}{2}, \frac{w_x}{2}\right]\right), z \sim U\left(\left[\frac{-w_z}{2}, \frac{w_z}{2}\right]\right)
\tag{124}
$$

On introduit également le faisceau gaussien (TM00) qui donne une meilleure description des faisceaux émis par la plupart des lasers. Nous négligeons cependant le terme de phase pour ne retenir que son profil en amplitude gaussien.

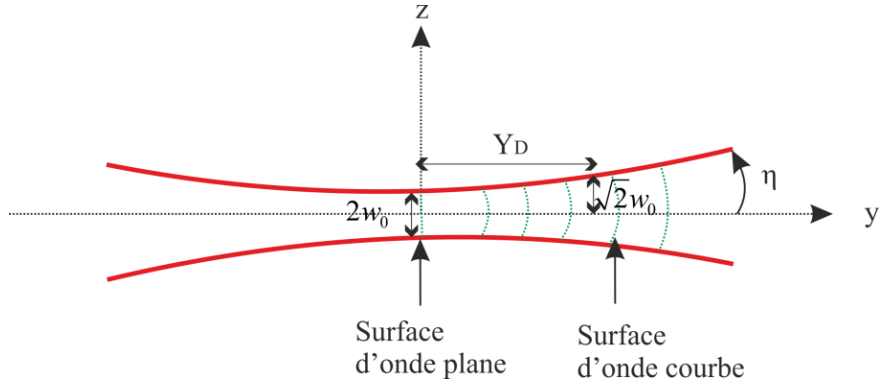

Figure 27 : Représentation géométrique d'un faisceau gaussien divergent avec les différents paramètres propres à ce dernier

Un tel faisceau se propageant selon l'axe $y$ a une intensité moyenne temporelle qui s'exprime par :

$$I(r,y) = I_0 \left( \frac{w_0}{w(y)} \right)^2 e^{\frac{-2r^2}{w(y)^2}} \qquad (125)$$

Avec $I_0$ l'intensité du champ électrique au centre du faisceau à l'origine, $w(y)$ la largeur du faisceau à la position $y$, $w_0$ la largeur au col du faisceau en $y=0$, $r$ la distance par rapport au centre du faisceau. On définit la largeur du faisceau selon $y$ par la relation :

$$w(y) = w_0 \sqrt{1 + \frac{y}{Y_R}} \qquad (126)$$

avec $Y_R = \pi \frac{w_0^2}{\lambda}$ la distance de Rayleigh.

L'angle de divergence d'un faisceau laser millimétrique étant très faible ($\eta \approx \lambda/\pi w_0 \approx 0.2$mrad pour $w_0 = 1$mm et $\lambda = 633$nm), on prendra comme hypothèse que le faisceau est collimaté, ce qui revient à négliger sa divergence. Ceci signifie que $w(y)$ est constant et $w(y) = w_0$, il vient alors :

$$I(r) = I_0 e^{\frac{-2r^2}{w_0^2}} \qquad (127)$$

Comme pour l'onde plane, le faisceau gaussien doit être discrétisé selon les coordonnées cartésiennes $(x, z)$ pour un faisceau se propageant selon l'axe y.

La méthode de Box-Muller [**Muller 1959**] consiste à générer, à partir de lois uniformes, des variables aléatoires qui suivent une loi normale centrée réduite. On dit que $X$ suit une loi normale d'écart-type $\sigma$ et d'espérance $\mu$ si sa fonction de densité est de la forme :

$$\phi(x) = \frac{1}{\sigma\sqrt{2\pi}} \exp\left(-\frac{1}{2}\left(\frac{x-\mu}{\sigma}\right)^2\right) \tag{128}$$

On note alors $X \sim N(\mu, \sigma^2)$. Pour $X$ suivant une loi normale centrée $(\sigma = 1, \mu = 0)$, on a :

$$\phi(x) = \frac{1}{\sqrt{2\pi}} \exp\left(\frac{-x^2}{2}\right) \tag{129}$$

La fonction de densité jointe de deux variables aléatoires, $X$ et $Z$, suivant chacune une loi normale centrée réduite, il vient que :

$$\phi(x,z) = \phi(x)\phi(z) = \frac{1}{2\pi} \exp\left(\frac{-(x^2+z^2)}{2}\right) \tag{130}$$

La méthode de Box-Muller consiste à écrire la relation en coordonnées polaires pour trouver une expression des variables $x$ et $z$ décrivant chacune une loi gaussienne. En effectuant le changement de variables :

$$\begin{aligned} x &= r\cos\theta \\ y &= r\sin\theta \end{aligned} \tag{131}$$

on obtient, pour les coordonnées polaires (indépendantes) $r$ et $\theta$ :

$$\phi(x,z)dxdz = r\exp\left(\frac{-r^2}{2}\right)dr \frac{1}{2\pi}d\theta \tag{132}$$

On voit que $\theta$ suit une loi uniforme sur $[0, 2\pi]$ permettant de déduire la fonction de répartition de $r$ :

$$f(r) = \int_0^r t\exp\left(\frac{-t^2}{2}\right)dt = 1 - \exp\left(\frac{-r^2}{2}\right) \tag{133}$$

Avec le générateur de Mersenne-Twister, on tire un nombre aléatoire $U_r$ de loi uniforme sur $[0,1]$ tel que $f(r) = U_r$, ce qui nous donne :

$$r = \sqrt{-2\ln U_r} \tag{134}$$

En tirant un second nombre aléatoire $U_\theta$ de loi uniforme dans $[0,1]$, on obtient pour la loi uniforme dans $[0,2\pi]$ :

$$\theta = 2\pi U_\theta \tag{135}$$

En repassant en coordonnées cartésiennes :

$$x = \sqrt{(-2\ln(U_r))}\cos(2\pi U_\theta)$$
$$z = \sqrt{(-2\ln(U_r))}\sin(2\pi U_\theta) \tag{136}$$

Grâce à l'expression du faisceau gaussien non divergent donnée dans l'expression (127), on en déduit que la fonction de densité jointe de x et z est donnée par :

$$\phi(x,z) = I_0 \exp\left(-2\frac{(x^2 + z^2)}{w_0^2}\right) \tag{137}$$

$I_0$ est fixée par les paramètres de la source et ne représente qu'une constante de normalisation des probabilités. La relation peut se reformuler de la manière suivante :

$$\phi(x,z) = \phi(x)\phi(z) = I_0 \exp\left(\frac{-2x^2}{w_0^2}\right)\exp\left(\frac{-2z^2}{w_0^2}\right) \tag{138}$$

La fonction de densité d'une loi normale étant $\phi(x) = \exp\left(-\frac{1}{2}((x-\mu)/\sigma)^2\right)\Big/\sigma\sqrt{2\pi}$, on peut en conclure que $x$ et $z$ sont distribuées selon deux lois normales centrées d'écart-type $w_0/2$. De plus, sachant que $N\left(0,(w_0/2)^2\right) = (w_0/2)N(0,1)$, la méthode de Box-Muller peut être utilisée pour générer x et z à partir de lois normales centrées réduites adaptées à notre cas :

$$x = \left(\frac{w_0}{2}\right)\sqrt{(-2\ln(U_r))}\cos(2\pi U_\theta)$$
$$z = \left(\frac{w_0}{2}\right)\sqrt{(-2\ln(U_r))}\sin(2\pi U_\theta) \tag{139}$$

Nous obtenons ainsi un faisceau gaussien à profil circulaire. Pour obtenir un faisceau gaussien à profil elliptique (astigmatisme des faisceaux émis par les diodes lasers par exemple), il suffit de choisir deux variances différentes pour $x$ et $z$, un tel faisceau étant défini par :

$$\phi(x,z) = I_0 \exp\left(-2\left(\frac{x^2}{w_x^2} + \frac{z^2}{w_z^2}\right)\right) \tag{140}$$

$x$ et $z$ devenant alors :

$$x = \left(\frac{w_x}{2}\right)\sqrt{\left(-2\ln\left(U_r\right)\right)}\cos\left(2\pi U_\theta\right)$$

$$z = \left(\frac{w_z}{2}\right)\sqrt{\left(-2\ln\left(U_r\right)\right)}\sin\left(2\pi U_\theta\right)$$

(141)

Des simulations ont été réalisées en FORTRAN 90 pour vérifier la validité de la discrétisation du faisceau gaussien. Pour ce faire, plusieurs critères doivent être respectés :

- la variable $x$ est distribuée selon une loi normale centrée d'écart-type $w_x/2$

- la variable $z$ est distribuée selon une loi normale centrée d'écart-type $w_z/2$

- le tracé de $z$ en fonction de $x$ est une ellipse d'axes $\left(w_x, w_z\right)$

La Figure 28 montre les deux types de profils laser simulés avec un faisceau (a) gaussien à profil circulaire et (b) gaussien à profil elliptique.

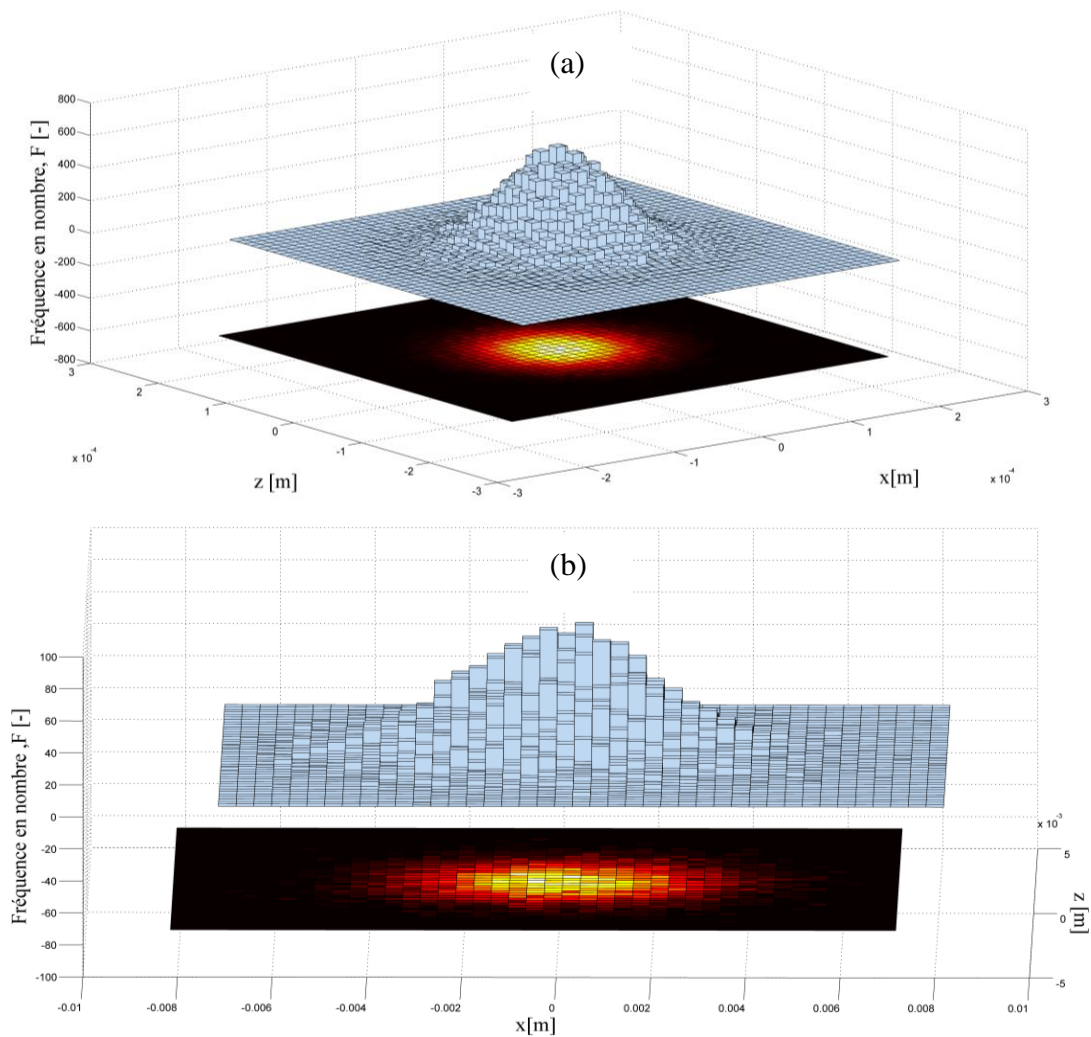

Figure 28 : Faisceaux gaussiens simulés avec la méthode de Box-Muller pour (a) un profil circulaire et (b) un profil elliptique

On peut également voir dans le tableau suivant, les paramètres nominaux $(w_x, w_z)$ utilisés pour la simulation des faisceaux de la Figure 28, ainsi que les vérifications faites sur les séries générées pour reproduire leurs profils $(\sigma_x, \sigma_z, \mu_x, \mu_z)$. On remarque le bon accord avec les paramètres nominaux. Ceci valide le modèle implémenté.

| Faisceau Gaussien | $w_x$ | $w_z$ | $\sigma_x$ | $\sigma_z$ | $\mu_x$ | $\mu_z$ |
| --- | --- | --- | --- | --- | --- | --- |
| | [mm] | [mm] | [mm] | [mm] | [mm] | [mm] |
| Profil circulaire (a) | 0.1 | 0.1 | 0.0499 | 0.0498 | 0.00001 | 0.0000002 |
| Profil elliptique (b) | 0.4 | 0.2 | 1.999 | 0.998 | 0.000003 | 0.000001 |

Tableau 1 : Récapitulatif des caractéristiques des faisceaux gaussiens générés par la méthode de Box-Muller

En plus du profil du faisceau laser incident, on peut préciser la direction (vecteur directeur des rayons lancés), la position ainsi que la polarisation (qui servira pour calculer les coefficients de Fresnel ainsi que les diagrammes de diffusion particulaire, voir dans la suite) de ce dernier.

## 3.4 Description de l'interaction avec des surfaces spéculaires

Durant sa propagation, le rayon peut interagir avec différents objets et notamment avec les différents systèmes dioptriques ou catadioptriques du montage optique. Il faut donc déterminer les intersections entre ce rayon et les surfaces spéculaires de ces objets, ainsi que le comportement résultant du rayon.

Afin de simuler la plupart des composants d'un montage optique classique, plusieurs éléments géométriques ont été traités : le plan, le cylindre, le trou optique, la sphère ainsi que des combinaisons comme la lentille sphérique plan-convexe. Pour chaque élément géométrique de l'expérience simulée, on doit calculer la distance du rayon avec ces derniers et garder la distance minimale pour savoir avec quel objet le rayon interagit en premier.

Dans tout ce qui suit, le rayon est localisé dans le repère du laboratoire par ses coordonnées $(x_0, y_0, z_0)$ et sa direction de propagation, par son vecteur directeur $\mathbf{d_p} = \left(\mathbf{d_p}(1), \mathbf{d_p}(2), \mathbf{d_p}(3)\right)^*$ où le symbole * indique l'opérateur transposition. La "droite de propagation" sera utilisée sous sa forme paramétrique :

$$\begin{cases} x = x_0 + t\mathbf{d_p}(1) \\ y = y_0 + t\mathbf{d_p}(2) \\ z = z_0 + t\mathbf{d_p}(3) \end{cases} \qquad \text{avec } t \in \mathbb{R}^+ \qquad (142)$$

L'exclusion des valeurs négatives de $t$ vient du fait qu'il faut conserver la direction de propagation du photon.

### 3.4.1 Plan infini/fini

Dans le repère du laboratoire, l'équation cartésienne d'un plan infini est donnée par :

$$ax + by + cz + d = 0 \qquad (143)$$

avec $a,b,c$ les composantes d'un vecteur normal au plan et $d$ un décalage par rapport à cette normale.

Si le photon interagit avec le plan, les équations (142) et (143) doivent avoir une solution commune, on injecte donc (142) dans (143) et on trouve que :

$$t\left(a\mathbf{d_p}(1) + b\mathbf{d_p}(2) + c\mathbf{d_p}(3)\right) + ax_0 + by_0 + cz_0 + d = 0 \qquad (144)$$

Il suffit alors de résoudre l'équation (144) pour trouver l'intersection entre la droite et le plan, en distinguant trois cas :

- l'équation n'a pas de solution, c.-à-d. que $\left(a\mathbf{d_p}(1) + b\mathbf{d_p}(2) + c\mathbf{d_p}(3)\right) = 0$ et que $ax_0 + by_0 + cz_0 + d \neq 0$, la trajectoire du rayon est parallèle au plan.

- l'équation admet tout nombre réel $t$ comme solution. Le rayon se propage dans le plan.

- l'équation admet une solution unique qui s'exprime par :

$$t = -\frac{ax_0 + by_0 + cz_0 + d}{\left(a\mathbf{d_p}(1) + b\mathbf{d_p}(2) + c\mathbf{d_p}(3)\right)} \qquad (145)$$

Le point d'intersection de la trajectoire du rayon avec le plan est donné par l'équation (142) et la distance est donnée par :

$$d_{plan} = |t|\sqrt{\mathbf{d_p}(1)^2 + \mathbf{d_p}(2)^2 + \mathbf{d_p}(3)^2} \qquad (146)$$

Pour le plan fini, la résolution est exactement la même, excepté que le point d'intersection trouvé doit appartenir à un certain domaine de l'espace.

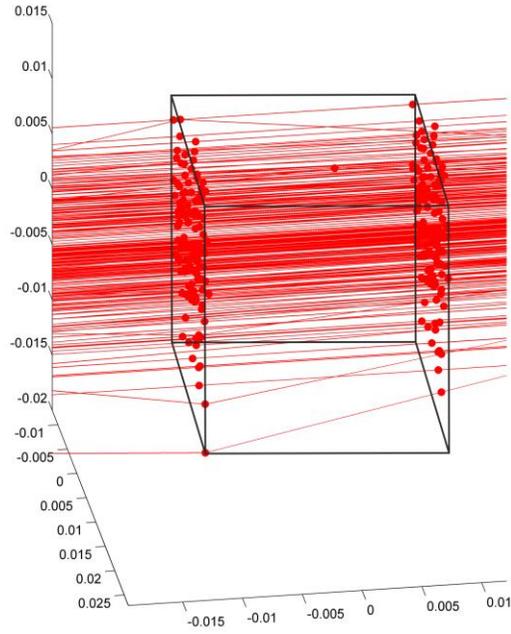

Figure 29 : Trajectoires de 100 rayons dans une cuve parallélépipédique rectangle composée de 4 plans parallèles. Les cercles rouges indiquent les points d'intersection. Pour les besoins de la figure, les parois ont été rajoutées "à la main"

On peut observer sur la Figure 29, une illustration du lancé de 100 rayons sur une cuve parallélépipédique rectangle. Le traitement pour ce cas passe par la résolution de 4 intersections (avec 4 plans) comme explicité par les équations ci-dessus.

### 3.4.2 Le cylindre

L'équation cartésienne d'un cylindre d'axe $z$, centré dans le repère du laboratoire est donnée par :

$$\left(x - x_c\right)^2 + \left(y - y_c\right)^2 - R_c^2 = 0 \quad z \in \left[z_{\text{inf}}, z_{\text{sup}}\right] \tag{147}$$

avec $R_c$ le rayon du cylindre, $\left(x_c, y_c\right)$ les coordonnées du centre du cylindre et $z_{\text{inf}}, z_{\text{sup}}$ les bornes de ce dernier selon l'axe $z$.

On procède comme précédemment, en injectant l'équation paramétrique de la droite (142) dans l'équation (147) :

$$\left(x_0 + t\mathbf{d_p}(1) - x_c\right)^2 + \left(y_0 + t\mathbf{d_p}(2) - y_c\right)^2 - R_c^2 = 0 \tag{148}$$

On développe l'équation (148) pour obtenir un polynôme du second degré en $t$ :

$$\left(\mathbf{d_p}(1)^2 + \mathbf{d_p}(2)^2\right)t^2 + 2\left((x_0 - x_c)\mathbf{d_p}(1) + (y_0 - y_c)\mathbf{d_p}(2)\right)t + \left((x_0 - x_c)^2 + (y_0 - y_c)^2 - R_c^2\right) = 0 \quad (149)$$

Pour faciliter et alléger la suite des calculs on pose :

$$\alpha t^2 + \beta t + \gamma = 0$$

$$\text{avec} \begin{cases} \alpha = \left(\mathbf{d_p}(1)^2 + \mathbf{d_p}(2)^2\right) \\ \beta = 2\left((x_0 - x_c)\mathbf{d_p}(1) + (y_0 - y_c)\mathbf{d_p}(2)\right) \\ \gamma = \left((x_0 - x_c)^2 + (y_0 - y_c)^2 - R_c^2\right) \end{cases} \quad (150)$$

La résolution de cette équation donne au final :

$$\begin{cases} t_1 = \dfrac{-\beta + \sqrt{\Delta}}{2\alpha} \\ t_2 = \dfrac{-\beta - \sqrt{\Delta}}{2\alpha} \end{cases} \text{avec } \Delta = \beta^2 - 4\alpha\gamma \quad (151)$$

On distingue alors trois cas :

- aucune solution (géométriquement parlant, $t_1 \in \mathbb{C}, t_2 \in \mathbb{C}$),

- une unique solution si $\Delta = 0$ et $t = -\beta/2\alpha$

- si la condition précédente n'est pas remplie on obtient deux solutions données par l'équation (151).

Dans le cas où l'équation admet deux solutions (qui peuvent être identiques), les distances du point de départ du photon au cylindre sont :

$$\begin{cases} d_{cyl1} = |t_1|\sqrt{\mathbf{d_p}(1)^2 + \mathbf{d_p}(2)^2 + \mathbf{d_p}(3)^2} \\ d_{cyl2} = |t_2|\sqrt{\mathbf{d_p}(1)^2 + \mathbf{d_p}(2)^2 + \mathbf{d_p}(3)^2} \end{cases} \quad (152)$$

On en déduit que le photon interagit avec le cylindre à la distance $d_{cyl} = \min\left(d_{cyl1}, d_{cyl2}\right)$ en $, x = x_0 + td_{cyl}(1), y = y_0 + td_{cyl}(2), z = z_0 + td_{cyl}(3)$ si $z \in \left[z_{\inf}, z_{\sup}\right]$.

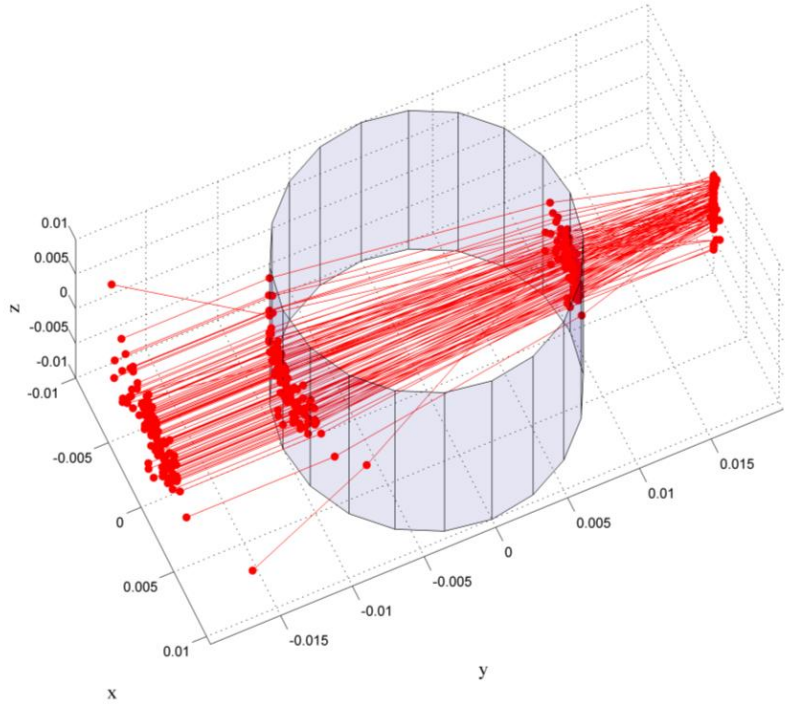

Figure 30 : Trajectoires de 100 rayons réfractés/réfléchis par un cylindre de verre

La Figure 30 montre une illustration d'un lancé de 100 rayons sur une cuve cylindrique, on voit que cette dernière concentre les rayons vers l'avant

### 3.4.3 La sphère

Dans le repère du laboratoire, l'équation cartésienne d'une sphère s'écrit :

$$\left(x-x_c\right)^2 + \left(y-y_c\right)^2 + \left(z-z_c\right)^2 - R_c^2 = 0 \tag{153}$$

Nous ne détaillerons pas ici la résolution de l'intersection entre la droite et la sphère puisque c'est exactement la même que pour le cylindre, excepté que :

$$\alpha t^2 + \beta t + \gamma = 0$$

$$\text{avec}\quad \begin{cases} \alpha = \left(\mathbf{d_p}(1)^2 + \mathbf{d_p}(2)^2 + \mathbf{d_p}(3)^2\right) \\ \beta = 2\left(\left(x_0 - x_c\right)\mathbf{d_p}(1) + \left(y_0 - y_c\right)\mathbf{d_p}(2) + \left(z_0 - z_c\right)\mathbf{d_p}(3)\right) \\ \gamma = \left(\left(x_0 - x_c\right)^2 + \left(y_0 - y_c\right)^2 + \left(z_0 - z_c\right)^2 - R_c^2\right) \end{cases} \tag{154}$$

La suite des étapes est donnée par les équations (151) et (152). Le photon interagit avec la sphère à la distance $d_{sphe} = \min\left(d_{sphe1}, d_{sphe2}\right)$ en $x = x_0 + td_{sphe}(1), y = y_0 + td_{sphe}(2), z = z_0 + td_{sphe}(3)$ .

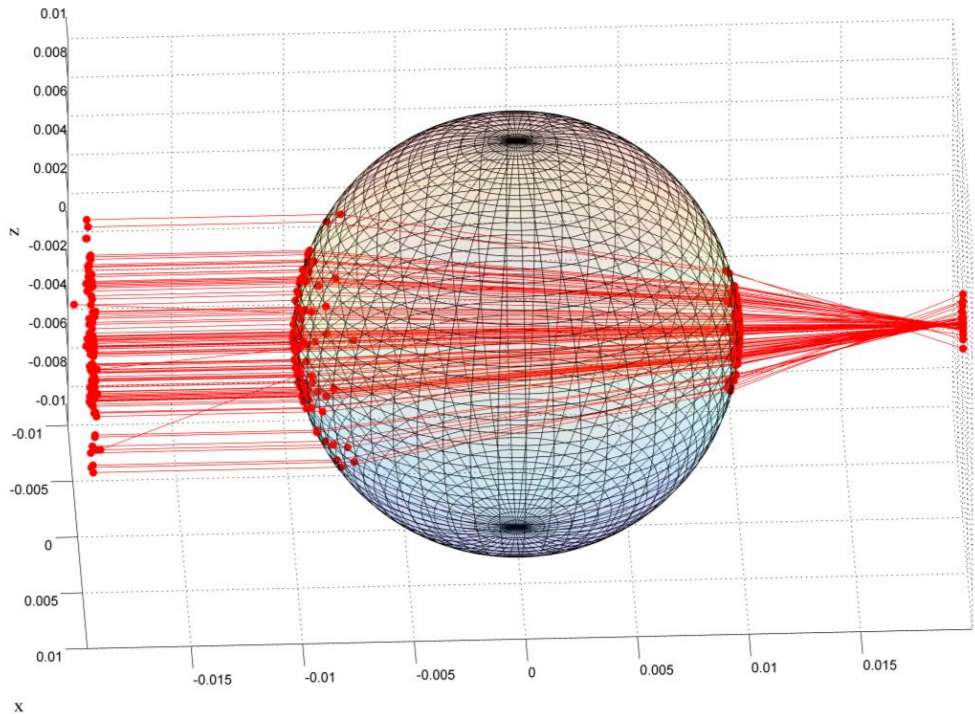

Figure 31 : Illustration de lancé de 100 rayons réfractés/réfléchis par une sphère de verre

Sur la Figure 31, 100 rayons ont été lancés et suivis dans leur propagation à travers une sphère. Comme dans le cas du cylindre, on remarque une zone de focalisation après le passage des rayons dans la sphère. Ce cas permet de valider, en partie, le cas de la lentille plan-convexe sphérique (formée à partir d'une sphère et d'un plan) qui sera présenté dans la suite.

### 3.4.4 Le trou optique (ou sténopée)

Le trou optique est un cas particulier par rapport aux précédents puisque physiquement celui-ci n'a pas d'existence. Il faut donc que ce dernier appartienne à un des objets précédemment définis pour avoir une existence physique. Le traitement choisi n'est probablement pas le plus approprié mais le principe est le suivant : on définit à quelle surface le trou appartient et lorsque le photon interagit avec cette dernière, on calcule s'il appartient également au trou optique. Si c'est bien le cas, le photon continue sa propagation sans changer de vecteur directeur.

### 3.4.5 La lentille plan convexe

La lentille plan-convexe sphérique est modélisée par deux dioptres : un plan et une sphère, ainsi que des bornes (diaphragme). Il faut calculer la distance du photon à la sphère et au plan, puis garder la distance minimale pour identifier le premier dioptre rencontré par le photon.

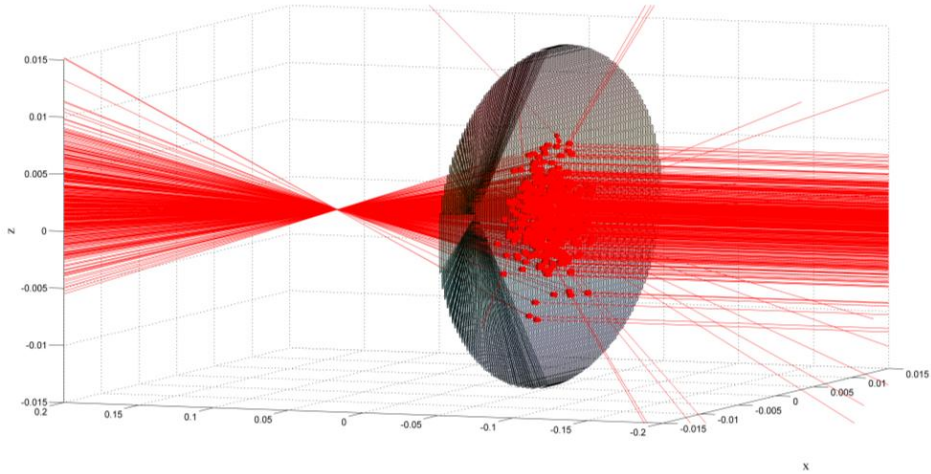

Figure 32 : Tracé de rayons par méthode de Monte-Carlo dans le cas d'une lentille plan-convexe sphérique de focale $f = 100mm$

La Figure 32 illustre le tracé de rayons se propageant de droite à gauche à travers une lentille plan-convexe sphérique. On distingue aisément le point de focalisation à la sortie de la lentille.

### 3.4.6 Direction de propagation et indice de réfraction

Pour l'ensemble $E$ des $k$ objets qui composent le montage optique, on calcule les $k$ distances euclidiennes $d_i$ et on retient la distance objet-photon $d_o$ minimale :

$$d_o = \min_{i=1,2\dots k}\left(d_i\right) \tag{155}$$

Cette étape permet de déterminer *in fine* les nouvelles coordonnées du photon, au point d'impact. Il faut ensuite modéliser le comportement du photon lors de l'interaction. A condition de négliger les effets ondulatoires, l'optique géométrique est encore le meilleur outil pour étudier la propagation et l'interaction de la lumière dans un environnement complexe (avec cellule cylindrique, miroirs, lentilles, supports mécaniques, etc…). Les lois de Snell-Descartes sous forme vectorielle ([**Wozniak 2012**],[**Ren 2011**]), en combinaison avec les coefficients de Fresnel en énergie $\left(r_{\perp,\parallel}{}^2\right)$, sont utilisées pour déterminer l'événement de diffusion après interaction (réflexion, réfraction) ainsi que la nouvelle direction de diffusion. Les expressions des coefficients de Fresnel ainsi que la forme des lois de Snell-Descartes vectorielles sont détaillées dans le **chapitre 2**.

Pour déterminer l'événement de diffusion, un nombre aléatoire $\chi$ est tiré et à chaque interaction du photon avec un dioptre, les coefficients énergétiques de Fresnel $0 \leq r_{\perp,\parallel}^2 \leq 1$ sont calculés, avec :

- si $r_{\perp,\parallel}^2 \leq \chi$ : le photon est réfracté (et il peut être absorbé),

- si $r_{\perp,\parallel}^2 > \chi$ : le photon est réfléchi.

La nouvelle direction de propagation est calculée avec les lois de Snell-Descartes. Cependant, la normale à la surface rencontrée doit être définie dans le bon sens, pour cela on impose que :

$$\mathbf{d_p} \cdot \mathbf{n} > 0 \tag{156}$$

avec $\mathbf{d_p}$ le vecteur directeur du photon et $\mathbf{n}$ la normale de la surface au point d'interaction. Une fois ce critère mis en place, comme dans le chapitre précédent, on calcule les angles de réflexion/réfraction :

$$\cos(i_1) = -\mathbf{d_p}.\mathbf{n} \tag{157}$$

$$\cos(i_2) = \sqrt{\left(1 - \left(\frac{m_1}{m_2}\right)^2 \left(1 - \cos(i_1)^2\right)\right)} \tag{158}$$

avec $m_1$ l'indice de réfraction du milieu 1 et $m_2$ l'indice de réfraction du milieu 2. On peut aussi remarquer que lorsque la radicande de l'équation (87) est négative, on est dans le cas de la réflexion totale.

Grâce à ces formules on peut exprimer le vecteur directeur du photon qui serait réfracté :

$$\mathbf{d_p} = \frac{n_1}{n_2}\mathbf{d_p} + \left(\frac{n_1}{n_2}\cos(i_1) - \cos(i_2)\right)\mathbf{n} \tag{159}$$

et celui du photon qui serait réfléchi :

$$\mathbf{d_p} = \mathbf{d_p} + 2\cos(i_1)\mathbf{n} \tag{160}$$

avec $n_1$ et $n_2$ la partie réelle de l'indice de réfraction $m_1$ et $m_2$ respectivement. Ces lois sont utilisées avec les parties réelles des indices de réfraction pour les raisons évoquées dans le **chapitre 2** sur les cas particuliers des lois de Snell-Descartes.

Il faut également, dans le code, définir correctement les indices de réfraction mis en jeu. En effet, sur une interface, comment savoir vers quel milieu le photon se propage? Pour cela on teste à quel objet appartient le point milieu du vecteur directeur du photon. C'est-à-dire que si on note $x_0, y_0, z_0$ les coordonnées actuelles du photon et $x, y, z$ ses coordonnées au point d'impact de la surface, alors on teste à quel élément appartient $\left\{ \left( x - x_0 \right)^2 + \left( y - y_0 \right)^2 + \left( z - z_0 \right)^2 \right\} / 4$.

### 3.4.7   Phase des rayons

La phase des rayons doit être intégrée au modèle pour prendre en compte les interférences entre les rayons. Et notamment, les interférences induites par certains composants optiques comme, par exemple, la cuve cylindrique contenant le système particulaire.

La prise en compte de la phase a été détaillée dans le modèle de Van de Hulst, lui-même décrit dans le chapitre précédent. Nous avons vu alors que plusieurs types de déphasages doivent être pris en compte. Parmi ceux-ci, on distingue le déphasage associé aux réflexions et réfractions (pris en compte dans les coefficients de Fresnel), celui associé au chemin optique et celui dû aux lignes focales (la courbure des fronts d'onde).

Nous ne prenons pas en compte ici les déphasages liés aux fronts d'onde, car le modèle de Van de Hulst ne s'applique qu'aux sphères. Nous n'aurons donc pas nécessairement toutes les structures d'ondulation mais, a minima, les principales.

Pour prendre en compte le déphasage lors des réflexions et réfractions aux interfaces, il suffit de sommer la phase des coefficients de Fresnel à chaque interaction. C'est-à-dire que :

$$\zeta_{Fresnel} = \sum_{i=1}^{n} phase(\varepsilon_{\perp,\parallel}^{i})$$ (161)

avec $n$, le nombre d'impacts sur les surfaces spéculaires.

Pour prendre en compte le déphasage dû au chemin optique, il suffit de sommer les distances parcourues dans chacun des milieux :

$$\sigma_{\perp,\parallel} = \sum_{i=1}^{m} \frac{2\pi d_i n_i}{\lambda}$$ (162)

avec $d_i$ la distance parcourue dans le milieu i d'indice de réfraction réel $n_i$ et m le nombre de milieux traversés.

Dans notre code, le déphasage total pris en compte vaut donc :

$$P = \sum_{i=1}^{n} phase(\varepsilon_{\perp,\parallel}^{i}) + \sum_{i=1}^{m} \frac{2\pi d_i n_i}{\lambda} \qquad (163)$$

### 3.4.8 Propagation dans un milieu homogène absorbant

Lorsque le rayon se propage dans un milieu homogène absorbant, il peut être considérablement atténué avant sa sortie. Généralement la loi de Beer-Lambert (voir par exemple [**Yuhan 2012**]) est utilisée dans ce cas-là et s'exprime par :

$$T = e^{-4\pi k_1 d_0 / \lambda} \qquad (164)$$

avec T la transmission, $k_1$ la partie imaginaire de l'indice de réfraction du milieu, $d_0$ la distance parcourue dans le milieu et $\lambda$ la longueur d'onde du photon dans le milieu.

Pour intégrer cette contribution dans le modèle de Monte-Carlo, on met en place une procédure stochastique simple [**Kirk 1992**] qui consiste à tirer un nombre aléatoire $\chi$ avec :

$$\begin{aligned} &\text{-si } T \leq \chi : \text{ le photon est absorbé,} \\ &\text{-si } T > \chi : \text{ le photon sort du milieu.} \end{aligned} \qquad (165)$$

On répète les étapes décrites dans les paragraphes précédents jusqu'à l'arrivée du photon dans le milieu particulaire, ou sur une zone photosensible, ou bien encore si ce dernier sort du domaine de calcul.

## 3.5 Description de l'interaction avec le milieu particulaire

Comme nous l'avons déjà évoqué, lorsque les contributions ondulatoires sont prédominantes, comme pour les systèmes nano et micro-particulaires, les modèles physiques et électromagnétiques ( Lorenz-Mie, Debye, T-Matrice, Airy par exemple [**Bohren 1998, Onofri 2012**] restent les meilleures solutions pour décrire avec précision les propriétés d'absorption et de diffusion de la lumière. Dans le milieu particulaire, on doit utiliser un ou plusieurs de ces modèles pour traiter l'interaction du photon incident avec les particules diffusantes. Cependant, une description au préalable du milieu particulaire et de ses principales caractéristiques doit être réalisée afin de prédire le comportement de la lumière dans ce dernier.

### 3.5.1 Description du milieu particulaire

Afin de décrire le milieu particulaire, un certain nombre de paramètres doivent être définis. Parmi ceux-ci, il y a l'indice de réfraction complexe de la phase fluide (milieu continu) $m_i$, la concentration volumique en particules du milieu diphasique $C_v$, la distribution en taille des particules, leur indice de réfraction complexe $m_p$ et *in fine*, leurs propriétés optiques (diagrammes de diffusion, sections efficaces).

Il faut également définir le profil en concentration de l'écoulement. Cependant, comme nous l'avons déjà indiqué, on prend ici comme hypothèse que l'écoulement est homogène (tailles, concentrations et vitesses). Il s'agit d'une hypothèse forte, valable pour les colloïdes mais pas nécessairement pour les écoulements en lit fluidisé par exemple. A terme, cette hypothèse pourrait être levée sans trop de difficultés mais ceci demandera des informations supplémentaires sur le système étudié.

La distribution granulométrique des particules peut être monodisperse, c'est-à-dire que toutes les particules ont la même taille. La distribution, en forme de Dirac, dépend d'un unique paramètre : le diamètre des particules. Le cas polydisperse, plus réaliste, est modélisé par une distribution à deux paramètres. En pratique, nous n'utilisons ici que la distribution log-normale (du fait notamment de sa grande utilisation dans le domaine des sprays et poudres), mais d'autres distributions pourraient également être utilisées : gamma, Rosin Rammler,... Dans ce cas, les deux paramètres sont le diamètre moyen et un écart-type. La distribution en nombre $n(D)$ s'écrit :

$$n(D) = \frac{1}{Ds\sqrt{2\pi}} \exp\left[ -\frac{(\ln(D) - \mu)^2}{2s^2} \right]$$  (166)

où $\mu$ et $s$ représentent respectivement la moyenne et l'écart-type du logarithme du diamètre $D$. Ces grandeurs sont reliées au diamètre moyen $\overline{D}$ et à l'écart-type $\sigma_D$ par :

$$s = \sqrt{\ln\left[ \left( \frac{\sigma_D}{\overline{D}} \right)^2 + 1 \right]}$$

$$\mu = \ln(\overline{D}) - \frac{s^2}{2}$$  (167)

Le type de distribution retenu est la distribution en nombre $n(D)$ mais on aurait pu utiliser une distribution surfacique $s(D)$ ou volumique $v(D)$ [**Onofri 2012**]. Dans tous les cas ces distributions doivent respecter :

$$\int_0^\infty n(D)dD = \int_0^\infty s(D)dD = \int_0^\infty v(D)dD = 1$$  (168)

La figure suivante montre l'allure de différentes distributions dont le diamètre moyen est de 200µm.

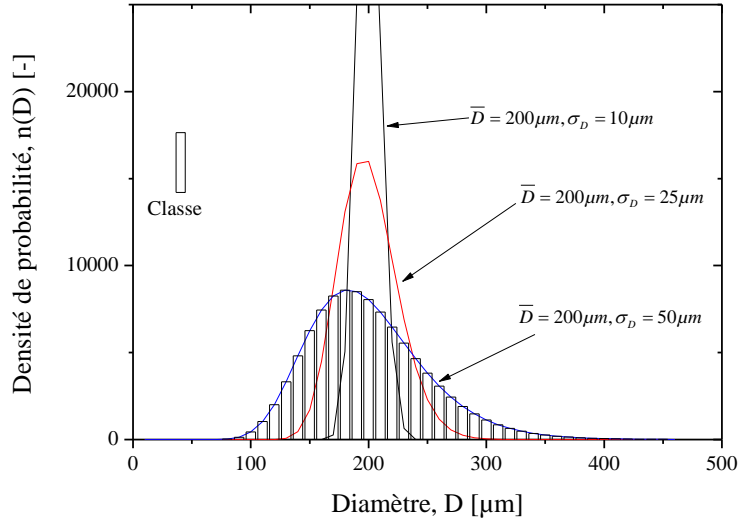

Figure 33 : Exemple de lois log-normales pour différents diamètres moyens et écarts-type

Ce type de distribution étant infini, il est nécessaire de les tronquer. Nous avons fixé le diamètre minimum à $D = 0$. Pour le diamètre maximum, nous recherchons celui correspondant à une densité de probabilité égale à 0.001 fois le maximum de la distribution (représente 99.9% de la population). Ce dernier est déterminé pour le mode $M_{\log Norm}$ :

$$M_{\log Norm} = e^{\mu - s^2},$$ (169)

ainsi, on peut en déduire la valeur maximale de la distribution en ce point:

$$n(M_{\log Norm}) = \frac{e^{\frac{s^2}{2} - \mu}}{s\sqrt{2\pi}}$$ (170)

Ce qui donne pour la définition des bornes de la distribution granulométrique:

$$\forall D \in \mathbb{R}^+, \begin{cases} n(D) = \dfrac{1}{Ds\sqrt{2\pi}} \exp\left[ -\dfrac{(\ln(D) - \mu)^2}{2s^2} \right] \ si \ n(D) \geq \dfrac{e^{-\frac{s^2}{2} - \mu}}{1000 s\sqrt{2\pi}} \\ n(D) = 0 \ sinon \end{cases}$$ (171)

Cette distribution est par la suite discrétisée en classes de taille dont la largeur peut être fixée à l'initialisation du programme.

La capacité d'une particule à diffuser ou à absorber est introduite via la section efficace d'extinction $C_{ext}$ qui est la somme de la section efficace de diffusion $C_s$ et de la section efficace d'absorption $C_a$. Comme présenté dans le **chapitre 2**, ces grandeurs sont principalement calculées à l'aide de la théorie de Lorenz-Mie.

Dans le cas d'une distribution polydisperse, on introduit une section efficace moyenne d'extinction qui s'exprime par :

$$\overline{C}_{ext} = \int_0^\infty C_n C_{ext}(D) n(D) dD, \tag{172}$$

où la concentration en nombre $C_n$ (en particules par m$^3$) est reliée à la concentration en volume $C_v$ par :

$$C_v v(D) = \frac{\pi D^3}{6} C_n n(D), \tag{173}$$

soit en intégrant de part et d'autre de l'équation (173) :

$$C_n = \frac{6}{\pi} \frac{C_v}{\int_0^\infty D^3 n(D) dD}. \tag{174}$$

### 3.5.2   Description de l'interaction rayon/particule

Grâce à ces grandeurs, on peut déterminer si, une fois dans le milieu particulaire, le rayon va rencontrer une particule. Pour cela, on utilise la notion de libre parcours moyen qui s'exprime par :

$$\Lambda = \frac{1}{C_n \overline{C}_{ext}} \tag{175}$$

On tire alors un autre nombre aléatoire $\chi \sim U\big(]0,1]\big)$. Si statistiquement, après avoir parcouru la distance $d_0 = -\Lambda \log(\chi)$, le rayon est toujours dans la suspension, alors il va interagir avec une particule en :

$$\begin{cases} x = x_0 - \Lambda \log(\chi) \mathbf{d_p}(1) \\ y = y_0 - \Lambda \log(\chi) \mathbf{d_p}(2) \\ z = z_0 - \Lambda \log(\chi) \mathbf{d_p}(3) \end{cases} \tag{176}$$

Il faut alors déterminer l'évènement de diffusion associé à cette interaction (diffusion ou absorption), voir l'algorithme de la Figure 26. Pour cela, on introduit, de manière classique, l'albédo moyen du système particulaire. Cet albédo est le rapport de la section efficace de diffusion moyenne sur la section efficace d'extinction moyenne. Cette grandeur est inférieure ou égale à 1 :

$$a = \frac{\overline{C_s}}{\overline{C_{ext}}}, \quad (177)$$

avec $\overline{C_s}$ la section efficace de diffusion moyenne (calculée également par la théorie de Lorenz-Mie). On tire alors un nouveau nombre aléatoire $\chi$ et :

$$si \ a < \chi \ \text{alors le photon est absorbé}$$
$$si \ a \geq \chi \ \text{alors le photon est diffusé} \quad (178)$$

S'il y a absorption, le photon ne contribue pas directement au diagramme de diffusion, voir l'algorithme de la Figure 26. Dans le cas contraire, il faut calculer une nouvelle direction de propagation pour le photon.

En coordonnées sphériques, cette nouvelle direction de diffusion est caractérisée par les angles de diffusion $\theta$ et $\varphi$. Lorsque la polarisation est négligée (autre hypothèse importante), l'angle azimutal est distribué selon une loi uniforme :

$$\varphi = 2\pi\chi \quad (179)$$

Pour obtenir l'angle de diffusion polaire $\theta$ après interaction avec une particule, il faut calculer son diagramme de diffusion et le transformer en densité de probabilités cumulées selon l'angle polaire $\theta \in [0, \pi]$. Pour ce faire, on peut utiliser une approche simplifiée reposant sur la fonction de phase du type de celle de Henyey-Greenstein [**Henyey 1941**] par exemple, ou bien utiliser une théorie rigoureuse comme la théorie de Lorenz-Mie. Avec le développement des capacités informatiques, le choix de la théorie de Lorenz-Mie n'est plus un problème.

La partie du programme calculant les diagrammes de diffusion et sections efficaces est directement inspirée des codes de calcul développés par Barber et Hill [**Barber 1990**]. Le programme a été transposé en Fortran 90, réorganisé en sous-fonctions et les calculs sont réalisés en double précision pour être appelés par le programme principal.

En suivant le même raisonnement que pour calculer la section efficace d'extinction moyenne, on obtient l'intensité moyenne :

$$\overline{I}(\overline{D}, \sigma_D, \theta, \varphi) = \int_0^\infty C_n I(D, \theta, \varphi) n(D) dD \quad (180)$$

On transforme ensuite cette intensité moyenne en une fonction de densité cumulée pour en déduire la probabilité du photon d'être diffusé dans la direction $\theta_0 \in [0, \pi]$ :

$$P(\theta_0) = \frac{\int_0^{\theta_0} \overline{I}(\theta) d\theta}{\int_0^{\pi} \overline{I}(\theta) d\theta} \tag{181}$$

On tire de nouveau un nombre aléatoire, pour en déduire l'angle polaire de diffusion du photon :

$$\theta = \{\theta_0 \mid P(\theta_0) = \chi\} \tag{182}$$

Le principe de cette méthode est schématisé dans la Figure 34.

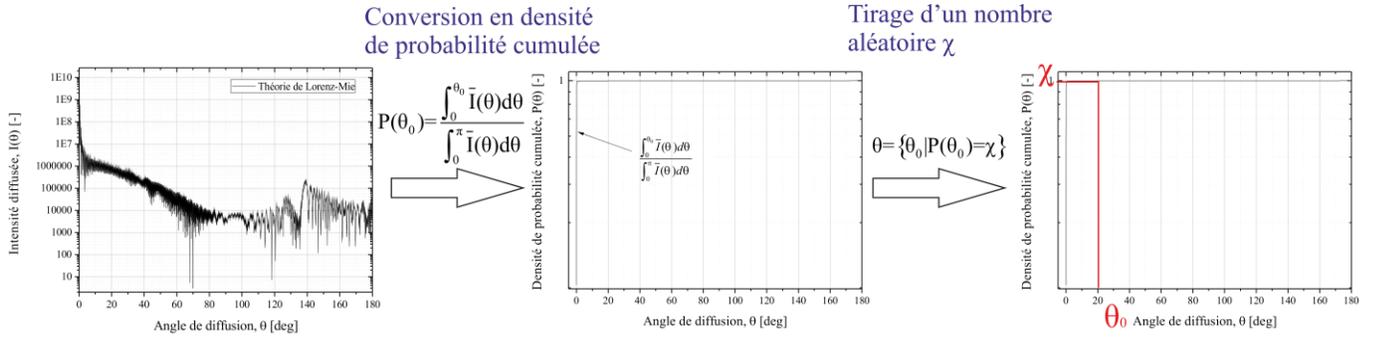

Figure 34 : Principe de calcul probabiliste de l'angle de diffusion polaire

Selon la dimension du problème, le diagramme de diffusion selon l'angle $\theta$ doit être transformé. En effet, en 2D, l'élément de surface élémentaire est $I(\theta)d\theta$ alors qu'en 3D l'élément de surface élémentaire est $I(\theta)\sin(\theta)d\theta d\varphi$.

Les diagrammes de diffusion ont été calculés avec la théorie de Lorenz-Mie pour un certain nombre d'angles $\theta$, pour plus de précision on passe par une interpolation linéaire dans l'intervalle $\Delta\theta = \theta_2 - \theta_1$ :

$$\theta = \min\left(P(\theta_1) - \chi\right)\Delta\theta + \frac{\chi - P(\theta_1)}{P(\theta_2) - P(\theta_1)}\Delta\theta \tag{183}$$

On doit alors calculer la nouvelle direction de diffusion dans le repère du laboratoire.

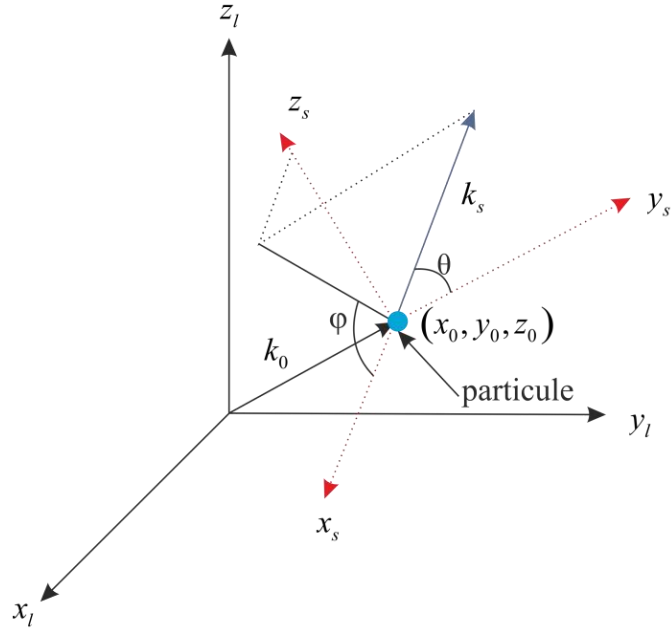

Figure 35 : Position du photon $(x_0, y_0, z_0)$ dans le système de coordonnées du laboratoire $(x_l, y_l, z_l)$, direction de diffusion $\mathbf{k_s}$ et système de coordonnées associé $(x_s, y_s, z_s)$

Sur la Figure 35, le repère $(x_l, y_l, z_l)$ représente le repère du laboratoire et le repère $(x_s, y_s, z_s)$ représente le repère local de diffusion au point d'interaction.

On note par $\mathbf{k_0}$ la direction de diffusion avant interaction avec la particule et $\mathbf{k_s}$ la direction de diffusion après l'interaction. On note $\mathbf{k_s}$ la base locale de diffusion [**Mul 2011**] avec :

$$\mathbf{k_s} = \cos(\theta)\mathbf{y_s} + \sin(\theta)\big(\cos(\varphi)\mathbf{x_s} + \sin(\varphi)\mathbf{z_s}\big) \tag{184}$$

Pour se ramener au repère du laboratoire, on utilise la transformation[**Mul 2004**] :

$$\mathbf{y_s} = \frac{\mathbf{k_0}}{\|\mathbf{k_0}\|}; \mathbf{x_s} = \frac{\mathbf{y_s} \times \mathbf{z_l}}{\|\mathbf{y_s} \times \mathbf{z_l}\|}; \mathbf{z_s} = \frac{\mathbf{x_s} \times \mathbf{y_s}}{\|\mathbf{x_s} \times \mathbf{y_s}\|} \tag{185}$$

On reproduit toutes les étapes antérieures jusqu'à l'arrivée sur un détecteur.

### 3.5.3 Description de la détection

Si le détecteur est considéré comme parfait (c.-à-d. que la partie complexe de son indice de réfraction est infinie), on somme simplement les photons arrivant sur les zones photosensibles (nombres complexes, ou entiers, suivant que l'on prend en compte, ou non, leurs phases). Si ce n'est pas le cas (efficacité quantique inférieure à l'unité), comme précédemment, les coefficients de Fresnel sont calculés pour savoir si le photon est effectivement détecté (absorbé) ou bien réfléchi. Dans le second cas, on calcule de nouveau les lois de Snell-Descartes vectorielles pour déterminer la nouvelle direction de propagation du photon qui repart vers le milieu particulaire.

L'intérêt d'utiliser la méthode de Monte-Carlo pour modéliser la réponse des photodétecteurs organiques réside dans le fait que, selon la partie du diagramme de diffusion que l'on souhaite caractériser (diffraction vers l'avant, arc-en-ciel,...), on peut aisément tester différentes configurations expérimentales. Cette optimisation consiste à prédire grâce au code de Monte-Carlo, la forme et le nombre de photodétecteurs nécessaires pour obtenir une information pertinente sur le système particulaire à partir du signal optique.

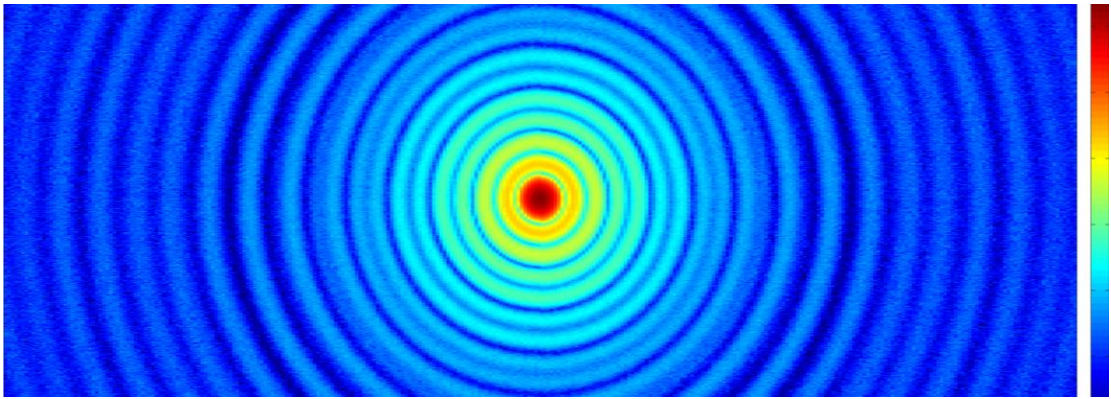

Figure 36 : Figure de diffraction calculée par méthode de Monte-Carlo pour un jet de billes de 100µm dans l'air observée derrière une lentille plan-convexe en configuration de Fourier

Pour cela, nous avons mis en place deux types de sorties au programme de Monte-Carlo développé. La première sortie consiste à produire un diagramme de diffusion en intégrant les photons sur des surfaces photosensibles préalablement définies (formes, distances, indices). La seconde sortie permet de générer une "carte de diffusion" globale, c'est-à-dire qu'on récupère la distribution des photons distribués sur une surface cylindrique (que l'on déplie pour les représentations graphiques). Ceci permet d'optimiser "rapidement" la forme et le nombre de photodétecteurs (sortie n°2), puis de raffiner le calcul dans un deuxième temps (sortie n°1). La Figure 36 montre une carte, générée à partir de la sortie 2, de la distribution des photons dans la zone de diffraction d'un jet de billes de verre de diamètre 100µm observée derrière une lentille plan-convexe (configuration de Fourier). A travers cet exemple, il est clair que, pour analyser cette portion du diagramme de diffusion, il faut produire des photodétecteurs organiques dont la forme est celle d'anneaux (ou de portions d'anneaux) circulaires.

## 3.6   Implémentation numérique

Le code est écrit en FORTRAN 90 et parallélisé (en termes de CPU) en utilisant la librairie libre MPI (Message Passing Interface). La parallélisation dans le cas du lancé de rayons est trivial puisque chaque rayon est indépendant par rapport aux autres, ainsi aucune communication entre processeurs n'est à implémenter. Ce cas se rapproche très fortement du cas asymptotique dans le sens où le temps de calcul est divisé par le nombre de processeurs (en réalité ce cas n'est jamais atteignable).

Le code est développé sur Linux (distribution Ubuntu) en utilisant le compilateur classique gfortran sans utilisation de librairies Fortran particulières, ce qui en fait un code portable sur d'autres OS. L'algorithme général a déjà été présenté sur la Figure 26.

Pour aider l'utilisateur, un "MakeFile" est implémenté afin de compiler proprement les différents fichiers du programme (création des fichiers objets, des liens entre les fichiers, de l'exécutable...).

Le temps d'exécution du programme varie énormément selon les cas étudiés (diffraction, diffusion simple, diffusion multiple...). Par exemple, pour une description du diagramme de diffusion en champ lointain en régime de diffusion simple, quelques 200 millions de rayons simulés suffisent à obtenir un résultat satisfaisant en 2 heures. Par contre le même cas en régime de diffusion multiple pour le même nombre de rayons peut prendre, selon la concentration, jusqu'à 100 fois plus de temps! C'est pour ce type de résultat que la parallélisation du code joue un rôle primordial. L'utilisateur doit donc correctement cerner ses besoins avant de simuler certains cas, par exemple si une autre alternative est utilisable (Théorie de Lorenz-Mie par exemple).

Une des difficultés majeures pour la programmation du code de Monte-Carlo est la création d'un langage pour que l'utilisateur puisse rentrer les données d'entrée au programme. En effet, comment distinguer les types d'éléments que veut ajouter l'utilisateur (sphère, plan...), les indices de réfraction associés à ces éléments, la polarisation du faisceau incident... De nouveau, l'utilisateur est confronté à un "pré-calcul" afin de remplir le fichier d'entrée au programme.

Pour finir, les différents post-traitements possibles sont traités par des programmes Matlab externes qui permettent de créer les cartes d'intensités ainsi que les diagrammes de diffusion.

## 3.7   Validation du code pour des cas du PAT

Le code développé doit être validé sur des cas de référence, mais ceci sont peu nombreux. La théorie de Lorenz-Mie constitue clairement un de ces cas. Cependant, dans le contexte du PAT, les simulations effectuées doivent reproduire des conditions de mesures bien plus complexes. Dans le temps imparti pour ce travail, nous avons essayé de procéder par étapes.

Le système est composé d'une cuve cylindrique (de rayon $R_r$, d'épaisseur $e_r$ et d'indice $m_r$) contenant des particules sphériques (de diamètre moyen $D_p$, d'écart-type $\sigma_p$, d'indice $m_p$, de concentration volumique $N_p$) dans un écoulement considéré comme uniforme.

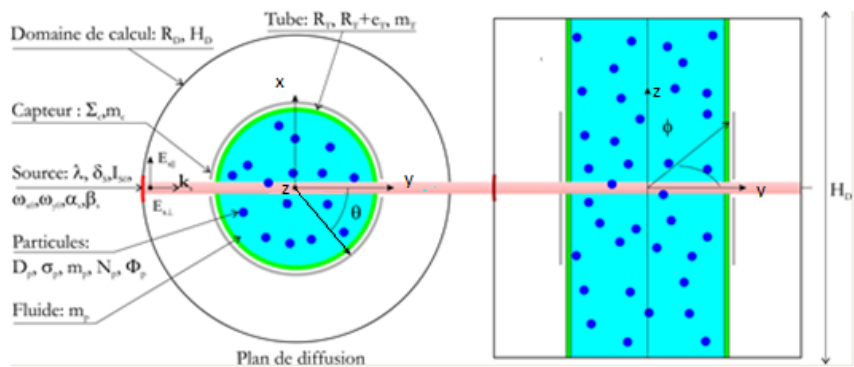

Figure 37 : Schéma de principe du système-type étudié numériquement

L'indice de réfraction du fluide est $m_f$. Une couronne de photodétecteurs organiques conformables (d'ouverture $\Sigma_c$, d'indice de réfraction $m_c$) est placée autour de cette cuve. Un faisceau laser collimaté éclaire la cuve parallèlement à un de ses diamètres. La couronne de photodétecteurs est percée de trous optiques pour laisser passer le faisceau en entrée et en sortie.

Les résultats numériques présentés dans ce qui suit comparent les prédictions de notre code Monte-Carlo avec ceux de la théorie de Lorenz-Mie, à grande distance des particules (on parle de *remote sensing approximation,* le signal optique est dans ce cas peu sensible à la position relative des particules dans la cuve), puis au "voisinage/proche" des particules (configuration réelle quand la détection n'est pas au foyer de lentille).

### 3.7.1   Cas idéal : approximation de détecteurs lointains

On présente ici les résultats obtenus pour le cas où les détecteurs sont placés à grande distance de l'échantillon. Les simulations ont été réalisées pour un nuage dilué de gouttes d'eau, de diamètre $D = 200\mu m$, placé dans l'air. De manière classique, les intensités ont été normalisées par le pic de diffraction en $I(\theta = 0)$ et un offset d'une décade à été introduit pour plus de clarté, voir la Figure 38.

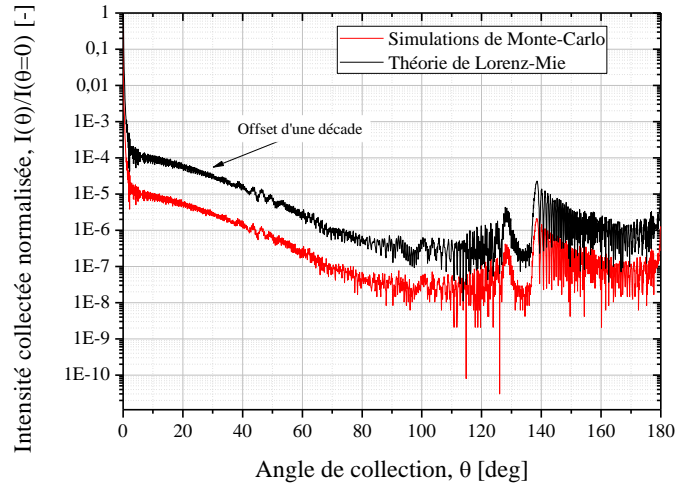

Figure 38 : Comparaison des diagrammes de diffusion obtenus par méthode de Monte-Carlo et par la théorie de Lorenz-Mie dans le cas d'un nuage de gouttes d'eau de diamètre $D = 200\mu m$.

On observe un accord quasi-parfait entre les deux approches. On remarquera la bonne description des arcs-en-ciel primaire et secondaire (autour des 140°), celle du pic de diffraction (petits angles), les oscillations basses et hautes fréquences qui sont caractéristiques de la taille des particules. Le code semble donc fonctionner de manière satisfaisante pour ce cas de référence.

On simule le même cas que précédemment mais avec une distribution polydisperse de diamètre moyen $\overline{D} = 200\mu m$ et d'écart-type relatif $\sigma_D / \overline{D} = 50\%$. La Figure 39 montre que l'écart-type tend à gommer les oscillations des hautes fréquences du diagramme. Cela peut se comprendre de manière " intuitive " dans la mesure où, nous avons superposé ici des diagrammes de diffusion dont la fréquence d'oscillation est différente pour chaque classe de taille. Il s'en suit un filtrage passe-bas qui appauvrie le signal optique.

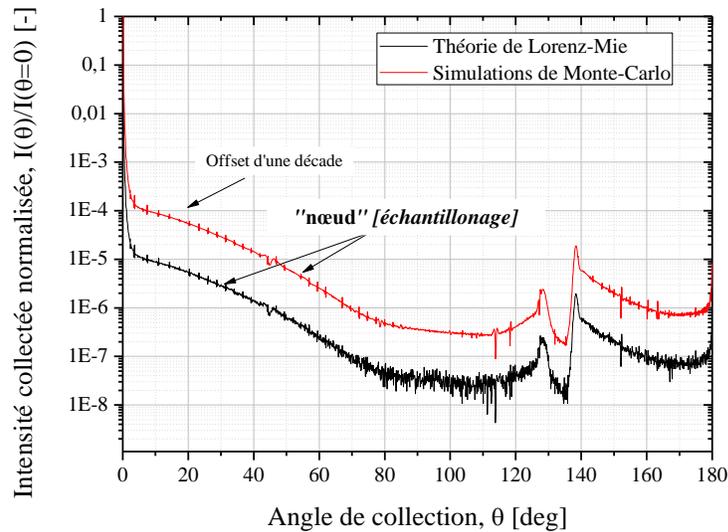

Figure 39 : Diagramme de diffusion d'une polydispersion, optiquement diluée, de gouttes d'eau dans de l'air (diamètre moyen $\overline{D} = 200\mu m$ et d'écart-type relatif $\sigma_D / \overline{D} = 50\%$ )

On peut néanmoins toujours identifier les arcs-en-ciel primaires et secondaires même si la structure d'ondulation a disparu. Les petites oscillations résiduelles observées avec la méthode de Monte-Carlo sont attribuées à la largeur excessive des classes utilisées pour discrétiser la distribution des diamètres (taille des classes 0.1µm). La sous-estimation des fluctuations au voisinage de 90°, ou dans la zone sombre d'Alexander, est dû à la faiblesse du signal dans ces zones et donc, *in fine*, au trop faible nombre de photons tirés (200 millions) dans le cas présent.

Le principal avantage de la méthode de Monte-Carlo est bien évidemment que cette dernière permet de prendre en compte certains effets négligés par des théories comme celle de Lorenz-Mie, et notamment, les effets dus à la cuve. La Figure 40 montre qu'en présence d'une cuve cylindrique (d'indice $m_r = 1.51$, de rayon $R_r = 5cm$ et d'épaisseur 1mm, entourant le système monodisperse de la Figure 38, les structures cohérentes du diagramme de diffusion (diffraction, arcs-en-ciel,...) sont brouillées. Le second arc-en-ciel a par exemple totalement disparu et le premier arc-en-ciel primaire est considérablement "atténué".

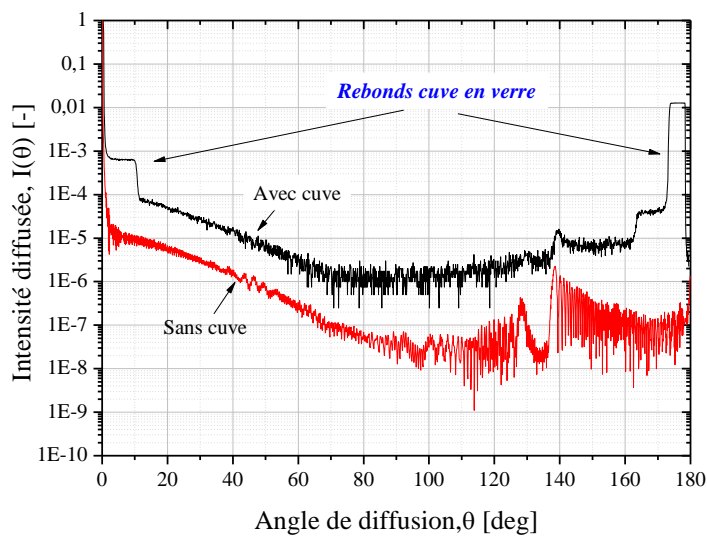

Figure 40 : Comparaison des diagrammes de diffusion avec et sans cuve de verre (cuve avec système particulaire de la Figure 37)

On observe également une série de plateaux d'intensité, vers l'avant (autour de 0°) et vers l'arrière (autour de 180°). Ces derniers perturbent fortement la zone de diffraction dont les oscillations sont perdues. L'apparition de ces plateaux s'explique par les réflexions multiples des photons, d'avant en arrière, sur les parois de la cuve. L'intensité du premier plateau vers l'arrière est de l'ordre de 5% de celle du premier plateau vers l'avant. L'intensité globale du diagramme de diffusion obtenu en présence de la cuve est plus importante que sans cuve. Ceci s'explique par une diffusion accrue du système particulaire liée aux multiples réflexions internes des photons.

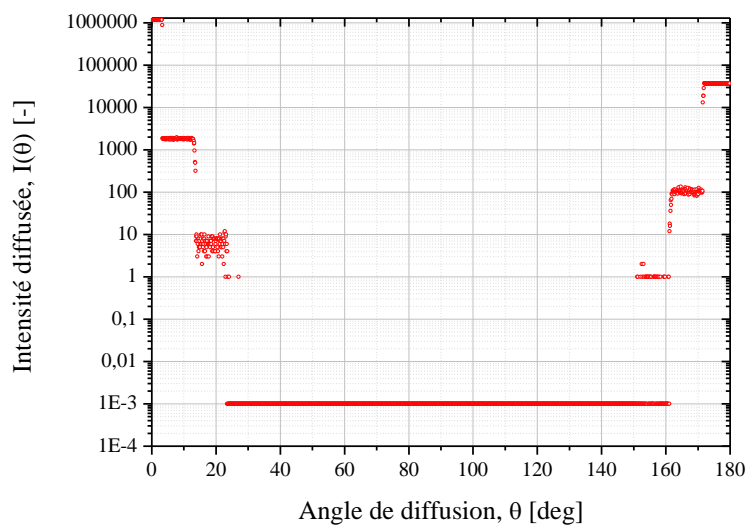

Figure 41 : Intensité diffusée par une cuve de verre d'indice $m_r = 1.51$ et de rayon $R_r = 10cm$ renfermant uniquement de l'air

La Figure 41 permet d'estimer la contribution de la cuve aux diagrammes de diffusion. On constate également que la largeur angulaire des différents plateaux augmente continûment avec le nombre de réflexions. La Figure 42 montre que la largeur angulaire du premier plateau (au voisinage de $\theta = 0$) augmente avec l'inverse du rayon de la cuve et la largeur du faisceau laser incident.

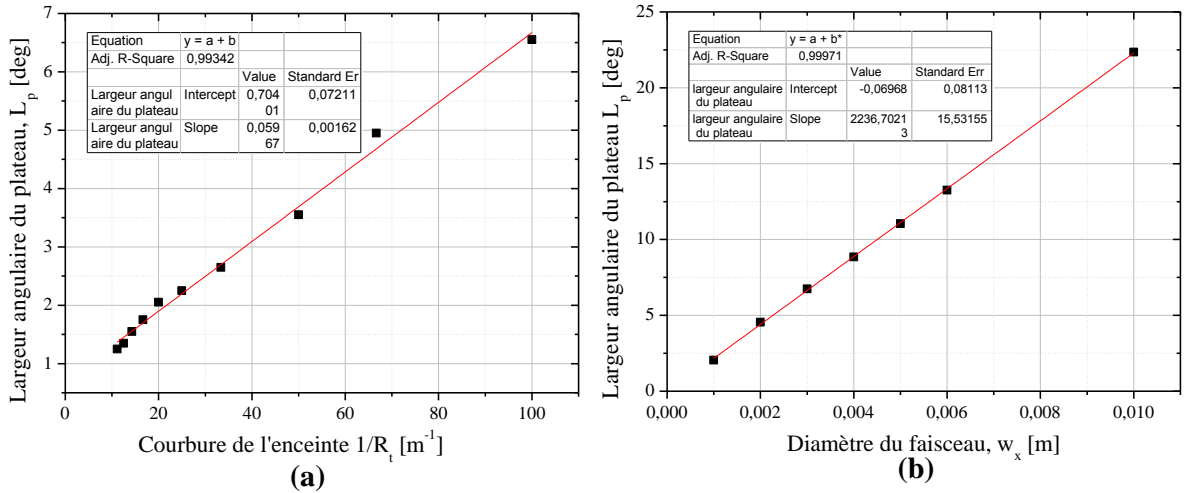

**(a)** **(b)**

Figure 42 : Étude paramétrique de l'influence sur la taille des plateaux angulaires de (a) la courbure de la cuve cylindrique et (b) du diamètre du faisceau

Sur la Figure 42 (a), la relation entre la courbure de la cuve et la largeur du plateau primaire semble linéaire. On observe qu'une cuve de petite taille (cas du rayon de courbure élevé) entraine des plateaux de taille plus conséquente, perturbant d'avantage le diagramme de diffusion final. En effet, plus le diamètre de la cuve est grand, plus on se rapproche d'une surface localement plane. Pour ces simulations, un faisceau gaussien circulaire de diamètre $w = 0.2mm$ a été utilisé. Sur la Figure 42 (b), la relation entre le diamètre du faisceau gaussien incident et la largeur angulaire du premier plateau est de nouveau linéaire. Pour conclure sur ce point, disons que pour accéder aux propriétés de diffusion des particules dans une cuve cylindrique, il vaut mieux minimiser la taille du faisceau et minimiser la courbure de la surface de la cuve. D'autres paramètres secondaires interviennent, comme les indices de la cuve et du milieu fluide, l'épaisseur de la paroi de la cuve, etc…

La validation en 3D de notre code de calcul est plus ardue. La Figure 43 présente le diagramme de diffusion 3D d'un nuage monodisperse de gouttes d'eau de diamètre $D = 2\mu m$ dans l'air.

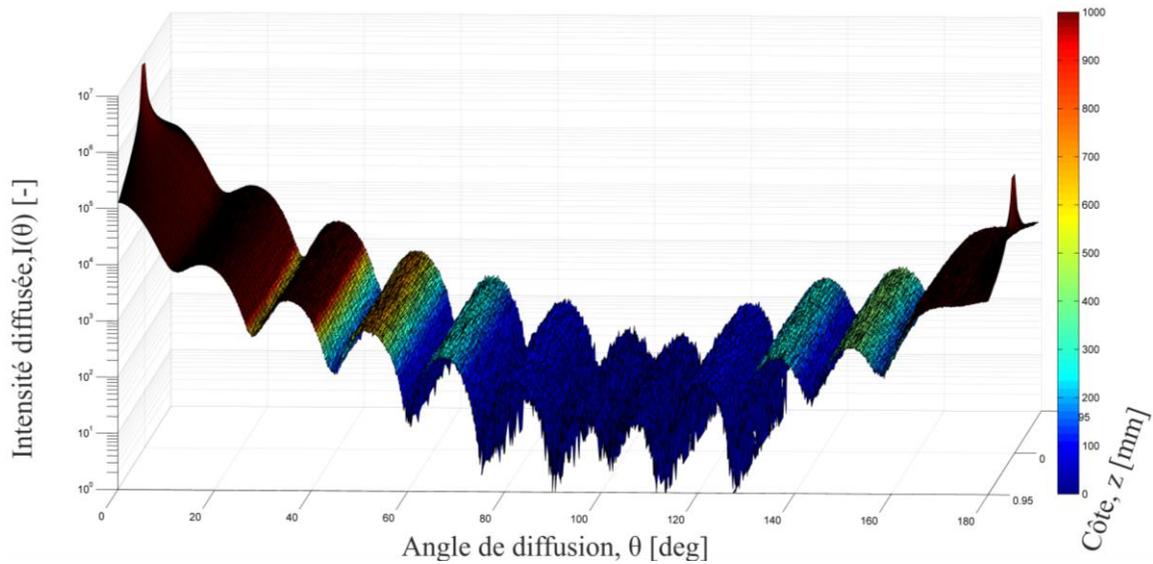

Figure 43 : Diagramme de diffusion 3D d'un nuage dilué de gouttes d'eau monodisperses de diamètre $D = 2\mu m$ (placées dans l'air)

On reconnait aisément l'allure du diagramme de diffusion d'une goutte d'eau de $D = 2\mu m$ dans l'air quelque soit la côte z. On remarque une décroissance de l'intensité selon $z$ lorsqu'on s'éloigne du plan de diffusion ($z = 0$) en un profil gaussien. En régime de diffusion simple, cette décroissance est directement liée à la décroissance de l'intensité du faisceau laser selon $z$ (profil gaussien circulaire avec $w_x = w_z = 0.2mm$).

Pour finir on peut observer le diagramme de diffusion en 3D d'un nuage de gouttes monodisperses de diamètre $D = 20\mu m$ dans l'air, contenu dans une cuve en verre dont le diamètre externe est de 20mm. Comme pour le cas 2D, on observe des plateaux successifs dus aux réflexions des photons sur les parois en verre. Le profil gaussien selon z est particulièrement clair avec cette représentation.

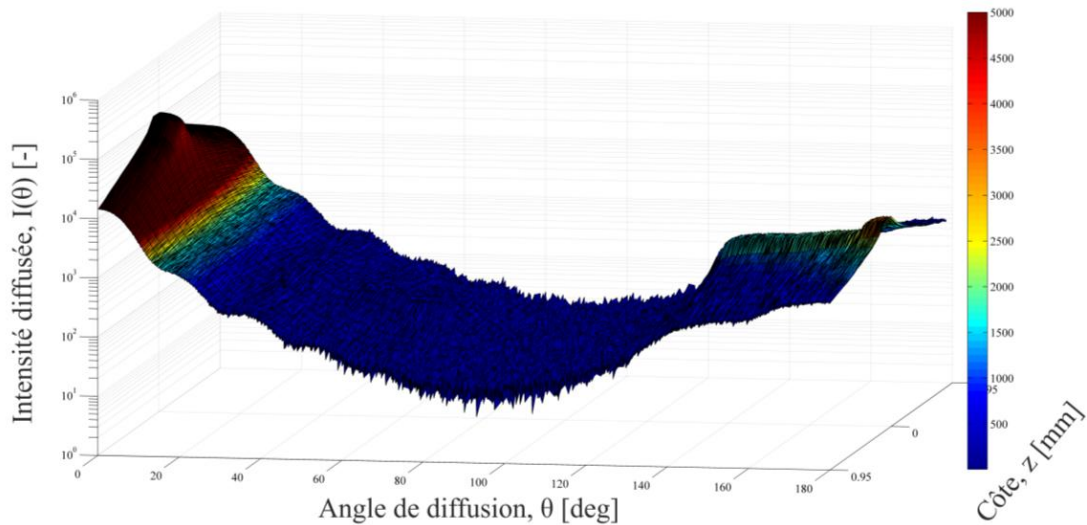

Figure 44 : Diagramme de diffusion 3D d'un nuage de gouttes d'eau monodisperses de diamètre $D = 20\mu m$ contenue dans une cuve en verre de diamètre externe égal à 20mm

Sans recours à des optiques, l'approximation de détecteur lointain n'est généralement pas valable dans des conditions de laboratoire. Cependant, dans le cadre du PAT, l'utilisation d'optiques de collection (lentilles, miroirs) - rigides - n'aurait pas beaucoup de sens. En effet, un des principaux avantages des photodiodes organiques réside dans leur conformabilité. L'utilisation d'optiques rigides reviendrait donc à renoncer à cet avantage. En plus de cela, le prix du montage serait largement supérieur au prix des photodiodes et l'alignement deviendrait complexe.

### 3.7.2 Cas réel : détecteurs au voisinage de l'échantillon

En milieu industriel confiné, la feuille de photodétecteurs est nécessairement à proche distance du milieu particulaire, entourant celui-ci, derrière une canalisation ou une cuve transparente (cf. Figure 37). Il est nécessaire dans cette configuration d'étudier la réponse optique du système afin de voir si une inversion est possible.

Sur la Figure 45, on peut observer le diagramme de diffusion obtenu à faible distance de détection ($R_d = 25mm$) dans le cas d'un nuage de gouttes d'eau monodisperses placé dans l'air et de diamètre moyen $\overline{D} = 200\mu m$ - sans prise en compte des effets de cuve. Ce cas est comparé à la réponse obtenue quand la détection est située à très grande distance des gouttes.

A courte distance de détection, on constate une disparition des structures cohérentes. Ceci s'explique par le fait qu'à un angle de collection correspond différents angles de diffusion (le volume de mesure est de taille non négligeable par rapport à la distance particule/détection).

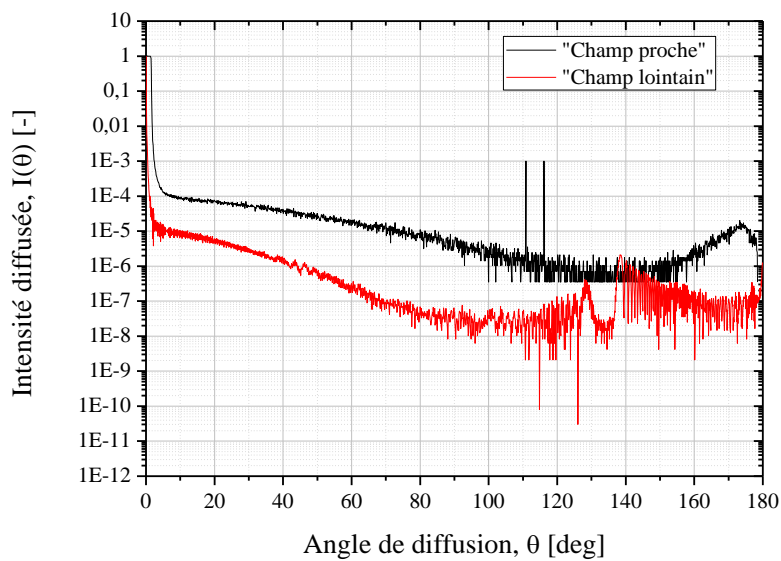

Figure 45 : Diagrammes de diffusion d'un nuage de gouttes d'eau monodisperses de diamètre moyen $\overline{D} = 200\mu m$ dans de l'air : détection à courte et grande distances

Si on rajoute les effets de la cuve, voir la Figure 46, toutes les informations granulométriques classiques semblent perdues (diffraction, arc-en-ciel...). Il semble donc quasiment impossible d'inverser un tel diagramme de diffusion.

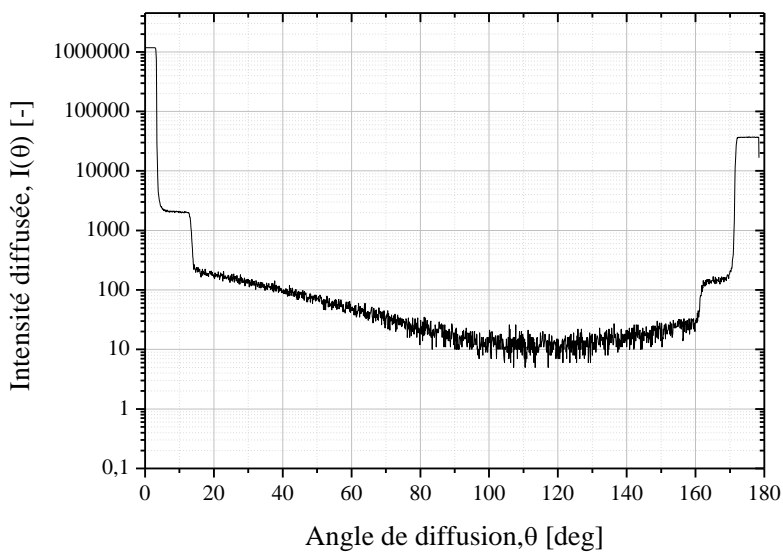

Figure 46 : Diagramme de diffusion d'un nuage de gouttes d'eau monodisperses de diamètre moyen $\overline{D} = 200\mu m$ dans de l'air - détection à courte distance et effets de cuve

Après ces quelques exemples, l'exploitation de mesures à courte distance semble totalement impossible dans une perspective PAT.

# Chapitre 4

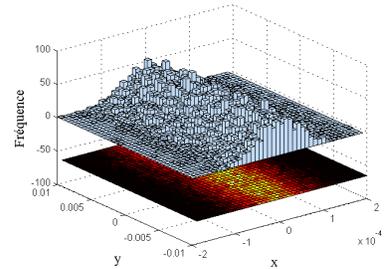

## Optimisation numérique de configuration en milieux dilués

Classiquement, on distingue trois régimes de diffusion à mesure que, de manière simplifiée, la concentration du milieu particulaire augmente :

- la diffusion simple qui, pour chaque photon incident, met en jeu au plus une interaction entre le photon et le milieu particulaire [**Wang 2002**], voir la Figure 47. Ce régime de diffusion est le cas de référence de la quasi-totalité des granulomètres optiques et de nombreux travaux comme ceux évoqués dans le **chapitre 2**

- la diffusion multiple pour laquelle une quantité non négligeable de photons issus de la source interagissent successivement avec plusieurs particules du milieu [**Hulst 1980**] avant leur détection. Dans ce régime, chaque particule diffuse de manière indépendante des autres. On peut donc traiter chaque interaction photon-particule comme une diffusion simple.

- la diffusion dépendante [**Maret 1987**] devient significative lorsque la distance inter particulaire est si faible que les particules diffusent de manière collective, comme des agrégats.

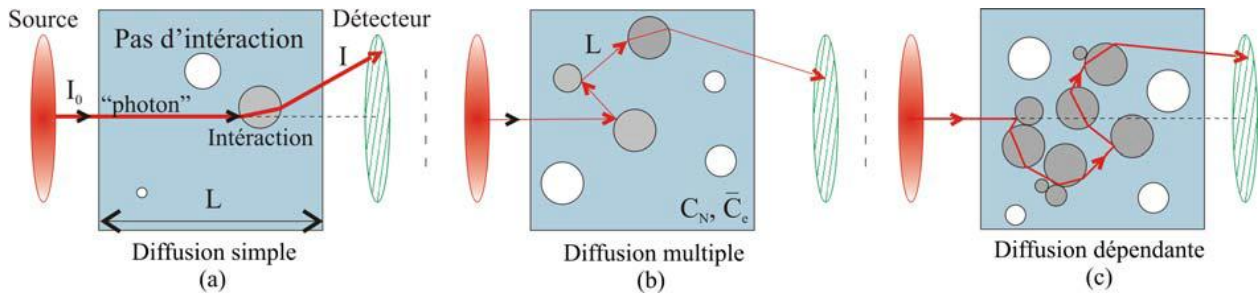

Figure 47 : Illustration des différents régimes de diffusion de la lumière par un milieu particulaire de densité optique croissante : (a) diffusion simple, (b) diffusion multiple et (c) diffusion dépendante [**Onofri 2012**]

Il est très difficile de fournir un critère fiable et universel pour déterminer le régime de diffusion. En effet, ce régime dépend de nombreux paramètres du milieu à caractériser (concentration, volume, taille des particules...), mais également du dispositif de détection (position et ouverture angulaire, volume de mesure observé,...) ou encore du dispositif d'émission (longueur d'onde, polarisation,...) [**Onofri 2012**]. Un critère, parfois simpliste, pour identifier les conditions de diffusion simple, repose sur la transmission minimale pour laquelle le milieu peut être considéré comme dilué. Nous avons choisi un critère "raisonnable" quant à la transmission du faisceau dans le milieu :

$$T = \exp(-C_n \overline{C}_{ext} L) \geq 95\% \tag{186}$$

avec $C_n$ pour la concentration en nombre de particules du milieu, $\overline{C}_{ext}$ le coefficient d'extinction correspondant et $L$ la largeur caractéristique du milieu. Certains auteurs considèrent qu'une transmission supérieure à 60% assure un régime de diffusion simple [**Jones 1999**] dans la mesure où, s'il y a de la diffusion multiple elle reste négligeable devant la diffusion simple. Mais, nous nous répétons, ce critère dépend énormément de la configuration optique du système de mesure.

Le code de Monte-Carlo présenté dans le **Chapitre 3** permet d'étudier l'impact de différents paramètres sur le régime de diffusion. Par exemple, la Figure 48 présente les effets de la concentration volumique en billes de verre $C_v$ (voir le **Chapitre 3** pour la relation avec $C_n$) sur le nombre d'interactions rayons/particules. On constate que, pour cette plage de concentrations et ce système, le nombre moyen d'interactions évolue quasi-linéairement avec la concentration (Figure 48 (b)), tandis que la distribution du nombre d'interactions s'élargie très rapidement avec la concentration en particules du milieu (cf. Figure 48 (a)). Pour une concentration volumique de l'ordre de $C_v = 10\%$, on compte en moyenne près de 800 interactions par photon détecté. Ceci n'est pas sans répercussions sur les propriétés optiques du milieu et les temps de calcul.

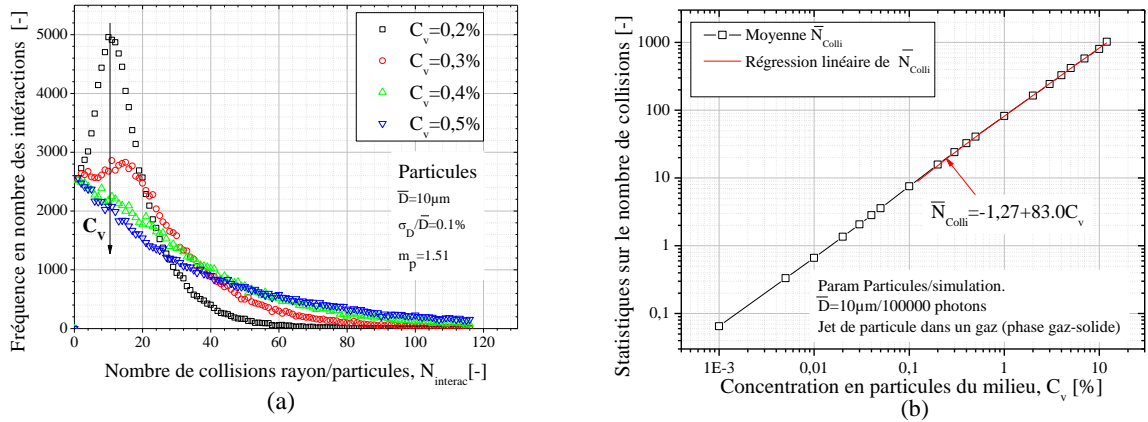

Figure 48 : Statistiques, pour des concentrations $C_v$ croissantes, sur (a) le nombre d'interactions photon/particule par photon et (b) évolution du nombre moyen d'interactions par photon. Milieu particulaire : jet cylindrique de billes de verre dans l'air (diamètre moyen $\overline{D} = 10 \mu m$ et écart- type relatif $\sigma_D / \overline{D} = 0.1\%$ )

Pour rappel, nos simulations sont menées avec des distributions granulométriques à deux paramètres, une loi log-normale (classique en granulométrie), pour laquelle les paramètres d'itération privilégiés sont le diamètre moyen $\overline{D}$ et l'écart-type relatif $\sigma_D/\overline{D}$. Le critère énoncé dans l'équation (186) est validé en utilisant ces deux paramètres. Pour cela on crée des cartes de validité (1 ≡ critère vérifié, 0 ≡ critère infirmé) en fonction de ces derniers et pour différentes concentrations volumiques, voir la Figure 49. On observe que le domaine de validité de l'équation (186) diminue lorsque la concentration volumique augmente et que ce sont les plus "petites" particules qui posent problème. En effet, dans un système diphasique donné et à concentration volumique fixée, le nombre de petites particules est plus important que le nombre de grosses particules. On aura donc moins de mal à résoudre le problème de diffusion de la lumière pour des particules plus grosses et une distribution polydisperse (les distributions log-normales ont tendance à s'élargir vers les plus grosses particules).

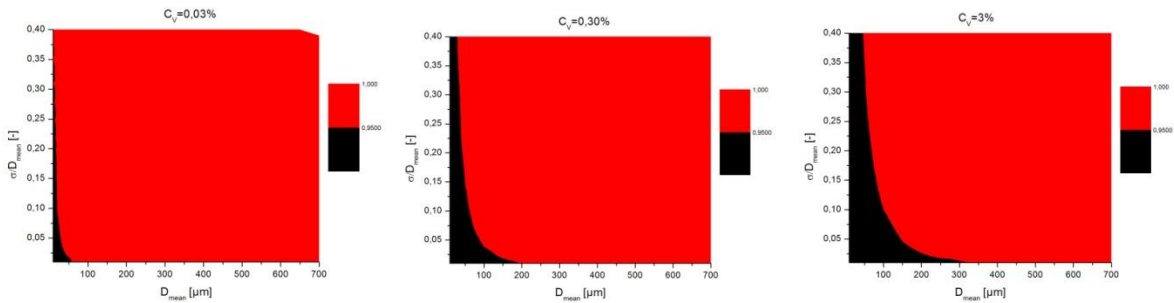

Figure 49 : Évolution du domaine de validité du critère de diffusion simple avec la taille, l'écart-type et la concentration volumique du milieu particulaire.

Ce chapitre ne traite que des milieux optiques dilués, dans la limite de la validité de l'équation (186). Nous y proposons et étudions les deux configurations optiques intéressantes pour la granulométrie du milieu considéré et qui mettent à profit les potentialités des photodétecteurs organiques. Ces configurations sont développées et optimisées grâce au code de Monte-Carlo et différents modèles analytiques. La première repose sur le principe des systèmes multi-angulaires opérant sur une large plage (néphélomètre ou bien MALS pour *Multi-Angle Light-Scattering* en anglais), la seconde est basée sur l'analyse d'une zone réduite du diagramme de diffusion : la zone de diffraction.

## 4.1   Configuration néphélométrique (ou MALS)

Dans le paragraphe **3.7**, nous avons étudié la diffusion de la lumière par un écoulement particulaire confiné dans une cuve cylindrique (éprouvette, cellule, tube d'analyse). Cette géométrie est typique des exigences du PAT et des conditions de mesures dans le milieu industriel (en laboratoire l'utilisation de cuves rectangulaires est plus répandu). Nous avons déjà montré que dans cette configuration et pour des photodétecteurs très peu distants de la cuve, le signal de diffusion obtenu semble très pauvre en informations granulométriques (cf. **3.7**). L'inversion de ces signaux semble extrêmement difficile, voire impossible. Nous avons donc intérêt à nous ramener à des conditions proches des configurations expérimentales classiques et des conditions décrites par des théories comme celle de Lorenz-Mie. En pratique, une de ces conditions fondamentales est celle dite de "détecteur lointain" (*remote sensing approximation* en anglais) qui permet d'obtenir des mesures résolues angulairement.

Pour être dans ce cas de figure, la solution la plus répandue consiste à se placer à l'infini *via* un montage de Fourier (la détection est dans le plan image d'une lentille convergente par exemple). Cependant, l'une des principales caractéristiques des OPS étant leur conformabilité, il n'y aurait pas de sens à utiliser une méthode employant des composants rigides (lentilles, miroirs, ...). De ce fait, et pour concevoir un système simple et à coût réduit, nous avons opté pour l'utilisation de trous optiques, ou "sténopés" [**Young 1971**], au lieu des lentilles. Des trous et fentes optiques sont donc utilisés, ils permettent de limiter la largeur angulaire du champ observé.

Cette solution n'est évidemment pas idéale, il nous faut par exemple, résoudre certains problèmes comme le "*cross-talking*" optique entre les différents détecteurs tout en maximisant les flux collectés et le nombre de détecteurs. En tenant compte des contraintes connues sur la taille de la feuille de photodétecteurs et la taille maximale acceptable pour un premier prototype, nous avons débuté par l'étude d'un montage à trous circulaires. Sur cette base, nous nous sommes dirigés vers l'élaboration d'un prototype à fentes optiques plus complexes, validé par des simulations de Monte-Carlo et des diagrammes de diffusion générés par la théorie de Lorenz-Mie.

### 4.1.1 Conception

L'imagerie par un trou circulaire est connue depuis déjà longtemps, sous les termes de "*caméra obscura*" ou de "sténopé". Pour cette étude, le système présenté sur la Figure 50 est composé d'une enceinte cylindrique de diamètre $D_c$ ( $D_c/2 < R_0 < 390/\pi \; [mm]$ ) renfermant un milieu diphasique uniforme et homogène, en écoulement ou non. Les particules y sont distribuées aléatoirement. Une couronne de filtres optiques est placée autour de cette enceinte à la distance $R_0$. Cette dernière est percée de trous optiques circulaires de diamètre curviligne $L_0$. Les photodiodes sont accolées sur la surface externe d'une seconde couronne qui est elle-même percée de trous optiques de diamètre curviligne équivalent à $L_d$. Chaque photodiode de diamètre $D_{opd}$ est donc placée en face de deux trous optiques. Les contraintes mécano-optiques imposées sont les suivantes $L_0 \geq 500 \mu m$, avec $D_{opd} \geq 500 \mu m$. La couronne de détecteurs est placée à une distance $R_d$ du centre de la cuve.

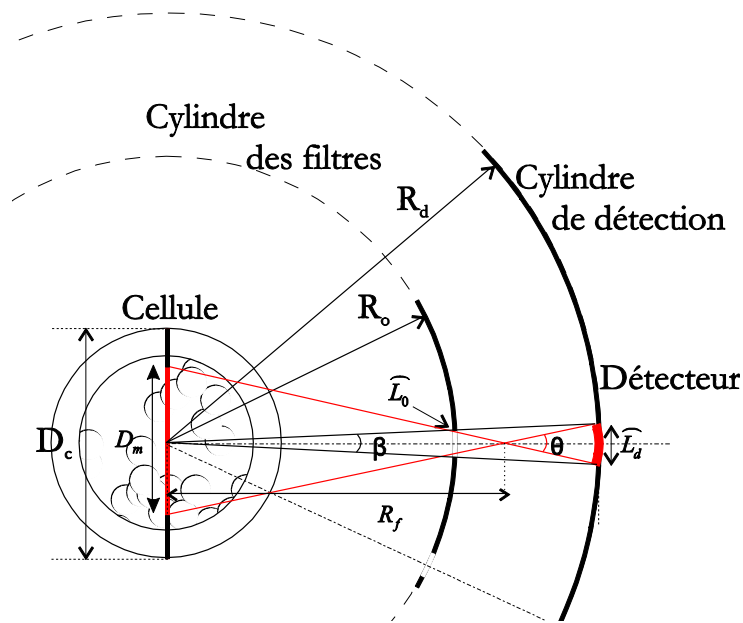

Figure 50 : Géométrie du prototype à filtres optiques circulaires

Dans ce système, le nombre de paramètres à déterminer est relativement important. En effet, il nous faut obtenir la distance du centre de la cuve à la couronne de filtre $R_0$, la distance du centre de la cuve au cylindre de détection $R_d$, le nombre de trous/détecteurs, leurs emplacements ainsi que leurs tailles. Pour résoudre ce problème, on cherche à déterminer $R_d, R_0$ en fonction de $\delta\theta$ en imposant les valeurs de $D_c, L_d$ et $L_0$.

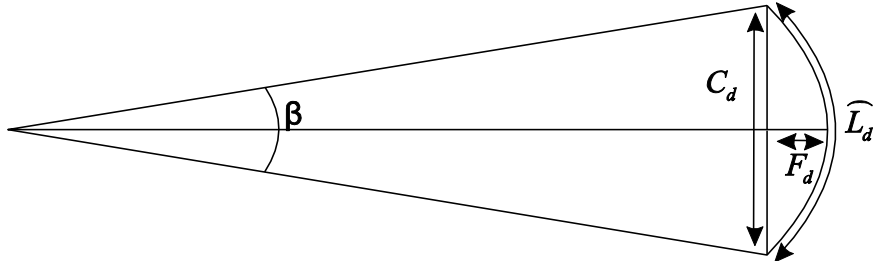

Figure 51 : Représentation des différents paramètres caractérisant un arc de cercle

Soit $L_d$ l'arc de cercle de rayon $R_d$ sous-tendant l'angle $\beta$ (voir la Figure 51), il vient la relation suivante :

$$L_d = \beta R_d \Rightarrow \beta = \frac{L_d}{R_d} \tag{187}$$

La corde $C_d$ de cet arc s'exprime par :

$$C_d = 2R_d \sin\left(\frac{\beta}{2}\right) = 2R_d \sin\left(\frac{L_d}{2R_d}\right) \tag{188}$$

On utilise le théorème de Pythagore pour calculer la flèche $F_d$ associée à cet arc : $R_d{}^2 = (R_d - F_d)^2 + (C_d/2)^2$, soit $|R_d - F_d| = \sqrt{R_d{}^2 - (C_d/2)^2}$. Cependant comme $R_d \geq F_d$ et $C_d = 2R_d \sin\left(\frac{L_d}{2R_d}\right)$, il vient que $F_d = R_d - \sqrt{R_d{}^2 - \left(2R_d \sin\left(L_d/2R_d\right)/2\right)^2}$ puis

$F_d = R_d - R_d\sqrt{1 - \sin^2\left(L_d/2R_d\right)}$.

On obtient finalement :

$$F_d = R_d\left[1 - \cos\left(\frac{L_d}{2R_d}\right)\right] \tag{189}$$

On raisonne de la même manière pour le trou. Soit $L_0$ l'arc de cercle de rayon $R_0$ sous-tendant l'angle $\beta$, la corde $C_0$ de cet arc s'exprime par :

$$C_0 = 2R_0 \sin\left(\frac{L_0}{2R_0}\right) \tag{190}$$

On utilise le théorème de Pythagore pour calculer la flèche $F_0$ associée à cet arc :

$$R_0{}^2 = (R_0 - F_0)^2 + \left(\frac{C_0}{2}\right)^2 \text{ avec } F_0 = R_0\left[1 - \cos\left(\frac{L_0}{2R_0}\right)\right] \tag{191}$$

On peut alors exprimer la tangente de l'angle $\delta\theta$ :

$$\tan\left(\frac{\delta\theta}{2}\right) = \frac{D_m}{2R_f} = \frac{C_0}{2(R_f - R_0 + F_0)} = \frac{C_d}{2(R_d - R_f - F_d)} \tag{192}$$

On exprime $R_d$ grâce à la relation (192) :

$$R_d - R_f - F_d = \frac{C_d}{2\tan\left(\dfrac{\delta\theta}{2}\right)} = R_f + F_d + \frac{C_d}{2\tan\left(\dfrac{\delta\theta}{2}\right)} \tag{193}$$

Or, d'après la relation (192), on a également $R_f = D_m \Big/ 2\tan\left(\dfrac{\delta\theta}{2}\right)$. Avec les relations (188) et

(189), il vient : $F_d = R_d\left[1 - \cos\left(L_d / 2R_d\right)\right]$ et $C_d = 2R_d \sin\left(L_d / 2R_d\right)$. On réinjecte ces relations dans

l'équation (193), ce qui donne :

$$R_d = \frac{D_m}{2\tan\left(\dfrac{\delta\theta}{2}\right)} + R_d\left[1 - \cos\left(\frac{L_d}{2R_d}\right)\right] + \frac{2R_d \sin\left(\dfrac{L_d}{2R_d}\right)}{2\tan\left(\dfrac{\delta\theta}{2}\right)} \tag{194}$$

On obtient finalement :

$$D_m = 2R_d\left[\cos\left(\frac{L_d}{2R_d}\right)\tan\left(\frac{\delta\theta}{2}\right) - \sin\left(\frac{L_d}{2R_d}\right)\right] \tag{195}$$

Le calcul est le même pour $R_0$ :

$$D_m = 2R_0\left[\cos\left(\frac{L_0}{2R_0}\right)\tan\left(\frac{\delta\theta}{2}\right) + \sin\left(\frac{L_0}{2R_0}\right)\right] \tag{196}$$

On égale les équations (195) et (196) pour se débarrasser de $D_m$ :

$$2R_d\left[\cos\left(\frac{L_d}{2R_d}\right)\tan\left(\frac{\delta\theta}{2}\right)-\sin\left(\frac{L_d}{2R_d}\right)\right]=2R_0\left[\cos\left(\frac{L_0}{2R_0}\right)\tan\left(\frac{\delta\theta}{2}\right)+\sin\left(\frac{L_0}{2R_0}\right)\right] \tag{197}$$

Dans la limite des petits angles, on utilise une approximation au premier ordre pour simplifier l'équation (197) qui devient : $R_d\delta\theta - L_d \simeq R_0\delta\theta + L_0$ .

On en déduit la relation suivante :

$$R_d - R_0 = \frac{L_0 + L_d}{\delta\theta} \tag{198}$$

La Figure 52 montre que plus la différence $(R_d - R_0)$ est petite, plus la résolution angulaire de la mesure sera grande. En se fixant une précision angulaire raisonnable, on peut trouver des couples $(R_d - R_0)$ de solutions techniquement réalisables. Si on ne considère qu'une seule zone photosensible, cette simple résolution suffit. Cependant, pour un système MALS comportant plusieurs détecteurs, une optimisation de la position de ces derniers est nécessaire afin d'éviter le "cross-talking" (un même détecteur collecte des flux passant par des trous différents).

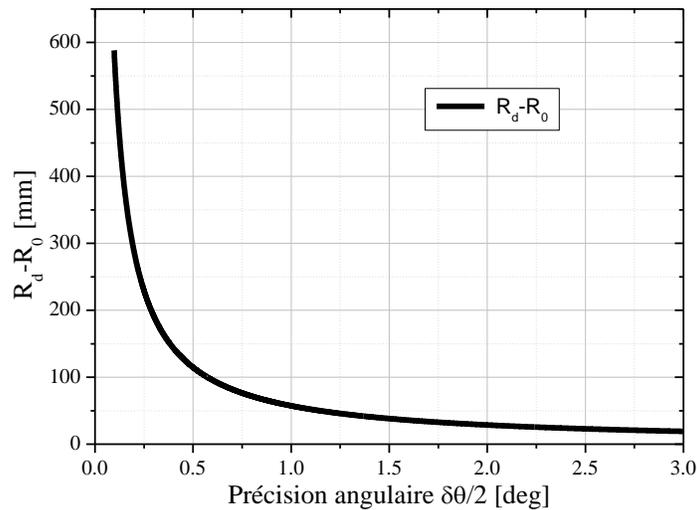

Figure 52 : Evolution de la différence entre la distance de détection $R_d$ et celle de la couronne de filtres $R_0$ en fonction de la précision angulaire recherchée

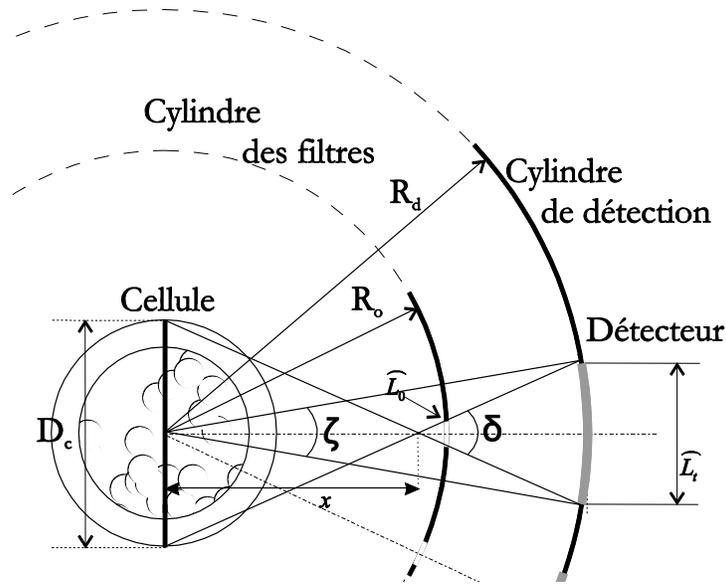

Figure 53 : Paramètres géométriques liés à la taille du halo sur la couronne de détection

Pour cela, il faut calculer la taille de la tâche $L_t$ en procédant comme précédemment. Soit $L_t$ l'arc de cercle de rayon $R_d$ sous-tendant l'angle $\zeta$, la corde $C_t$ de cet arc s'exprime par :

$$C_t = 2R_d \sin\left(\frac{L_t}{2R_d}\right) \tag{199}$$

On utilise le théorème de Pythagore pour calculer la flèche $C_t$ associée à cet arc :

$$F_t = R_d \left[1 - \cos\left(\frac{L_t}{2R_d}\right)\right] \tag{200}$$

On peut alors exprimer la tangente de l'angle $\delta$ :

$$\tan\left(\frac{\delta}{2}\right) = \frac{D_c}{2x} = \frac{C_t}{2(R_d - x - F_t)} = \frac{C_0}{2(R_0 - x - F_0)} \tag{201}$$

Cette relation permet d'obtenir : $(R_d - x - F_t)D_c = C_t x$, dont on tire :

$$x = \frac{(R_d - F_t)D_c}{C_t + D_c} \tag{202}$$

En reproduisant ces calculs avec le dernier membre de l'équation (201) :

$$x = \frac{(R_0 - F_0)D_c}{C_0 + D_c} \tag{203}$$

On peut alors égaler les équations (202) et (203) :

$$\frac{(R_0 - F_0)D_c}{C_0 + D_c} = \frac{(R_d - F_T)D_c}{C_T + D_c} \qquad (204)$$

Après développement et simplification, on a :

$$L_t(R_0 - F_0) + F_t(C_0 + D_c) = R_d(C_0 + D_c) - D_c(R_0 - F_0) \qquad (205)$$

Or $x_0$ et $L_0$ au premier ordre nous donne : $F_0 = 0$ et $C_0 = L_0$. On rappelle également grâce aux relations (199) et (200) que $C_t = 2R_d \sin\left(L_t / 2R_d\right)$ et $F_t = R_d\left[1 - \cos\left(L_t / 2R_d\right)\right]$. D'où l'on tire après quelques calculs :

$$2R_d R_0 \sin\left(\frac{L_t}{2R_d}\right) - R_d(L_0 - D_c)\cos\left(\frac{L_t}{2R_d}\right) + D_c R_0 = 0 \qquad (206)$$

Le but étant de diminuer la taille des taches, on cherche à obtenir : $\frac{L_t}{2R_d} \simeq 0$, ce qui nous amène à un développement au premier ordre de l'équation (206) :

$$2R_d R_0\left(\frac{L_t}{2R_d}\right) - R_d(L_0 - D_c) + D_c R_0 = 0 \qquad (207)$$

D'où :

$$L_t = \frac{(L_0 + D_c)R_d}{R_0} - D_c \qquad (208)$$

Le système d'équations est donc le suivant :

$$\begin{cases} R_d - R_0 \simeq \dfrac{L_0 + L_d}{2\tan\left(\dfrac{\delta\theta}{2}\right)} \\[4mm] L_t \simeq \dfrac{(L_0 + D_c)R_d}{R_0} - D_c \end{cases} \qquad (209)$$

Pour résoudre ce dernier, des choix sont nécessaires : si on opte pour une ouverture angulaire $\delta\theta$ faible, on dégrade (augmente) la taille de la tâche et vice-versa. La résolution la plus simple, consiste à imposer une précision angulaire $\delta\theta$. De cette dernière découle la valeur de $R_d - R_0$. On doit ensuite trouver les couples de valeurs $(R_d, R_0)$ qui minimisent la tâche $L_t$, c'est-à-dire minimiser le rapport $R_d / R_0$. Notons q la valeur telle que $R_d - R_0 = q$, on a alors :

$$\min\left(\frac{R_d}{R_0}\right) = \min\left(\frac{q + R_0}{R_0}\right) = 1 + q\min\left(\frac{1}{R_0}\right) = 1 + \frac{q}{\max(R_d - q)} \qquad (210)$$

Minimiser le rapport $R_d/R_0$ revient donc à maximiser $R_d$. On prendra donc systématiquement $R_d = 390/\pi$.

L'évolution de la précision angulaire ainsi que le nombre de détecteurs en fonction de la différence $(R_d - R_0)$, peuvent maintenant être quantifiés, voir le Tableau 2. Il apparait qu'augmenter la précision angulaire, revient à augmenter la taille des halos lumineux et *in fine* à dégrader la densité en détecteurs indépendants du système.

| $R_d - R_0$ [mm] | Ouverture angulaire $\delta\theta$ [deg] | Taille de la tache $L_t$ [mm] [deg] | Nombre de détecteurs minimum [-] |
|---|---|---|---|
| **114** | 0.501 | 234.48 108 | 1 |
| **100** | 0.572 | 85.91 39.7 | 4 |
| **90** | 0.636 | 54.76 25.3 | 7 |
| **80** | 0.716 | 37.77 17.4 | 10 |
| **70** | 0.818 | 27.07 12.5 | 14 |
| **57** | 1 | 18.17 8.4 | 21 |
| **50** | 1.146 | 14.35 6.63 | 27 |
| **10** | 11.421 | 3.101 1.43 | 125 |

Tableau 2 : Évolution des différents paramètres du système en fonction de l'écart $(R_d - R_0)$ entre la distance de détection et la position des filtres optiques

La résolution angulaire du prototype est fixée à $\pm 0.5°$ (i.e. $\delta\theta = 1°$), tandis que le cahier des charges impose une taille minimale des capteurs organiques conformables de 500 µm. Les paramètres d'entrée du prototype sont donc : $L_0 = 0.5mm, L_d = 0.5mm, D_c = 20mm, \delta\theta = 1°$. On en déduit alors que $R_0 \simeq 66.85mm, R_d = 390/\pi$ et $L_t \simeq 18.15mm \simeq 8.37°$. Avec ces paramètres, le nombre minimal de détecteurs est donc de 21 détecteurs pour un secteur de 180° (les calculs ont été réalisés dans le cas le plus défavorable : un angle de collection de 90°). Pour augmenter le nombre de détecteurs indépendants, on peut dans un premier temps observer la Figure 54. Cette dernière montre l'évolution de la largeur angulaire de la tache avec l'angle de diffusion. Plus on se rapproche des "petits" angles, plus la taille du halo lumineux diminue. On vérifie également que la largeur angulaire de la tâche à 90° a bien la valeur attendue (8.4°).

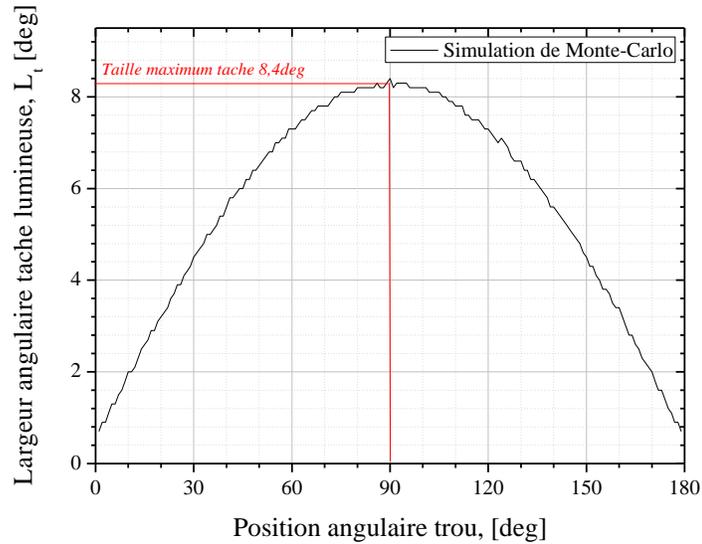

Figure 54 : Évolution de la largeur angulaire du halo lumineux en fonction de l'angle de détection

Il nous faut optimiser l'emplacement et le nombre de photodétecteurs sur la couronne de détection. L'espacement entre les trous/détecteurs dépend de la taille angulaire de la tache. La symétrie de cette courbe par rapport à 90° et son aspect général permettent de supposer qu'une approximation par une parabole pourrait être suffisante. On approxime cette courbe par une parabole de sommet $(90;8.4)$ passant par le point $(1;0.7)$. Connaissant l'équation d'une parabole dans le plan $ax^2 + bx + c$ on en déduit le système d'équations suivant :

$$\begin{cases} 90^2 a + 90b + c = 8.4 \\ 180a + b = 0 \\ a + b + c = 0.7 \end{cases} \tag{211}$$

La deuxième équation du système (211) est obtenue par dérivation au point de tangence $(90;8.4)$. La résolution de ce système nous donne le polynôme de degré 2 dont les coefficients sont :

$$\begin{cases} a = -\dfrac{1383}{1425780} \\ b = \dfrac{1386}{7921} \\ c = \dfrac{749949}{1425780} \end{cases} \tag{212}$$

La Figure 55 permet de comparer l'allure du polynôme obtenu et les simulations Monte-Carlo. Ce polynôme n'est pas le polynôme passant par le plus de points mais celui maximisant la taille des taches. C'est pour cela que les 3 points d'interpolation retenus sont les maxima et minima de la fonction (avec 10% d'erreur en moyenne). Grâce à cette approximation, la taille de la tâche à tous les angles est connue (quoique légèrement surestimée). A partir de ces données, on peut mettre en place un algorithme simple pour déterminer les positions angulaires optimales des photodétecteurs.

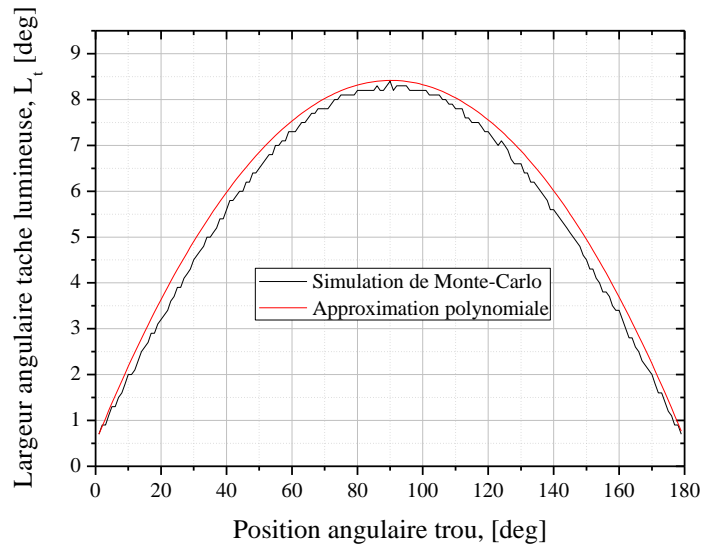

Figure 55 : Comparaison entre le polynôme d'ordre 2 approximant l'évolution de la largeur des halos à la simulation

On décide d'une position angulaire pour le premier détecteur $\theta_{init}$. Connaissant la taille de la tache $L_t$ (taille minimale 0.5°), la position angulaire de la zone photosensible suivante $\theta_{next}$ est recherchée. L'algorithme est le suivant :

---

**Algorithme d'optimisation de l'emplacement angulaire des photodétecteurs**

---

1:      $N_{detect} \leftarrow 1$
2:      $Position(N_{detect}) \leftarrow \theta_{init}$
3:      **Tant que** $\theta_{init} \leq 180°$ **faire**
4:          **Si** $\theta_{init} + L_t(\theta_{init})/2 - \left(\theta_{next} - L_t(\theta_{next})/2\right) < 0$ **alors**
5:              $\theta_{next} \leftarrow \theta_{next} + 0.5°$
6:          **sinon**
7:              $N_{detect} \leftarrow N_{detect} + 1$
8:              $\theta_{init} \leftarrow \theta_{next}$
9:              $Position(N_{detect}) \leftarrow \theta_{init}$
10:         **fin si**
11:     **fin tant que**

---

Le système ainsi conçu est composé de 41 points de mesure, répartis sur un secteur de près de 180°, avec une résolution angulaire de $\theta \pm 0.5°$. La Figure 56 permet d'observer pour chaque détecteur la largeur du halo lumineux (en noir) ainsi que l'intensité collectée (en bleu). Ces résultats sont obtenus par simulation Monte-Carlo pour des particules nanométriques ( $D = 1nm$ ) et sous une polarisation perpendiculaire. On s'aperçoit que la taille des tâches lumineuses est comme prévue maximale à 90° et minimale pour les très petits, et les très grands, angles. L'absence de recouvrement du signal entre les différentes zones de détection est à souligner. De même, l'intensité collectée est minimale vers 90° et maximale aux extrémités du diagramme de diffusion.

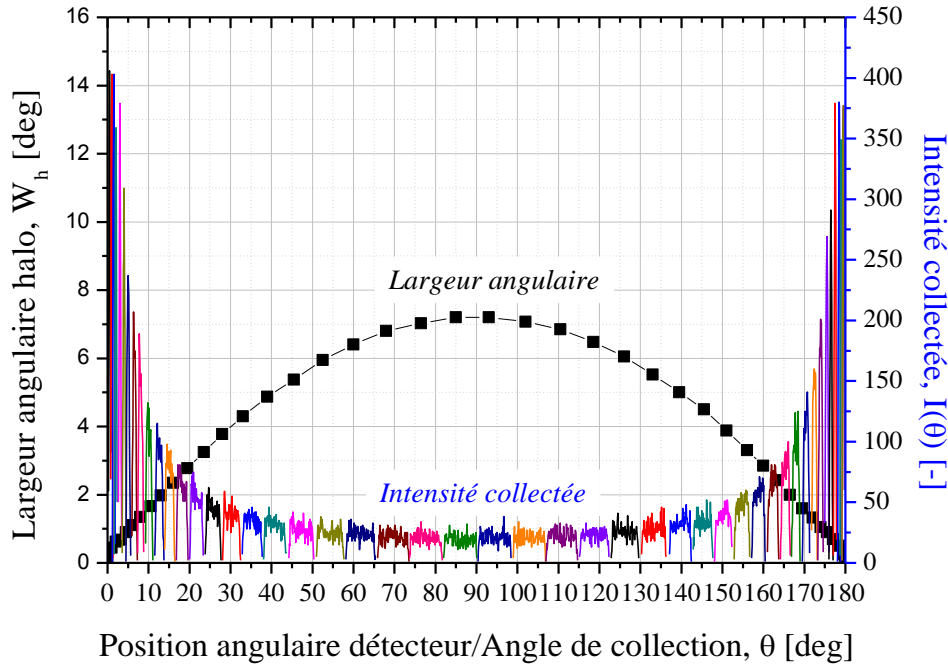

Figure 56 : Évolution de la taille du halo lumineux avec la position angulaire des détecteurs

La validité des calculs de dimensionnement est confirmée par l'estimation de certains paramètres via les simulations Monte-Carlo :

- la taille de la tache lumineuse à 90° ( $L_t \simeq 8.37°$ ) est en accord avec les résultats de la Figure 54.

- il en est de même pour le volume de mesure observé à 90° dont la valeur est donnée par l'équation (195) ou (196) : $D_m \simeq 1.66mm$

- l'ouverture angulaire du détecteur à 90° est bien de l'ordre de $\delta\theta \simeq 1°$ .

- pour évaluer la forme et les dimensions du volume de mesure, nous avons modifié le code Monte-Carlo de façon à sauvegarder la position des photons qui passent par le trou à 90° et atteignent le détecteur placé à 90° (les calculs précédents ont été réalisés dans cette configuration et ne s'appliquent pas aux autres angles). D'après cette simulation (Figure 57), la largeur du volume de mesure est de l'ordre de $D_m = 1.58mm$ selon l'axe $y$, ce qui est très proche du résultat attendu d'autant plus que le nombre de photons est ici un peu trop faible pour décrire de manière satisfaisante le volume de mesure. Pour l'autre dimension, selon $x$, nous retrouvons le diamètre imposé pour le faisceau laser incident : $w_x = 0.2mm$.

- pour évaluer la précision angulaire réellement obtenue, le code Monte-Carlo a été modifié de façon à conserver la mémoire de l'angle suivant lequel chaque photon ayant traversé le trou à 90° a été détecté. La Figure 58 montre la distribution des angles en question. De forme gaussienne, centrée sur 90° et de largeur de 1°, elle confirme la pertinence de notre démarche d'optimisation.

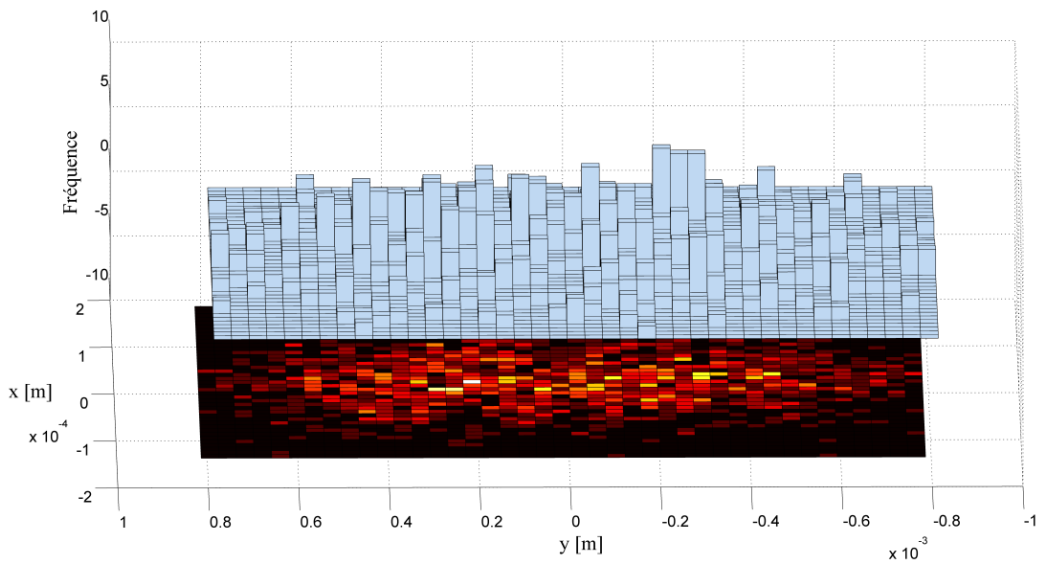

Figure 57 : Visualisation du volume de mesure simulé par méthode de Monte-Carlo pour le détecteur situé à 90°

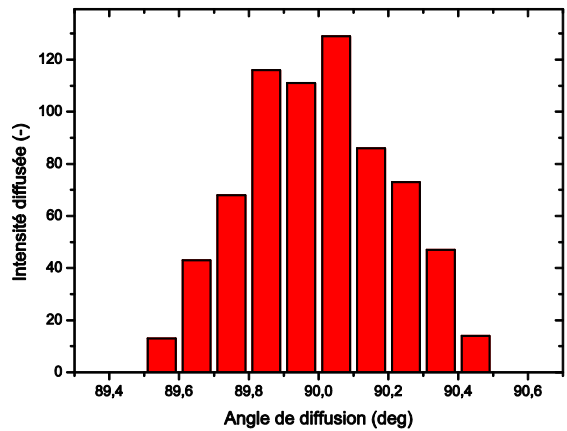

Figure 58 : Distribution des angles de diffusion des photons détectés par le détecteur placé à 90°

Cette configuration peut générer des problèmes de flux à certains angles de collection. En effet, le trou de 0.5mm de diamètre limite considérablement le flux du signal optique collecté. En régime de diffusion simple, pour des particules de taille micronique qui diffusent très peu de lumière à 90°, le seuil de détectabilité peut être atteint. Nous avons donc étudié l'impact de la taille du trou optique de filtrage sur le dimensionnement du système. Les résultats correspondants sont présentés dans le tableau suivant avec, en rouge, le cas retenu. On voit que le nombre de détecteurs décroit très rapidement avec l'augmentation du diamètre des trous. Nous avons décidé de ne pas modifier la taille des trous circulaires (une solution alternative sera présentée plus loin).

| $\phi_0$ **(mm)** | $R_d - R_0$ [mm] | **Taille de la tache $L_t$** **[mm deg]** | **Nombre détecteurs minimum [-]** |
|---|---|---|---|
| **0.5** | 57.29 | 18.15 8.37 | 21 |
| **0.6** | 63.02 | 22.57 10.43 | 17 |
| **0.8** | 74.48 | 32.08 14.82 | 12 |
| **1** | 85.94 | 48.42 22.37 | 8 |
| **1.2** | 97.40 | 78.82 36.42 | 4 |

Tableau 3 : Étude de l'influence de la taille du trou optique sur le nombre de zones photosensibles minimales

Le prototype de mesure que nous venons d'optimiser est schématisé dans la Figure 59. Il permet la mesure de l'intensité diffusée par le système particulaire contenu dans un tube en verre de diamètre 20mm. Un plus petit tube réduirait la taille des halos lumineux et augmenterait le nombre de points de mesures. Ce prototype filtre la lumière diffusée grâce à des trous optiques, de diamètre $L_0 = 0.5mm$, répartis sur une couronne concentrique de rayon $R_0 = 67mm$. La détection est assurée par une seconde couronne, de diamètre 124mm sur laquelle sont répartis 41 photodétecteurs circulaires de diamètre $L_d = 0.5mm$.

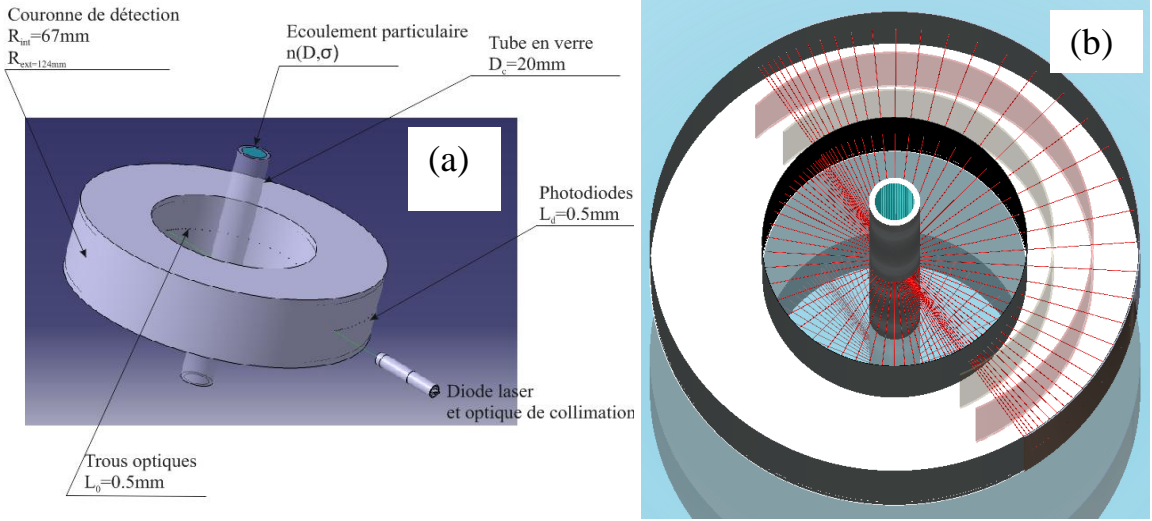

Figure 59 : Ebauches (sous CATIA) du prototype à trous de filtrage circulaires

## 4.1.2 Problème du volume de mesure observé

Les systèmes MALS [**Wang 2002, Burr 2007**] soulèvent certaines difficultés quant au traitement de l'intensité collectée, voir la Figure 60. Cette dernière dépend en effet fortement de l'angle de collection, même en l'absence de particules,...

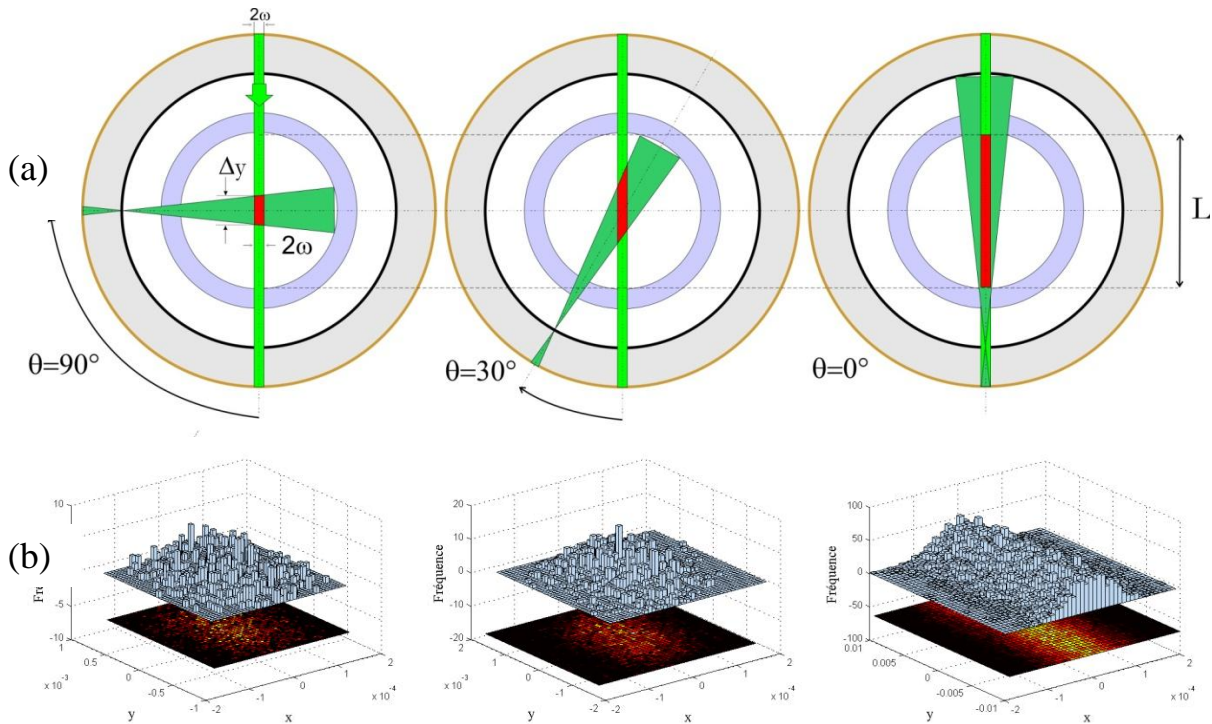

Figure 60 : (a) Illustration de la dépendance du volume de mesure avec l'angle de diffusion et (b) simulation par le modèle de Monte-Carlo de la forme de ce dernier

On remarque que pour les angles proches de 0°, le volume de mesure ne varie pas. En effet, on observe l'intégralité du faisceau lumineux aux petits angles. Le volume de mesure diminue lorsqu'on augmente l'angle d'observation jusqu'à 90° qui représente le cas le plus défavorable (plus petit volume de mesure). Une loi géométrique approchée de l'évolution du volume de mesure en fonction de l'angle d'observation peut être tirée des paramètres détaillés sur la Figure 61.

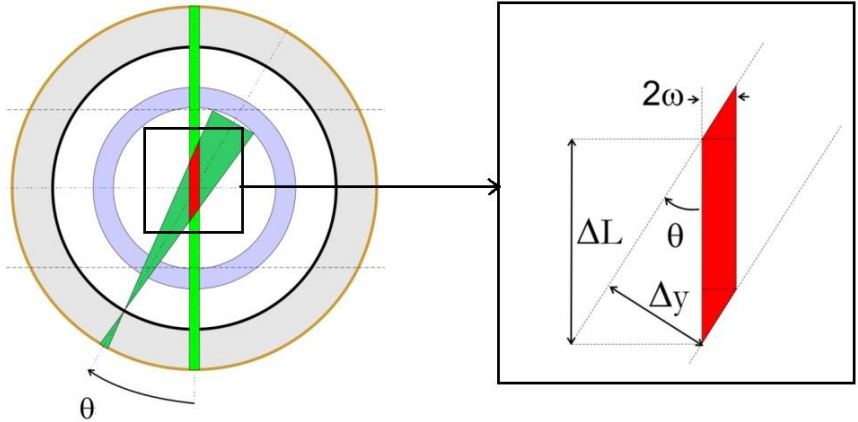

Figure 61 : Représentation des différents paramètres mis en jeu pour le calcul du volume de mesure

Sur la Figure 61, $w$ représente le rayon du faisceau gaussien, $\theta$ l'angle d'observation, $\Delta y$ la largeur du champ d'observation (constante selon les angles) et $\Delta L$ un des paramètres recherchés. L'aire $A$ d'un parallélogramme (représenté en rouge sur la Figure 61) s'exprime avec nos paramètres par :

$$A = 2w\Delta L \qquad (213)$$

Le sinus de l'angle $\theta$ donne accès à la grandeur $\Delta L$ :

$$\Delta L(\theta) = \frac{\Delta y}{\sin(\theta)} \qquad (214)$$

Le volume de mesure $V$ pour un faisceau gaussien à section circulaire de rayon $w$ est donc donné par :

$$V(\theta) = 4w^2 \frac{\Delta y}{\sin(\theta)} \qquad (215)$$

Pour un faisceau gaussien à section elliptique de petit axe $2w_x$ et de grand axe $2w_z$, le volume de mesure s'exprime par :

$$V(\theta) = 4w_x w_z \frac{\Delta y}{\sin(\theta)} \qquad (216)$$

Cependant, l'observation de la Figure 60 montre qu'à 0°, et ce jusqu'à un certain angle, la cuve est vue en totalité. Cet angle est obtenu lorsque le sinus de l'angle $\theta$ dépasse 1, c'est-à-dire :

$$\theta \leq \sin^{-1}\left(\frac{\Delta y}{D_c}\right) \qquad (217)$$

avec $D_c$ le diamètre de la cuve. Finalement, dans le cas général du faisceau gaussien à section elliptique, le volume de mesure est :

$$\begin{cases} V(\theta) = 4w_x w_z \dfrac{\Delta y}{\sin(\theta)} \text{ si } \theta \leq \sin^{-1}\left(\Delta y/D_c\right) \\ V(\theta) = 4w_x w_z D_c \text{ sinon} \end{cases} \qquad (218)$$

Sur la Figure 62, on observe la courbe d'évolution du volume de mesure à partir de la loi détaillée précédemment avec une normalisation par la valeur maximale $V(\theta) = 4w_x w_z D_c$

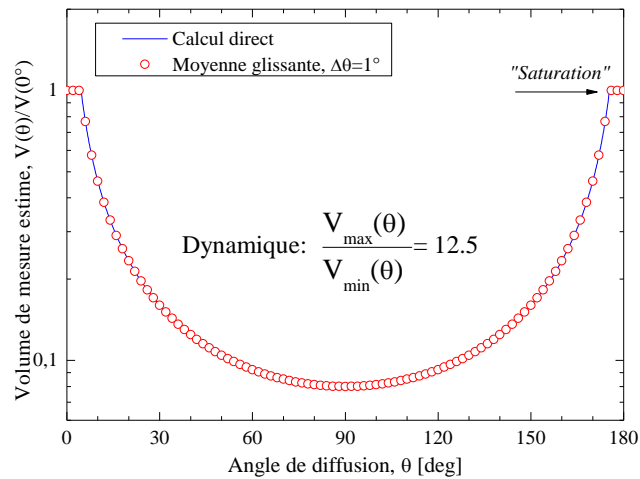

Figure 62 : Évolution du volume de mesure selon l'angle de collection pour notre montage

Dans notre configuration, il faudra donc couvrir une dynamique du signal de l'ordre de la décade. Il faudra également prendre en compte cette évolution du volume de mesure dans le calcul de l'intensité collectée.

### 4.1.3 Expression de l'intensité et normalisation

Les hypothèses énumérées précédemment permettent d'exprimer l'intensité moyenne du signal collecté comme suit :

$$\overline{I}(\theta) \propto C_n V_m(\theta) \int_0^\infty I(\theta, D) n(D) dD \qquad (219)$$

avec $C_n$ la concentration en nombre, $V_m$ la fonction d'évolution du volume de mesure avec l'angle d'observation (cf. Figure 62), $I$ l'intensité diffusée et $n$ la distribution des particules en nombre. Pour simplifier le problème en recherchant une quantité dépendant uniquement de la distribution en taille des particules $n(D)$, différentes étapes de normalisation et/ou de calibration doivent être suivies. En premier lieu, le volume de mesure dont la connaissance nous importe peu ne doit plus apparaître dans l'expression de l'intensité moyenne. Pour cela, on procède à une normalisation par l'intensité collectée avec des particules de Rayleigh, qui lorsqu'elles sont suffisamment petites (de l'ordre de 1nm) et observées sous une polarisation perpendiculaire, ont un diagramme de diffusion parfaitement isotrope. Toutes les variations résiduelles de l'intensité collectée sont dans ce cas imputables à celles du volume de mesure.

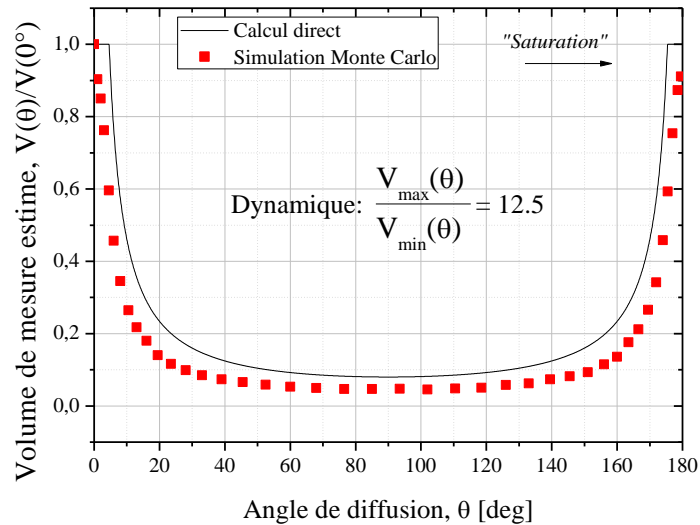

Figure 63 : Diagramme de diffusion obtenu par simulation de Monte-Carlo pour le montage à filtres optiques précédemment optimisé. Les particules ont un diamètre de 0.1nm.

La Figure 63 montre le résultat d'une simulation de Monte-Carlo dans les conditions précédemment citées (suspension aqueuse de nano billes de silice). On constate que la forme globale du diagramme de diffusion obtenue tend à décrire de manière exacte le volume de mesure explicité sur la Figure 61, on peut donc écrire :

$$I_{nano}(\theta) \propto g C_{nano} V_m(\theta) I_{nano}^{Mie}(\theta) \qquad (220)$$

avec pour $C_{nano}$ une constante liée à la concentration en nanoparticules et $g$ le gain du photodétecteur à l'angle $\theta$. En normalisant l'intensité collectée sur les détecteurs :

$$\frac{\bar{I}(\theta)}{I_{nano}(\theta)} \propto C_n \int_0^\infty I(\theta, D) n(D) dD \qquad (221)$$

Il est ensuite primordial de faire abstraction de la concentration en nombre $C_n$. Nous avons choisi une normalisation par le pic de diffraction en $\theta = 0°$, même si un autre angle est techniquement possible et même recommandé (nous en rediscuterons par la suite), soit :

$$\bar{I}(\theta \approx 0) \propto C_n \int_0^\infty I(\theta \approx 0, D) n(D) dD \qquad (222)$$

En normalisant l'équation (221) par l'intensité en $\theta = 0°$, on obtient :

$$\frac{\bar{I}(\theta)}{I_{nano}(\theta) I(\theta \approx 0)} \propto \int_0^\infty I(\theta, D) n(D) dD \qquad (223)$$

Ainsi à indice de réfraction connu, il ne reste donc plus que les paramètres de la distribution à déterminer par inversion de l'intensité mesurée et normalisée.

## 4.1.4 Prototype amélioré : vers des filtres optiques plus complexes

Le premier prototype présenté utilise des petits trous de diamètre identique, avec les problèmes de flux qui en découlent. Pour augmenter ces flux tout en comprimant la dynamique des signaux détectés, il nous faut concevoir des filtres de forme plus complexe.

On rappelle que dans notre repère, et en négligeant les effets de polarisation, le diagramme de diffusion d'une particule dépend uniquement de l'angle de diffusion $\theta$ [**Mul 2011**], voir la Figure 64. Le diagramme de diffusion est donc axisymétrique par rapport à l'axe du faisceau.

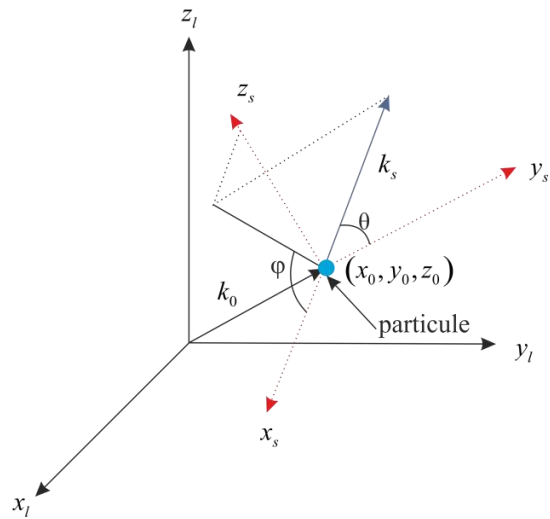

Figure 64 : Repère du laboratoire : $(x_l y_l z_l)$, position de la particule: $(x_s y_s z_s)$, vecteur d'onde du rayon incident $\mathbf{k_0}$, angle de diffusion $\theta$ et angle azimutal $\varphi$ avec $\varphi \in [0, 2\pi]$ (la polarisation n'est pas prise en compte)

Par ailleurs comme les détecteurs conformables ne peuvent être courbés que suivant une dimension, les surfaces de détection de référence sont « cylindriques ». Pour de telles surfaces, les masques optiques correspondent à l'intersection entre un cône (obligatoirement circulaire dans le cas de particules sphériques) et un cylindre (circulaire pour assurer une distance de mesure constante dans le plan de diffusion). La Figure 65 schématise, à gauche, l'intersection (2) du cône de section circulaire (1) et d'un cylindre, et à droite, l'extraction de la surface d'intersection correspondante (en forme de "chips" (2)). La forme de cette dernière diffère notablement de celle de la surface d'intersection d'un plan et d'un cône (1).

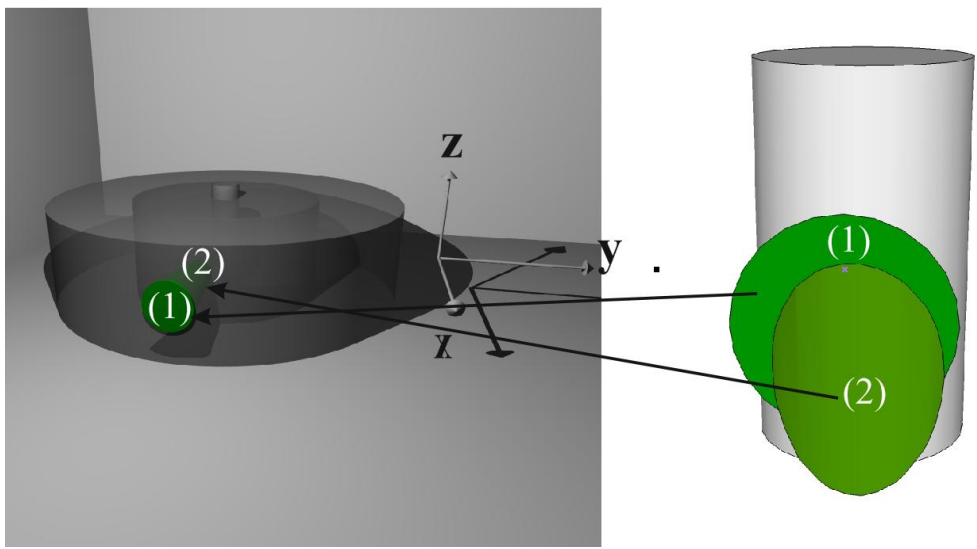

Figure 65 : Représentation de la surface d'intersection (2) entre un cône d'axe y et un cylindre d'axe z (de section circulaire (1)). A gauche la géométrie générale du problème et à droite la forme de la surface d'intersection

Pour décrire la forme de ces surfaces, il est nécessaire de déterminer l'équation de cette intersection. Le montage ne reproduisant pas parfaitement les conditions de détecteur lointain, il est nécessaire de se fixer une résolution angulaire type. La surface recherchée est donc l'intersection entre le cylindre de détection et la différence entre les cônes d'ouvertures angulaires $\theta + \partial\theta/2$ et $\theta - \partial\theta/2$ avec comme auparavant $\partial\theta/2 = \pm 0.5°$ pour la précision angulaire.

Dans un premier temps, on cherche uniquement à résoudre l'intersection entre le cylindre et le cône d'ouverture angulaire $\theta - \partial\theta/2$, la résolution pour le cône d'ouverture $\theta + \partial\theta/2$ étant exactement la même. Pour ce problème, une technique classique consiste à injecter une équation paramétrique dans une équation cartésienne. Ceci, afin de trouver l'équation paramétrique (ou cartésienne quand cela est possible) de la surface recherchée. Dans le repère du laboratoire, l'équation paramétrique du cône d'ouverture angulaire $\theta - \partial\theta/2$ s'écrit :

$$\begin{cases} x_l = h\cos(\alpha)\tan(\theta - \partial\theta/2) \\ y_l = h \\ z_l = h\sin(\alpha)\tan(\theta - \partial\theta/2) \end{cases} \quad \text{avec } \left\{\alpha \in [0, 2\pi], h \in [0, H]\right\} \tag{224}$$

avec $H$ la hauteur du cône de diffusion. Dans ce même repère, l'équation cartésienne d'un cylindre d'axe $z_l$ et de rayon $R_c$ est la suivante :

$$x_l^2 + y_l^2 = R_c^2 \tag{225}$$

Avec les équations (224) et (225), il vient :

$$h = \sqrt{\frac{R_c^2}{1 + \cos^2(\alpha)\tan^2(\theta - \partial\theta/2)}} \tag{226}$$

Finalement :

$$\begin{cases} x = \sqrt{\dfrac{R_c^2}{1 + \cos^2(\alpha)\tan^2(\theta - \partial\theta/2)}}\cos(\alpha)\tan(\theta - \partial\theta/2) \\[2ex] y = \sqrt{\dfrac{R_c^2}{1 + \cos^2(\alpha)\tan^2(\theta - \partial\theta/2)}} \\[2ex] z = \sqrt{\dfrac{R_c^2}{1 + \cos^2(\alpha)\tan^2(\theta - \partial\theta/2)}}\sin(\alpha)\tan(\theta - \partial\theta/2) \end{cases} \quad \text{avec } \left\{\alpha \in [0, 2\pi], \theta < \pi/2\right\} \tag{227}$$

La surface à décrire est donc comprise entre la courbe donnée par le système d'équations (227) et la courbe décrite par le système d'équations (228), qui représente l'intersection entre le cylindre de détection et le cône d'ouverture angulaire $\theta + \partial\theta/2$ :

$$\begin{cases} x = \sqrt{\dfrac{R_c^{\ 2}}{1+\cos^2(\alpha)\tan^2(\theta+\partial\theta/2)}}\cos(\alpha)\tan(\theta+\partial\theta/2) \\[2mm] y = \sqrt{\dfrac{R_c^{\ 2}}{1+\cos^2(\alpha)\tan^2(\theta+\partial\theta/2)}} \\[2mm] z = \sqrt{\dfrac{R_c^{\ 2}}{1+\cos^2(\alpha)\tan^2(\theta+\partial\theta/2)}}\sin(\alpha)\tan(\theta+\partial\theta/2) \end{cases} \qquad \text{avec } \left\{\alpha\in\left[0,2\pi\right],\theta<\pi/2\right\} \qquad (228)$$

Ces formules sont valables jusqu'à $\theta = 90°$, les cônes étant symétriques par rapport à 90°, pour les angles supérieurs à 90° il vient :

$$\begin{cases} x = \sqrt{\dfrac{R_c^{\ 2}}{1+\cos^2(\alpha)\tan^2(\pi-(\theta+\partial\theta/2))}}\cos(\alpha)\tan(\pi-(\theta+\partial\theta/2)) \\[2mm] y = \sqrt{\dfrac{R_c^{\ 2}}{1+\cos^2(\alpha)\tan^2(\pi-(\theta+\partial\theta/2))}} \\[2mm] z = \sqrt{\dfrac{R_c^{\ 2}}{1+\cos^2(\alpha)\tan^2(\pi-(\theta+\partial\theta/2))}}\sin(\alpha)\tan(\pi-(\theta+\partial\theta/2)) \end{cases} \qquad \left\{\alpha\in\left[0,2\pi\right],\theta\in\left[\pi/2,\pi\right]\right\} \qquad (229)$$

Comme nous le verrons dans ce qui suit, ces équations définissent des "anneaux" complexes. La validité des équations précédentes a été testée à l'aide de simulations de type Monte-Carlo, dans le cas simple d'un volume de mesure cylindrique sans cuve et pour des particules placées dans l'air. Pour visualiser les résultats obtenus, nous avons établi une carte du pourcentage des photons diffusés dans la direction : $\theta_{\text{detect}}\in\left[\theta-0.5°,\theta+0.5°\right]$ avec $\theta_{\text{detect}}$ l'angle sous lequel un photon est détecté dans le repère cylindrique ( $r_l$, $\theta_{\text{detect}}$, $z_l$ ). La surface de détection, cylindrique, est ensuite « dépliée » pour obtenir un plan. Pour ce faire, on pose :

$$\begin{cases} r = \sqrt{x^2+y^2} \\ \theta_{\text{detect}} = \arctan(y/x) \end{cases} \qquad (230)$$

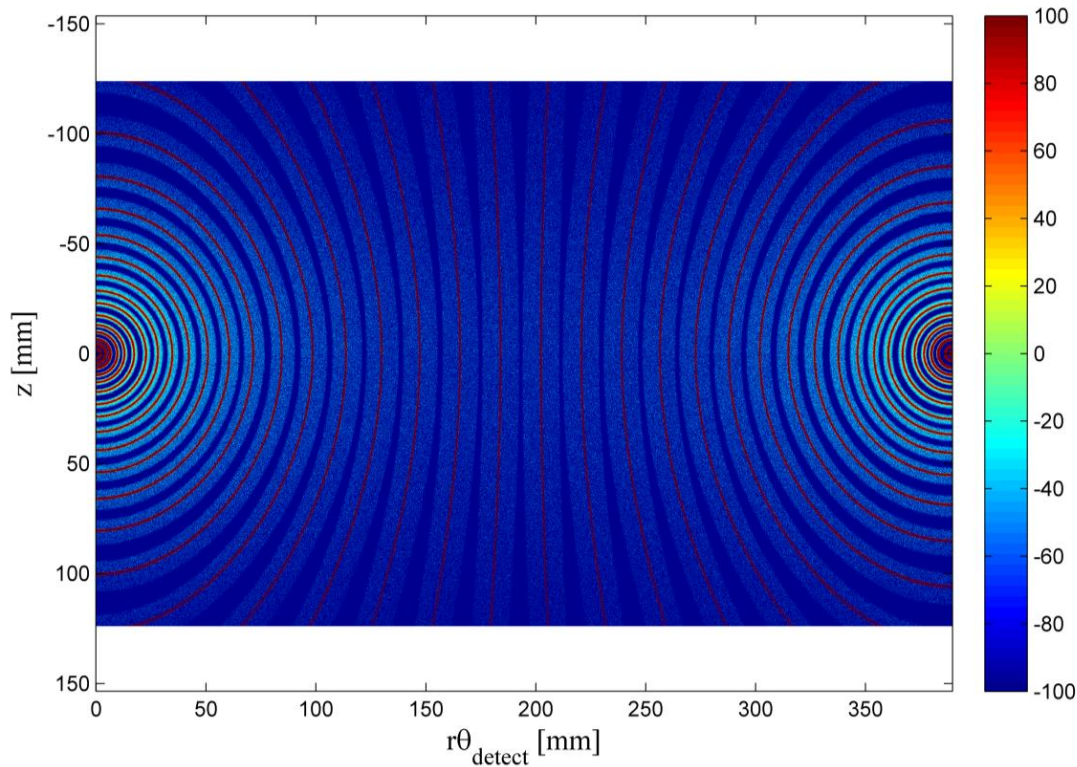

Figure 66 : Carte conditionnelle obtenue après dépliage de la zone cylindrique de détection. Le codage en couleur correspond au pourcentage des photons détectés et qui respectent le critère de résolution de ±0.5° (particules de Rayleigh)

La Figure 66 montre, en dégradé de couleurs, la carte dépliée sur un plan pour le pourcentage des photons collectés qui respectent le critère de résolution angulaire imposé. A noter que pour des raisons numériques, la surface de détection a été décomposée en pixels. Les principaux paramètres du calcul sont $\theta_{\det ect} \in [\theta - 0.5°, \theta + 0.5°]$, $R_c = 124 mm$, billes de silice nanométriques. Pour une meilleure compréhension, la valeur -100% est affectée par défaut aux zones (pixels) où aucun photon n'a été collecté. Le cas 0% correspond à des zones où aucuns des photons collectés ne respectent le critère angulaire. Bien évidemment, le cas 100% (clairement visualisé par des "lignes" rouges) correspond aux zones de détection où tous les photons impactant respectent le critère angulaire. Ce sont ces zones rouges qui doivent être utilisées pour la conception des zones photosensibles.

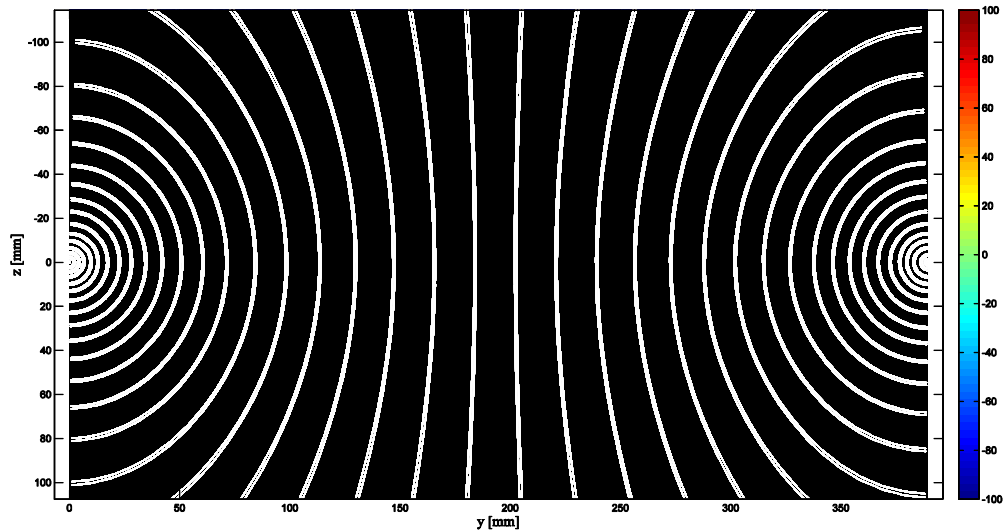

Figure 67 : Vérification de la forme des anneaux - Comparaison entre la simulation Monte-Carlo et le modèle paramétrique avec $y = r\theta_{\text{detect}}$

La Figure 67 compare ces zones rouges, obtenues à l'aide des simulations Monte-Carlo, avec les courbes « cylindro-coniques » décrites par les équations (227), (228) et (229). On remarque tout de suite l'excellent accord obtenu. Les courbes rouges sont parfaitement encadrées par les "cylindro-coniques" (en blanc), et ceci, quelque soit la hauteur $z_l$. On notera cependant que, selon le milieu particulaire à caractériser, le design optimal des zones photosensibles peut évoluer. En effet, si on observe la carte conditionnelle de la Figure 68, calculée pour des particules en suspension dans l'eau et dans un tube en verre, la forme et la taille des zones rouges respectant le critère de résolution fixé évoluent. Cela signifie que dans l'état, un masque spécifique doit être créé à chaque fois que le milieu fluide change.

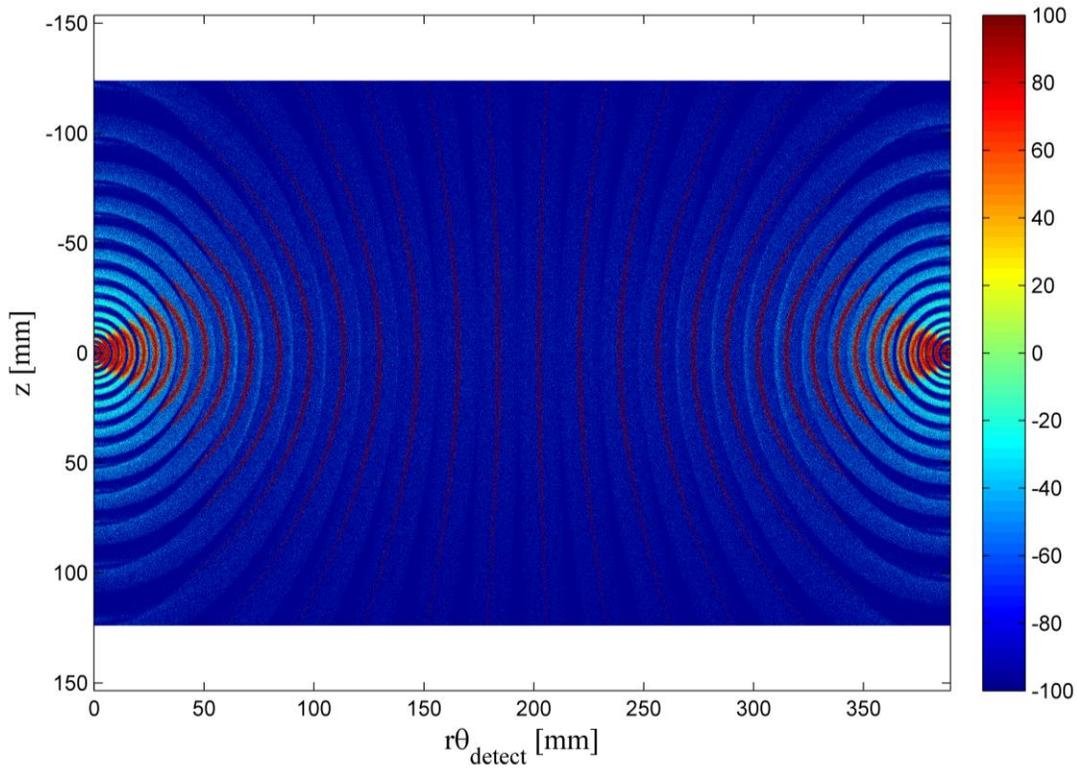

Figure 68 : Carte conditionnelle pour le pourcentage de photons détectés respectant le critère de résolution de $\pm 0.5°$ - cas d'une suspension aqueuse de particules

La Figure 69 montre la carte en intensité correspondante (échelle logarithmique de base e). Pour cette simulation, la distribution en taille des particules est monodisperse ($D = 1\mu m$). Plusieurs constatations peuvent être faites :

- comme nous l'avions déjà montré en 2D avec les trous optiques circulaires : l'intensité collectée est minimale et la largeur des halos est maximale vers 90°,

- par ailleurs les diagrammes de diffusion collectés sont déformés par la projection sur le cylindre : la largeur des halos lumineux dans le plan de diffusion (défini par $z_i = z = 0$) est inférieure à celle dans le plan perpendiculaire contenant l'axe z (c.-à-d. le plan d'équation $x_i = 0$).

- la largeur des halos est également variable : minimale dans le plan de diffusion et maximale dans le plan vertical.

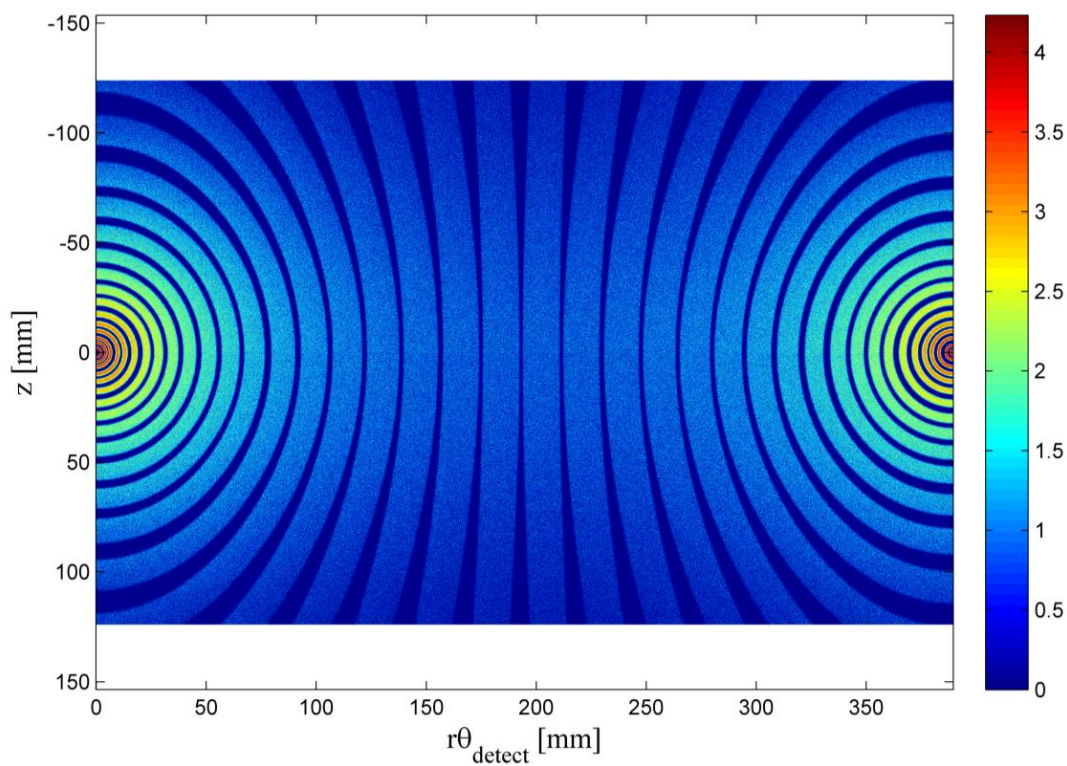

Figure 69 : Carte de l'intensité lumineuse collectée sur le cylindre de détection déplié

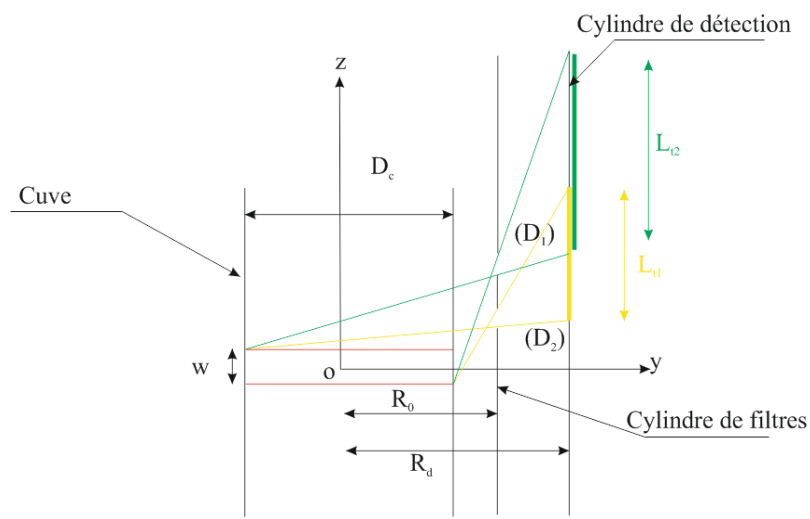

Figure 70 : Modèle géométrique expliquant l'évolution des paramètres des halos de lumière selon la côte z (par simplicité, le plan $x_l = 0$ est utilisé)

Un modèle géométrique simplifié permet d'expliquer ce dernier effet, ainsi que la variation d'intensité observée qui est imputable à une épaisseur optique variable du nuage particulaire, voir la Figure 70. La tache lumineuse $L_{r1}$ (en jaune) apparait clairement plus petite que le halo (en vert) $L_{t2}$. Plus les trous sont positionnés à une côte élevée, plus la droite (D$_1$) a une pente importante, ceci apparaissant déjà très clairement à la Figure 69. La Figure 71 fournit des données quantitatives sur la dépendance selon z de la taille et du centre des halos. Enfin la variation des coefficients de Fresnel aux interfaces milieu/verre et verre/air se rajoute à ses effets.

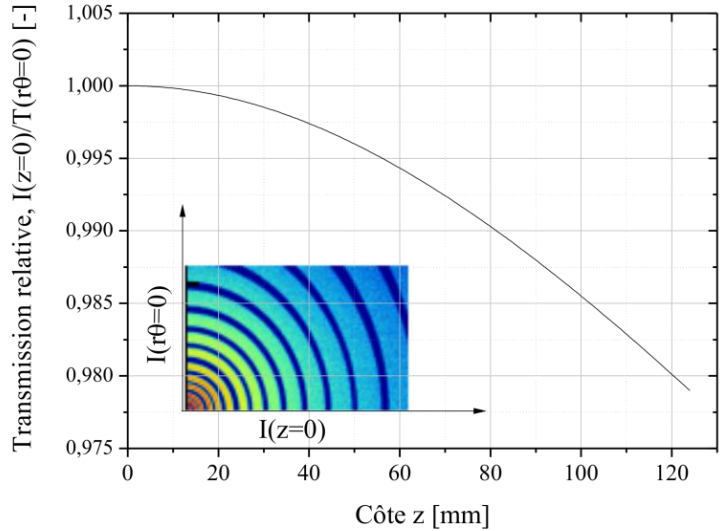

Figure 71 : Évolution du rapport des intensités collectées dans le plan de diffusion et le long de l'axe du cylindre

En régime de diffusion simple, on s'attend également à ce que l'intensité sur un même anneau diminue très légèrement quand $z$ augmente du fait de la variation de l'épaisseur du milieu particulaire traversé. Dans le plan de diffusion, cette largeur est minimale, avec $L \equiv D_c$ (cf. équation (186)), alors que pour les détecteurs dans le plan $x_l = 0$ on a : $L \equiv D_c / \cos\left(\tan^{-1}\left(z/R\right)\right)$.

Pour un même anneau, le rapport entre son intensité dans le plan de diffusion et perpendiculairement à celui-ci vaut :

$$R_l = I^{1/\cos\left(\tan^{-1}\left(z/R_d\right)\right)}$$

(231)

La Figure 71 montre que cette atténuation reste faible, puisqu'elle n'excède pas 2% pour nos paramètres.

La Figure 72 est identique à la Figure 69 à ceci près qu'elle a été obtenue pour des particules de diamètre $D=100\mu m$. On peut remarquer plusieurs choses :

- la dynamique est beaucoup plus importante, de sorte que certains anneaux (lobes de Mie) semblent avoir disparu alors qu'ils sont en pratique plus nombreux,

- des variations importantes de l'intensité apparaissent au sein même de certains anneaux, y compris dans le plan de diffusion. Ces variations sont dues à l'anisotropie "naturelle" des diagrammes de diffusion (la largeur angulaire de certains lobes est inférieure à la résolution angulaire du système). Les variations hors de ce plan sont "amplifiées" par des effets parasites : réflexion totale sur les parois de la cuve, variation des coefficients de Fresnel,…

- de même, un élargissement des anneaux est visible dans le plan de diffusion par rapport au cas sans cuve et sans eau. Cet effet est simplement dû à la cuve qui se comporte comme une lentille cylindrique,

- une légère atténuation de l'intensité le long des anneaux, dans le plan vertical, a été mise en évidence numériquement . Elle est très certainement due aux variations des coefficients de Fresnel et à un effet guide d'onde pour les grands angles de diffusion (réflexion totale sur et entre les parois de la cuve).

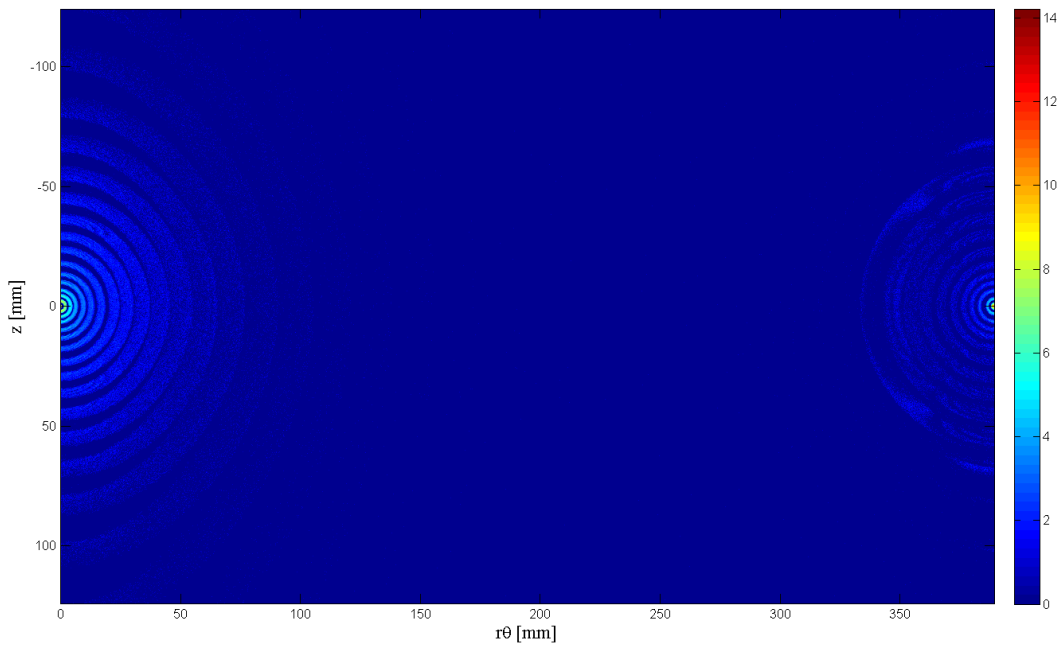

Figure 72 : Carte d'intensité sur le cylindre de détection pour des particules de diamètre $D = 100\mu m$ avec prise en compte de la cuve en verre et du milieu contenant les particules (eau)

A partir de ces calculs analytiques ainsi que des simulations numériques, un masque (de filtres) optique est créé, le masque des photodiodes organiques étant homothétique à ce dernier.

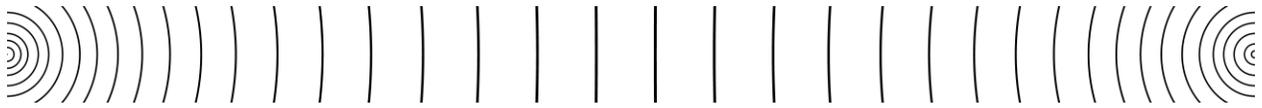

Figure 73 : Image (au format bitmap 1200dpi) du masque (de filtres) optique obtenu après optimisation du système - Les zones noires représentent la forme des photodiodes organiques devant être conçues par le CEA

Ce dernier permet, en plus d'augmenter le flux autour des 90°, de compenser la dynamique du signal (petites surfaces près de 0° et grandes autour de 90°). Cependant cette évolution des surfaces photosensibles est à prendre en compte dans l'expression de l'intensité moyenne collectée. En effet, la surface des photodiodes précédentes étant constante, l'intensité collectée présentait donc simplement un offset par rapport à la valeur attendue. Pour déterminer la valeur de ces surfaces $S(\theta)$ selon l'angle de collection $\theta$, plusieurs approches sont mises en place. L'approche par intégration mathématique des cylindro-coniques étant extrêmement complexe, nous nous sommes focalisés sur deux autres méthodes numériques.

La première méthode consiste à simuler par le code de Monte-Carlo le diagramme de diffusion obtenu pour un nuage monodisperse de particules nanométriques dans l'air, pour la polarisation perpendiculaire et sans filtrage optique (et sans cuve). Ainsi le volume de mesure disparaît et les variations du diagramme de diffusion sont directement liées à l'évolution surfacique des détecteurs. Ces résultats sont reportés en rouge sur la Figure 74 .

La seconde méthode consiste à transformer les cylindro-coniques (décrites par des équations paramétriques) en surfaces pixélisées, voir la Figure 73. Le principe de l'algorithme développé sous Matlab est tout d'abord de créer une carte pixélisée haute résolution (1200dpi) où tous les pixels sont initialisés à 1. S'il y a intersection entre la courbe décrite mathématiquement et un pixel, ce dernier est mis à 0 (les zones photosensibles optimales apparaissent donc en noir). Ces deux premières étapes permettent seulement de déterminer le contour des surfaces recherchées, il faut alors mettre tous les pixels compris entre les deux courbes pixélisées à 0. La Figure 74 propose une comparaison du résultat issu du calcul numérique des surfaces (en noir) à celui obtenu avec le code de méthode de Monte-Carlo (en rouge). Ces deux méthodes présentent un excellent accord.

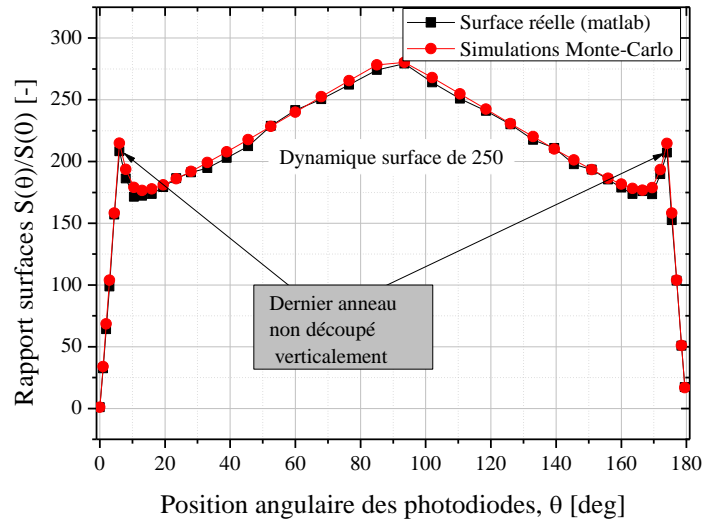

Figure 74 : Évolution de la surface des photodiodes selon l'angle de collection $\theta$ par méthode de Monte-Carlo et par simulation avec Matlab

Ces rapports de surface (Figure 74) peuvent être utilisées pour corriger les diagrammes de diffusion bruts des effets de l'évolution angulaire des surfaces photosensibles et du volume de mesure. On obtient ainsi un diagramme expérimental qui approche le diagramme théorique recherché :

$$\frac{\bar{I}(\theta)}{I_{nano}(\theta)I(\theta \approx 0)S(\theta)} \propto \int_0^\infty I(\theta,D)n(D)dD \tag{232}$$

On peut également chercher à optimiser les surfaces photosensibles pour améliorer la dynamique du système du granulomètre. Pour ce faire, il faut notamment réduire la dynamique en intensité des signaux optiques et donc ajuster les surfaces photosensibles. La Figure 75 montre un résultat obtenu, pour une configuration avec eau, lorsque seule la hauteur des anneaux est ajustée. Dans ce dernier cas, pour des raisons de design imposées par le CEA, l'évolution de cette hauteur est simplement linéaire avec l'angle (croissante de 0 à 90° puis décroissante de 90° à 180°). En fait, une évolution de type "logarithmique" des surfaces photosensibles aurait été plus indiquée. De même, la symétrie gauche-droite du masque de la Figure 75 est liée aux contraintes de fabrication des détecteurs photo-organiques.

Pour réaliser des masques optiques, les contraintes sont beaucoup plus faibles. Différentes solutions existent, néanmoins, comme ceux-ci doivent être morphologiquement identiques aux capteurs photo-organiques nous avons produit uniquement des masques dont le design est conforme à la Figure 75.

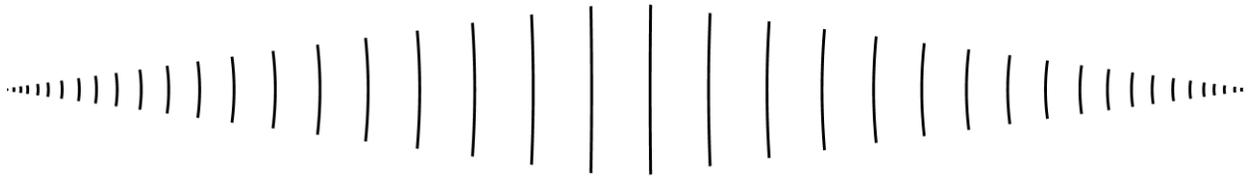

Figure 75 : Masque optique optimisé verticalement

### 4.1.5 Validation du montage

Pour réaliser des comparaisons entre le modèle de Monte-Carlo et la théorie de Lorenz-Mie, il faut appliquer un filtre passe-bas aux diagrammes obtenus avec cette dernière. Il s'agit d'une méthode simple, voire simpliste, pour prendre en compte l'ouverture angulaire du détecteur avec la théorie de Lorenz-Mie. La Figure 76 montre un résultat typique obtenu pour des détecteurs dont l'ouverture angulaire totale est de 1° et des billes de verre monodisperses de $D = 100\mu m$ L'accord est globalement assez bon, sauf pour les angles proches de 90° où le bruit statistique de la méthode de Monte-Carlo est important (intensité faible rime avec évènements peu probables). La méthode employée pour la normalisation des diagrammes influe également beaucoup sur l'appréciation globale de l'accord. La Figure 77 montre les résultats obtenus pour des particules plus petites. L'accord semble bien meilleur pour ces particules dont la dynamique des diagrammes est plus faible. Néanmoins, nous insistons sur le fait que dans la Figure 76, c'est la normalisation qui pose problème, pas le code de Monte-Carlo.

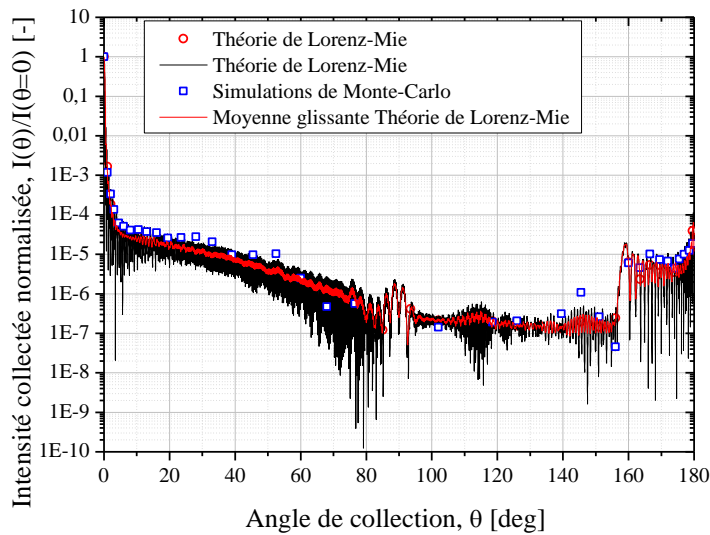

Figure 76 : Illustration des différentes étapes pour comparer les résultats obtenus pour le prototype et la théorie de Lorenz-Mie. Cas pour un jet de billes monodisperses, dans l'air, de diamètre $D = 100\mu m$

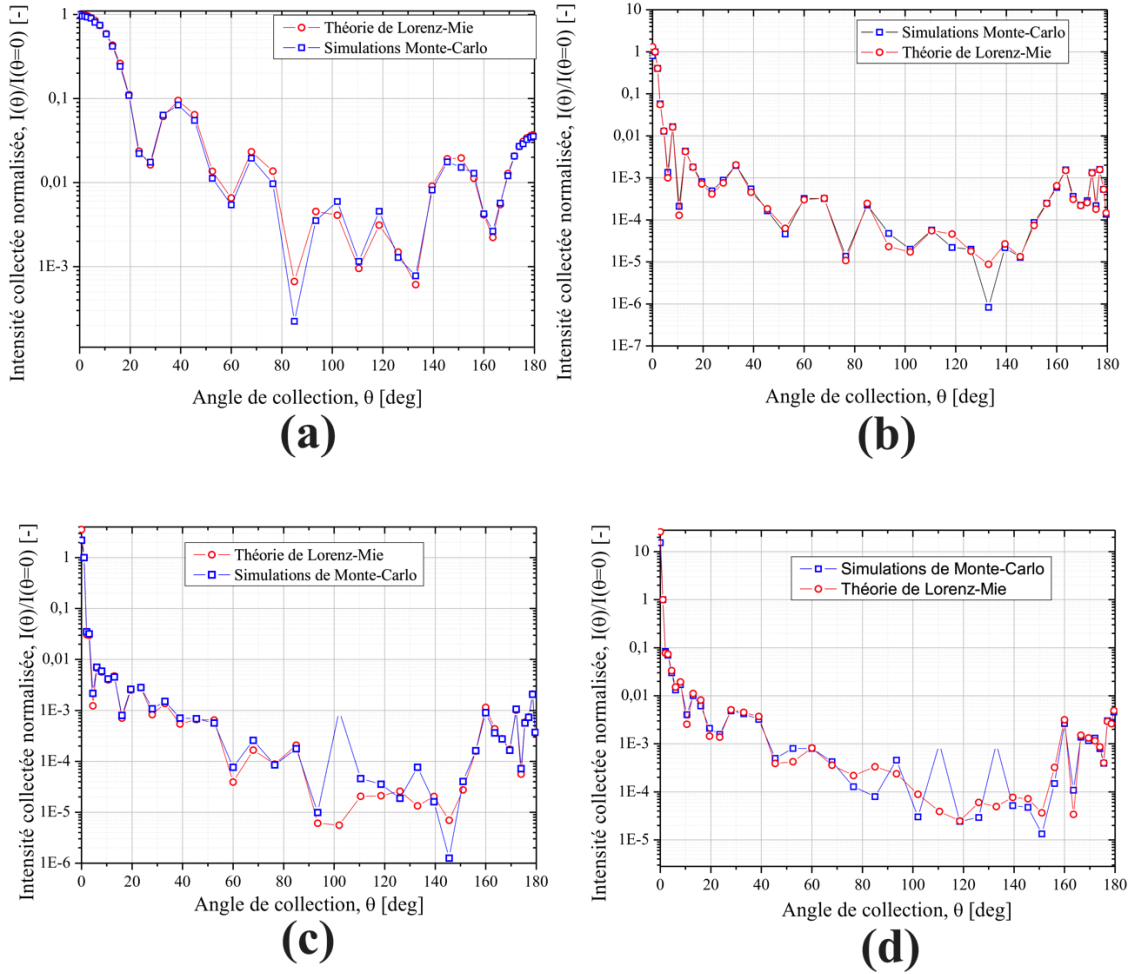

Figure 77 : Comparaisons des diagrammes de diffusion obtenus pour le prototype développé pour la méthode de Monte-Carlo et la théorie de Lorenz-Mie. Cas d'un jet de billes de verre monodisperses dans l'air pour (a) $D = 1\mu m$, (b) $D = 10\mu m$, (c) $D = 20\mu m$ et (d) $D = 30\mu m$

La Figure 78 propose une comparaison des prédictions de la théorie de Lorenz-Mie et de la méthode de Monte-Carlo dans le cas plus réaliste d'une suspension aqueuse de billes de verre monodisperses ($\sigma_D / \overline{D} = 0.1\%$) en conduite cylindrique.

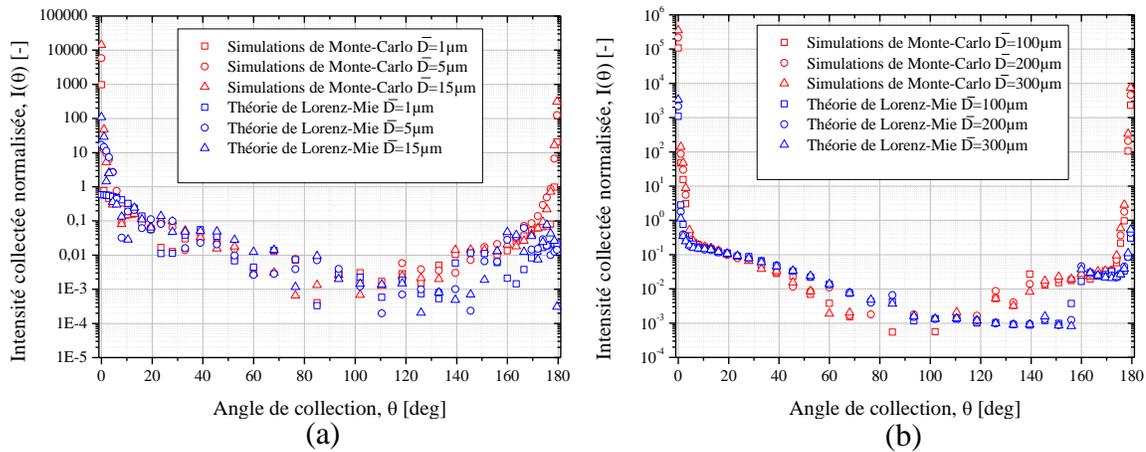

Figure 78 : Comparaison des diagrammes de diffusion obtenus par simulation de Monte-Carlo pour le montage précédent et les diagrammes de diffusion obtenus par la théorie de Lorenz-Mie pour (a) des petites particules et (b) des grosses particules. Le paramètre d'itération est le diamètre moyen (ordre 0) et les distributions sont monodisperses ( $\sigma_D / \overline{D} = 0.1\%$ )

Les simulations numériques et la théorie de Lorenz-Mie semblent en bon accord, à l'exception de l'arrière (au-dessus de 120°) et de l'avant (en-dessous de 10°) du diagramme de diffusion. On peut aisément relier ce phénomène à l'existence de réflexions parasites sur les parois externe et interne de la cuve cylindrique pour l'arrière du diagramme et au faisceau direct pour l'avant du diagramme. Ces réflexions augmentent l'intensité diffusée vers l'arrière : réflexion spéculaire et même diffraction par les particules. Ces effets ne sont bien évidemment pas pris en compte par la théorie de Lorenz-Mie.

A l'issue de cette partie, nous concluons que le modèle de Monte-Carlo, la méthode de correction du volume de mesure, le design des masques et des capteurs photo-organiques sont validés et donnent satisfaction. Notre conception permet bien de retrouver les résultats de la théorie de Lorenz-Mie, c'est-à-dire de minimiser certains effets de la cuve et d'être quasiment dans l'approximation du détecteur lointain. Mais ceci n'est vrai que sous certaines conditions (diffusion simple, couples d'indices, distances, dimensions...) et pour certaines régions angulaires (en dehors de la zone de rétrodiffusion et de la diffusion vers l'avant). Dans le cas favorable, la théorie de Lorenz-Mie doit être utilisée à la place de la méthode de Monte-Carlo (très gourmande en temps de calculs).

### 4.1.6 Inversion des diagrammes

Pour inverser les mesures des systèmes MALS, plusieurs approches sont possibles [**Xu 2002**]. Nous n'évoquons ici que des pistes compatibles avec les contraintes du PAT, c'est-à-dire des méthodes simples, et dans la mesure du possible, originales.

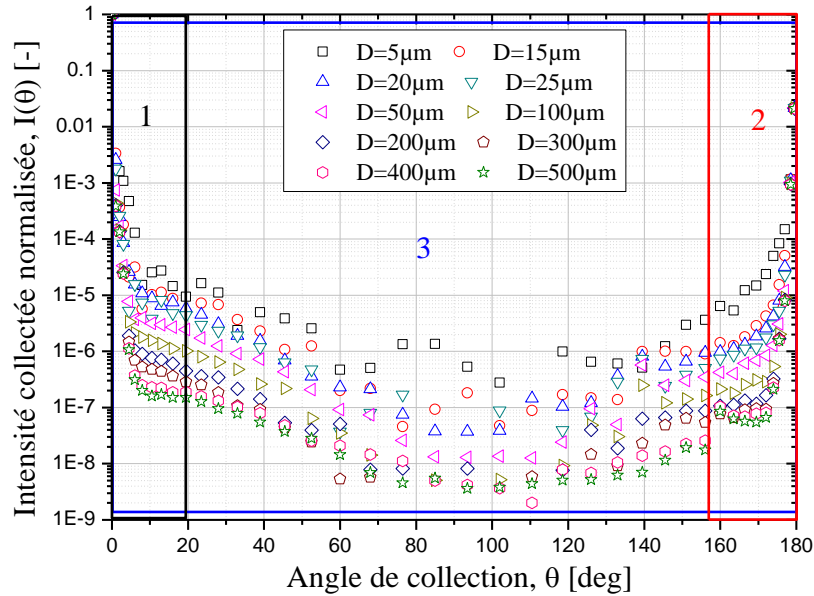

Figure 79 : Illustration de quelques techniques d'inversion de diagrammes de diffusion

On peut utiliser une des trois zones du diagramme de diffusion "expérimental" (généré avec les simulations Monte-Carlo donc "expérimental numérique") pour trouver un phénomène directement relié à la nature intrinsèque des particules, voir la Figure 79. Sur les premiers degrés du diagramme (zone 1, en noir), le phénomène de diffraction est visible. Son motif intense et facilement modélisable dans le cadre de l'optique physique est largement utilisé pour caractériser les systèmes particulaires [**Xu 2002**]. Malheureusement, en raison des limites technologiques actuelles des OPS, la densité des capteurs (nombre par unité de surface ou longueur) n'est pas encore suffisante pour caractériser de manière satisfaisante cette région.

Une autre méthode classique consiste à caractériser la rétrodiffusion ou diffusion de côté (zone 2 en rouge dans la Figure 79). Dans cette zone du diagramme, on peut par exemple observer des phénomènes singuliers comme l'arc-en-ciel (gouttes et particules solides) [**Bohren 1998**] ou la diffusion critique (bulles). On peut voir sur le diagramme de diffusion de la Figure 79 que la hauteur relative du "plateau" arc-en-ciel est directement reliée à la taille des particules, alors que sa position angulaire absolue dépend essentiellement de leur indice de réfraction.

La dernière méthode, qui permet d'utiliser de manière optimale les photodiodes organiques, repose sur l'utilisation du diagramme de diffusion dans son intégralité à travers une fonction d'asymétrie (zone 3, en bleu).

### 4.1.6.1 Méthode du rapport d'intensité

L'arc-en-ciel est un phénomène très étudié en tant que singularité, comme pour le développement de méthodes inverses de caractérisation des milieux particuliers (voir par exemple, le cas des sprays libres [**Beeck 1996**]). Cependant, tous les travaux expérimentaux reposent sur une analyse (au centième de degré près, avec une caméra CCD haute résolution) de la position angulaire des premières franges d'Airy, voire du diagramme complet avec une inversion algébrique [**Onofri 2012**]. La résolution angulaire de notre système est bien trop limitée pour ce type d'analyse.

*4.1.6.1.1 Cas d'un ratio arc-en-ciel / diffusion avant*

Nous proposons ici d'utiliser un simple ratio d'intensité entre deux régions angulaires très distinctes. Sur la Figure 79, dans la zone de rétrodiffusion, on remarque un minimum dans le "plateau" correspondant approximativement à la région du premier arc-en-ciel (de 160- 170°) de billes de verre dans l'air. La Figure 80 montre l'évolution du rapport entre l'intensité correspondant approximativement au minimum local de la zone arc-en-ciel et une intensité mesurée vers l'avant à $\theta = 0°$ (c'est-à-dire une mesure du pic de diffraction et du faisceau direct). Ce rapport, $\Re_1(\theta) = I(\theta = 172°) / I(\theta = 0°)$, sur lequel repose cette première méthode, est donné pour différents diamètres moyens de la distribution granulométrique et différentes transmissions $T$ du faisceau direct. Son évolution avec le diamètre peut être décomposée en deux cas distincts.

Dans la zone 1, c'est-à-dire pour les "petites" particules, la relation entre le rapport d'intensité $\Re_1$ et le diamètre des particules ne présente pas de comportement asymptotique marqué. Ceci s'explique par le fait que, pour ces particules, la structure basse fréquence des diagrammes de diffusion évolue très rapidement avec la taille. La résolution de notre système est en quelque sorte trop importante pour lisser les lobes de Mie. Globalement, les tendances entre la théorie de Lorenz-Mie et les simulations de Monte-Carlo sont assez similaires avec cependant des pentes plus marquées pour la théorie de Lorenz-Mie que pour les simulations Monte-Carlo (qui prennent en compte les réflexions de la cuve et le faisceau direct), et ce d'autant plus que l'on augmente la transmission. En fait, pour les petites particules, l'augmentation de la transmission et donc de la pollution relative des diagrammes par les réflexions parasites, entraine une diminution de la sensibilité du rapport $\Re_1$ au diamètre. De fait, l'exploitation de ce rapport pour la granulométrie des petites particules semble inenvisageable.

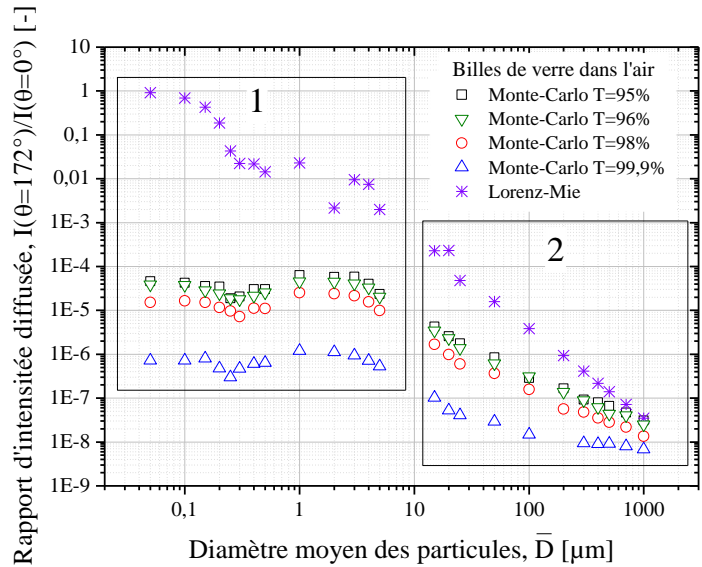

Figure 80 : Évolution du rapport entre l'intensité minimum du plateau arc-en-ciel et l'intensité au voisinage du pic de diffraction pour différents diamètres moyens et transmissions

Dans la zone "2", c'est-à-dire dire les "grosses" particules, on constate en revanche que le rapport d'intensité suit une loi de puissance en $\overline{D}^\gamma$ offrant une bonne sensibilité avec le diamètre moyen. La Figure 81 se focalise sur l'évolution de ce rapport pour la zone 2, en fournissant une estimation des pentes (et coefficients de corrélation). On constate que pour des transmissions "pas trop faibles", les pentes obtenues avec la méthode de Monte Carlo sont très voisines ($\gamma$ entre -1.11 et -1.15) quelque soit la transmission. La théorie de Mie prévoie également un comportement asymptotique mais de pente différente. Pour une transmission très élevée ($T = 99.9\%$ dans notre exemple), le milieu particulaire est tellement dilué que les réflexions parasites polluent complètement le signal de diffusion avant (voir la Figure 82) qui devient, *in fine*, inexploitable. Ce cas limite ne sera plus discuté dans ce qui suit.

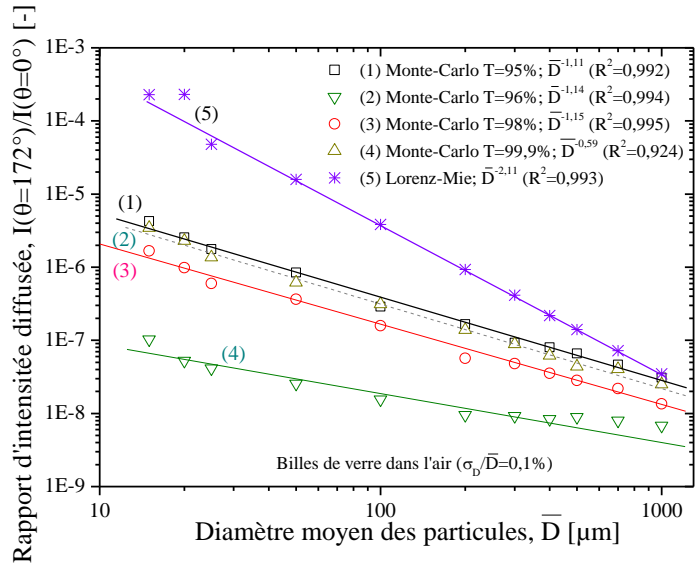

Figure 81 : Évolution du rapport $\Re_1$ pour des billes de verre dans l'air en fonction de leur diamètre moyen et pour différentes transmissions

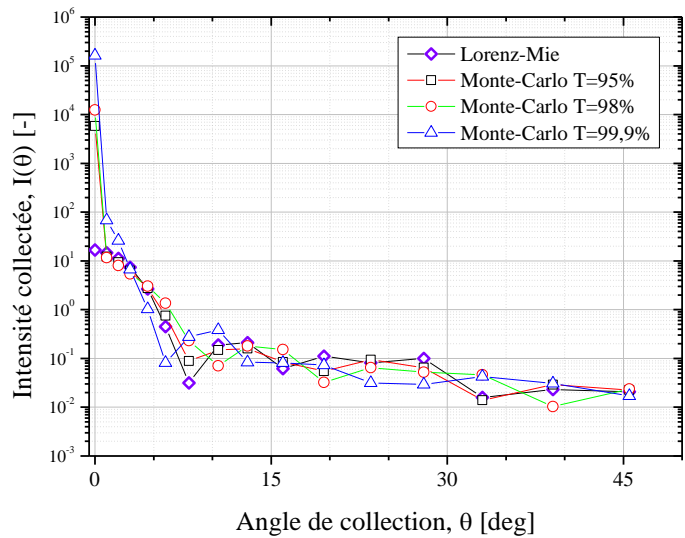

Figure 82 : Comparaison des diagrammes de diffusion de billes de verre ($\overline{D} = 5\mu m$) calculés pour différentes valeurs de la transmission, avec la théorie de Lorenz-Mie et le code Monte-Carlo

Si la transmission (et donc la concentration du milieu) semble très légèrement modifier l'exposant de la loi de puissance, son effet le plus significatif porte sur l'offset de la loi obtenue. La Figure 81 montre très clairement que cet offset $\eta(T)$ dépend du taux de transmission du faisceau, de sorte que le rapport d'intensité mesuré est plutôt de la forme :

$$\Re_1(D) \approx \beta D^\gamma + \eta(T) \tag{233}$$

En pratique, pour déduire le diamètre moyen des particules il faut donc mesurer le taux de transmission du faisceau et le rapport $\mathfrak{R}_1$, de même que disposer d'un modèle permettant d'estimer l'exposant $\gamma$, le coefficient multiplicateur $\beta$ et la fonction $\eta(T)$. Les deux premières quantités sont mesurables avec notre prototype et les trois dernières peuvent être calculées avec le code de Monte-Carlo.

A partir de calculs comme ceux présentés par la Figure 81, on peut estimer à l'aide d'une régression linéaire (en échelle log-log) : $\beta, \gamma$ pour différentes valeurs de la transmission et donc $\eta(T)$. En mesurant le ratio $\mathfrak{R}_1(D)$ et le taux de transmission $T$ on peut donc directement estimer le diamètre moyen :

$$D_{inv} \approx \left( \frac{\mathfrak{R}_1(D) - \eta(T)}{\beta} \right)^{1/\gamma} \tag{234}$$

La Figure 83 propose une comparaison des diamètres nominaux $D_{th}$ et ceux obtenus $D_{inv}$ avec cette méthode. Pour ce faire les mesures "numériques" sont simplement interprétées comme étant des mesures expérimentales. On constate que les estimations sont globalement correctes compte tenu de la simplicité de la méthode, même si l'erreur relative moyenne est de 10%, avec certaines fluctuations atteignant près de 20%. L'utilisation de courbes de régression plus complexes permettrait certainement d'améliorer ces premiers résultats, surtout pour les plus grosses particules (D>600µm).

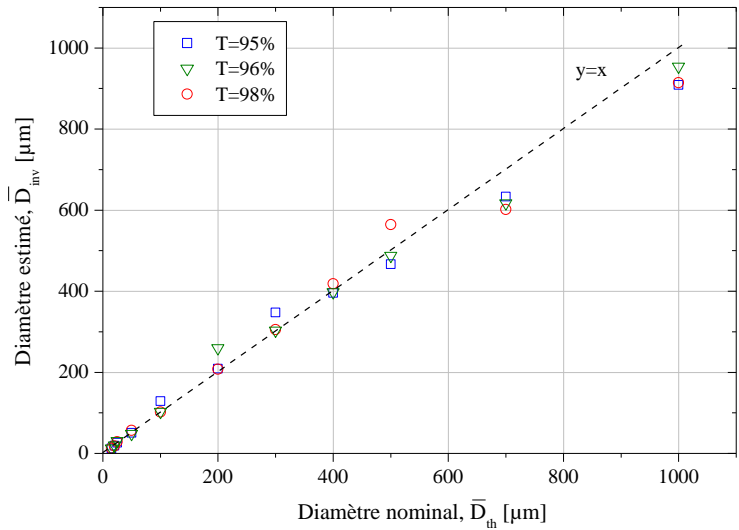

Figure 83 : Évolution du diamètre estimé par inversion par rapport au diamètre en entrée du programme et calcul de l'erreur relative de la méthode d'inversion sur l'estimation du diamètre moyen des particules pour la zone "2"

*4.1.6.1.2  Cas d'un ratio arc-en-ciel / rétrodiffusion*

Nous avons vu que le rapport $\Re_1(\theta)$ est fortement contaminé par le faisceau direct via le taux de transmission. Nous avons recherché d'autres ratios moins sensibles à cet effet. La Figure 84 montre l'évolution du rapport d'intensité $\Re_2(\theta) = I(\theta = 172°)/I(\theta = 179.5°)$. $\Re_2$ propose une comparaison entre l'intensité du signal arc-en-ciel et celle du signal diffusé vers l'arrière. Ce dernier est composé de la rétrodiffusion du système particulaire (au sens de Mie), et d'une fraction de la diffraction vers l'avant qui a été redirigée vers l'arrière par la cuve et une inévitable contribution des réflexions de la cuve. A noter que du fait de la courbure du cylindre la diffraction rétrodiffusée est en quelque sorte étalée angulairement, ce qui n'est pas une mauvaise chose compte tenu de la résolution angulaire très limitée de notre prototype. Son comportement est assez similaire à celui du rapport $\Re_1(\theta)$ et les erreurs relatives sont sensiblement du même ordre de grandeur, voir la Figure 85 et la Figure 86. Ce rapport permet néanmoins de limiter les accès optiques requis (un seul hublot peut suffire à la mesure).

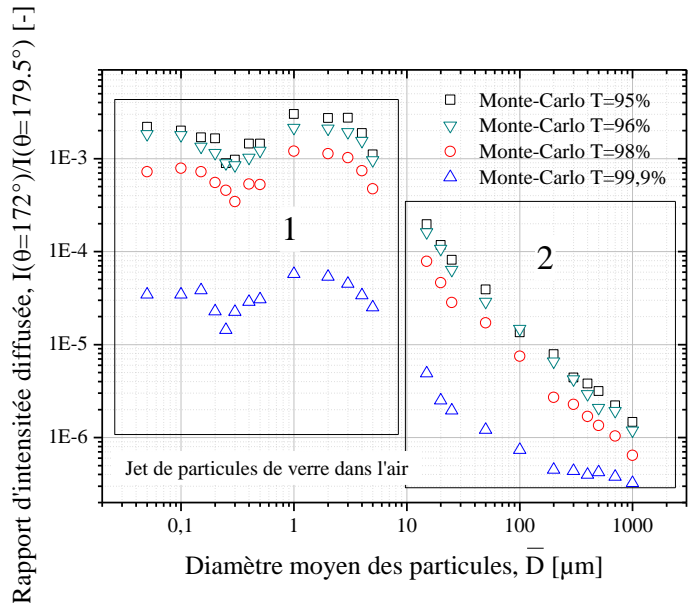

Figure 84 : Évolution du rapport d'intensité I(172°)/I( 179.5°) avec le diamètre moyen de billes de verre dans l'air et pour différentes transmissions

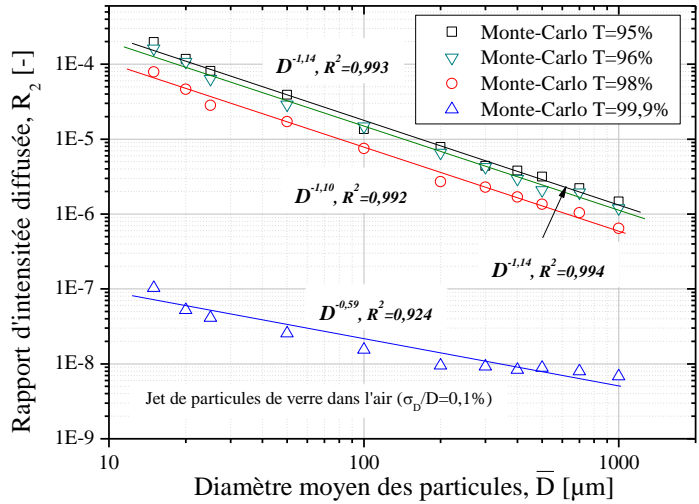

Figure 85 : Évolution du rapport d'intensité $\Re_2 = I(172°)/I(179.5°)$ avec le diamètre moyen

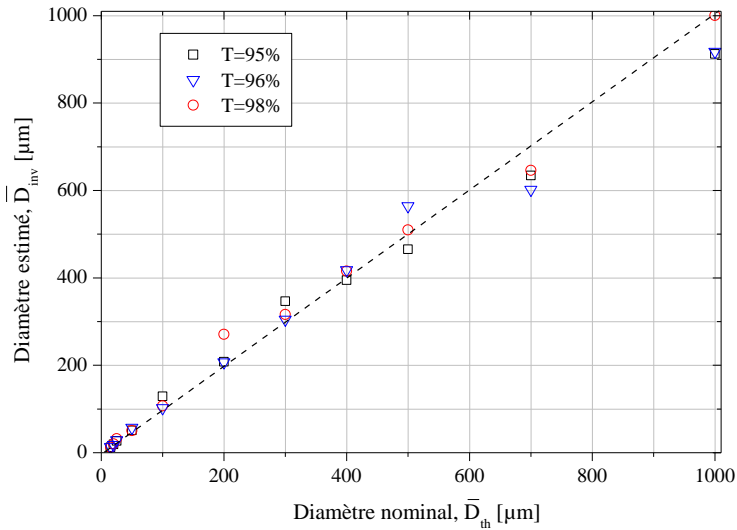

Figure 86 : Évolution du diamètre estimé avec le rapport d'intensité $\Re_2$.

### 4.1.6.1.3 Éléments d'interprétation sur les lois de puissance observées

L'origine des lois de puissance observées est difficile à interpréter précisément du fait des réflexions parasites directes et indirectes. Néanmoins, dans le cas d'un jet libre (sans cuve) et du rapport $\Re_1$, on peut distinguer trois contributions possibles: Airy, Fraunhofer et optique géométrique.

La théorie de Fraunhofer (cf. **chapitre 2**) donne l'amplitude du champ diffusé dans la zone de diffraction :

$$S_d(\theta) = x^2 \frac{J_1(x\sin(\theta))}{x\sin(\theta)} \tag{235}$$

avec $J_1(x\sin(\theta)) = \sum_{p=0}^{\infty} \frac{(-1)^p}{p!(1+p)!}\left(\frac{x\sin(\theta)}{2}\right)^{2p+1}$ et $x$ le paramètre de taille de la particule. Au voisinage de $\theta \approx 0°$ le développement limité de l'équation (235) vaut :

$$\lim_{\theta \to 0}\left(x^2 \frac{J_1(x\sin(\theta))}{x\sin(\theta)}\right) = \frac{x^2}{2} \tag{236}$$

L'intensité au voisinage de $\approx 0°$ est alors proportionnelle à :

$$I(\theta \approx 0) \propto D^4 \tag{237}$$

Pour l'arc-en-ciel, la théorie bidimensionnelle d'Airy indique une dépendance de l'intensité diffusée avec le diamètre en $I(\theta \approx \theta_{ac}) \propto D^{4/3}$. Soit, en prenant en compte le second rayon de courbure des particules :

$$I(\theta \approx \theta_{ac}) \propto D^{7/3} \tag{238}$$

Si, dans le cadre de l'optique physique, $\Re_1$ est bien un simple rapport entre l'intensité au voisinage de l'angle d'arc-en-ciel et celle au voisinage du pic de diffraction, il vient les relations en lois de puissance suivantes :

$$\Re_1 \propto D^{-8/3} \approx D^{-2.66} \tag{239}$$

$$\Re_1 \propto D^{-5/3} \approx D^{-1.66} \tag{240}$$

Force est de constater que les valeurs de $\gamma$ trouvées ici diffèrent fortement des valeurs numériques obtenues pour le système particulaire en cuve ($\gamma = $-1.11 à -1.15, voir la Figure 81).

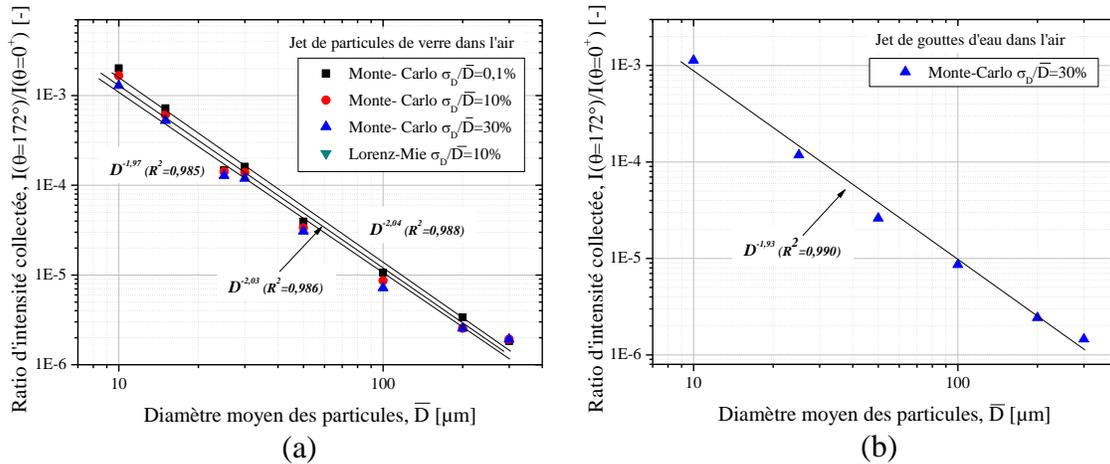

(a)                                                                (b)

Figure 87 : Évolution du ratio entre l'intensité minimum du plateau arc-en-ciel et la zone de diffraction pour différents diamètres moyens et écarts-types - sans prise en compte du faisceau direct - pour (a) un jet de particules de verre et (b) pour des gouttelettes d'eau

La Figure 87 présente les résultats obtenus, sans cuve, pour (a) des particules de verre (indice de réfraction, $m \approx 1.51$) et (b) des gouttelettes d'eau (indice de réfraction, $m \approx 1.33$) dans l'air. On remarque que, sans cuve, la pente de la loi de puissance est en $\gamma \approx -2$ avec une très faible dépendance vis-à-vis de l'écart-type de la distribution granulométrique et de l'indice des particules. D'un point de vue pratique, le fait que $\Re_1$ soit très peu sensible à l'écart-type est à la fois un atout et un inconvénient. C'est un atout car cela permet une estimation directe du diamètre moyen, avec peu de moyens informatiques. Mais c'est également un inconvénient si la largeur de la distribution est importante pour le problème traité. Dans tous les cas, pour un jet ou écoulement libre, le rapport $\Re_1$ présente une bien meilleure sensibilité au diamètre moyen et vraisemblablement une plus grande robustesse.

La valeur $\gamma \approx -2$ peut être expliquée par le fait qu'à 172°, même pour des billes de verre, on se situe à l'extrême limite de la zone du premier arc-en-ciel. Comme nous l'avons montré dans le **chapitre 2** avec le modèle hybride, l'optique géométrique décrit relativement bien cette région du diagramme de diffusion. Or, en négligeant la divergence et les interférences (de haute fréquence pour des grosses particules), l'optique géométrique pure prévoit une dépendance de l'intensité diffusée en $I(\theta) \propto D^2$. En remplaçant dans l'équation (239) la contribution d'Airy par celle de l'optique géométrique, il vient $\Re_1 \propto D^{-2}$, soit $\gamma \approx -2$. Le rapport d'intensité que nous utilisons serait donc un rapport entre l'efficacité de diffusion d'une singularité (la diffraction dans le cas de $\Re_1$) et d'une composante géométrique. Nos recherches bibliographiques indiquent que ce concept est original. Nous pensons qu'il pourrait être étendu au delà des limites imposées par les OPS actuels.

### 4.1.6.2 Méthode du paramètre d'asymétrie

Cette méthode utilise le diagramme de diffusion dans son intégralité. Le facteur d'asymétrie [**Irvine 1964**], noté $g$, quantifie l'anisotropie d'un diagramme de diffusion en comparant son intégrale à celle de sa projection sur l'axe de propagation de l'onde incidente, voir l'équation (241) :

$$g(D) = \frac{\int_0^\pi I(\theta, D)\cos(\theta)d\theta}{\int_0^\pi I(\theta, D)d\theta} \qquad (241)$$

avec :

- $g \to 0$ pour des particules diffusant de manière isotrope (particules de Rayleigh, petites devant la longueur d'onde), voir la Figure 88,

- $g \to 1$ pour les particules grandes devant la longueur d'onde et qui diffractent beaucoup la lumière,

- $g \to -1$ pour des particules diffusant uniquement vers l'arrière (un peu comme un miroir plan).

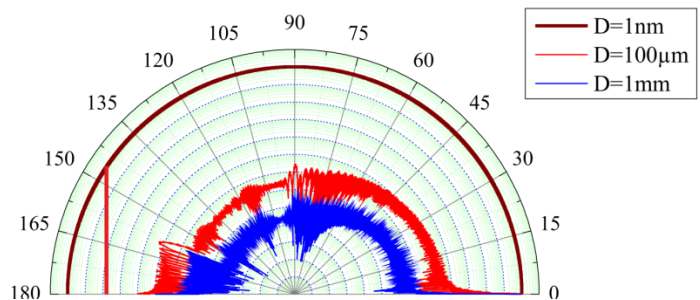

Figure 88 : Représentation polaire de trois diagrammes de diffusion et projection sur l'axe

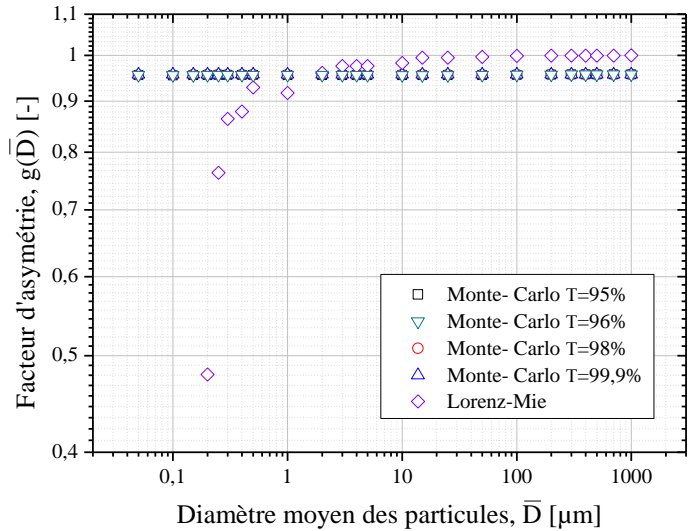

Figure 89 : Évolution du facteur d'asymétrie en fonction du diamètre moyen

La Figure 89 propose une comparaison des résultats de simulations obtenus par la théorie de Lorenz-Mie ou par le code de Monte-Carlo, dans la gamme probable de valeurs prises par le facteur d'asymétrie par un écoulement de billes de verre en cuve. On constate malheureusement que le facteur d'asymétrie est quasi indépendant du diamètre moyen des particules avec $g(D) \sim 1$. Ce sont de nouveau l'élargissement angulaire du faisceau direct et les réflexions parasites générées par la cuve qui expliquent ce comportement totalement aberrant.

Avec une détection légèrement au-dessus du plan de diffusion, la collection du faisceau direct et des réflexions spéculaires peut être évitée. Cette solution, testée au laboratoire, semble donner de très bons résultats [**Montet 2014**]. Nous avons néanmoins choisi ici de montrer, voir la Figure 90, l'évolution de $g$ lorsque la borne inférieure d'intégration, $\theta_1 > 0°$ (avec ici $\theta_1 = 1°$), est suffisamment importante pour rejeter le faisceau direct :

$$\left| g^*(D) \right| = \left| \frac{\int\limits_{\theta_1 > 0}^{\pi} I(\theta, D) \cos(\theta) d\theta}{\int\limits_{\theta_1 > 0}^{\pi} I(\theta, D) d\theta} \right| \tag{242}$$

Par commodité, nous avons choisi de représenter la valeur absolue du facteur d'asymétrie $g$ ainsi modifié, noté $\left| g^*(D) \right|$. Pour les particules dont le diamètre moyen est tel que $\overline{D} < 10 \mu m$, on observe une dépendance forte et monotone de $\left| g^*(D) \right|$ avec le diamètre, comme attendu. La

remontée de $\left|g^*(D)\right|$ pour $\overline{D}$>10μm s'explique par la diffraction particulaire renvoyée vers l'arrière par les réflexions de la cuve.

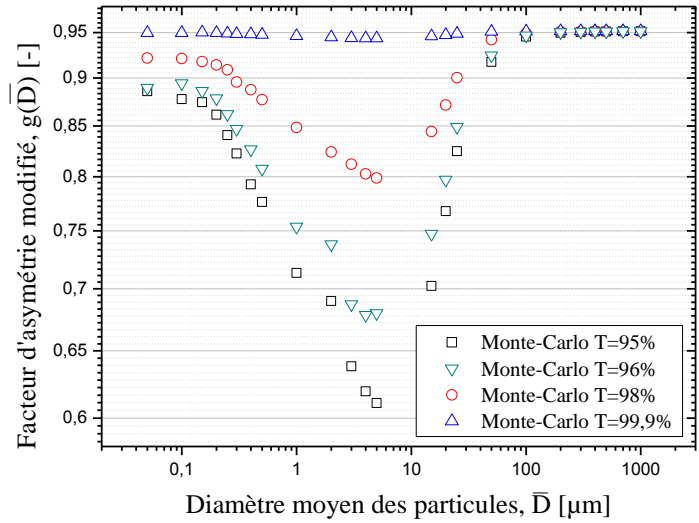

Figure 90 : Évolution du facteur d'asymétrie modifié en fonction du diamètre moyen et pour plusieurs transmissions

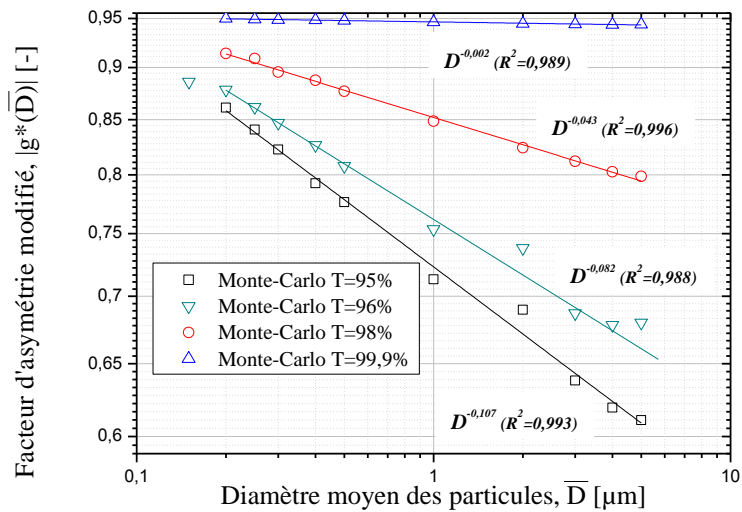

Figure 91 : Évolution du facteur d'asymétrie modifié en fonction du diamètre moyen pour plusieurs transmissions pour des diamètres entre 0.15μm et 5μm

La Figure 91 montre que sur la gamme $\overline{D}$=0.3μm-10μm, $\left|g^*(\overline{D})\right|$ évolue comme une loi de puissance. L'exposant est cependant très faible et l'offset est fortement dépendant de la transmission. La Figure 92 (a) propose une comparaison entre le diamètre nominal et celui estimé à partir des paramètres des lois de puissance de la Figure 91, une mesure de la transmission et une mesure de $\left| g^*(\overline{D}) \right|$. On constate que l'accord diamètre nominal-diamètre estimé est globalement bon. La Figure 92 (b) montre l'erreur relative correspondante qui est inférieure à 10%.

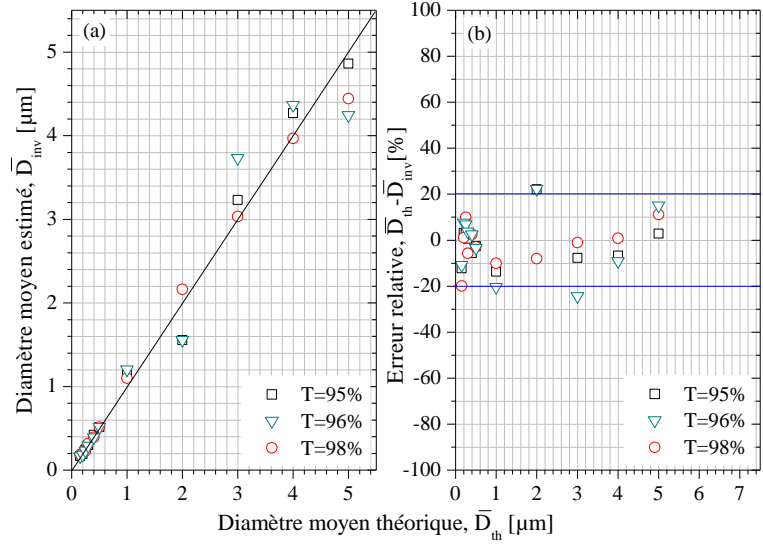

Figure 92 : Méthode de facteur d'asymétrie modifié : (a) Estimation du diamètre moyen et (b) de l'erreur relative correspondante.

### 4.1.6.3 Analyse et inversion des diagrammes avec une méthode algébrique et la théorie de Lorenz-Mie

Du fait du design du prototype, les prédictions de la théorie de Lorenz-Mie coïncident assez bien avec celles de Monte-Carlo lorsque l'on s'éloigne des zones parasitées par les effets de cuve. La Figure 78 montre que cette plage angulaire se situe entre 10° et 120°. Dans celle-ci, on peut donc légitimement utiliser la théorie de Lorenz-Mie pour inverser les données expérimentales et donc calculer des matrices d'inversion. L'intérêt principal de la théorie de Lorenz-Mie est ici de permettre des calculs beaucoup plus rapides.

L'équation (232) décrit le diagramme expérimental mesuré par notre prototype pour un angle $\theta_i$. En régime de diffusion simple, l'intensité moyenne diffusée par le milieu dans cette direction angulaire s'écrit après discrétisation :

$$I\left(\theta_i\right) = C_n \int_0^\infty I\left(\theta_i, D\right) n\left(D\right) dD = C_n \sum_{j=1}^N \mathbf{I}_{i,j}^{Mie} \mathbf{N}_j \tag{243}$$

où $n(D)$ est la distribution granulométrique en nombre que nous cherchons à déterminer, $\mathbf{I}_{i,j}^{Mie}$ l'intensité diffusée à l'angle $\theta_i$ par une particule de diamètre $D_j$ et qui est calculée avec la théorie de Lorenz-Mie, $\mathbf{N}_j$ le vecteur inconnu. La matrice de diffusion $\mathbf{I}^{Mie}$ est dans notre cas une matrice N x M =22 x 200 où chaque terme s'exprime par :

$$\mathbf{I}_{i,j}^{Mie} = \int_{D_j}^{D_{j+1}} I\left(\theta_i, D\right) dD \tag{244}$$

Ce problème peut être écrit sous forme matricielle de la manière suivante :

$$\mathbf{I} = \mathbf{I}^{Mie}\mathbf{N} \tag{245}$$

où $\mathbf{I}$ est le vecteur intensité mesurée (c'est-à-dire le diagramme de diffusion). Déterminer $\mathbf{N}$ revient à résoudre un problème inverse. Pour ce faire, nous utilisons ici une méthode dite "des moindres carrés avec solution non négative". $\mathbf{N}$ est identifié en minimisant la norme $\left\|\mathbf{I}^{Mie}\mathbf{N} - \mathbf{I}\right\|$ :

$$Min_{N>0} \left\|\mathbf{I}^{Mie}\mathbf{N} - \mathbf{I}\right\|_2^2 \tag{246}$$

où $\left\|.\right\|_2$ représente la norme L2 (c'est-à-dire, la norme euclidienne). En pratique, cette minimisation est effectuée à partir de la fonction *Lsqnonneg* implémentée sous Matlab.

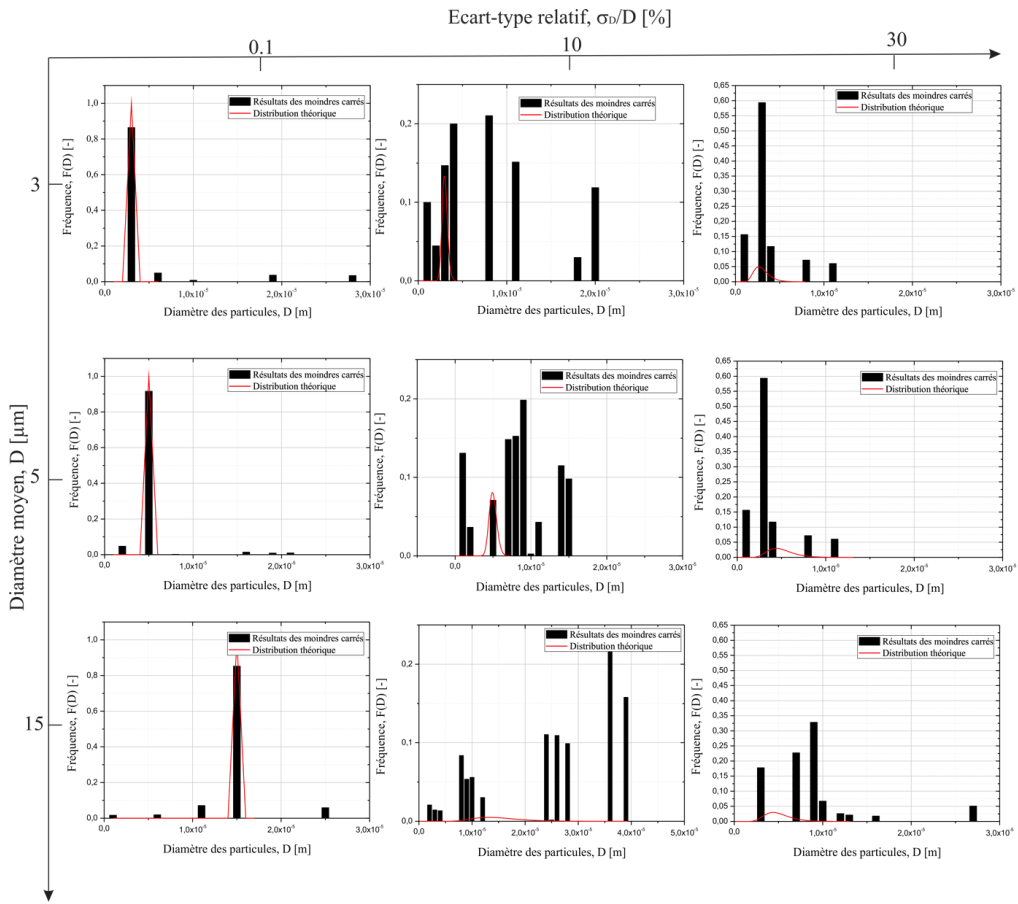

Figure 93 : Reconstruction de distribution granulométrique pour le prototype néphélométrique utilisant une méthode des moindres carrés non négative

La Figure 93 montre les résultats de quelques tests d'inversion des simulations Monte-Carlo (assimilées à des expériences réelles) avec un noyau calculé à l'aide de la théorie de Lorenz-Mie. La méthode semble fonctionner pour les distributions quasi-monodisperses ($\sigma_D / \overline{D} = 0.1\%$), même si quelques artefacts sont visibles. En revanche, les résultats deviennent catastrophiques dès lors que l'écart-type devient significatif. Il y a plusieurs explications à cela. La première est que sur l'intervalle 10-120° la morphologie des diagrammes de diffusion des grosses particules devient quasi-indépendante de leur taille (c'était déjà une des conclusions du **chapitre 2**) à l'exception des lobes de Mie (qui, pour les grosses particules, sont rapidement lissés du fait de la polydispersion de l'échantillon). La seconde explication est qu'il n'y a que 18 détecteurs sur l'intervalle 10-120°. Par ailleurs et paradoxalement, les résultats des simulations Monte-Carlo sont probablement plus bruités par les fluctuations statistiques (on sait que les méthodes inverses sont très sensibles au bruit) que ne l'est un signal expérimental. Et enfin le problème n'est pas régularisé, c'est-à-dire qu'aucune contrainte n'a été imposée sur la forme de la solution recherchée, à l'exception de son caractère positif [**Hansen 2010**].

## 4.2   Configuration diffractométrique

La configuration étudiée ici repose sur l'analyse de la figure de diffraction [**Xu 2002**] en respectant les contraintes de fabrication actuelles des photodétecteurs organiques. Comme expliqué auparavant, cette zone est très difficilement caractérisable avec les OPS produits par le CEA. Pour pallier en partie ce problème, nous utilisons ici une lentille de Fourier tout en conservant l'atout de conformabilité des OPS.

### 4.2.1  Design

Le design repose sur une lentille plan-convexe sphérique de focale $f = 100 mm$, de diamètre $D_l = 51.5 mm$ et d'indice de réfraction $n_l = 1.515$, voir la Figure 94. La cuve, de base carrée, permet de limiter l'étalement angulaire du faisceau direct qui pollue totalement la zone de diffraction.

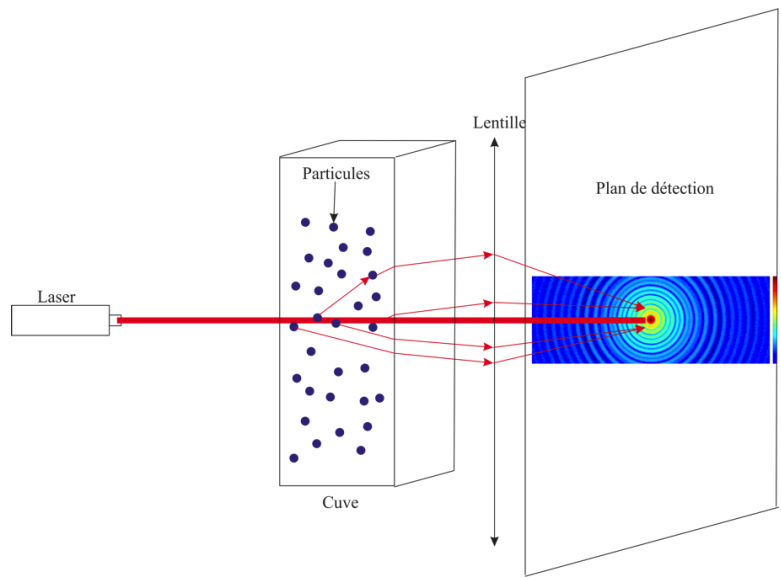

Figure 94 : Schéma de principe du montage diffractométrique

Dans la zone de diffraction, la dynamique et la dépendance angulaire du signal optique sont très importantes. Les zones photosensibles sont donc conçues pour compenser ces deux effets. De manière classique [**Xu 2002**] elles sont petites et resserrées aux petits angles, et elles sont larges et espacées aux grands angles. Pour la variation de taille avec l'angle de diffusion nous avons opté pour une loi exponentielle croissante. L'augmentation de la densité angulaire des OPS aux petits angles est beaucoup plus problématique car leur taille physique ne peut être en deçà de 1mm$^2$ avec un espacement minimal de 1 mm (**cf. Chapitre1**).

La Figure 95 présente le design conçu. Les contraintes de dynamique et d'encombrement font que celui-ci présente nécessairement des similitudes avec les capteurs silicium qui équipent certains diffractomètres laser [**Heffels 1998**]. Dans cette figure, les zones photosensibles sont représentées en blanc. Les connecteurs et pistes électriques sont indiqués en rouge. On remarque l'évolution exponentielle des surfaces, que ce soit en hauteur ou en largeur. Un trou circulaire, difficilement visible sur ce schéma, est percé au centre des branches de façon à laisser passer le faisceau direct. Cette configuration permet d'obtenir 4 fois plus de points de mesure qu'une simple impression en longueur. Tous les films de photodiodes sont identiques à l'exception des découpages, du positionnement du trou pour le faisceau direct et des trous de fixation/alignement. Ces derniers sont positionnés légèrement différemment de façon à faire varier les angles de détection. Le lecteur trouvera plus de détails sur la procédure utilisée pour générer les branches "hélicoïdales" dans l'Annexe 2. .

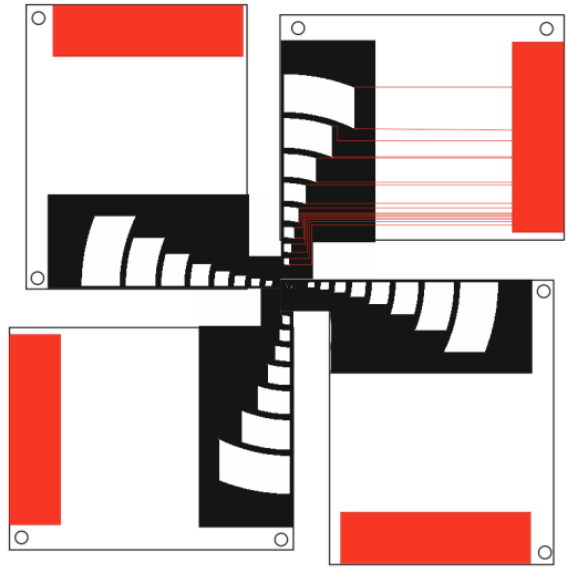

Figure 95 : Design du masque de photodiodes organiques pour la caractérisation de particules en configuration diffractométrique.

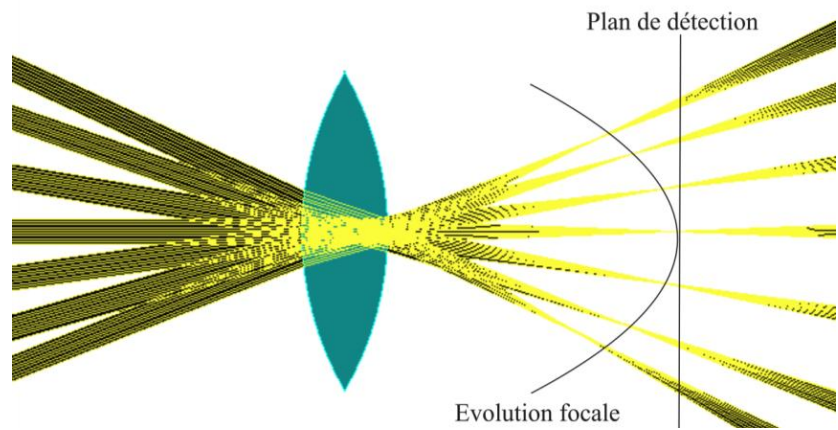

Figure 96 : Evolution de la position du point de focalisation d'une lentille ( $D_l = 51.5mm$, $n_l = 1.515$, focale sur l'axe optique: 100mm) avec l'angle d'incident des rayons (Image générée avec le logiciel Pintar InterArcive Physics VirtuaLab Optics program)

On sait que, quand on collecte la lumière sur une grande ouverture angulaire, on doit à minima prendre en compte l'évolution de la focale effective de la lentille avec l'angle de collection, voir la Figure 96. Grâce au code Monte-Carlo, il est possible de suivre l'évolution du point focal de la lentille avec l'angle d'incidence du faisceau laser, en faisant varier le vecteur directeur des rayons incidents ainsi que la position de la source laser. Faute de temps, la localisation de ce point focal a été simplement extraite graphiquement. La Figure 97 montre les résultats obtenus pour une focale de $f = 100mm$, d'ouverture $N = 3.2$. La lentille étant sphérique, la courbure dans le plan XoY est la même que pour le cas présenté sur la Figure 98. En fait, cette forme particulière de la surface de focalisation peut être modélisée par les polynômes de Zernike [**Zernike 1934**]. Ces derniers jouent un rôle très important en optique [**Born 1999**] et notamment pour la caractérisation des différentes aberrations optiques classiques (aberrations sphériques, comas...). Le troisième polynôme de Zernike est utilisé pour caractériser les erreurs de focalisation [**Liang 1994**], voir la Figure 97, avec :

$$R_2^0(r) = 2r^2 - 1 \qquad (247)$$

Pour la définition des autres polynomes de Zernike, se reporter à l'Annexe 3.

La conformabilité des OPS peut permettre de compenser cette courbure du plan de focalisation. Il suffit pour cela de les adosser à un "berceau métallique" dont la courbure est égale à celle de la surface de focalisation. Avec le design de la Figure 95 et un berceau à deux bras (en forme de croix courbe), on pourrait compenser les aberrations selon deux directions perpendiculaires.

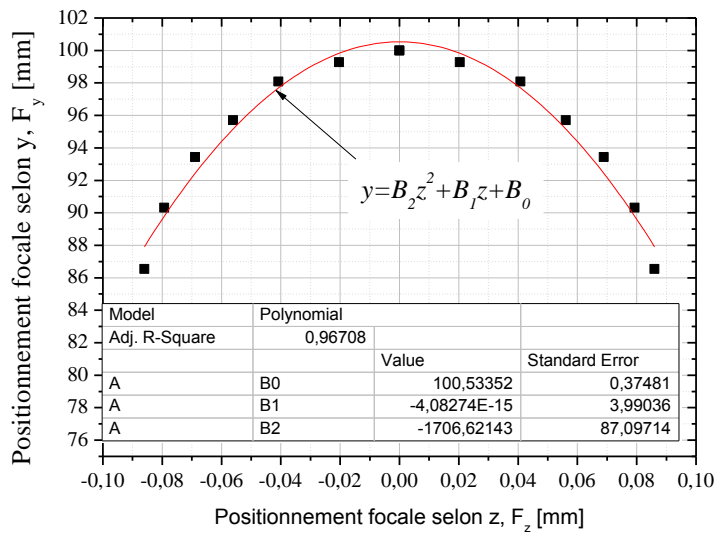

Figure 97 : Approximation de l'évolution de la focale par le troisième polynôme de Zernike

## 4.2.2 Évaluation numérique des diagrammes et premières tendances

Le design des photodiodes présenté sur la Figure 95 est trop complexe pour être représenté par des équations cartésiennes. Nous avons donc fait le choix de passer par une carte d'intensité pixellisée, c'est-à-dire une surface de détection discrétisée (sortie du programme numéro 2). La réponse de chaque détecteur est obtenue par l'intégration sélective des pixels correspondants dans la carte fournie par le code Monte-Carlo.

Deux normalisations sont utilisées successivement pour calculer l'intensité moyenne collectée par un détecteur, dont le centroïde de sa surface $S(\theta)$ est en $\theta$. Elles visent à supprimer les paramètres influant sur l'intensité collectée sans apporter d'informations sur la granulométrie de l'échantillon analysé, à savoir $C_n$ et $S(\theta)$. La Figure 98, qui montre l'évolution de $S(\theta)$, confirme que l'évolution des surfaces des OPS suit bien une loi exponentielle. On constate que pour cette plage angulaire d'une vingtaine de degrès, le rapport des surfaces permet de compresser la dynamique des signaux de près de deux ordres de grandeur.

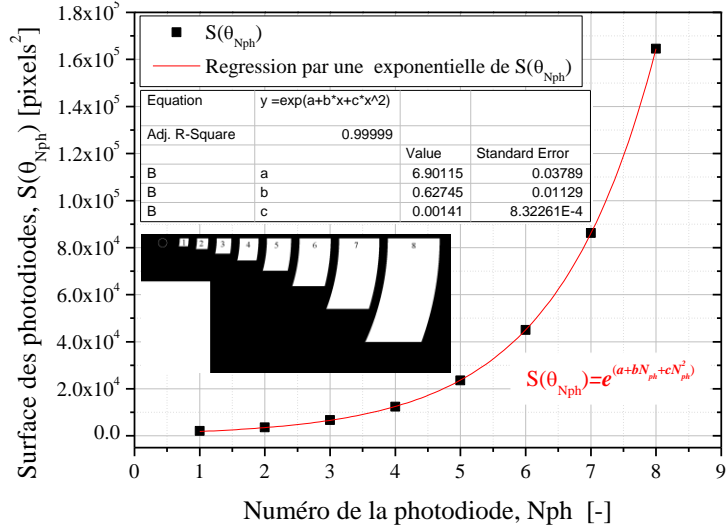

Figure 98 : Évolution de la surface des détecteurs du système diffractométrique

On peut maintenant simuler les diagrammes de diffraction mesurés par les 32 photodiodes du prototype de la Figure 95. Les simulations sont faites ici pour une cuve de spectrophotométrie classique, de 1 cm de côté, des billes de verre de différentes granulométries immergées dans l'eau. Dans le cas de la Figure 99 le milieu est monodisperse ($\sigma_D / \overline{D} = 0.1\%$), alors que dans le cas de la Figure 101 il est polydisperse ($\sigma_D / \overline{D} = 10\%$). Pour faciliter les comparaisons, les diagrammes ont été normalisés selon l'équation :

$$I_{Nph}(\theta) = \frac{\overline{I}(\theta)}{S(\theta) \sum_{i=1}^{N} I_i} \qquad (248)$$

avec $I_{Nph}$ l'intensité moyenne sur une photodiode, $N$ le nombre total de photodiodes et $I_i$ l'intensité sur la photodiode i. Pour le cas monodisperse et pour les plus faibles diamètres ($\overline{D} < 100\mu m$), on distingue encore les oscillations ou "lobes" de diffraction. Pour les diamètres plus importants, la résolution angulaire est trop faible pour en permettre l'observation.

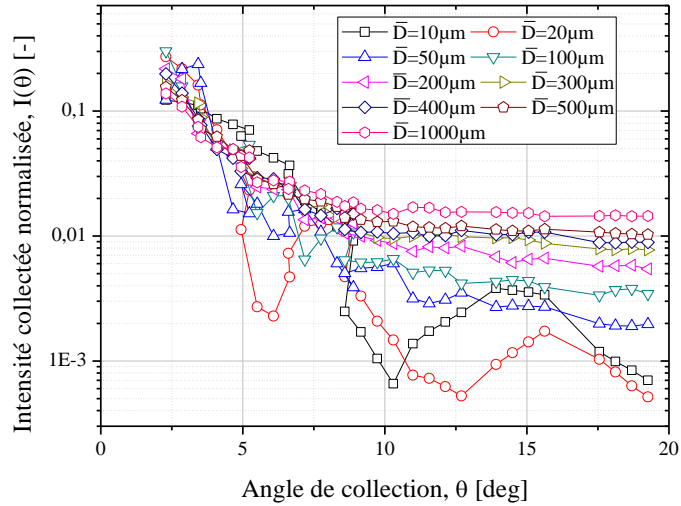

Figure 99 : Diagrammes de diffusion obtenus avec le prototype diffractométrique dans le cas d'une suspension aqueuse de billes de verre monodisperses

Tous les diffractomètres actuels utilisent des méthodes d'inversion algébriques [**Xu 2002**] du type de celle que nous avons implémentée dans la section précédente. Dans le cadre du PAT et pour faire simple, nous nous limitons à l'étude du ratio d'intensité :

$$\Re_3 = \sum_{i=16}^{32} I_i \bigg/ \sum_{i=1}^{32} I_i \qquad (249)$$

La Figure 100 montre que l'évolution de $\Re_3$ avec le diamètre moyen est du type loi de puissance lorsque $\overline{D} \geq 10\mu m$. Comme nous l'avons déjà montré, ce type de loi peut être utilisé pour estimer directement le diamètre moyen des particules.

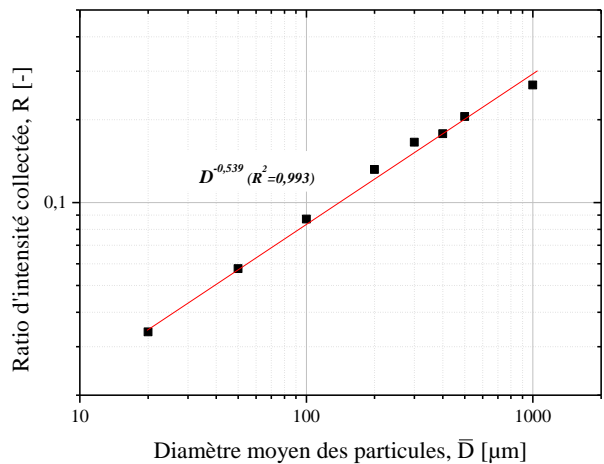

Figure 100 : Évolution d'un ratio d'intensité collectée avec le diamètre moyen des particules pour la configuration diffractométrique

Pour une distribution polydisperse les diagrammes de diffusion n'oscillent plus, quelque soit le diamètre. Malheureusement, comme on peut le voir par exemple pour $\overline{D}=10\mu m$ et $\overline{D}=100\mu m$, la polydispersion induit une uniformisation de l'allure des diagrammes. La recherche d'un ratio évoluant en loi de puissance n'est donc plus pertinente.

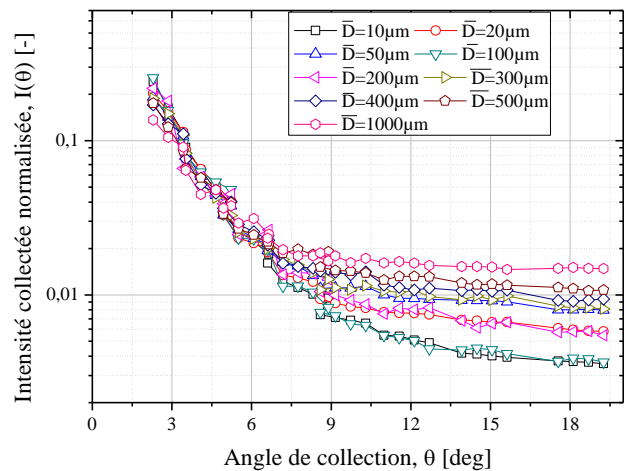

Figure 101 : Diagrammes de diffusion obtenus avec le prototype diffractométrique dans le cas d'une suspension aqueuse de billes de verre polydisperses

## 4.3 Conclusion

Force est de constater que, malgré nos optimisations, la très faible résolution angulaire du système néphélométrique et les réflexions parasites de la cuve cylindrique ne permettent pas une estimation précise du diamètre moyen des particules. Pour pallier la faible résolution angulaire du système, nous avons proposé l'utilisation d'un simple rapport d'intensité qui, pour certaines gammes de tailles, donne des résultats satisfaisants du point de vue des applications PAT. Pour minimiser la pollution du signal de diffusion par les réflexions parasites et le faisceau direct, la solution proposée consiste à réaliser la détection hors du plan de diffusion conventionnel [**Montet 2014**] et d'augmenter un peu le diamètre de la cuve de mesure (l'étalement angulaire du faisceau direct s'en trouvant réduit). La méthode du rapport d'intensité, vue comme une comparaison entre l'efficacité des effets singuliers et des effets géométriques, donne de meilleurs résultats sans cuve, c'est-à-dire pour des écoulements libres dont l'intérêt est évident même s'il sort un peu du cadre du PAT. La configuration diffractométrique utilise la conformabilité des OPS pour corriger certaines aberrations optiques. Néanmoins, ici encore, la très faible résolution angulaire du système et le très faible nombre de détecteurs, ne permettent pas de proposer une solution totalement satisfaisante.

# Chapitre 5

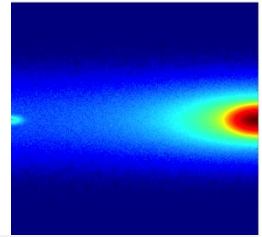

## Etude qualitative de configurations en milieux denses

Dans le chapitre précédent, différentes méthodes de mesure ont été proposées pour caractériser, dans le cadre du PAT, des milieux particulaires optiquement dilués. Cependant, force est de constater que de nombreux procédés industriels mettent en jeu des milieux particulaires optiquement denses, voire très denses. Un même procédé peut donc conduire à des changements de régimes de diffusion : de la diffusion simple jusqu'à la diffusion multiple et même jusqu'à la diffusion dépendante (cf.**chapitre 2**). La solution la plus communément utilisée pour rester en régime de diffusion simple est de procéder à des prélèvements et dilutions automatiques, mais cette solution n'est pas compatible avec les objectifs du PAT (cf. **chapitre 1**).

La caractérisation optique (fine) des milieux denses est bien plus complexe que celle des milieux dilués. En effet, même si la diffusion multiple [**Hulst 1980**] est assimilable à une succession de phénomènes de diffusions simples, elle restreint de manière drastique notre capacité à caractériser un système particulaire donné. L'impact du brouillard sur la sécurité routière, voir par exemple [**Zhang 2011**], illustre parfaitement le problème : les feux de route produisent une lumière rétrodiffusée, qui en plus d'empêcher le conducteur de s'orienter, l'éblouit..

La diffusion multiple de la lumière a été largement étudiée dans différents domaines : en météorologie pour le suivi des émissions de polluants [**Hansen 1970, Hansen 1974**] avec le LIDAR notamment [**Bissonnette 1995, Hespel 2007**], en chimie, pour la mesure de la concentration de suspensions [**Bergougnoux 1995**] ou de fronts de sédimentation [**Mengual 1999**], en biologie, pour la caractérisation de cellules et tissus biologiques [**Mourant 1998**],...

Ces recherches ont conduit au développement de différentes techniques de laboratoire pour mesurer, notamment, la taille moyenne de particules présentes dans des suspensions. Ainsi, la DWS, pour "*Diffusive Wave Spectroscopy*" [**Pine 1988, Scheffold 2002**] est sans contexte la technique la plus aboutie dans ce domaine. Elle permet de caractériser des suspensions denses *via* l'analyse du mouvement brownien des particules. On peut la combiner à la technique DLS (pour "*Dynamic Light Scattering*" [**Berne 2000**]) dont elle est issue. L'association des deux permet alors d'analyser des milieux denses et dilués. La DWS est une technique inverse qui nécessite, dans certains cas, l'utilisation d'un code Monte-Carlo dépendant du temps [**Hespel 2007, Calba 2008**]. En dépit de leurs succès, ces deux techniques demeurent cependant limitées aux particules de taille inférieure à quelques microns. Par ailleurs, les mesures sont longues (plusieurs minutes) et sensibles au bruit (de par la technique d'inversion). Ajoutons que la résolution temporelle exigée par la DLS et la DWS (µs à ms) est très supérieure à celle des OPS actuels (dizaine de ms, voire plusieurs secondes pour minimiser le bruit).

Une autre approche consiste à étudier le cône de rétrodiffusion cohérent [**Albada 1985, Wolf 1988**]. Observé au voisinage du faisceau d'éclairage (déviation angulaire de quelques milli-radians), le phénomène sous-jacent résulte d'interférences constructives liées à des chemins inverses équiprobables [**Chaneliere 2004**]. En utilisant une méthode d'inversion *ad-hoc* [**Blum 1977, Ishimaru 1982**], cette méthode permet de remonter au libre parcours moyen des particules [**Ishimaru 1978**] et donc, au diamètre moyen de celles-ci. Cependant, la résolution angulaire exigée par cette approche est incompatible avec les exigences du PAT et la résolution angulaire qui peut être atteinte avec les OPS actuels.

Une autre méthode classique, que l'on pourrait qualifier de plus "intuitive", consiste à étudier les propriétés de la tache de rétrodiffusion produite par le système particulaire, et plus spécifiquement son profil en intensité [**Mengual 1999, Mengual 1999, Bordes 2003**] ou son état de polarisation [**Raković 1998**]. Si l'on reprend notre exemple initial (celui du brouillard), il s'agit d'étudier les propriétés du halo lumineux vu par le conducteur automobile. Pour illustrer notre propos, la Figure 102 montre la tache de rétrodiffusion produite par un jet cylindrique dense de billes de verre dans l'air. Le régime de diffusion multiple étant atteint, on observe très clairement la tache de rétrodiffusion sur la droite de la figure, alors que la tache de diffusion vers l'avant (essentiellement le faisceau direct et la diffraction en diffusion simple) est à peine visible sur la gauche de l'image. Contrairement aux deux autres approches citées précédemment, les dimensions spatiales de la tache de rétrodiffusion sont compatibles avec la résolution actuelle des OPS.

Toutes les raisons évoquées précédemment nous ont conduit à proposer dans ce chapitre des premiers résultats obtenus par analyse des propriétés de la tache de rétrodiffusion. Pour cela, et afin d'économiser du temps de calcul, la configuration néphélométrique du chapitre précédent est utilisée.

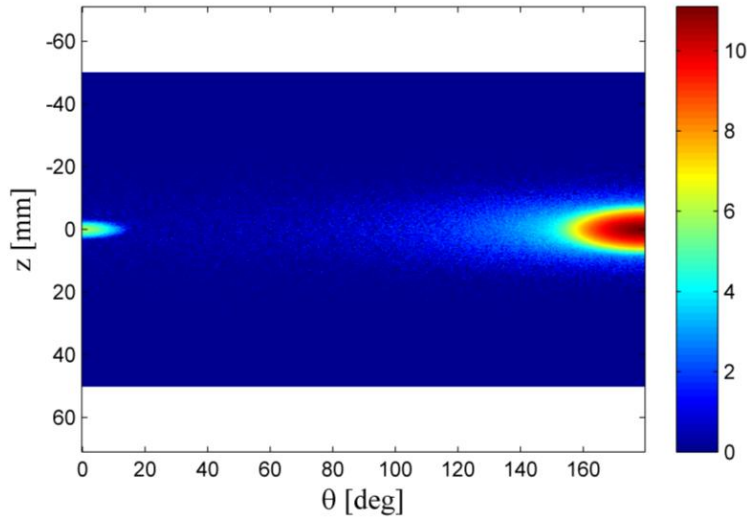

Figure 102 : Exemple de tache de rétrodiffusion (à droite, $\theta \to 180°$) produite par la focalisation d'un laser sur un milieu dense de billes de verre dans l'air. La tache de diffusion vers l'avant (à gauche, $\theta \to 0°$) est à peine visible. Les couleurs symbolisent l'intensité diffusée (bleu : minimale, rouge : maximale)

## 5.1 Configuration néphélométrique et hypothèses

Cette configuration correspond à celle présentée dans le chapitre précédent (**paragraphe 4.1**) : tube cylindrique (cuve, distance par rapport à la couronne de filtres, distance de détection...). Le système particulier étudié est une distribution log-normale quasi-monodisperse ($\sigma_D / \overline{D} = 0.1\%$) de billes de verre dans l'air. On se limite à l'étude numérique du signal restitué pas les OPS en fonction de la concentration volumique $C_v$ et du diamètre moyen $\overline{D}$ des particules.

Le traitement de l'intensité collectée diffère du chapitre précédent dans la mesure où la concentration volumique et le volume de mesure $V_m$ ne peuvent plus être supprimés par une simple normalisation. En effet, l'intensité collectée par un détecteur identifié par sa position angulaire $\theta$ et sa surface $S(\theta)$ n'est plus proportionnelle à la concentration volumique $C_v$ en particules du milieu. De même, la forme et les dimensions du volume de mesure $V_m$ varient de manière très complexe [**Haeringen 1990**] avec l'angle de diffusion $\theta$, la distribution granulométrique $n(D)$ et la concentration $C_v$. Dans ce qui suit, la réponse brute de chaque détecteur est seulement normalisée par sa surface de détection $S(\theta)$. Dans notre analyse, en supposant que la cuve, la géométrie du dispositif expérimental, le type (forme, composition) des particules demeurent inchangés pour une étude donnée, l'intensité collectée dépend explicitement de trois paramètres : la concentration volumique $C_v$, le diamètre moyen $\overline{D}$ des particules et l'écart-type $\sigma_D$ de la distribution granulométrique $n(D)$. Cette étude repose sur l'analyse de 100 diagrammes de diffusion obtenus avec 10 concentrations volumiques différentes et 10 diamètres moyens différents, mais à écart-type fixé. Comme souligné précédemment, les calculs dans cette configuration sont particulièrement longs. Aussi, nous avons arbitrairement limité le nombre d'interactions rayon/particules à 5000. Une simple étude de sensibilité a d'ailleurs montré que cette simplification a peu d'influence sur les simulations menées.

Avant de présenter et de discuter les résultats de cette série de simulations, il est à noter qu'à notre connaissance aucun article portant sur l'analyse de la tache de rétrodiffusion ne propose réellement de solution permettant de remonter à $\sigma_D$ par cette méthode.

En guise d'illustration, la Figure 103 présente les résultats obtenus pour trois concentrations volumiques : $C_v = 1\%$, $15\%$ et $40\%$. On constate que plus les particules sont petites, plus les diagrammes de diffusion particulaires présentent une intensité rétrodiffusée importante (voir par exemple la Figure 103 (a)).

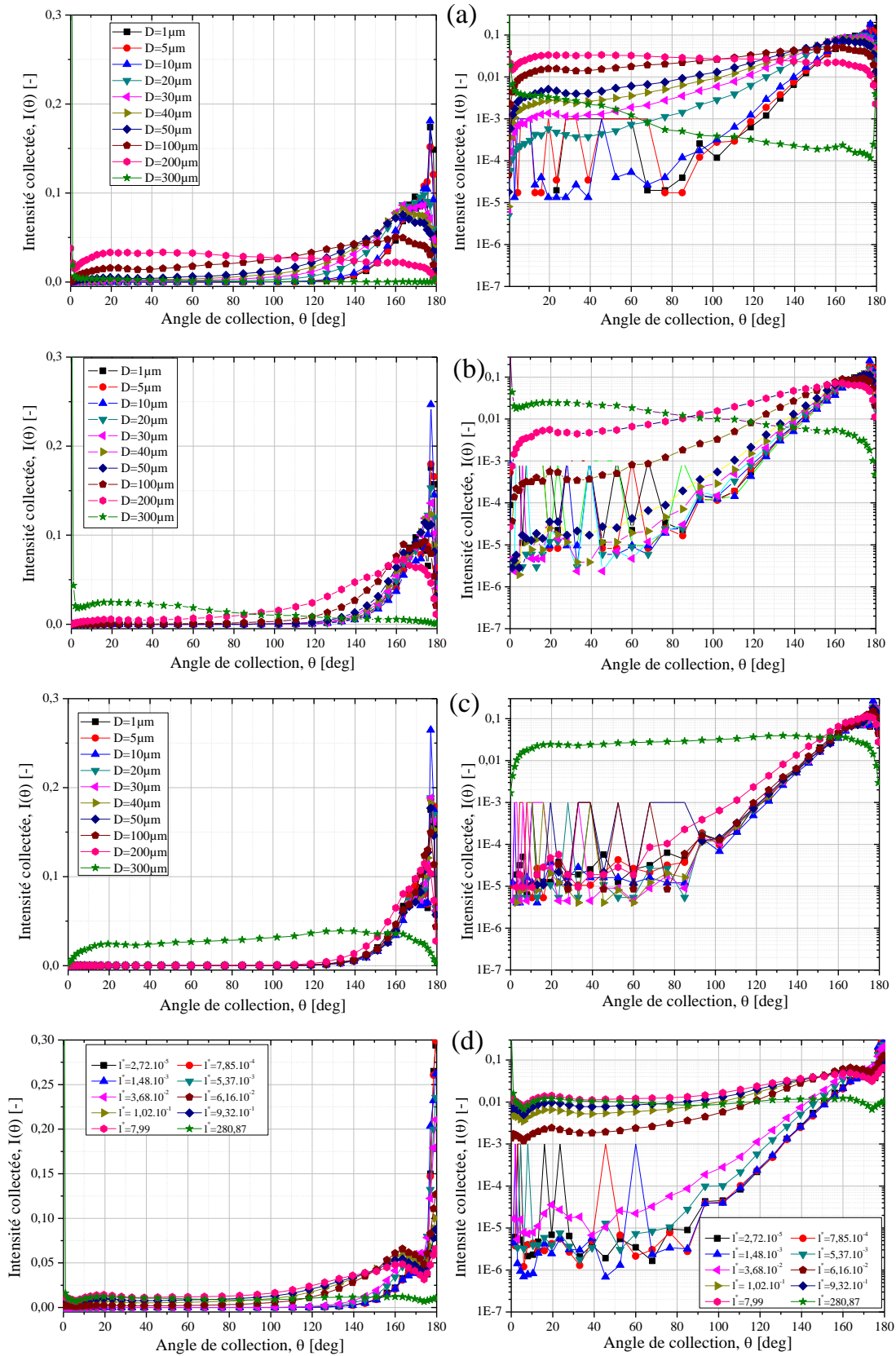

Figure 103 : Réponses simulées pour différents diamètres moyens et concentrations volumiques de (a) $C_v = 1\%$ , (b) $C_v = 15\%$ et (c) $C_v = 40\%$ .et (d) réponses pour différents libre parcours moyen $l^*$ (en échelle linéaire et log)

En complément sur la Figure 103(d), l'évolution du signal en fonction d'un libre parcours moyen $l^*$ est présentée. Cette grandeur est plus classiquement utilisée dans la littérature pour la diffusion multiple de la lumière [**Mengual 1999**] et s'exprime par :

$$l^* = \frac{l}{1-g} \qquad (250)$$

avec $l$ , le libre parcours moyen définie par l'équation (175) et g définie par l'équation (241).

Pour certains diamètres moyens, le diagramme de diffusion présente un pic principal vers l'avant, indiquant que le principal mode de diffusion est celui de la diffusion simple (par exemple la Figure 103 (a) pour $D = 300\mu m$ ). On note par ailleurs que l'intensité rétrodiffusée augmente avec la concentration volumique. Le maximum est atteint pour des particules dont le diamètre est de l'ordre de $\approx 10\mu m.$ , ce qui peut apparaître comme étonnant en première approche. Notre interprétation est double : d'une part, le diagramme des petites particules est moins anisotrope (elles répartissent spatialement davantage l'énergie rétrodiffusée) et d'autre part les plus grosses particules tendent à diminuer la diffusion multiple (et *in fine* l'intensité rétrodiffusée). Il semble alors que d'après la Figure 103 les particules de $\approx 10\mu m$ sont à l'intersection de ces deux tendances.

La Figure 103 , indique donc qu'il est très difficile de trouver une tendance pouvant être exploitée pour élaborer une procédure d'inversion de ces diagrammes de diffusion. La nature cylindrique du système particulaire, les effets de la cuve, la faible résolution angulaire des OPS et la quasi inexistence d'un régime asymptotique établi expliquent cela. On remarque néanmoins sur la Figure 103 que dans la zone du diagramme entre $\approx 130°$ et $\approx 160°$, sa pente semble évoluer à la fois avec le diamètre moyen mais également avec la concentration. Ceci est confirmé par la Figure 104 qui montre que l'intensité du diagramme de diffusion semble être reliée à l'angle de collection par une « loi de puissance ». Si l'on estime la pente de ces diagrammes, on obtient la Figure 105. Cette dernière montre, en courbes de niveau, le taux de variation angulaire de l'intensité diffusée par le milieu particulaire en fonction de son diamètre moyen et de sa concentration volumique. Il est à noter que pour tous les cas traités ici, les coefficients de régression sont très proches de 1 (même pour les particules de diamètre 300µm qui ne sont plus en régime de diffusion multiple). On voit qu'à concentration fixée et sur une certaine plage de diamètres, on peut déduire le diamètre moyen de manière biunivoque et directe, et vice-versa. Une étude plus poussée permettrait peut-être d'affiner, d'étendre et de quantifier cette méthode d'estimation des paramètres recherchés.

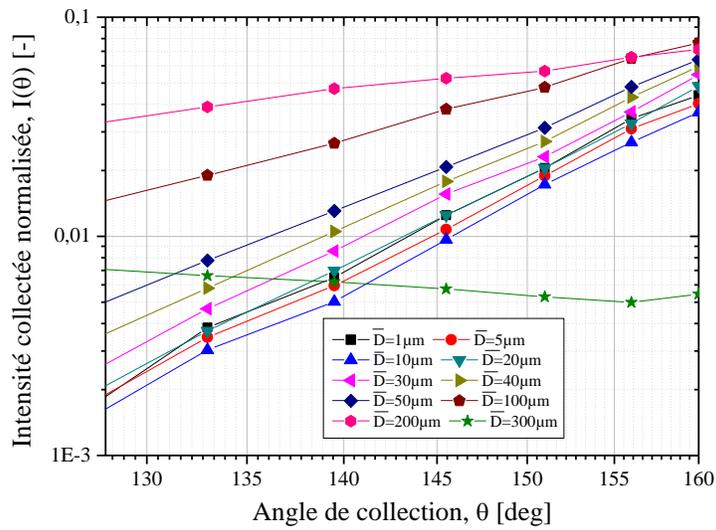

Figure 104 : Diagrammes de diffusion (échelles log-log) d'un jet de particules de verre dans l'air pour différents diamètres moyens - concentration volumique fixée à $C_v = 10\%$

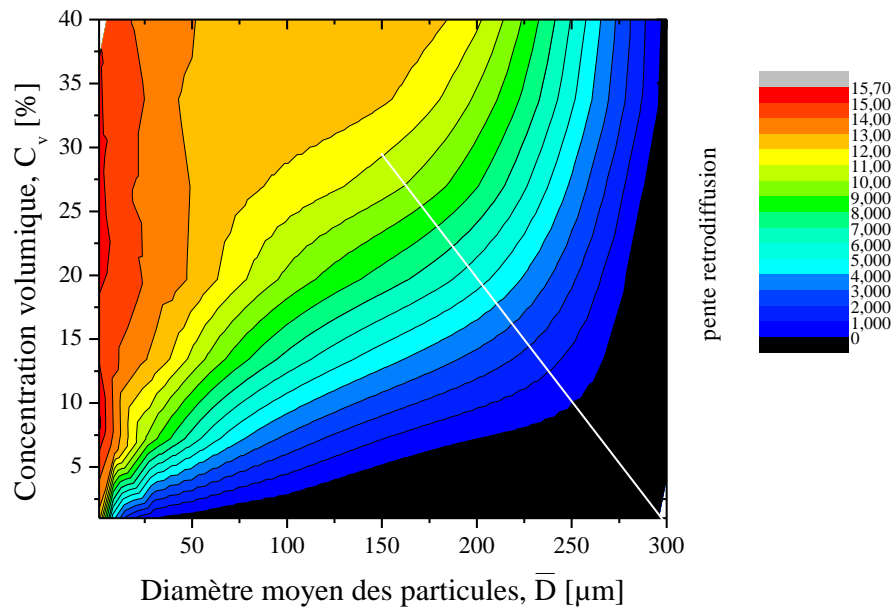

Figure 105 : Étude de l'influence du diamètre moyen et de la concentration volumique sur le taux de variation angulaire de l'intensité diffusée, dans la zone 130-165°

## 5.2 Configuration en détecteur proche (prototype Indatech)

Une étude complémentaire dans le cadre du projet OPTIPAT a été menée pour tester un prototype conçu par la société INDATECH. Ce prototype s'inspire de celui proposé dans le paragraphe précédent.

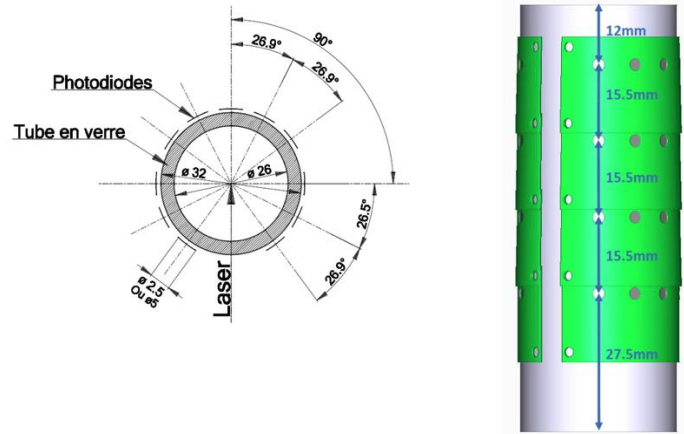

Figure 106 : Prototype de mesure en champ proche conçu dans le cadre du PAT (INDATECH et IUSTI)

Il est composé d'un tube de verre dans lequel s'écoule un milieu particulaire. Quatre barrettes de 11 photodiodes sont enroulées autour du cylindre de verre. Les positions des différentes photodiodes sont précisées sur la Figure 106.

Pour cette configuration, différentes "cartes de détection" (comme sur la Figure 102) ont été générées par la méthode de Monte-Carlo. Une étude paramétrique a été menée pour simuler la réponse du prototype en fonction du diamètre moyen, de la concentration ainsi que de l'écart-type de la distribution log-normale de billes de verre dans l'eau. La Figure 107 présente plusieurs résultats caractéristiques en fonction du diamètre moyen et de la concentration volumique. Il s'agit de cas limites où les transitions sont claires.

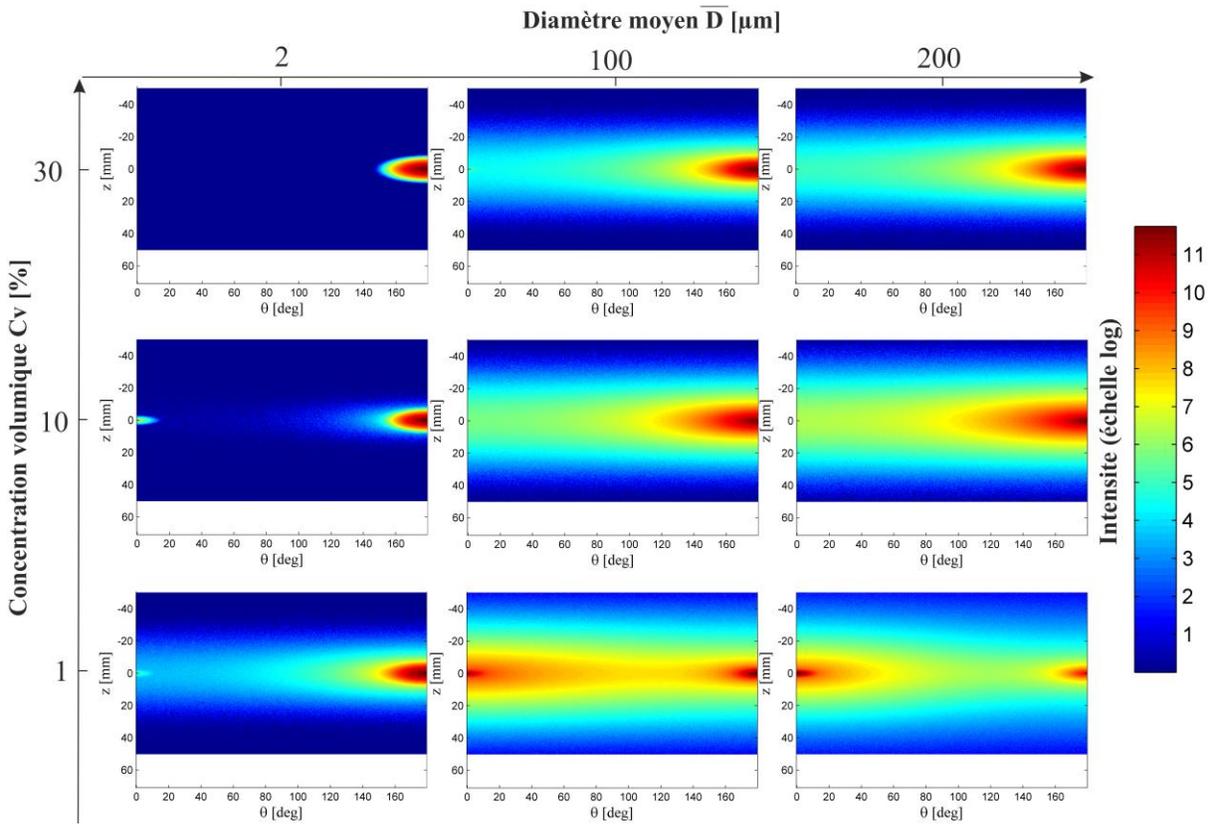

Figure 107 : Exemples de cartes générées pour le prototype INDATECH : l'abscisse correspond à la demi-circonférence du prototype et l'ordonnée à la hauteur de collection vis-à-vis du plan de diffusion où se propage le faisceau laser initial

On constate que l'effet de la concentration volumique sur la tache de rétrodiffusion est de diminuer sa taille au profit de son intensité maximale. Ceci confirme bien le fait que la diffusion multiple augmente avec la fraction volumique en particules. À concentration fixée, lorsque le diamètre des particules augmente, l'effet est inverse. Ainsi, pour une concentration volumique de $C_v = 1\%$ et lorsque le diamètre moyen augmente, l'extension de la tache de rétrodiffusion diminue au profit de la diffusion vers l'avant. A partir de $\approx 200\mu m$, la tache de rétrodiffusion semble même quasiment disparaitre. La Figure 108 illustre l'influence de l'écart-type relatif de la distribution granulométrique sur la tache de rétrodiffusion. Ce paramètre joue un rôle semblable à celui du diamètre moyen; ce qui s'explique par le fait qu'avec une distribution log-normale, l'augmentation de l'écart-type entraine un élargissement préférentiel de la distribution granulométrique vers les grosses particules.

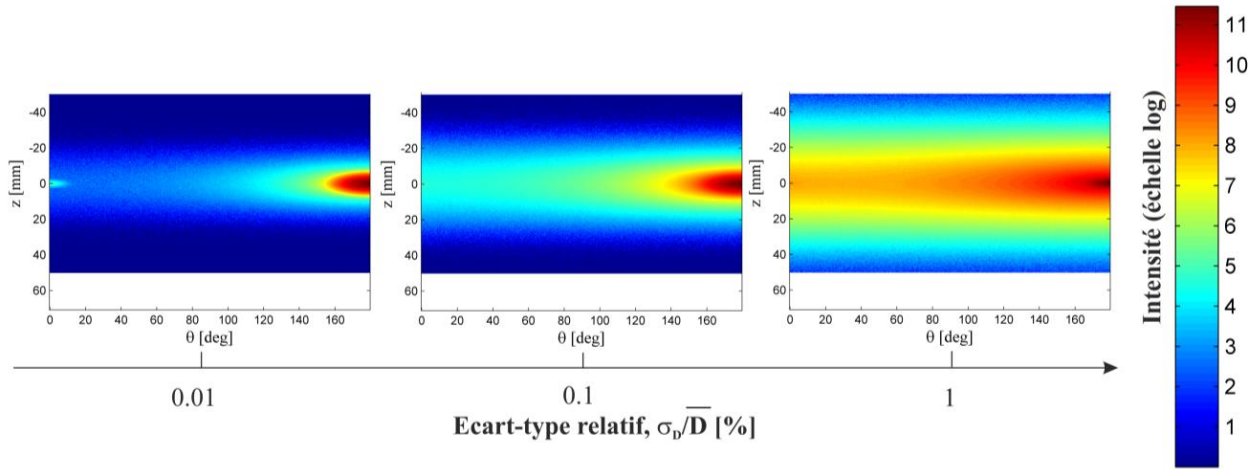

Figure 108 : Influence de l'écart-type relatif de la distribution granulométrique sur la tache de rétrodiffusion, concentration volumique fixée à 1%

En complément de ces cartes, un programme Matlab de post-traitement des données a été développé pour positionner de manière optimale les photodiodes à partir de la sortie n°2 de notre code de calcul (cf. **chapitre 3**). L'utilisateur peut y préciser les coordonnées de la photodiode ainsi que son rayon (INDATECH utilise des photodiodes circulaires). Le programme restitue l'intensité que collecterait cette dernière. Une interface graphique simplifiée aide l'utilisateur dans ses choix, voir la Figure 109 pour l'interface et l'Annexe 4. pour plus de détails sur le code.

Les signaux ainsi modélisés ont été utilisés par Indatech et Ondalys pour mener une étude de sensibilité reposant sur un algorithme génétique.

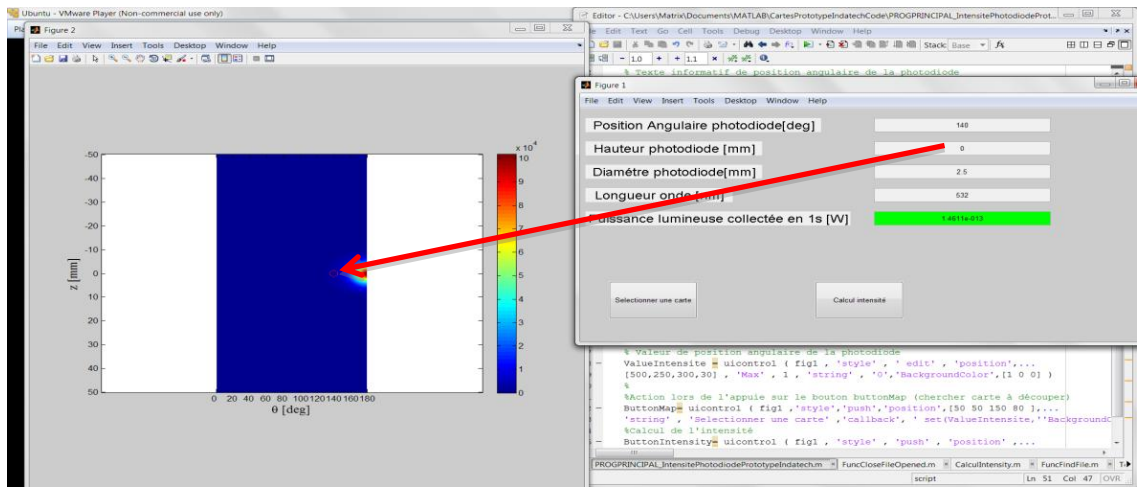

Figure 109 : Copie d'écran du programme de post-traitement des cartes d'intensité simulées pour le prototype Indatech

Afin d'estimer la réponse du prototype en régime de diffusion multiple (forte concentration volumique du système particulaire), la Figure 110 montre l'évolution du profil horizontal en intensité (z=0) en fonction de la concentration volumique. Il apparait clairement que, sur la plage de concentrations retenue, la diffusion multiple et la configuration "détecteurs proches" tendent à homogénéiser et lisser les diagrammes de diffusion.

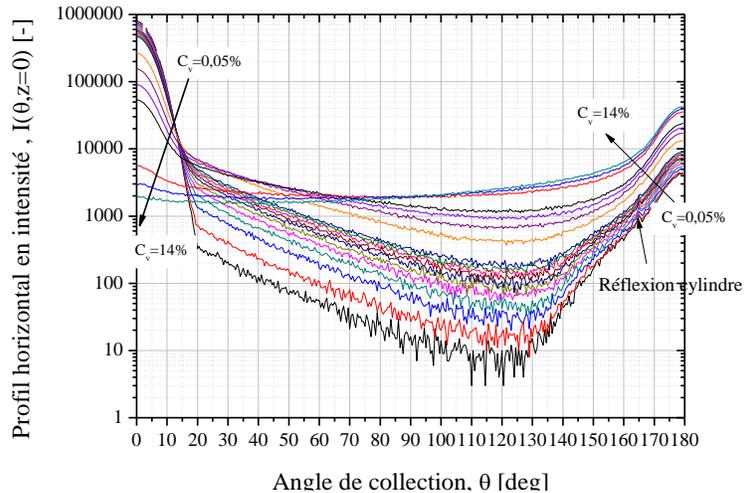

Figure 110 : Diagrammes de diffusion obtenus par méthode Monte-Carlo pour le montage de la Figure 106 : un jet de billes de verre de 100µm placées dans l'eau, à différentes concentrations volumiques $C_v$

Comme nous l'avons déjà indiqué, les caractéristiques de la tache de rétrodiffusion a d'ores et déjà fait l'objet de nombreuses études. On citera notamment les travaux de Mengual et ses collaborateurs [**Mengual 1999, Buron 2004**]. Ces derniers relient les propriétés du milieu particulaire aux caractéristiques de la tache observée sous un angle de $\theta = 135°$. La Figure 111 (a) montre le profil d'intensité de cette dernière, en fonction de la côte $z$ et pour différentes concentrations volumiques. On remarque que ces profils s'apparentent à des distributions canoniques (de type lorentziennes). La Figure 111 (b) présente des grandeurs caractéristiques tirées de ces profils. De manière surprenante, on constate que, quelque soit la grandeur utilisée pour décrire le profil d'intensité de la tache de rétrodiffusion (intensité maximale, moyenne ou écart-type de la distribution), celle-ci évolue quasi-linéairement avec la concentration volumique du milieu.

La Figure 112 montre l'évolution de ces mêmes profils en fonction du diamètre moyen et de l'écart-type de la distribution log-normale utilisée pour modéliser la distribution granulométrique. La Figure 112 indique que dans cette région angulaire, l'intensité diffusée par les particules de diamètre $D=1µm$ est extrêmement faible. Cela vient probablement du fait que, pour ces particules, on est en régime de diffusion dépendante. En effet, d'après Snabre et Arhaliass [**Snabre 1998**], ce régime apparait lorsque :

$$\frac{\pi D}{\lambda} < 20 \qquad (251)$$

Or, dans notre étude, nous avons $\pi D / \lambda = 5.9$. Nous serions donc en régime de diffusion dépendante et ceci, même pour les particules de diamètre D=2µm. Pour les autres cas, l'augmentation du diamètre moyen des particules semble élargir la courbe et augmenter l'intensité mais, avec seulement "deux points de mesure", il est très compliqué d'en tirer une conclusion. Le franchissement de la limite fixée par l'équation (251) implique également que nos simulations de type Monte-Carlo, qui ne prennent pas en compte la diffusion dépendante, doivent être regardées avec une extrême prudence.

Néanmoins, pour des particules plus grosses, la Figure 112 (b) indique que la variation de l'écart-type relatif semble influer à la fois sur la valeur de l'intensité maximale et sur la forme du profil (c'est-à-dire du diagramme de diffusion). De plus, les courbes correspondant aux écart-types relatifs de 10% et 30% semblent adopter un comportement inverse aux autres. Nous supposons ici encore qu'un changement de régime de diffusion a lieu et que les deux dernières courbes correspondent à un régime de diffusion simple majoritaire.

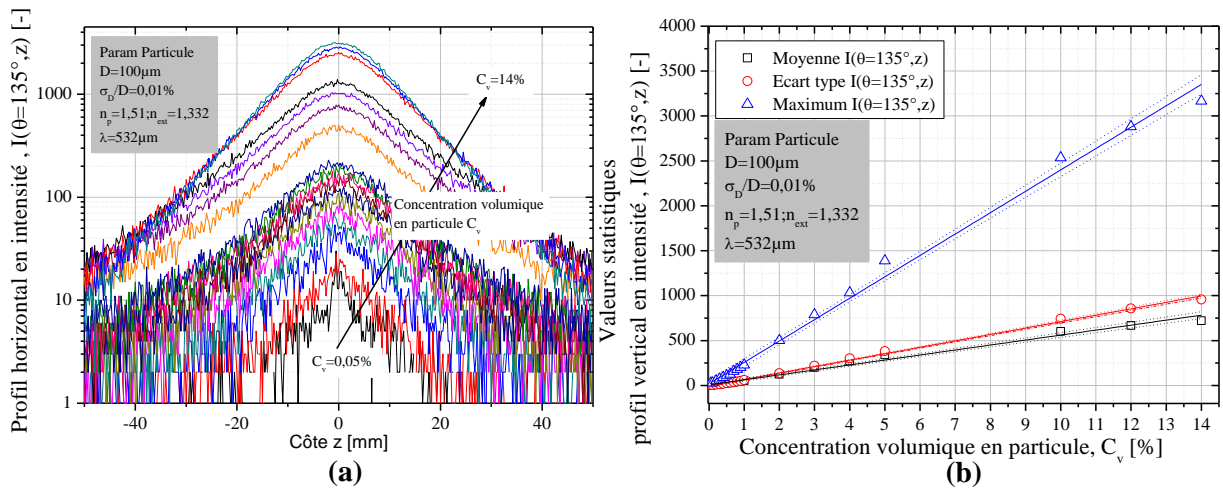

Figure 111 : (a) Profils d'intensité de la tache de rétrodiffusion observée à $\theta = 135°$ pour différentes concentrations volumiques et (b) évolution des caractéristiques (moyenne, écart type et maximum) de ce profil

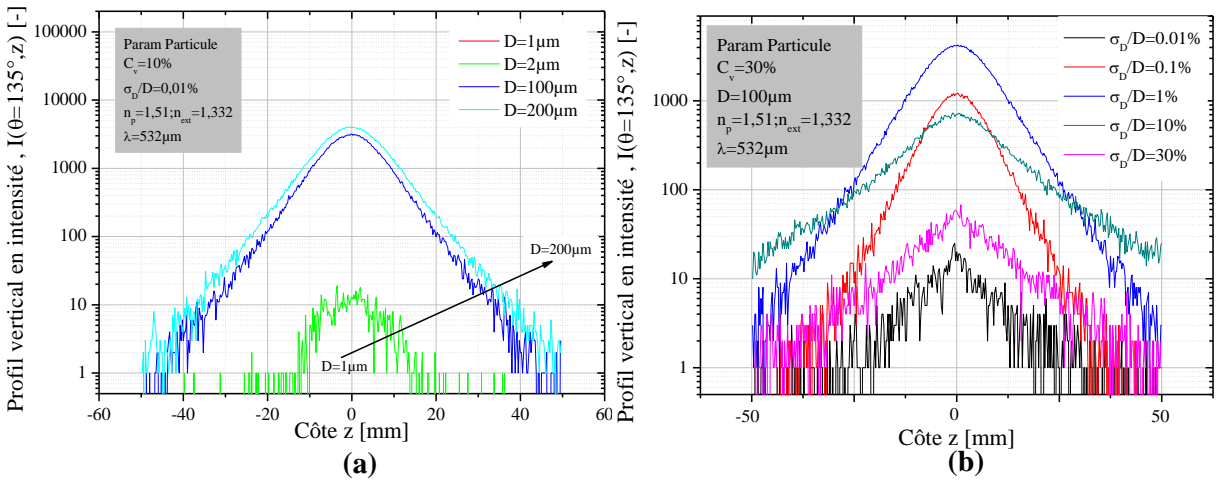

Figure 112 : Influence sur l'allure du profil vertical en intensité (a) du diamètre moyen et (b) de l'écart-type relatif, de la distribution Log-Normale

Nous avons étudié l'évolution des 4 premiers moments d'une distribution lorentzienne s'ajustant au mieux aux profils simulés, pour différents paramètres granulométriques. Malheureusement, aucune tendance probante n'a été dégagée de cette étude. Contrairement au cas de la concentration volumique, en diffusion multiple, nous n'avons pas encore pu identifier une signature simple des effets du diamètre et de l'écart type.

# Chapitre 6

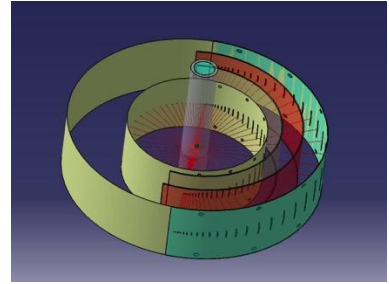

## Tests expérimentaux

Les chapitres 4 et 5 proposent quelques solutions ou quelques pistes de caractérisation des milieux dilués et des milieux denses. Ces solutions et les comportements sous-jacents ont été testés et étudiés numériquement grâce au code de Monte-Carlo implémenté durant cette thèse. Dans le cadre du projet OPTIPAT, une validation expérimentale s'avère nécessaire. Nos travaux expérimentaux se sont concentrés sur le développement de la configuration néphélométrique présentée dans le 4.1, principalement pour deux raisons. La première est que cette configuration s'est révélée être la plus prometteuse à l'issue de l'étude numérique. La seconde est que les premières barrettes de photodétecteurs organiques ne nous sont parvenues que dans les derniers mois de cette thèse et qui plus est, en étant non conformes. Pour pallier ce problème, une première expérience de validation a été conçue à partir d'un système goniométrique et d'une caméra CCD. Le principe de fonctionnement de ce montage multi-angles et de son logiciel de pilotage sont décrits dans la première partie de ce chapitre. Dans la seconde partie, les bases de l'expérience avec photodiodes organiques sont détaillées. Dans une troisième partie, une étude annexe d'un montage développé au laboratoire est faite pour valider le code de Monte-Carlo.

## 6.1 Montage goniométrique avec caméra CCD

Le banc optique de mesure est présenté sur la Figure 113. Dans celui-ci, la détection est assurée par une caméra CCD classique (DMK23U274 de *The Imaging Source*: 1600x1200 pixels, 8/12bits, monochrome) solidaire d'un berceau goniométrique de précision (*Micro*

*Contrôle*). Ce dernier permet à la caméra de tourner autour du système particulaire à analyser qui est contenu dans une cuve cylindrique en verre standard pour le PAT (*Bene-Inox*, hauteur $H = 90mm$, diamètre intérieur $D_{int} = 26mm$, diamètre extérieur $D_{ext} = 32$). Le faisceau laser incident, gaussien circulaire et de polarisation parallèle, est produit par un laser Helium-Néon rouge ($\lambda = 632.8$nm) de 5 mW de puissance et de diamètre $2w_0 = 0.8mm$. La caméra collecte la lumière diffusée par le système particulaire dans le plan de diffusion. Une roue à filtres, équipée de filtres de densité, permet d'atténuer ou non la puissance nominale du faisceau incident sur la cuve.

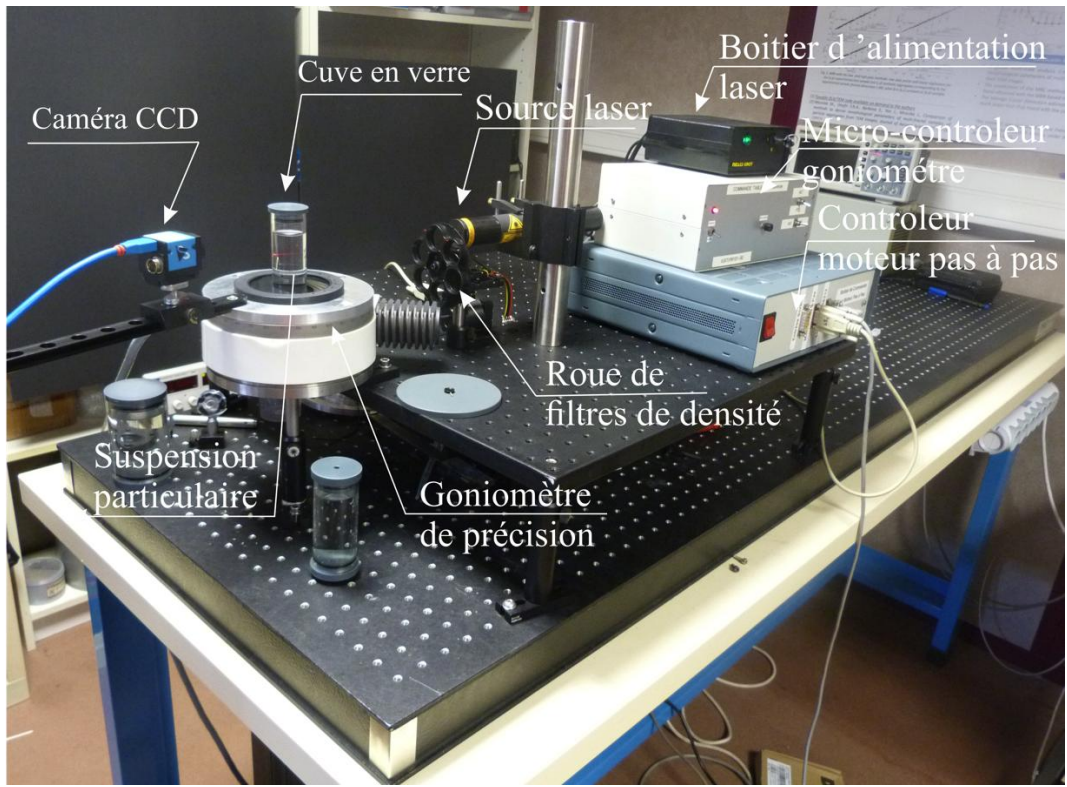

Figure 113 : Photographie du montage goniométrique avec caméra CCD

Pour faciliter la reproductibilité du centrage du système et du positionnement du faisceau, le goniomètre est équipé d'un dispositif de centrage à base de diaphragmes, voir la Figure 114. Ces derniers permettent le centrage de "gros" tubes (diaphragmes *Thorlabs* D75SZ, diamètre $D = 75mm$ ) ou "petits" tubes capillaires (*Thorlabs* D5S, diamètre $D = 1mm$). Ces pièces sont solidaires du rail optique de la caméra.

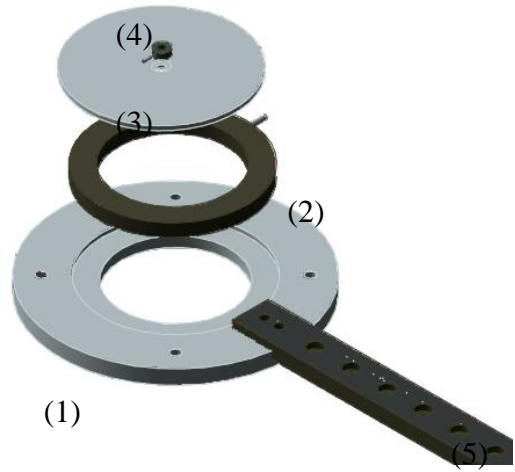

Figure 114 : Schéma du système de centrage de cuves cylindriques et capillaires (figure et conception mécanique S. Noel). On distingue les pièces d'adaptation au goniomètre ou diaphragme (1 ; 3); le diaphragme de centrage de la cuve cylindrique (2), le diaphragme de centrage des capillaires (4) et le rail optique de la camera (5)

### 6.1.1 Pilotage de la caméra

Le programme de pilotage de la caméra est capable de calculer de manière automatique l'intensité collectée pour toutes les positions angulaires du goniomètre. Pour ce faire nous utilisons actuellement une caméra 8/12bits (8bits pour le signal, le reste est utilisé pour sous-traiter le bruit de fond électronique). Le principal intérêt de cette caméra est sa compacité, son coût et la possibilité de la programmer en VB.NET via une bibliothèque de fonctions implémentées par le constructeur.

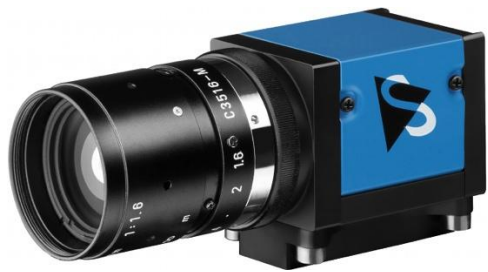

Figure 115 : Photographie de la caméra utilisée sur le banc goniométrique

#### 6.1.1.1 Dynamique des signaux

La programmation de la caméra doit permettre de couvrir une large dynamique en intensité des signaux, tout en évitant la saturation des pixels ou des rapports signal sur bruit trop faibles. Dans une même image, la dynamique des signaux à mesurer peut être estimée en calculant le rapport d'intensité $I_{max}(\theta) / I_{min}(\theta)$ au moyen de la théorie de Lorenz-Mie. La Figure 116 présente cette grandeur calculée pour l'ouverture angulaire (sans objectif) de la caméra, soit $\approx 3.2°$ à 124mm.

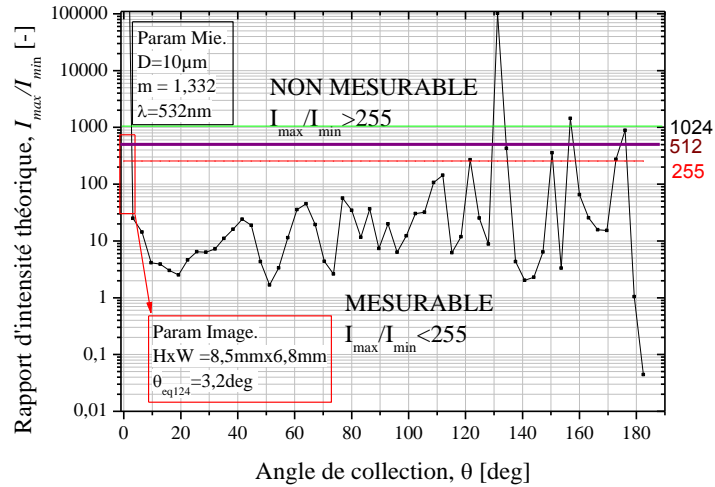

Figure 116 : Évolution angulaire du rapport théorique $I_{max}(\theta) / I_{min}(\theta)$ obtenu par utilisation de la théorie de Lorenz-Mie

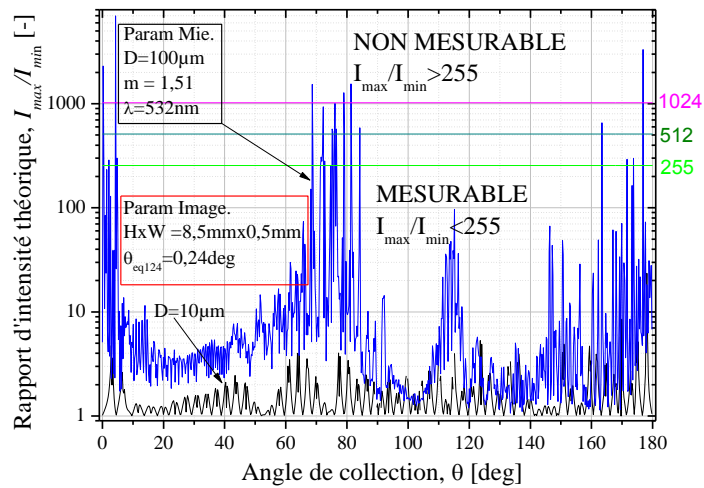

Figure 117 : Évolution angulaire du rapport théorique (théorie de Lorenz-Mie) $I_{max}(\theta) / I_{min}(\theta)$ obtenu pour une ouverture angulaire réduite à $\approx 0.24°$

On constate que pour des particules de diamètre $\bar{D} = 10 \mu m$, la dynamique du signal est bien trop importante pour être mesurée avec une caméra CCD 8 bits, et même une caméra 16bits, notamment autour de l'angle d'arc-en-ciel et de la diffraction. Pour pallier ce problème, on peut réduire l'ouverture angulaire de la caméra en utilisant un masque optique. Ainsi, avec une fente optique de 500µm de large (taille des trous optiques du montage néphélométrique du **chapitre 4**), on réduit l'ouverture angulaire de la caméra à $\approx 0.24°$. Dans ce cas, la Figure 117 indique que le diagramme de diffusion d'une particule de diamètre $\bar{D} = 10 \mu m$ devient localement mesurable par une caméra 8bits.

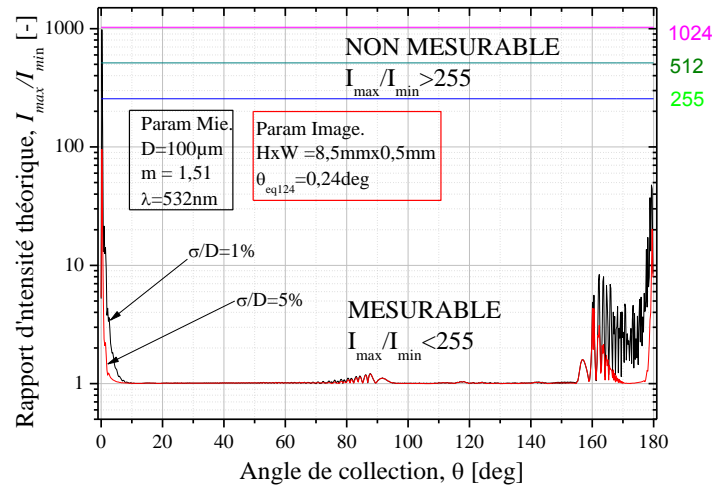

Figure 118 : Représentation de la dynamique du signal $I_{max}(\theta) / I_{min}(\theta)$ dans une même image pour des particules de diamètre moyen $\bar{D} = 100 \mu m$ et des écarts-types variables

Néanmoins, la dynamique de particules de plus gros diamètre (par exemple 100μm, voir la courbe en bleu sur la Figure 117) devient vite trop importante, malgré l'ouverture angulaire réduite de la caméra. Ceci implique donc que plus le diamètre des particules augmente, plus l'ouverture angulaire de la caméra doit être réduite. Toutefois, on remarque que, dans le cas d'un milieu polydisperse, voir la Figure 118, l'écart-type relatif ($\sigma_D / \bar{D} = 1\%$) de la distribution granulométrique tend à limiter la dynamique des signaux en lissant les "résonances de Mie".

### 6.1.2 Procédures d'enregistrement et reconstitution

Pour caractériser un diagramme complet, comme le diagramme à certains angles, il faut augmenter la dynamique du système de détection, par exemple en faisant varier le gain et le temps d'exposition du CCD. Le calcul de l'intensité collectée passe alors essentiellement par trois étapes : auto-ajustement des paramètres d'enregistrement de la caméra, acquisition de plusieurs images pour augmenter le rapport signal sur bruit et reconstruction de l'histogramme des intensités.

La première étape d'auto-ajustement consiste à configurer les paramètres de la caméra pour éviter la saturation tout en conservant le maximum d'informations sur le signal (l'image). Par défaut, le gain de la caméra est réduit au minimum pour éviter d'augmenter inutilement le ratio signal sur bruit. Le seul paramètre de réglage restant est donc le temps d'exposition $t_{\exp}$. Au démarrage, le temps d'exposition est réglé à sa plus faible valeur de façon à éviter une saturation (et, in fine, une destruction) du capteur. On calcule l'histogramme des intensités. Si aucun pixel n'est saturé (c'est-à-dire qu'aucun pixel n'a une intensité de 255), on augmente graduellement le temps d'exposition et l'on recalcule l'histogramme des intensités. Lorsqu'un pixel atteint le niveau 255, on se ramène au cas non saturé antérieur en diminuant d'une unité de temps (c.-à-d. d'un facteur deux pour cette caméra) le temps d'exposition.

La deuxième étape consiste à acquérir et à moyenner plusieurs images afin de minimiser l'importance du bruit du capteur et pour moyenner les fluctuations de l'écoulement. Les images sont sauvegardées sur disque pour le post-traitement (étape 3).

La troisième étape constitue la partie la plus délicate de la procédure. Il s'agit "d'étirer" l'histogramme des intensités de $[0,255]$ à $[0,1024]$ niveaux, grâce, de nouveau, au temps d'exposition. On produit un premier histogramme des intensités avec le réglage obtenu dans l'étape 2. Il est possible que dans ce cas-là, l'histogramme présente un pic en 0. Ce dernier peut indiquer une absence "totale" de lumière comme des niveaux d'intensité non mesurables. Si ce n'est pas le cas, l'histogramme est optimal et aucun réglage supplémentaire n'est nécessaire pour calculer l'intensité finale. Dans le cas contraire (présence d'un pic pour les intensités inférieures ou égales à 0), on effectue une mesure complémentaire en augmentant d'une unité le temps d'exposition. La partie $[128,255]$ de l'histogramme précédent est alors décalée sur l'intervalle $[256,510]$. Le doublement du temps d'exposition fait que seules les valeurs paires des niveaux sont connues. Les niveaux impairs sont obtenus par interpolation linéaire :

$$Histo(2j+1) = Histo(2j) + (Histo(2j+2) - Histo(2j)) / 2, \ \forall j \in \mathbb{N} \qquad (252)$$

avec *Histo(j)* pour le nombre de pixels ayant un niveau d'intensité *j*.

On calcule le nouvel histogramme en concaténant les valeurs sur $[0,255]$ et $[256,510]$. On répète l'opération précédente en doublant le temps d'exposition et en sauvegardant les valeurs $[256,510]$ de l'histogramme précédent dans les valeurs $[512,1020]$ du nouvel histogramme. En procédant à une nouvelle interpolation linéaire on sauvegarde les valeurs $[128,255]$ vers $[256,510]$, et ainsi de suite. On obtient finalement un histogramme sur $[0,1020]$ niveaux dont on déduit l'intensité totale collectée par le capteur CCD pour un angle de diffusion donné :

$$I = \sum_{j=0}^{1020} j \text{Histo}(j) \qquad (253)$$

La Figure 119 montre une copie d'un des écrans du logiciel de pilotage et de reconstruction des diagrammes de diffusion.

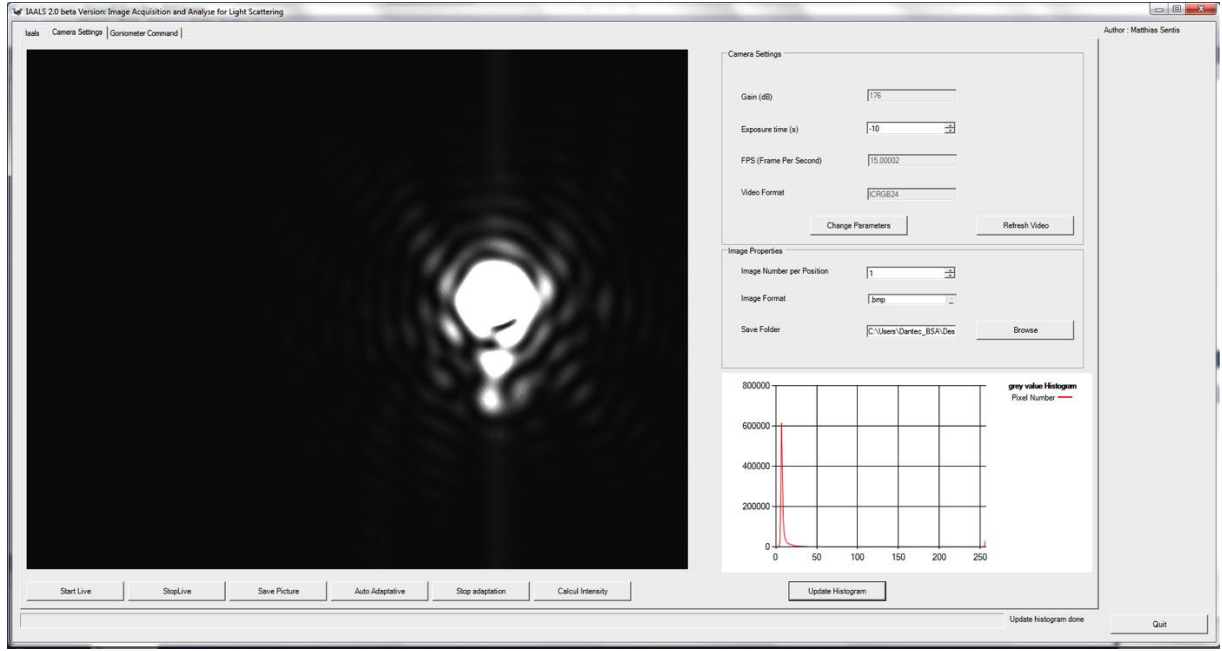

Figure 119 : Fenêtre du programme de pilotage de la caméra développé en VB.NET avec visualisation d'un cas saturé et son histogramme 8 bits

L'algorithme décrit ci-dessus pourrait être utilisé pour étendre encore la dynamique du système mais, sans informations détaillées sur les capacités du capteur à gérer la saturation, nous avons opté pour la prudence en nous limitant à une numérisation sur $\approx 10$ bits. Cet algorithme pourrait bien évidemment être utile pour étendre la dynamique de caméras 12, 14 ou 16 bits.

### 6.1.3 Pilotage du goniomètre

Le goniomètre permet de positionner précisément la caméra à un angle de diffusion donné, voir la Figure 113. Ce dernier peut être piloté manuellement, ou via un microcontrôleur communiquant par port série (RS 232). Un langage élémentaire et un logiciel ont été développés spécifiquement pour piloter par ordinateur le goniomètre, voir la Figure 120.

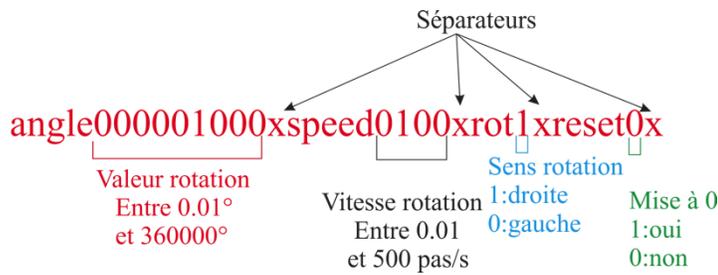

Figure 120 : Exemple d'instruction envoyée par le port série au microcontrôleur du goniomètre

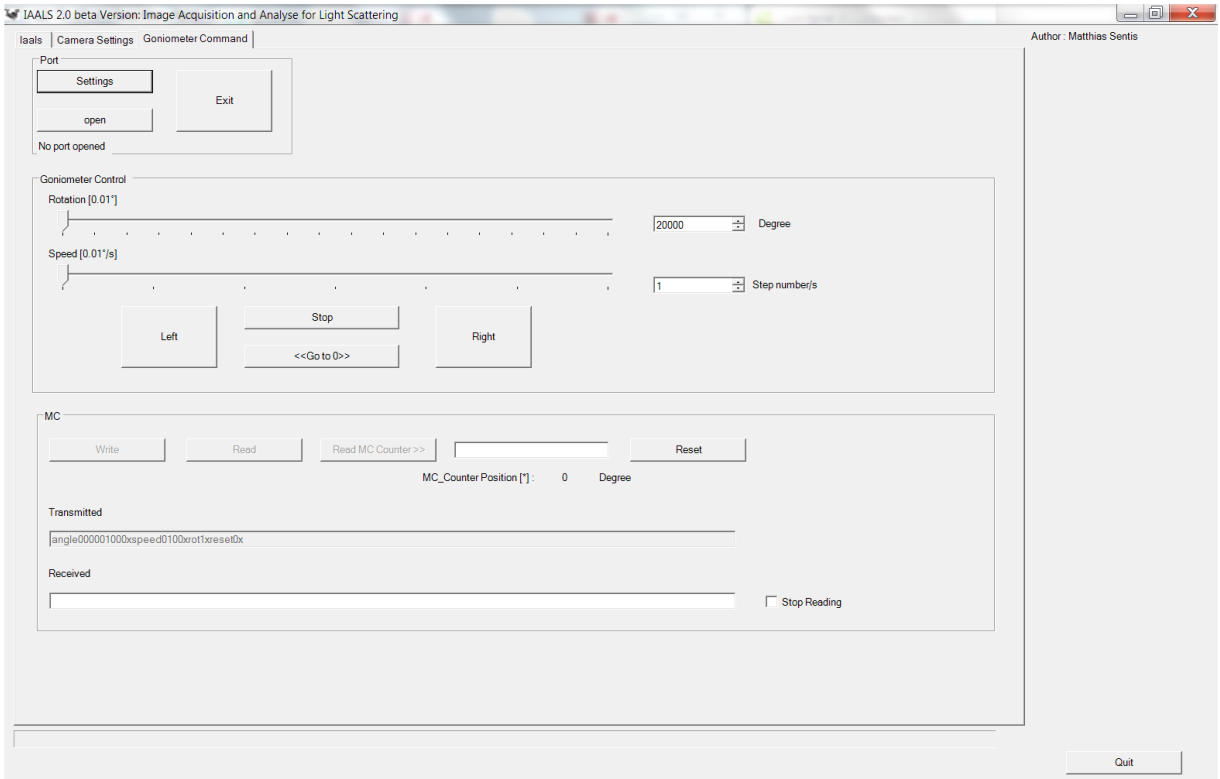

Figure 121 : Fenêtre du programme de pilotage du goniomètre (développé en VB.NET)

La Figure 121 montre une copie d'écran du logiciel de commande du goniomètre développé durant cette thèse. Ce programme en VB.NET est compatible avec le logiciel de pilotage de la caméra. Il permet de choisir le port série utilisé et ses paramètres *(Baud rate, Data Bits, parity...)*, la vitesse et la plage de rotation du goniomètre, etc…, mais aussi de lire les positions réellement atteintes par le goniomètre, de faire coïncider le zéro du goniomètre et du montage optique, etc…

### 6.1.4 Programme final et exemple de résultats

Un programme maître–ordonnant les tâches a été réalisé dans le but d'automatiser les mesures. L'utilisateur précise dans un fichier d'entrée les différentes positions angulaires $V_{angle}$ pour lesquelles il veut obtenir l'intensité. Grâce aux interfaces graphiques précédentes (Figure 119 et Figure 121) l'utilisateur peut également préciser tous les paramètres des différents systèmes. Une fois ceux-ci explicités, le déroulement des instructions est détaillé dans l'algorithme suivant :

**Algorithme de construction du diagramme de diffusion**

1:      **Pour i de 1 à** $Nb_{angle}$ **faire**
2:          **Si** i = 1 **alors**
3:              Vrot ← Vangle$[i]$
4:          **sinon**
5:              Vrot ← Vangle$[i]$ − Vangle$[i-1]$
6:          **fin si**
7:          *Tourner*( $V_{rot}$, Direction)
8:          $t_{exp}$ ← AdaptAuto()
9:          *Intensite* ← *CalculIntensite*()
10:         *EcrireFichier*($V_{angle}$, $t_{exp}$, *Intensite*)
11:     **fin pour**

avec $Nb_{angle}$ le nombre de points de mesure, $V_{rot}$ la valeur angulaire de la rotation, *Tourner* la fonction qui donne l'ordre au goniomètre de tourner, *AdaptAuto*() la fonction d'auto-calibration de la caméra, *CalculIntensite* la fonction qui retourne l'intensité à partir de l'histogramme étiré et *EcrireFichier* la fonction qui écrit les résultats dans un fichier texte.

Le fichier de « *résultats* » contient l'enregistrement de trois grandeurs : la valeur de l'angle de collection $\theta$, le temps d'exposition et l'intensité collectée. L'intensité finale mesurée pour l'angle $\theta$ est donnée par :

$$I(\theta) = \frac{Intensite}{t_{exp}} \qquad (254)$$

Ce banc optique et cette procédure d'acquisition peuvent avoir de nombreuses applications. La principale est bien sûr de produire des diagrammes de diffusion afin de caractériser la taille de particules en écoulement. Les quatre figures suivantes montrent une confrontation entre des expériences réelle et numérique (Monte-Carlo), dans le cas d'une suspension aqueuse, quasi-mono disperse, de particules sphériques de latex ( $\bar{D} = 3\mu m$, écart-type absolu : $\sigma_D = 0.065\mu m$ ). Tous les autres paramètres du montage (faisceau laser, cuve,...) sont les mêmes que pour les simulations Monte-Carlo du **chapitre 4.**

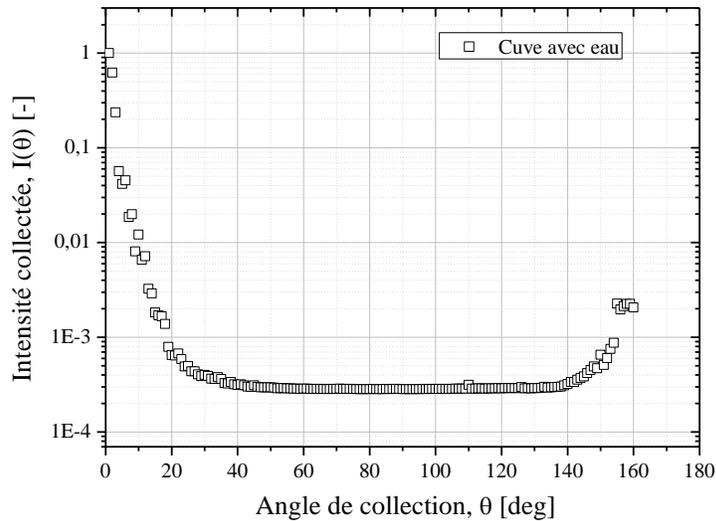

Figure 122 : Diagramme de diffusion expérimental de la cuve avec eau

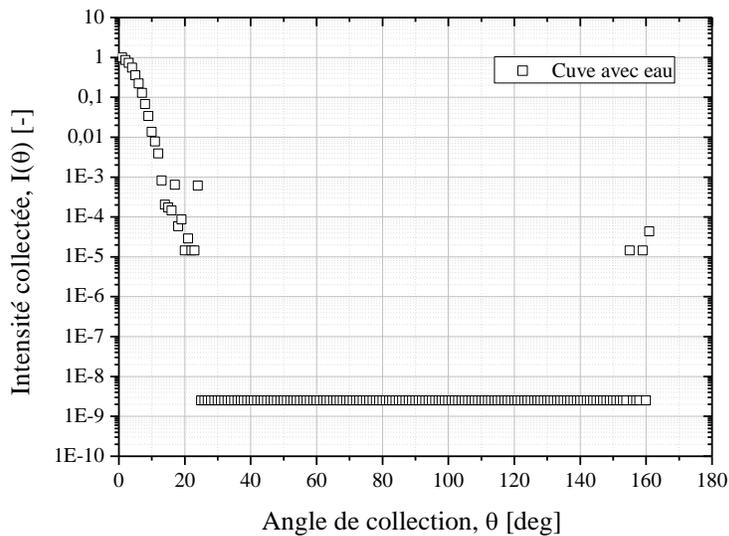

Figure 123 : Diagramme de diffusion de la cuve avec eau simulé avec le code de Monte-Carlo

Les premières mesures réalisées portent sur le cas de la cuve remplie d'eau, voir la Figure 122. On remarquera tout d'abord la forte dynamique en intensité des diagrammes avec près de 4 ordres de grandeur. L'intensité est maximale au voisinage $\theta = 1°$ (premier point de mesure pour éviter le faisceau direct), c'est-à-dire au voisinage du pic de diffraction et du faisceau direct. L'intensité décroît alors pour atteindre sa valeur minimale autour des 20°. Dans cette plage angulaire, pour ce montage, les mesures sont donc perturbées par les réflexions du faisceau incident. Un "plateau" apparait également autour des 160°, ce dernier rend compte des réflexions du faisceau direct vers l'arrière.

Les simulations avec le code Monte-Carlo en Figure 123 semble confirmer une pollution du signal optique par les réflexions parasites dans la cuve sur les mêmes gammes angulaires du diagramme de diffusion.

Aux autres angles, l'intensité collectée ne varie plus mais ne passe pas par zéro pour l'expérience. En théorie, ceci n'est pas possible en l'absence de particules dans le milieu, comme l'indique d'ailleurs le code Monte-Carlo, voir la Figure 123. Nous attribuons cela, au bruit de la caméra pour les temps d'exposition élevés.

La Figure 124 compare les simulations de Monte-Carlo et les résultats expérimentaux pour des particules de latex de diamètre moyen $\bar{D} = 95.5 nm$. Pour ces expériences, un filtrage par trous optiques est réalisé de manière à reproduire le prototype néphélométrique présenté dans le **chapitre 4** qui est également utilisé pour les simulations de Monte-Carlo. On constate un assez bon accord des prédictions des simulations de Monte-Carlo avec l'expérience 1 et 2 avec cependant un léger décroché sur les premiers degrés. Le "saut" soudain pour les expériences laisse présagé une erreur dans le calcul de l'intensité, soit à l'auto ajustement, soit plus probablement à la conversion du temps d'exposition.

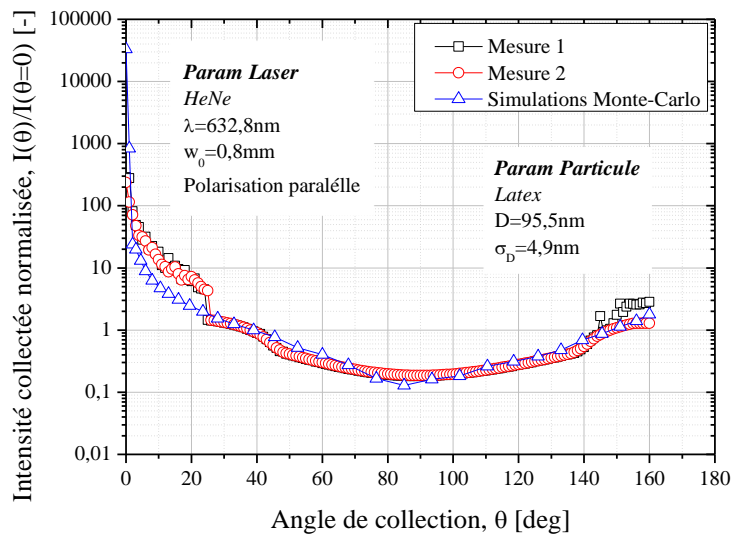

Figure 124 : Diagrammes de diffusion de particules de latex de diamètre moyen $\bar{D}$ =95.5nm et d'écart-type $\sigma_D = 4.9 nm$ dans de l'eau : simulations de Monte-Carlo et expériences.

## 6.2  Configuration néphélométrique

Le masque optique, optimisé dans le **chapitre 4**, a été réalisé à l'aide d'une technologie utilisée en électronique classique pour l'élaboration des masques de gravure pour les Composants Montés en Surface (CMS). Ces masques " pour la  lithogravure et gravure sous UV " sont conçus par découpe de feuilles métalliques de 127µm d'épaisseur (parfaitement conformables). La précision de cette technologie mature est amplement suffisante pour les tailles et formes de nos fentes optiques. Pour atténuer les réflexions parasites sur cet élément, le masque a été anodisé noir (traitement chimique de surface). Le résultat de cette conception est montré sur la Figure 125 (1).

Nous avons également réalisé le design du film de photodiodes organiques, voir à ce sujet le **chapitre 4**. Celui-ci a été techniquement réalisé par le CEA-LITEN en deux films assemblés. La Figure 125 (2) montre une moitié de film avant assemblage. La décomposition en deux films s'est avérée nécessaire du fait des limites actuelles du procédé de fabrication et ce, malgré le cahier des charges initial du projet (cf. **Chapitre 1**). La livraison tardive du film photosensible et sa connectique électrique spécifique nous a interdit toute campagne de mesure durant cette thèse.

La Figure 125 montre néanmoins un schéma de principe du prototype néphélométrique. Il est composé d'une cuve en verre (contenant le milieu particulaire) centrée au milieu d'un dispositif mécanique maintenant de manière concentrique: le masque de filtres optiques (Figure 125 (1)), un filtre neutre de densité, un filtre dichroïque et le film de photodiodes organiques (Figure 125 (2)).

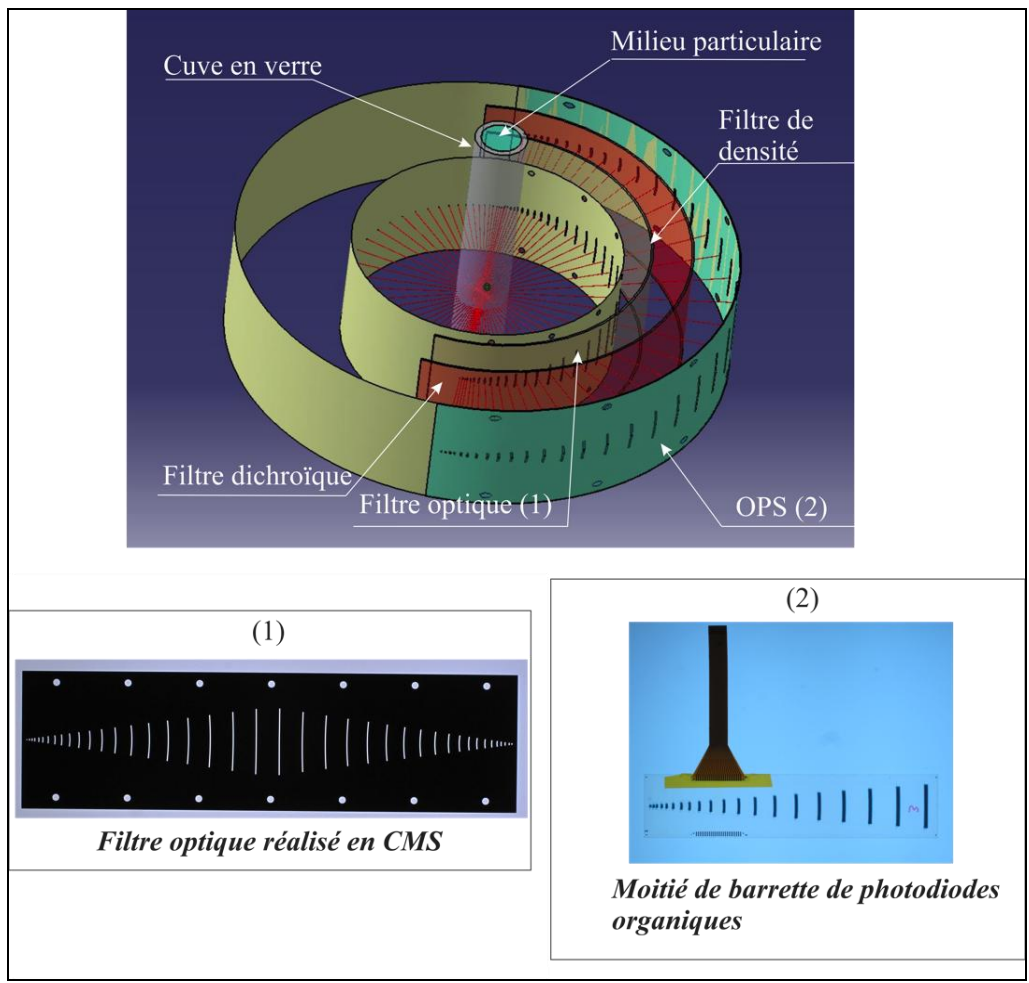

Figure 125 : Schéma du montage néphélométrique avec deux encarts sur son masque métallique de sténopés (1) et une des deux barrettes de photodiodes organiques (2)

L'électronique d'acquisition a été finalisée selon le schéma de fonctionnement de la Figure 126. Le film de photodiodes organiques réalisé par le CEA (a) est relié par une limande (b) à un mini-connecteur électrique (c). Des câbles blindés multibrins (d) acheminent les signaux électriques vers un boitier d'amplification (e). Ce dernier intègre 48 amplificateurs courant/tension *(Scitec)* avec un gain maximal de $10^3$ V/A et un offset réglable. Les signaux de tension sont acheminés via des câbles blindés vers un convertisseur (f) permettant d'attaquer simultanément les 16 entrées différentielles de chacune des trois cartes d'acquisition 16 bits/200kHz (*National instruments*, Ni 9205) (g) synchronisées entre elles et pilotées par un ordinateur (h) *via* un port USB. Un logiciel développé en C# permet de gérer l'acquisition simultanée des 48 voies à une cadence maximale de près de 1kHz (voir. Figure 127).

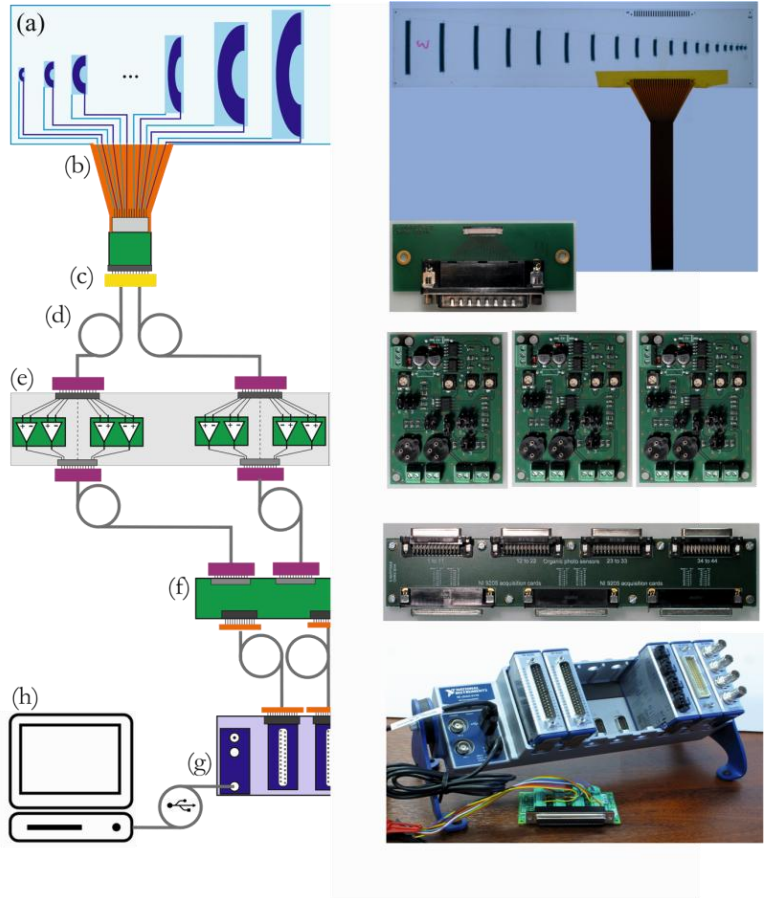

Figure 126 : Schéma du système électronique d'acquisition adapté aux photodiodes organiques (IUSTI)

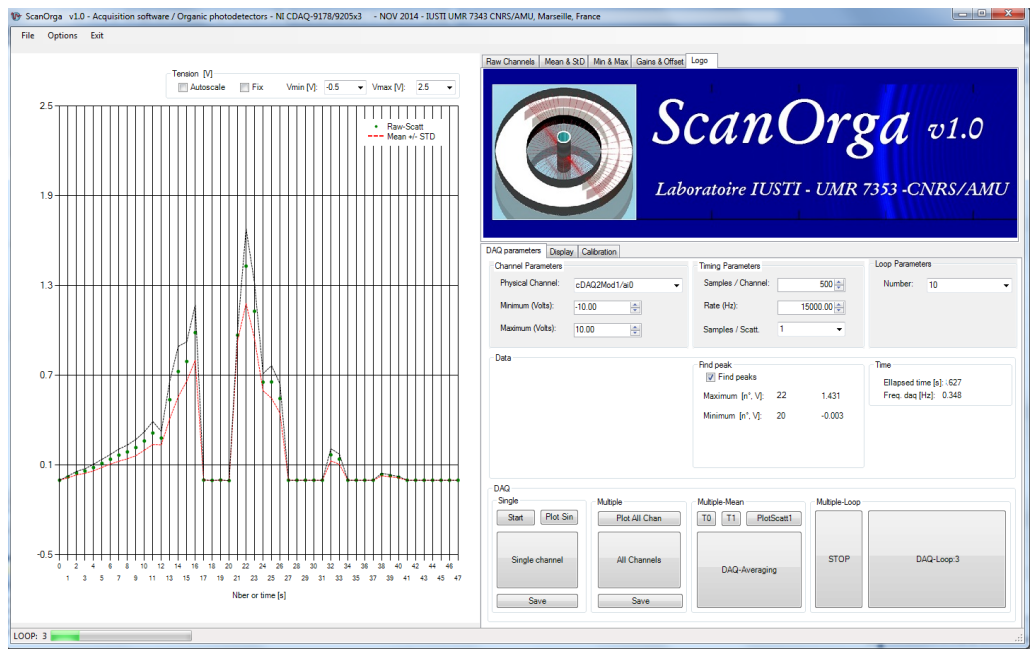

Figure 127 : Écran principal du programme d'acquisition synchrone développé en C#

La conception mécanique du montage néphélométrique présentée sur la Figure 125 est observable sur la Figure 128. Une diode laser de 30mW et λ=0.6328nm est focalisée sur un tube en verre contenant une suspension de particules de silice. Différentes rainures permettent de positionner les différents filtres, voir la Figure 125 avec, de l'intérieur vers l'extérieur : le masque CMS, un filtre de couleur et un polariseur linéaire. La lumière diffusée par le système est collectée sur le masque de photodiodes organiques fixé après le polariseur

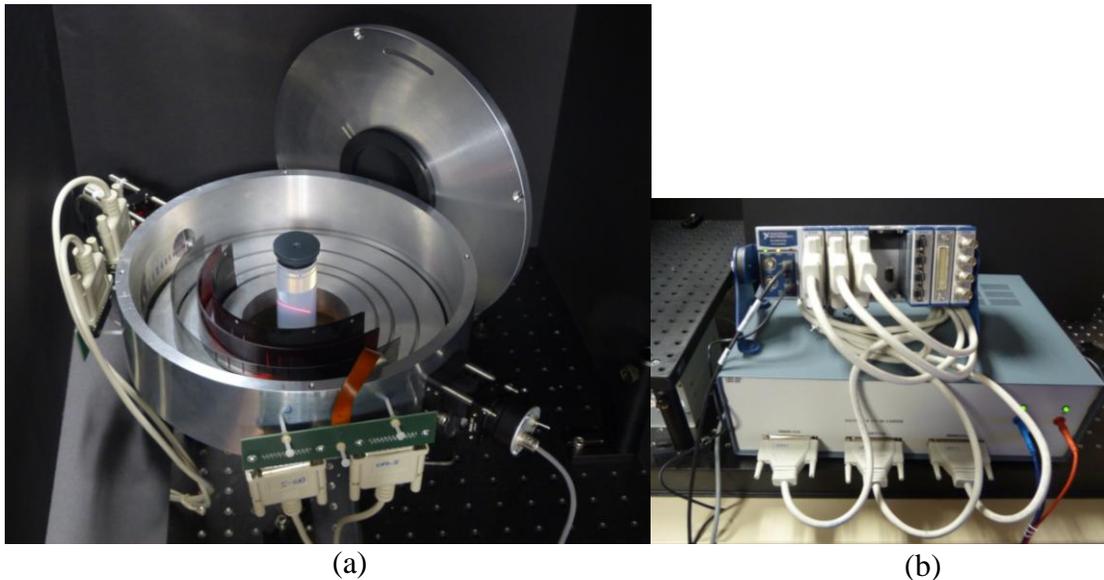

(a) (b)

Figure 128 : Réalisation mécanique du montage néphélométrique (a) et de l'électronique d'acquisition (boitier contenant l'électronique d'amplification et conversion courant-tension, cartes d'acquisition) (b)

Au terme de cette thèse, seules des mesures sur les capacités des photodiodes organiques ont pu être réalisés et notamment sur le temps de réponse et la linéarité avec le flux de ces dernières.

Sur la Figure 129, la réponse temporelle des photodiodes organiques est étudiée en hachant à différentes fréquences le faisceau généré par la diode laser (au moyen d'un chopper optique). A 2.3Hz, la réponse conserve une allure de créneau (rapport cyclique de 1) mais cette dernière devient légèrement triangulaire pour une fréquence de 15Hz puis parfaitement triangulaire pour 170Hz. En se fixant un critère raisonnable pour le taux acceptable de distorsion du signal, on retiendra que le temps de réponse des diodes est de l'ordre de 100ms.

Sur la Figure 130, la réponse des photodiodes organiques est mesurée lorsque le faisceau de la diode laser est atténué progressivement grâce à un filtre de densité variable. Avec les réglages actuels de l'électronique d'amplification, la réponse des photodiodes est linéaire pour une transmission décroissante de 100% à 40% (1 à 0.4).  En dessous de cette plage, les non linéarités deviennent très importantes. Le seuil de détection semble se situer autour de $I_{max}/I_{min}\# 1/40$, ce qui est relativement faible. D'après le CEA, le réglage de la tension de polarisation des diodes pourrait permettre d'étendre de manière significative la plage de linéarité des diodes.

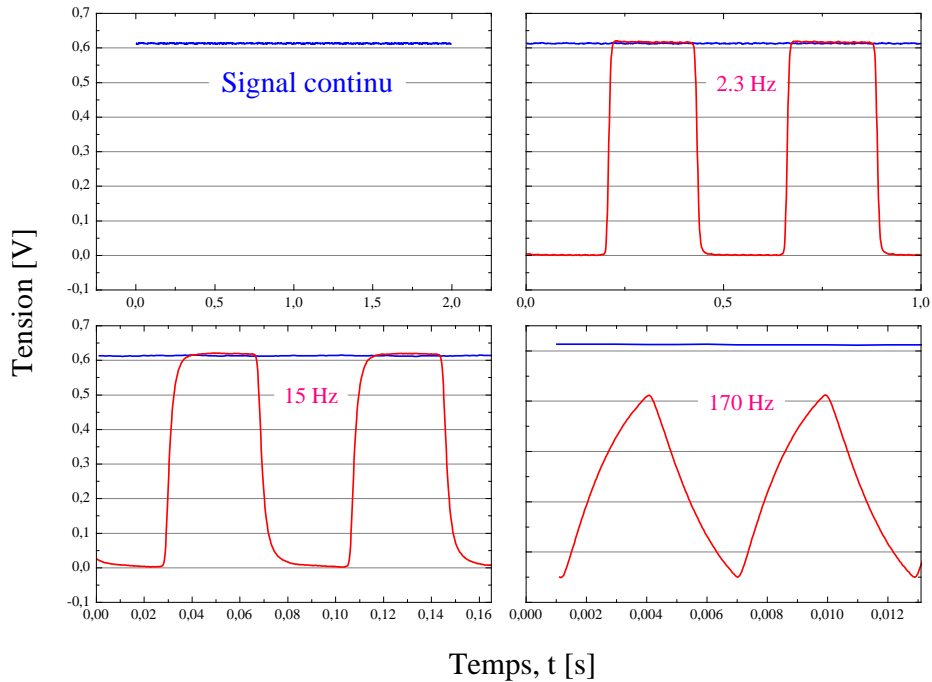

Figure 129 : Étude du temps de réponse des photodiodes organiques pour un signal de rapport cyclique unitaire

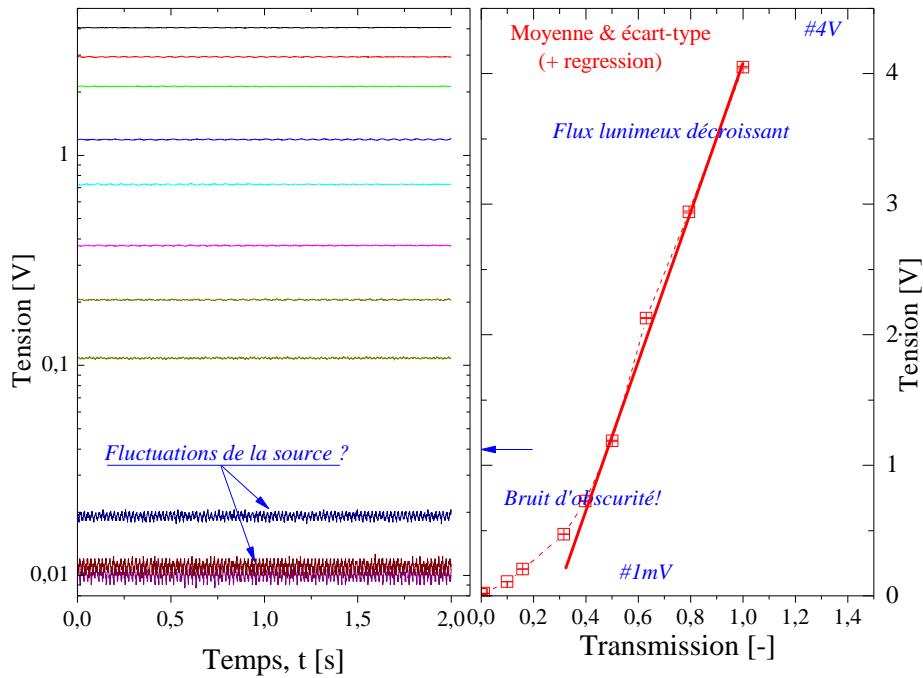

Figure 130 : Étude de la linéarité de la réponse des photodiodes organiques avec le flux lumineux

## 6.3 Expérience annexe

En parallèle du projet OPTIPAT, un autre banc multi-angle automatisé a été mis en place au laboratoire IUSTI pour caractériser des nanoparticules et leurs agrégats [**Montet 2014**]. Ce banc permet d'enregistrer des diagrammes de diffusion sur une large plage angulaire et avec une grande précision, voir la Figure 131. Un laser continu émettant à 407 nm (1) est focalisé sur une cuve cylindrique (ou rectangulaire) (8) contenant des nanoparticules. La lumière diffusée par ces nanoparticules est enregistrée par un photomultiplicateur (10). Ce dernier est positionné sur un berceau goniométrique de précision (9) de manière à pouvoir faire varier sa position autour de la cuve. Le goniomètre, de chez *Micos Gmbh*, est mis en mouvement par un moteur pas à pas lui-même piloté par un contrôleur *Pollux* interfacé à un PC. Deux photodiodes avalanches (APD, de chez *Thorlabs*) sont utilisées pour compenser à posteriori les fluctuations de puissance du laser (2) et pour mesurer l'intensité du faisceau transmis, c'est-à-dire le signal d'extinction (10).

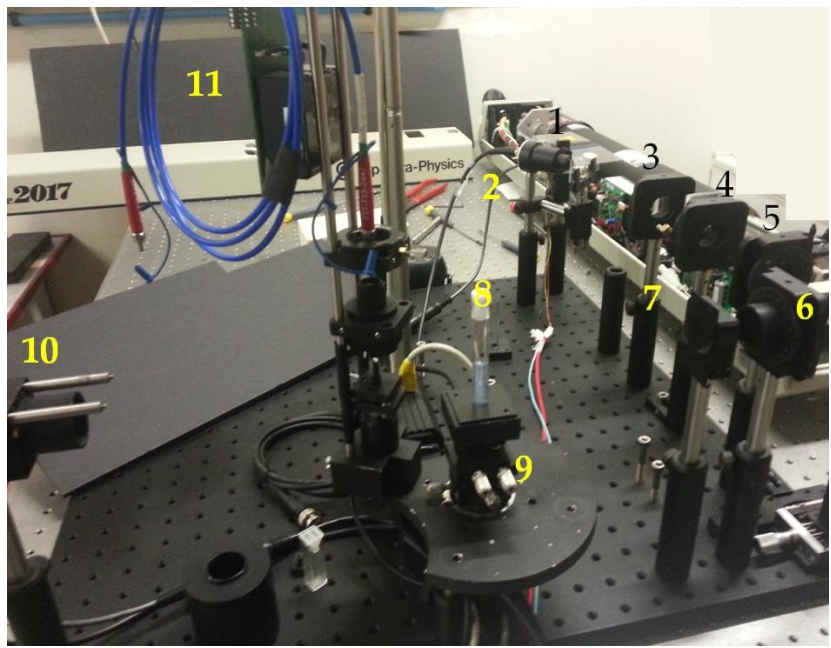

Figure 131 : Photographie du montage goniométrique

Comme l'indique la Figure 131, plusieurs optiques sont utilisées pour former le faisceau et notamment deux lentilles cylindriques (3-4) sont positionnées de manière à former une nappe laser horizontale. Deux polariseurs (5-6) sont également utilisés pour contrôler la polarisation du faisceau laser et celle de la lumière collectée. Enfin, une fente optique (7) est positionnée avant la cuve pour bloquer certaines réflexions parasites. Le signal diffusé est réfléchi par un prisme (9) vers une fibre optique de 10 à 100 µm de cœur reliée au photomultiplicateur. Trois optiques supplémentaires sont placées sur le montage : un filtre dichroïque centré sur la longueur d'onde laser (10), un diaphragme d'ouverture réglable (11) et une lentille permettant de focaliser le faisceau sur la fibre. Pour limiter les réflexions parasites, le miroir est positionné légèrement sous le plan de diffusion conventionnel. De fait, outre les méthodes inverses employées, ce dispositif expérimental comporte trois caractéristiques innovantes: la forme du faisceau (une nappe laser horizontale), la forme de la cuve (cylindrique), la détection (sous le plan de diffusion usuel, angle $\alpha$).

Les Figure 132 et Figure 133 proposent une comparaison entre les diagrammes expérimentaux et ceux simulés avec le code Monte-Carlo pour une suspension colloïdale de particules de latex de $\bar{D}$ =95.5nm, éclairée par un faisceau gaussien circulaire de polarisation parallèle. La lumière est collectée à $\alpha$ = 3° sous le plan de diffusion. Deux cas sont considérés pour les simulations.

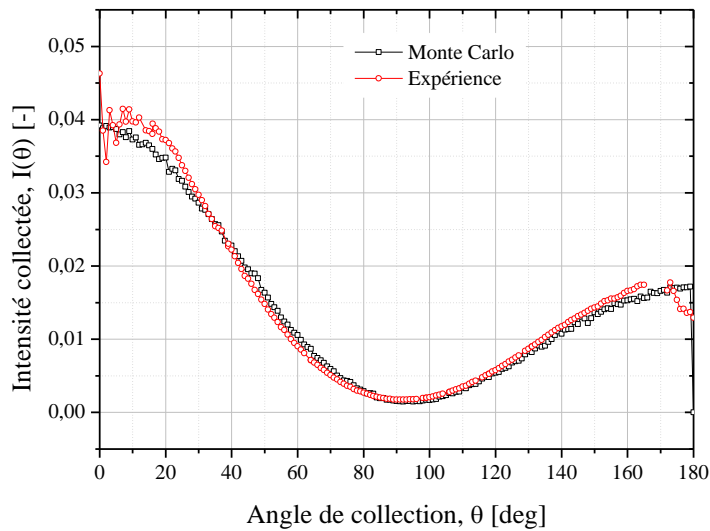

Figure 132 : Diagramme de diffusion expérimental (configuration néphélométrique) obtenu pour des particules de latex de 95.5nm et d'écart-type 4.9nm en polarisation parallèle

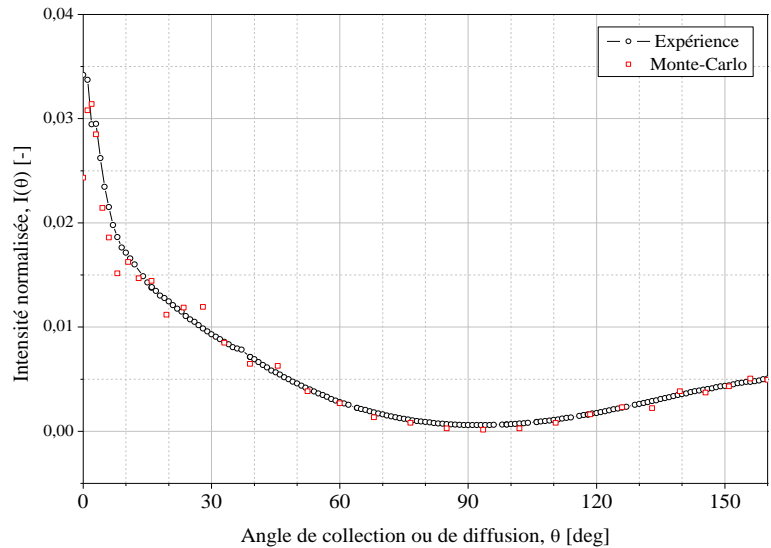

Figure 133 : Diagramme de diffusion pour des particules de latex de 95.5 nm - polarisation parallèle obtenue pour la configuration néphélométrique

Dans le premier cas, les conditions expérimentales sont simulées en se plaçant très loin de l'échantillon à caractériser, la détection sous le plan est prise en compte mais pas la lentille de collection. Le saut à 170° est un artefact du dispositif expérimental. On constate que l'accord entre expériences et simulations est plutôt bon à l'exception de l'avant du diagramme de diffusion. Nous attribuons cette différence aux nanorugosités de cuve qui ne sont pas pris en compte dans le modèle de Monte-Carlo.

La deuxième configuration correspond au prototype néphélométrique présenté dans le **chapitre 4** qui utilise des trous optiques pour le filtrage et qui comporte comme on peut le voir sur la Figure 133, 41 points de mesures. On constate encore une fois un bon accord entre l'expérience et les simulations mais tout de même moins bon que précédemment. En fait, le filtrage par trous optiques circulaires sous le plan de diffusion réduit considérablement le flux lumineux filtrée entrainant alors un bruit statistique important. En effet, à nombre de rayons constant par rapport à la configuration précédente, on constate une description beaucoup moins bonne du diagramme de diffusion.

## 6.4 Perspectives

Dès réception des barrettes de photodiodes assemblées, avec trous de centrage et connectique ad hoc, l'assemblage schématisé par la Figure 125 sera réalisé. Le système complet sera ensuite étalonné, notamment en procédant au réglage du gain et de l'offset des 48 étages amplificateurs. Une fois ceci réalisé, il faudra tester la fonctionnalité des éléments photosensibles (le rendement à la fabrication est de 90%), les effets de *cross-talking* entre détecteurs, puis leur réponse au flux et en temps. Il nous faudra ensuite réaliser des tests sur différents écoulements (suspensions, lit fluidisé).

# Chapitre 7

## Conclusions et perspectives

Ces travaux de thèse de doctorat se sont inscrits dans le cadre d'un projet de recherche multipartenaire et multidisciplinaire, avec des aspects académiques et des visées applicatives. Dans ce projet, la tâche incombant au laboratoire IUSTI est d'étudier dans quelle mesure les photodétecteurs organiques conformables peuvent permettre l'élaboration de solutions innovantes pour la granulométrie optique d'écoulements diphasiques et de suspensions en conduite cylindrique.

Dans un premier temps, nous avons dressé un panorama des enjeux et des nombreux aspects de ce projet : écoulements diphasiques et suspensions, caractéristiques et limites des granulomètres actuels, spécificités des photodétecteurs organiques. Sans être exhaustif, les principaux modèles et théories de diffusion de la lumière par des particules ont également été passés en revue, afin de juger de leur pertinence dans le cadre de ce projet et comme introduction au développement de modèles originaux. Il s'avère que ces outils ne s'appliquent qu'à de très petites particules (qui peuvent être complexes en forme et composition), ou bien à des particules strictement sphériques et homogènes. Pour repousser ces limites, le modèle hybride de Van de Hulst avec une méthode d'échantillonnage aléatoire a été implémenté. Les résultats obtenus avec ce modèle s'accordent parfaitement à ceux des théories électromagnétiques en dehors des zones singulières (diffraction, arc-en-ciel, diffusion critique,...). La théorie d'optique physique de Marston a également été étendue au cas de particules sphéroïdales. Ce travail théorique a permis de montrer, notamment, qu'au voisinage de l'angle critique, comme pour l'arc-en-ciel, la diffusion pouvait être assimilée à un phénomène de (semi-)diffraction. Le développement de ce nouveau modèle de diffusion constitue un pas important vers la généralisation du modèle hybride aux particules non sphériques à surface spéculaire.

Pour coupler les résultats des modèles de diffusion de particules avec ceux décrivant les autres phénomènes de diffusion rencontrés dans notre problème (réflexions et réfraction des parois), nous avons intégralement développé un modèle et un code Monte-Carlo. Le code parallélisé permet de prendre en compte : le profil et la propagation du faisceau laser, les effets des différents éléments du montage optique (cuve, lentille, forme des détecteurs), les caractéristiques granulométriques du milieu, les transitions entre régimes de diffusion simple et multiple,... Grâce à cet outil de simulation, nous avons pu concevoir un prototype de néphélomètre (c'est-à-dire un granulomètre multi-angle) utilisant au mieux la conformabilité et la complexité de forme des photodiodes organiques. Pour conserver cette conformabilité, le prototype est équipé de sténopés (ou masques optiques) en lieu et place des optiques de collection utilisées classiquement en granulométrie optique. L'optimisation globale de son design et de ses propriétés (taille du faisceau, résolution angulaire,...) a été faite pour des conditions de diffusion simple, c'est-à-dire pour des écoulements optiquement dilués. Du fait de sa conception, la réponse de ce prototype peut-être directement comparée (pour certaines zones angulaires) aux prédictions de la théorie de Lorenz-Mie. Nous avons pu ainsi valider le code Monte-Carlo, mais également utiliser la théorie de Lorenz-Mie pour accélérer certains calculs directs. Les limites de cette approche sont liées aux réflexions spéculaires complexes induites par la cellule de mesure cylindrique et la résolution encore trop faible de la technique d'impression des photodiodes organiques. Trois premières méthodes, encore à l'état embryonnaire, ont néanmoins été proposées pour inverser les diagrammes de diffusion en diffusion simple. La première repose sur la mesure d'un rapport entre l'intensité diffusée au voisinage d'une singularité et l'intensité diffusée dans une zone correctement décrite par l'optique géométrique. L'idée est ici, qu'asymptotiquement, l'intensité diffusée croit en $D^2$ dans les zones classiquement décrites par l'optique géométrique, alors qu'elle croit en $D^4$ pour la diffraction, en $D^{7/3}$ pour l'arc-en-ciel, etc… Cette méthode, permet avec une simple loi de puissance de déduire directement le diamètre moyen d'une population de particules sphériques sur une gamme relativement étendue : $\approx$ 20-1000µm. Sa précision n'excède pas 10% en moyenne, mais cela peut être suffisant pour de nombreux procédés. A noter que la mesure est quasi insensible à l'écart-type de la distribution (log-normale dans notre cas). La seconde méthode repose sur l'inversion d'un diagramme complet à l'aide d'une méthode algébrique. Plus couteuse numériquement, elle devrait permettre d'obtenir une résolution accrue sur la forme de la distribution granulométrique. Dans l'état, elle ne donne pas de résultats exploitables sauf pour les distributions monodisperses. La troisième méthode repose également sur une loi de puissance mais cette dernière est calée sur l'évolution du facteur d'asymétrie. De ce fait, elle est limitée à la gamme de taille $\approx$ 200nm-10µm. Ce premier prototype, à 41 zones de détection réparties sur la longueur d'une feuille A3, a également été testé pour des conditions de diffusion multiple, c'est-à-dire des milieux particulaires denses. Les résultats numériques obtenus ont permis de dégager des tendances mais pas de proposer une solution de mesure pour les milieux denses. Il y a plusieurs raisons à cela. Tout d'abord, il s'est avéré assez rapidement que la mesure en cuve cylindrique avec faible rayon de courbure entraine une très forte pollution du signal de rétrodiffusion par la réflexion spéculaire. De plus, ce type de cuve ne permet pas de se placer dans des conditions de milieu semi-infini. Ajoutons également que la très faible résolution temporelle et spatiale des photodétecteurs organiques actuels constitue une limite critique. D'un point de vue expérimental, nous avons développé un banc goniométrique utilisant une caméra CCD pour l'enregistrement des diagrammes de diffusion. Cette expérience a été conçue pour tester notre code Monte-Carlo et dans l'attente des premiers photodétecteurs organiques. Un procédé de fabrication des masques optiques a été testé avec succès. Il repose sur la découpe laser de masques CMS qui sont ensuite anodisés en noir pour limiter les réflexions parasites. Malheureusement, bien que nous ayons développé toute une chaîne d'amplification et de numérisation des signaux issus des photodétecteurs organiques, ces derniers nous ont été livrés trop tard pour que des résultats expérimentaux puissent être produits dans ce manuscrit.

Les perspectives de ce travail sont nombreuses, que ce soit au niveau théorique, numérique ou expérimental. Ainsi, les effets de polarisation liés aux particules, les particules non sphériques et les effets liés aux singularités (diffraction par les sténopés, arc-en-ciel produit par la cellule cylindrique, ...) ne sont pas encore pris en compte par notre code Monte-Carlo. Ces limites pourraient être en partie levées en utilisant le formalisme des matrices de Muller et des théories électromagnétiques *ad hoc*. Il faudrait par ailleurs améliorer la robustesse des méthodes d'inversion proposées. Les études menées en condition de diffusion multiple doivent être également approfondies. Au niveau expérimental, la validation du prototype néphélométrique paraît incontournable.

Il résulte de ces travaux que la conformabilité des détecteurs organiques est clairement un atout pour la granulométrie optique. Avec une augmentation significative de la densité d'intégration des surfaces photosensibles, cette technologie émergente pourrait bien servir de base au développement de solutions de mesure performantes et originales.

# Annexe 1. Analyse des poudres SANOFI

Ces placébos ont été analysés au microscope électronique à balayage (MEB – Philips XL30 SFEG STEM ) au CP2M (Centre Pluridisciplinaire de Microscopie électronique et de Microanalyse, Marseille).

La Figure 1 (a) présente à titre d'exemple un cliché MEB du « *placebo 150* ». Celle-ci met en évidence la complexité de forme et la diversité des tailles des particules de cet échantillon (et *in fine* de la difficulté de les caractériser optiquement, les théories et modèles développés par le IUSTI se basant sur l'hypothèse de sphéricité des particules).

Il semble également que celui-ci présente des agrégats de particules. Ces derniers peuvent résulter du transport (compactage), tout comme être naturellement constitutifs du placébo du fait de son procédé de fabrication. De plus, l'état de surface des particules a pu être altéré par l'impact du faisceau d'électrons du MEB (« arrachage de matière »). Ce cliché MEB n'est donc pas nécessairement représentatif de la structure et de la taille des particules du Placébo 150 en cours de fabrication.

Le logiciel open source d'analyse d'image *ImageJ* a été utilisé pour post-traiter et extraire la granulométrie des clichés MEB. Le post-traitement de chaque cliché passe par plusieurs étapes, à savoir le seuillage et la binarisation de l'image (Figure 1 (b)), la détection des contours de chaque particule (Figure 1 (c)), puis l'estimation de leur aire projetée, $A$ (i.e l'aire des contours détectés). On peut en déduire le diamètre équivalent sphérique de chaque particule $D_{Sph} = \sqrt{4A/\pi}$ et ainsi obtenir la distribution granulométrique (PSD) associée à chaque cliché, comme le montre la Figure 134.

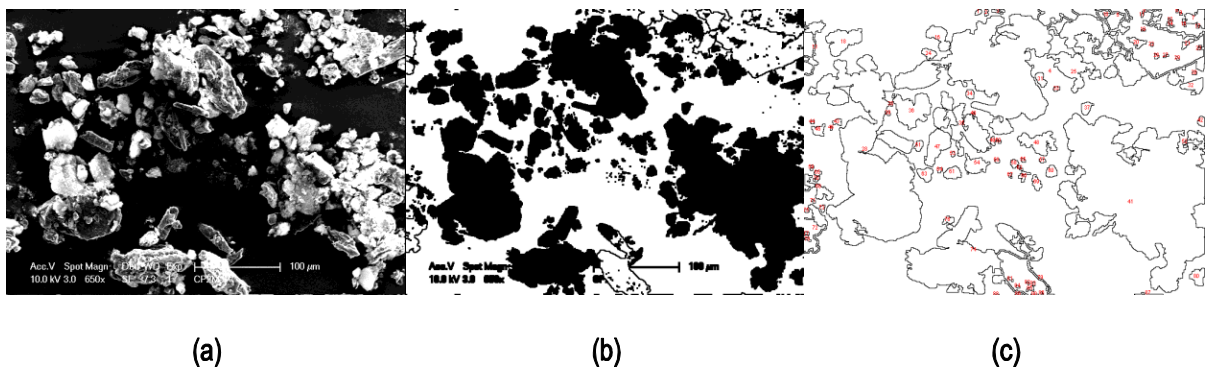

(a)  (b)  (c)

Figure 134 : (a) Image MEB d'un échantillon prélevé dans le sac de "placébo 150", (b) Binarisation de l'image (c) Détection des contours et numérotation des particules

## Distribution granulométrique du « placébo 150 »

Le diamètre équivalent sphérique des particules de la Figure 2 est essentiellement compris dans la gamme : [5µm – 25µm]. La PSD présente quelques pics dans les grandes tailles, correspondant aux quelques grosses particules (peut être agrégées).

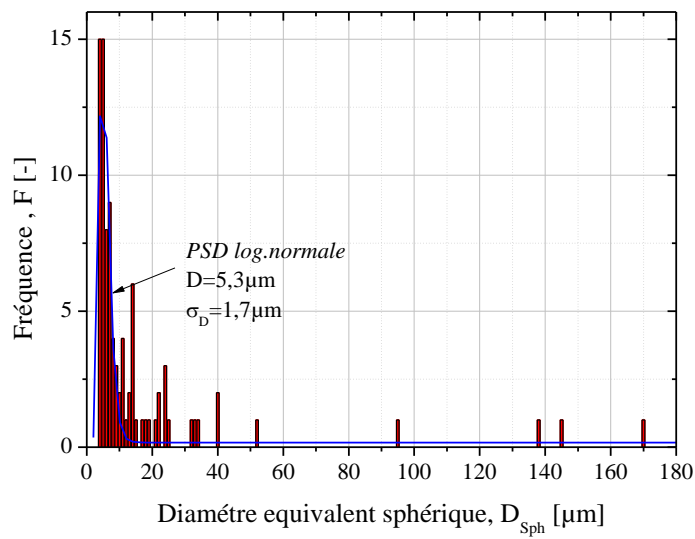

Figure 135 : Distribution en nombre du diamètre équivalent sphérique obtenue à partir de l'analyse de l'image MEB de la Figure 134

Une majorité des particules sont distribuées autour de la dizaine de microns. Les quelques pics observés pour les grandes tailles correspondent très probablement à de gros aggrégats.

## Distribution granulométrique du « placébo 850 »

Bien que visuellement les particules du *« placébo 850 »* semblent être de même nature que celles du *« placebo 150 »*, nous avons procédé à l'analyse d'un échantillon du *« placébo 850 »*.

La Figure 136 (a) présente un exemple de cliché MEB du *« Placebo 850 »*. *C*e dernier comporte un très grand nombre de particules dont la taille diffère sensiblement, de telle sorte qu'une analyse séparée des particules de grandes et petites tailles est nécessaire.

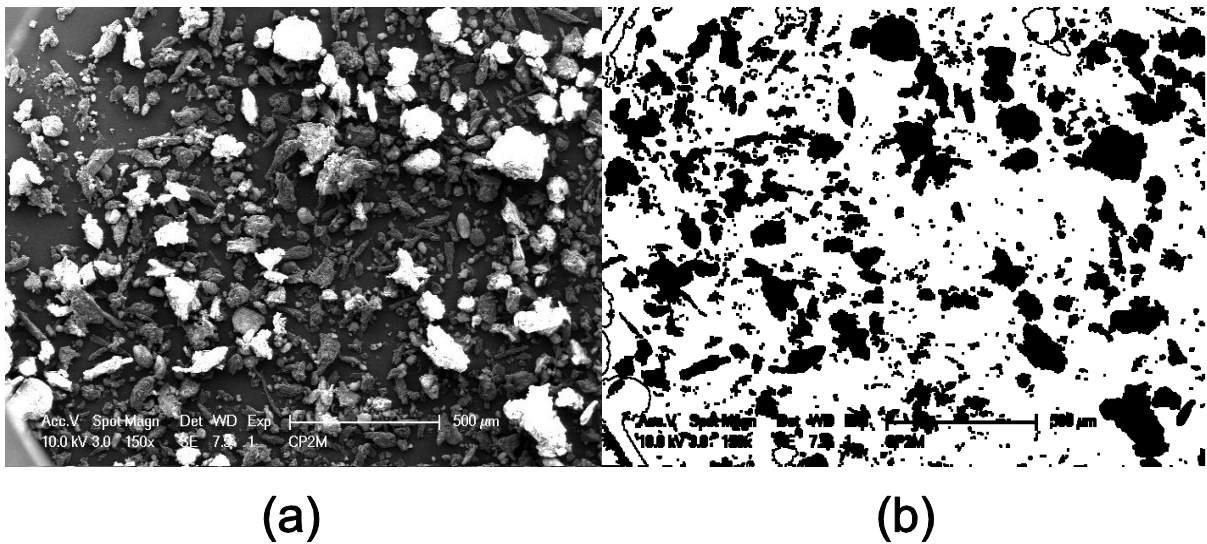

Figure 136 : (a) Image MEB d'un échantillon prélevé dans le sac de "placébo 850", (b) image binaire

La Figure 137 présente le contour associé à ces particules, tout en les discriminant selon leur taille (Figure 137 (a) pour les plus grosses et Figure 137 (b) pour les plus petites).

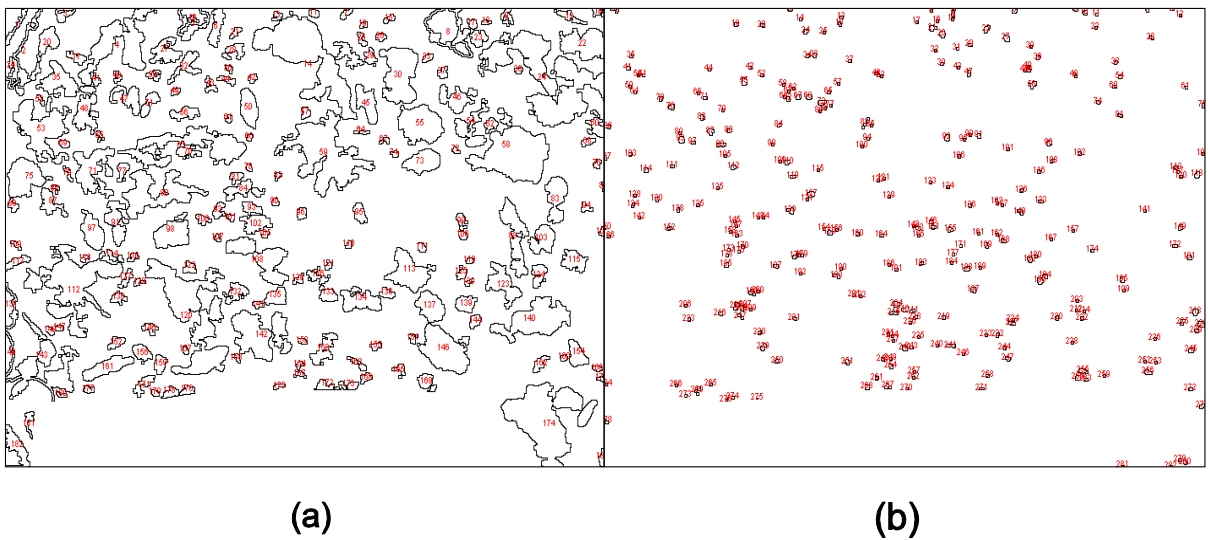

Figure 137 : Contour des deux populations de particules observées sur l'image Figure 136, (a) grosses et (b) petites particules

Les PSDs en nombre de $D_{Sph}$ associées sont reportées à la Figure 138. Les plus grosses particules ont un diamètre équivalent sphérique dans la gamme [25µm-250µm] avec un diamètre moyen de 24.3µm et un écart-type de 0.04µm (voir Figure 5 (a)), tandis que les particules de plus petite taille ont un diamètre équivalent sphérique dans [2µm- 22µm], avec un diamètre moyen de 11.2µm et un écart-type de 0.04µm

Les particules du « *Placebo 850* » semblent légèrement plus grosses, mais cela peut simplement résulter de la méthode de préparation des échantillons observés au MEB.

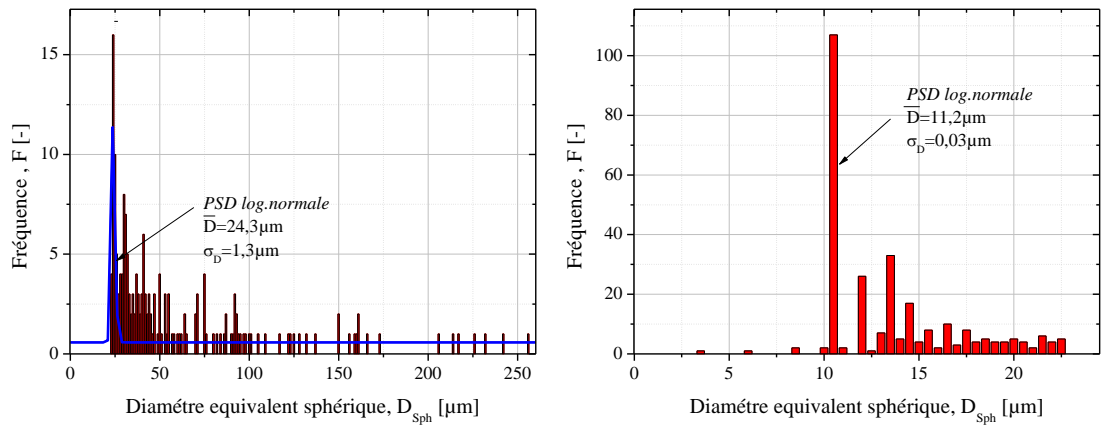

Figure 138 : PSD du diamètre équivalent sphérique des particules du *"Placebo 850"*. (a) PSDs des particules de la Figure 4 (a) et (b) PSDs des particules de la Figure 4 (b)

## Analyse des images enregistrées au microscope optique

L'étude précédente montre que la taille des particules est globalement largement supérieure à $D_{Sph} = 2\mu m$, de telle sorte que ces particules sont observables au microscope optique. Les analyses optiques ont été réalisées au laboratoire. Elles sont plus rapides à pratiquer et necessitent des conditions de préparation moins drastiques (pas de vide poussé).

Les clichés obtenus ont été analysés avec la méthode de traitement d'images précédente.

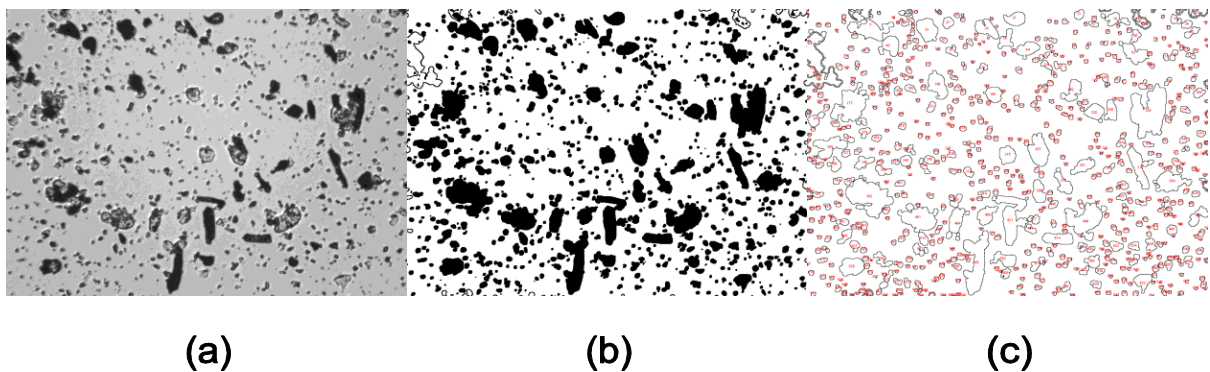

(a)           (b)           (c)

Figure 139 : (a) Image enregistré au microscope optique d'un échantillon prélevé dans le sac de 150g, (b) Binarisation de l'image (c) Détection des contours et numérotation des particules

La Figure 139 (a) montre que certaines particules sont transparentes, les particules étant à bse de sucre, on pourrait s'attendre à observer des crisataux. Les échantillons ont été observés au microscope optique en lumière polarisée. Aucuns changements significatifs de couleur n'a été observé avec notre dispositif, ce qui signife que l'échantillon ne contient pas ou peu de cristaux visibles.

La transparence des objets en microscopie optique fait que ceux-ci semblent de forme légérement moins complexe que les images MEB.

La Figure 140 présente la PSD de $D_{Sph}$ obtenue par analyse de la Figure 139 (a). Elle se situe dans la gamme [2µm-85µm], avec un diamètre moyen de 9.2 µm et un écart-type de 0.52µm. Ceci est en bon accord avec la distribution obtenue par analyse MEB.

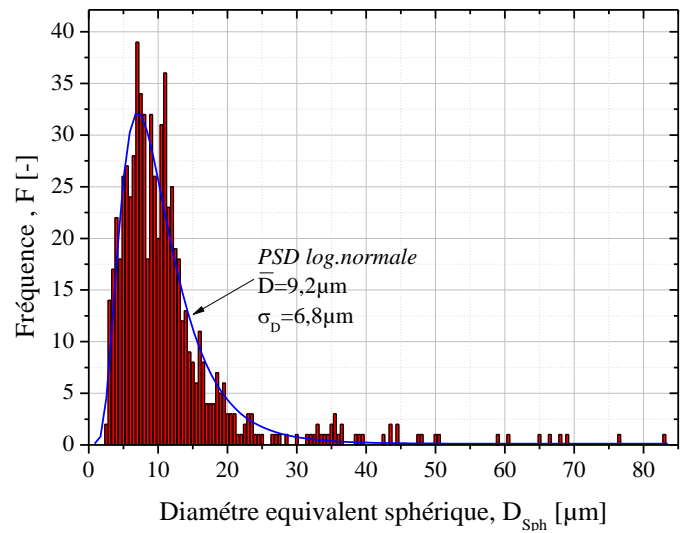

Figure 140 : PSD obtenue au microscope optique du *"Placebo 150"*

En conclusion, nos analyses des deux échantillons de placébos ont permis de souligner la diversité des tailles des particules et l'importance de la polydispersion, néanmoins elles se situent en moyenne dans une gamme de taille de 5µm à 50µm. De plus, les particules semblent difficiles à modéliser par des sphéres. La caractérisation optique de tels échantillons va donc être extrêmement complexes et ces placébos ne pourront en aucun cas servir de particules « modèle » pour tester nos prototypes de granulomètres utilisant des OPS.

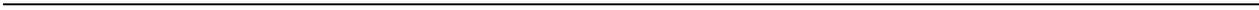

# Annexe 2. Code d'optimisation des surfaces photosensibles pour la mesure en configuration diffractométrique

```matlab
%%%%%%%%%%%%%%%%%%%%%%%%%%%%%%%%%%%%%%%%%%%%%%%%%%%%%%%%%%%%%%%%%%%%%%%%%%%%
%% Creation Branche masque photodiodes design diffraction %
%% Design final en assemblant 4 de ces branches %
%%%%%%%%%%%%%%%%%%%%%%%%%%%%%%%%%%%%%%%%%%%%%%%%%%%%%%%%%%%%%%%%%%%%%%%%%%%%
% Nettoyage des variables
clear all
% Définition taille des pixels pour carte 1200dpi
sizepix=25.4/1200
% Borne supérieur de la feuille de détection 40mmx16mm
borne=round(40/sizepix+1);
% Initialisation des coordonnées des pixels en x et y
for i=1:borne
 xpix(i)=(i-1)*sizepix;
end
borne=round(16/sizepix+1);
for i=1:borne
 ypix(i)=-1+(i-1)*sizepix;
end
% Initatialisation à 0 de la carte des pixels
I=zeros(int64(40/sizepix+1),int64(16/sizepix+1));
% Création du trous pour laisser passer le faisceau laser incident
r1=0.5;
theta=[0:0.0001:2*pi()];
x=r1.*cos(theta);
y=r1.*sin(theta);
taille=size(theta);
% Si intersection entre le cercle et un pixel, on met le pixel à 1
 for k=1:1:taille(2)
 ind1 =int64((x(k)+1)/sizepix+1);
 ind2= int64((y(k)+1)/sizepix+1);
 I(ind1,ind2)= 1;
 end
% Première zone photosensible respectant la distance minimale de 1mm avec
% le trou laser (ici décalage pour centrage+assemblage)
r2=3.5 ;
beta=log(10)/9;
alpha=1/exp(beta);
% Une branche comporte 8 zones photosensibles
for k=1:8
 % Largeur anneau évolue selon une exponentielle
 r1=r2+1;
 r2=r1+alpha*exp(k*beta);
 % La hauteur donc l'angle du cercle évolue selon une exponentielle
 beta2 =log(atan(14/r1)/atan(0.5/r1))/9;
 alpha2 = 2*atan(0.5/r1)/exp(beta2);
 % Borne de l'angle maximal du cercle
 thetamax=alpha2*exp(k*beta2);
 theta=[0:0.00001:thetamax];
 taille=size(theta);
 beta3 =log(atan(14/r2)/atan(0.5/r2))/9;
```

```matlab
alpha3 = 2*atan(0.5/r2)/exp(beta2);
thetamax=alpha3*exp(k*beta3);
theta2=[0:0.00001:thetamax-0.02];
taille2=size(theta2);
x=r1.*cos(theta);
y=r1.*sin(theta);
% Intersection pixel équation : pixel à 1
for k=1:1:taille(2)
ind1 =int64((x(k))/sizepix+1);
ind2= int64((y(k)+0.5)/sizepix+1);
I(ind1,ind2)= 1;
end
indminx1=double((x(1))/sizepix+1)
indminy1= double((y(1)+0.5)/sizepix+1);
indmaxx1=double((x(taille(2)))/sizepix+1)
indmaxy1= double((y(taille(2))+0.5)/sizepix+1);
x=r2.*cos(theta2);
y=r2.*sin(theta2);
for k=1:1:taille2(2)
ind1 =int64((x(k))/sizepix+1);
ind2= int64((y(k)+0.5)/sizepix+1);
I(ind1,ind2)= 1;
end
indminx2=double((x(1))/sizepix+1)
indminy2= double((y(1)+0.5)/sizepix+1);
indmaxx2=double((x(taille2(2)))/sizepix+1)
indmaxy2= double((y(taille2(2))+0.5)/sizepix+1);
a1=(double(indminy2)-double(indminy1))/(double(indminx2)-double(indminx1));
b1=double(indminy1)-a1*double(indminx1);
a2=(double(indmaxy2)-double(indmaxy1))/(double(indmaxx2)-double(indmaxx1));
b2=double(indmaxy2)-a2*double(indmaxx1);
for i=indminx1:0.1:indminx2
ind1=int64(i);
ind2=int64(a1*indminx2+b1);
I(ind1,ind2)= 1;
end
for i=indmaxx1:0.1:indmaxx2
ind1=int64(i);
ind2=int64(a2*indmaxx2+b2);
I(ind1,ind2)= 1;
end
end
imagesc(xpix,ypix,I');
I=logical(I');
imwrite(I,'DesignDiffractionCirc-Branche4.bmp');
axis equal
```

# Annexe 3. Définition basique des polynômes de Zernike [Zernike 1934]

Ces derniers jouent un rôle très important en optique [**Born 1999**] et notamment pour la caractérisation des différentes aberrations optiques classiques (aberrations sphériques, comas...).

Ils se décomposent en fonctions paires $Z_n^m$ et impaires $Z_n^{-m}$ avec

$$Z_n^m\left(r,\varphi\right) = R_n^m\left(r\right)\cos\left(m\varphi\right) \qquad (255)$$

$$Z_n^{-m}\left(r,\varphi\right) = R_n^m\left(r\right)\sin\left(m\varphi\right) \qquad (256)$$

où $m$ et $n$ des entiers naturels positifs ou nuls, $\varphi$ l'angle azimutal, $r$ la distance radiale avec $r \in \left[0,1\right]$ et $R_n^m$ les polynômes radiaux qui s'expriment par :

$$R_n^m\left(r\right) = \sum_{k=0}^{\frac{n+m}{2}} \frac{\left(-1\right)^k\left(n-k\right)!}{k!\left(\dfrac{n+m}{2}-k\right)!\left(\dfrac{n-m}{2}-k\right)!} r^{n-2k} \qquad (257)$$

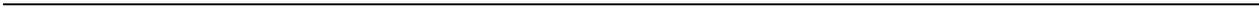

# Annexe 4. Code de traitement des cartes d'intensité en diffusion multiple

```matlab
%%%%%%%%%%%%%%%%%%%%%%%%%%%%%%%%%%%%%%%%%%%%%%%%%%%%%%%%%%%%%%%%%%%%%%%%%%
% Programme de calcul d'intensité collecté sur une photodiode %
% @Author : Matthias Sentis %
% @Mail : matthias.sentis@univ-amu.fr %
% @Version : 1.2 [06/01/2014] %
% % 
%En se basant sur le prototype proposé par Indatech, le code Fortran a %
%permis de fournir des cartes d'intensité (cylindre de détection déplié). %
%Ces cartes permettent à l'utilisateur (via une IHM triviale) de découper %
%dans ces cartes des photodiodes organiques circulaires de diamètre %
%variable et de position choisie et in fine obtenir une puissance %
%lumineuse théorique (aucun effet électronique n'a été pris en compte. %
%Les calculs ont été faits pour 100millions de photons(soit 3.7339e-011W %
%en 1s. %
%%%%%%%%%%%%%%%%%%%%%%%%%%%%%%%%%%%%%%%%%%%%%%%%%%%%%%%%%%%%%%%%%%%%%%%%%%
%
%Réinitialisation de l'environnement matlab (nettoyage des variables)
clear
%Ouverture d'une Fenetre graphique
fig1=figure;
%Positionnement de cette figure sur l'écran + taille
set( fig1 , 'position' , [ 100 , 100 , 1000 , 500 ]);
%
% Texte informatif de position angulaire de la photodiode
LabelPositionAngulaire = uicontrol( fig1 , 'style','text','position',...
[10,450,400,30] ,'string' , 'Position Angulaire photodiode[deg]' ,...
'fontsize' , 15 )
% Valeur de position angulaire de la photodiode
ValuePositionAngulaire = uicontrol ( fig1 ,'style',' edit' ,'position',...
[500,450,300,30] , 'Max' , 1 , 'string' , '0' )
%
% Texte informatif de position angulaire de la photodiode
LabelHauteur = uicontrol( fig1 , 'style' , 'text' , 'position' ,...
[10,400,300,30] ,'string' , 'Hauteur photodiode [mm]' , 'fontsize' , 15 )
% Valeur de position angulaire de la photodiode
ValueHauteur = uicontrol ( fig1 , 'style' , ' edit' , 'position',...
[500,400,300,30] , 'Max' , 1 , 'string' , '0' )
%
% Texte informatif de position angulaire de la photodiode
LabelDiam = uicontrol( fig1 , 'style' , 'text' , 'position' ,...
[10,350,300,30] ,'string' , 'Diamétre photodiode[mm]' , 'fontsize' , 15 )
% Valeur de position angulaire de la photodiode
ValueDiam = uicontrol ( fig1 , 'style' , ' edit' , 'position',...
[500,350,300,30] , 'Max' , 1 , 'string' , '2.5' )
%
% Texte informatif de position angulaire de la photodiode
LabelLongueurOnde = uicontrol( fig1 , 'style' , 'text' , 'position' ,...
[10,300,250,30] ,'string' , 'Longueur onde [nm]' , 'fontsize' , 15 )
% Valeur de position angulaire de la photodiode
ValueLongueurOnde = uicontrol ( fig1 , 'style' , ' edit' , 'position',...
```

```matlab
[500,300,300,30] , 'Max' , 1 , 'string' , '532' )
%
% Texte informatif de position angulaire de la photodiode
LabelIntensite = uicontrol( fig1 , 'style', 'text' , 'position' ,...
[10,250,450,30],'string','Puissance lumineuse collectée en 1s [W]',...
 'fontsize',15)
% Valeur de position angulaire de la photodiode
ValueIntensite = uicontrol ( fig1 , 'style' , ' edit' , 'position',...
[500,250,300,30] , 'Max' , 1 , 'string' , '0','BackgroundColor',[1 0 0] )
%
%Action lors de l'appuie sur le bouton buttonMap (chercher carte à découper)
ButtonMap= uicontrol ( fig1 , 'style','push','position',[50 50 150 80 ],...
'string' , 'Selectionner une carte' ,'callback',
set(ValueIntensite,''BackgroundColor'',[1 0
0]);FuncCloseFileOpened();[x,y,z]=FuncFindFile();')
%Calcul de l'intensité
ButtonIntensity= uicontrol ( fig1 , 'style' , 'push' , 'position' ,...
[400 50 150 80 ] ,'string' , 'Calcul intensité' ,'callback',...
'lambda= str2double(get(ValueLongueurOnde,''string''));
TestNumeriqueValeur(ValueLongueurOnde);ThetaPhotodiode=
str2double(get(ValuePositionAngulaire,''string''));
TestNumeriqueValeur(ThetaPhotodiode);Rdiode=
str2double(get(ValueDiam,''string''));
TestNumeriqueValeur(Rdiode);Zphotodiode=str2double(get(ValueHauteur,''string''
));
TestNumeriqueValeur(Zphotodiode);I=CalculIntensity(ThetaPhotodiode,Rdiode,Zpho
todiode,z,lambda);set (ValueIntensite , ''String'',num2str(I));
set(ValueIntensite,''BackgroundColor'',[0 1 0])')
%
% %%%%%%%%%%%%%%%%%%%%%%%%%%%%%%%%%%%%%%%%%%%%%%%%%%%%%%%%%%%%%%%%%%%%%%%%
% Programme de calcul d'intensité collecté sur une photodiode %
% @Author : Matthias Sentis %
% @Mail : matthias.sentis@univ-amu.fr %
% @Version : 1.2 [06/01/2014] %
% %
% Entrée : value (valeur à tester) %
% Sortie : aucune %
% %
%Teste seulement si les valeurs entrées par l'utilisateur sont numériques.%
%%%%%%%%%%%%%%%%%%%%%%%%%%%%%%%%%%%%%%%%%%%%%%%%%%%%%%%%%%%%%%%%%%%%%%%%
function []=TestNumeriqueValeur(value)
 %Si la valeur entrée n'est pas numérique
 if isnan(value)
 %Renvoie un message d'erreur et empêche la suite de l'éxécution
 errordlg('Valeur non numérique dans les paramétres','ERREUR','modal')
 return
 end
end
%%%%%%%%%%%%%%%%%%%%%%%%%%%%%%%%%%%%%%%%%%%%%%%%%%%%%%%%%%%%%%%%%%%%%%%%
% Fonction de recherche de fichier .fig %
% @Author : Matthias Sentis %
% @Mail : matthias.sentis@univ-amu.fr %
% @Version : 1.2 [06/01/2014] %
% %
% Entrées : Aucunes %
% Sortie : x (abcisse pixels), y (ordonnées pixels), I (intensité pour %
% chaque pixel. %
% %
%Fonction qui permet à l'utilisateur de sélectionner la carte avec laquel %
```

```matlab
%il veut travailler. %
%%%%%%%%%%%%%%%%%%%%%%%%%%%%%%%%%%%%%%%%%%%%%%%%%%%%%%%%%%%%%%%%%%%%%%%%%%%%

function [x,y,I]=FuncFindFile()
 %Ouverture de la boite de dialogue
 [filename, pathname] = uigetfile( ...
 { '*.fig', '*.fig' }, ...
 'Pick a file', ...
 'MultiSelect', 'on');
 %ouverture du fichier .fig selectionné
 fid = openfig(strcat(pathname,filename));
 % Recuperation des données de l'image
 [x,y,I]=getimage(fid);
 %close(fid)
end
%%%%%%%%%%%%%%%%%%%%%%%%%%%%%%%%%%%%%%%%%%%%%%%%%%%%%%%%%%%%%%%%%%%%%%%%%%%%
% Fonction de fermeture des fichiers .fig %
% @Author : Matthias Sentis %
% @Mail : matthias.sentis@univ-amu.fr %
% @Version : 1.3 [28/01/2014] %
% %
% Entrées : Aucunes %
% Sortie : Aucunes %
% %
%Fonction qui ferme toutes les figures courantes avant ouverture d'une %
%autre. Evite des confusions sur les figures lors du calcul de l'intensité%
%%%%%%%%%%%%%%%%%%%%%%%%%%%%%%%%%%%%%%%%%%%%%%%%%%%%%%%%%%%%%%%%%%%%%%%%%%%%
function []=FuncCloseFileOpened()
 %Récupération de toutes les figures ouvertes
 handles=findall(0,'type','figure')
 %Récupoération de la taille
 numfig = size(handles)
 %Si plus d'un fichier .fig est ouvert (le GUI est compté)
 if(numfig(1)>1)
 %On ferme toutes les figures
 for i=2:1:numfig(1)
 close(i)
 end
 end
end
%%%%%%%%%%%%%%%%%%%%%%%%%%%%%%%%%%%%%%%%%%%%%%%%%%%%%%%%%%%%%%%%%%%%%%%%%%%%
% Fonction de calcul d'intensité %
% @Author : Matthias Sentis %
% @Mail : matthias.sentis@univ-amu.fr %
% @Version : 1.2 [06/01/2014] %
% %
% Entrées : ThetaPhotodiode (position angulaire), Rdiode(Rayon de la %
% diode), Zphotodiode(hauteur photodiode),z (carte intensité), lambda( %
% longueur d'onde) %
% Sortie : Intensité calculée %
% %
%A partir des données entrées par l'utilisateur, ce programme somme %
%simplement la valeur des pixels appartenant à la photodiode pour ensuite%
%multiplier cette somme par l'énergie d'un photon hw. %
%%%%%%%%%%%%%%%%%%%%%%%%%%%%%%%%%%%%%%%%%%%%%%%%%%%%%%%%%%%%%%%%%%%%%%%%%%%%
%
function [I]=CalculIntensity(ThetaPhotodiode,Rdiode,Zphotodiode,z,lambda)
%Constante de Planck
h=6.626E-34;
```

```
%Célérité de ma lumière
c=299792458;
%rayon du cylindre de détection
r=16;
% Transformation du diamètre en rayon
Rdiode=Rdiode/2;
%Transformation position angulaire en distance
PositionDiodesSurCarte = ThetaPhotodiode*r*pi()/180;
%Initialisation de l'intensité à 0
 I=0;
 %Définition du maillage de la carte
 x= 0:1:359;
 y= 0:1:400;
 x=x./2;
 x = x.*(acos(-1.0)/180.0);
 x = x.*0.016;
 y=y.*250E-6-0.050;
 y=y.*1000;
 x=x.*1000;
 x=x';
 figure(2)
 hold on
 theta=[0:0.1:2*pi()+0.1];
 xdetect=Rdiode.*cos(theta)+PositionDiodesSurCarte;
 ydetect=Rdiode.*sin(theta)+Zphotodiode;
 plot(xdetect,ydetect,'red')
 % Pour tous les pixels, on test si ces derniers appartiennent à la diode
 % Si oui on somme la valeur de l'intensité
 for j=1:1:360
 for k=1:1:401
 if ((x(j)-PositionDiodesSurCarte)^2+(y(k)-Zphotodiode)^2<=Rdiode^2 )
 I = I+z(k,j);
 end
 end
 end
%Multiplication par l'énergie d'un photon
lambda=lambda*1E-9;
I=I*h*c/lambda;

end
```

# References

Abramowitz M et I Stegun (1964). Handbook of Mathematical functions with formulas, graphs and mathematical tables. New-York, Dover publications inc.

Airy G B (1838). "On the intensity of light in the neighbourhood of a caustic." Trans. Cambridge Philos. Soc **6**: 379-403.

Albada V, P Meint et A D Lagendijk (1985). "Observation of weak localization of light in a random medium." Physical review letters **55**(24): 2692-2695.

Albrecht H E, N Damaschke, M Borys et C Tropea (2003). Laser Doppler and Phase Doppler Measurement Techniques. Berlin, Springer.

Arca F, S F Tedde, M Sramek, J Rauh, P Lugli et O Hayden (2013). "Interface trap states in organic photodiodes." Scientific reports **3**.

Arnott W P et P L Marston (1991). "Unfolded optical glory of spheroids: Backscattering of laser light from freely rising spheroidal air bubbles in water." Appl. Opt. **30**(24): 3429-3942.

Auger J-C, V Martinez et B Stout (2007). "Absorption and Scattering Properties of Dense Ensembles of Non-spherical Particles." Journal of the Optical Society of America A **24**(11): 3508–3516.

Balboni M L (2003). "Process analytical technology." Pharmaceutical Technology.

Barber P W et S S Hill (1990). Light Scattering by Particles: Computational Methods. Singapore, World Scientific.

Barton J P, D R Alexander et S A Schaub (1988). "Internal and Near-Surface Electromagnetic Fields for a Spherical Particle Irradiated by a Focused Laser Beam." Journal of Applied Physics **64**(4): 1632-1639.

Beeck J P A J V et M L Riethmuller (1996). "Rainbow phenomena applied to the measurement of droplet size and velocity and to the detection of nonsphericity." Applied optics **35**(13): 2259-2266.

Beeck J P A V (1997). Rainbow phenomena : on development of a laser-based, non intrusive technique for measuring droplet size, temperature and velocity. Ph.D. Thesis Technical University of Eidoven ,Eindhoven, Pays-Bas.

Beeck J P A v et M L Riethmuller (1994). Simultaneous determination of temperature and size of droplets from rainbow using Airy theory. Proceedings of the 7th International Symposium Lisbon, Portugal, Springer.

Bergougnoux L (1995). Diagnostic optique pour la mesure de concentration de matières en suspension, Thèse de l'université de Provence, Marseille, France.

Berne B et R Pecora (2000). Dynamic light scattering: with applications to chemistry, biology, and physics. Mineola. New York, Courier Dover Publications.

Binek P (2007). Analysis of the Optically Dense Systems with Matrix T. Ph.D. thesis (in Polish), Wrocław University of Technology, Wrocław, Pologne.

Bissonnette L, P Bruscaglioni, A Ismaelli, G Zaccanti, A Cohen, Y Benayahu, M Kleiman, S Egert, C Flesia et P Schwendimann (1995). "Lidar multiple scattering from clouds." Applied Physics B **60**(4): 355-362.

Blum L et J S Høye (1977). "Mean spherical model for asymmetric electrolytes. 2. Thermodynamic properties and the pair correlation function." The Journal of Physical Chemistry **81**(13): 1311-1316.

Bohren C F et D R Huffman (1998). Absorption and Scattering of Light by Small Particles. Federal Republic og Germany, John Wiley and Sons.

Bordes C, P Snabre, C Frances et B Biscans (2003). "Optical investigation of shear-and time-dependent microstructural changes to stabilized and depletion-flocculated concentrated latex sphere suspensions." Powder technology **130**(1): 331-337.

Born M et E Wolf (1980). Principles of optics. Oxford, England, Pergamon press.

Born M et E Wolf (1999). Principles of optics : Electromagnetic theory of propagation, interference and diffraction of light. Cambridge, UK, Cambridge university press.

Briton J F (1989). simulations numériques de la diffusion multiple de la lumière par une méthode de Monte-Carlo, Thése de l'université de Rouen, Rouen, France.

Bruscaglioni P, A Ismaelli et G Zaccanti (1968). "Monte-Carlo calculations of Lidar returns:procedure an results." Applied physic **60**(4): 325-329

Buffon G L L (1733). "Mémoire sur le jeu du Franc Carreau." présenté à l'Académie des Sciences

Buron H, O Mengual, G Meunier, I Cayre et P Snabre (2004). "Optical characterization of concentrated dispersions: applications to laboratory analyses and on-line process monitoring and control." Polymer international **53**(9): 1205-1209.

Burr D W, K J Daun, O.Link, K A Thomson et G J Smallwood (2007). "Soot particle sizing by inverse analysis of multiangle elastic light scattering." NRC Publications Archive.

Buslenko N P, D I Golenko, Y A Shreider, I M Sobol et V G Sragovich (1966). The Monte Carlo Method: The Method of Statistical Trials. Oxford, Pergamon press.

Calba C (2008). Interaction entre une impulsion lumineuse ultra-brève et un nuage dense de particules: simulations numériques et expériences, Thése de l'université de Rouen, Rouen, France.

Chandrasekhar S (1960). Radiative transfert. New york, Dover publications.

Chaneliere T (2004). <u>Diffusion multiple coherente avec atomes froids de strontium: Effet de la saturation sur la retrodiffusion coherente-Piege magneto-optique sur raie etroite</u>, thèse de l'université Nice Sophia Antipolis, Nice, France.

Chang S C, J M Jin, J Jin et S Zhang (1996). <u>Computations of Special Functions</u>. New-York, John Wiley & Sons.

Collins D G, W G Blttner, M B Wells et H G Horak (1972). "Monte Carlo Calculations of the Polarization Characteristics of the Radiation Emerging from Spherical-Shell Atmospheres." <u>Applied optics</u> **11**(11): 2684-2696.

Davis G E (1955). "Scattering of light by an air bubble in water." <u>J. Opt. Soc. Am. A</u> **45**(7): 572-572.

De Beer T, C Bodson, B Dejaegher, B Walczak, P Vercruysse, A Burggraeve, A Lemos, L Delattre, Y V Heyden et J P Remon (2008). "Raman spectroscopy as a process analytical technology (PAT) tool for the *in-line* monitoring and understanding of a powder blending process." <u>Journal of pharmaceutical and biomedical analysis</u> **48**(3): 772-779.

Debye P (1909). "Der Lichtdruck auf Kugeln von beliebigem Material." <u>Ann. Phys</u> **4**(30): 58.

Desvignes F (1992). Radiométrie. Photométrie. <u>Techniques de l'ingénieur Optique instrumentale</u>. E. t. d. l'ingénieur. Paris, France, Techniques de l'ingénieur.

DeVoe H (1964). "Optical Properties of Molecular Aggregates. I. Classical Model of Electronic Absorption and Refraction." <u>Journal of Chemical Physics</u> **41**(2): 393-400.

Doicu A, T Wriedt et Y A Eremin (2006). <u>Light Scattering by Systems of Particles</u>. Berlin, Heidelberg, Springer.

Draine B T et P J Flatau (1994). "Discrete Dipole Approximation for Scattering Calculations." <u>Journal of the Optical Society of America A</u> **11**(4): 1491-1499.

El-Hagrasy A S et J K Drennen (2006). "A Process Analytical Technology approach to near-infrared process control of pharmaceutical powder blending. Part III: Quantitative near-infrared calibration for prediction of blend homogeneity and characterization of powder mixing kinetics." <u>Journal of pharmaceutical sciences</u> **95**(2): 422-434.

Fiedler-Ferrari N, H M Nussenzweig et W J Wiscombe (1991). "Theory of near-critical-angle scattering from a curved interface." <u>Phys. Rev. A</u> **43**(2): 1005-1005.

Fresnel A (1816). "Oeuvres." <u>Ann. Chim et Phys</u> **1**: 89-129.

Gessner H (1936). <u>L'Analyse mécanique : tamisage, sédimentation, lévigation</u>. Paris, Dunod.

Gouesbet G (2004). "Debye Series Formulation for Generalized Lorenz-Mie Theory with the Bromwich Method." <u>Part. Part. Syst. Charact.</u> **20**(6): 382-386.

Gouesbet G, B Maheu et G Grehan (1988). "Light Scattering From a Sphere Arbitrarily Located in a Gaussian Beam, Using a Bromwich Formulation." <u>Journal of the Optical Society of America A</u> **5**(9): 1427-1443.

Grace J R, T Wairegi et T H Nguyen (1976). "Shapes and velocities of single drops and bubbles moving freely through immiscible liquids." Trans. Inst. Chem. Eng. **54**(a): 167-173.

Gulink M (1943). "Sur la précision des analyses granulométriques par tamisage." Bulletin de la Société Belge de géologie Paléontologie hydrologie: 206-213.

Haeringen W et D Lenstra (1990). Analogies in Optics and Micro Electronics. Dordrecht, Kluwer.

Hansen J et J Pollack (1970). "Near-infrared light scattering by terrestrial clouds." Journal of the Atmospheric Sciences **27**(2): 265-281.

Hansen J et D Travis (1974). "Light scattering in planetary atmospheres." Space Science Reviews **16**(4): 527-610.

Hansen P C (2010). Discrete inverse problems: insight and algorithms. Philadelphia (PA), USA.

Heffels C M G, R Polke, M Radle, N Sachweh, M Schafer et N Scholtz (1998). "Control of Particulate Processes by Optical Measurement Techniques." Part. and Part. Syst. Charact. **15**(5): 211-218.

Henyey L C et J L Greenstein (1941). "Diffuse radiation in the galaxy." Astrophys. **93**: 70–83

Hespel L, M Barthelemy, N Riviere, A Delfour, L Mees et G Grehan (2007). "Temporal scattering of dense scattering media under ultra short laser light illumination: Application for particle sizing." ICHMT DIGITAL LIBRARY ONLINE **14**.

Hovenac E A et J A Lock (1992). "Assessing the contribution of surface waves and complex rays to far-field scattering by use of the Debye series." J. Opt. Soc. Am. A **9**(5): 781-795.

Hulst H C (1957). Light Scattering by Small Particles. New York, Courier Dover Publications.

Hulst H C (1980). Multiple light scattering : tables, formulas and applications. USA, Academic press, inc.

Hulst H C (1981). Light Scattering by small particles. New York, Courier Dover Publications.

Huygens C (1690). Traité de la lumière.

Irvine W (1964). "Light scattering by spherical particles : radiation pressure, assymetry factor, and extinction cross section." Journal of the Optical Society of America (JOSA) **55** (1): 16-21.

Ishimaru A (1978). "Diffusion of a pulse in densely distributed scatterers." JOSA **68**(8): 1045-1050.

Ishimaru A et Y Kuga (1982). "Attenuation constant of a coherent field in a dense distribution of particles." JOSA **72**(10): 1317-1320.

Jones A R (1999). "Light scattering for particle characterization." Progress in Energy and Combustion science **25**(1): 1-53.

Keller J B (1961). "Geometrical Theory of Diffraction." <u>Journal of the Optical Society of America</u> **52**(2): 116-130.

Kirk J T O (1992). "Monte Monte carlo modelling of the performance of a reflective tube absorption meter." <u>Appl optic</u> **31**(30): 6463-6468.

Kovalenko S A (2001). "Descartes-Snell law of refraction with absorption." <u>Semiconductor Physics, Electronics & Optoelectronics</u> **4**(3): 214-218.

Langley D S et P L Marston (1984). "Critical scattering of laser light from bubbles in water: measurements, models, and application to sizing bubbles " <u>Appl. Opt.</u> **7**(23): 1044-1054.

Lemaitre P, E Porcheron, A Nuboer, P Brun, P Cornet, J Malet, J Vendel et G Grehan (2004). <u>Développement de la réfractométrie arc-en-ciel global pour mesurer la température de gouttes en chute libre</u>. 9ème Congrès Francophone de Vélocimétrie Laser, IKV,ULB, AFVL.

Liang J, B Grimm, S Goelz et J F Bille (1994). "Objective measurement of wave aberrations of the human eye with the use of a Hartmann-Shack wave front sensor." <u>J. Opt. Soc. Am. A</u> **11**(7): 1949-1957.

Lock J A (2003). "Role of the tunneling ray in near-critical-angle scattering by a dielectric sphere." <u>J. Opt. Soc. Am. A</u> **20**(3): 499-507.

Lorenz L (1890). "Lysbevaegelsen i og uden for en af plane Lysbolger belyst Kugle. Det Kongelige Danske Videnskabernes Selskabs Skrifter." <u>6. Raekke, 6. Bind</u>: 1-62.

Lötsch H K V (1971). "Beam displacement at total reflection: the Goos-Hänchen effect." <u>Optick</u> **32**.

Maheu B (1988). <u>High fidelity simulation and modelisation of heat tranfer between a turbulent flow and a wall</u> Thése de l'université de Rouen, Rouen, France.

Maret G et P E Wolf (1987). "Multiple light scattering from disordered media. The effect of Brownian motion of scatterers." <u>Zeitschrift für Physik B Condensed Matter</u> **37**(4): 409-413.

Marsaglia G (1995) "DIEHARD Statistical Tests." <u>http://stat.fsu.edu/~geo/diehard.html</u>.

Marston P L (1979). "Critical scattering angle by a bubble: physical optics approximation and observations." <u>J. Opt. Soc. Am. A</u> **69**(9): 1205-1211.

Marston P L et D L Kingsbury (1981). "Scattering by a bubble in water near the critical angle: interference effects." <u>J. Opt. Soc. Am. A</u> **71**(2): 192-196.

Martin P A (2006). <u>Multiple scattering: interaction of time-harmonic waves with N obstacles</u>. Cambridge, Cambridge University Press.

Matsumoto M et Y Kurita (1994). "Twisted gfsr generators ii." <u>ACM Transactions on Modeling and Computer Simulation (TOMACS)</u> **4**(3): 254-266.

Matsumoto M et T Nishimura (1998). "Mersenne Twister: A 623-dimensionally equidistributed uniform pseudo-random number generator." ACM Transactions on Modeling and Computer Simulation **7**(1): 3-30.

Mengual O, G Meunier, I Cayre, K Puech et P Snabre (1999). "Characterisation of instability of concentrated dispersions by a new optical analyser: the TURBISCAN MA 1000." Colloids and Surfaces A: Physicochemical and Engineering Aspects **152**(1): 111-123.

Mengual O, G Meunier, I Cayre, K Puech et P Snabre (1999). "TURBISCAN MA 2000: multiple light scattering measurement for concentrated emulsion and suspension instability analysis." Talanta **50**(2): 445-456.

Metropolis N (1987). "The Beginning of the Monte Carlo Method " Los Alamos Science **15**: 125-130.

Metropolis N et S Ulam (1949). "The Monte Carlo Method." Journal of the American Statistical Association **44**( 247): 335-341.

Mie G (1908). "Beiträge zur Optik Trüber Medien, Speziell Kolloidaler Metallösungen." Annalen der Physik **330**(3): 377–445.

Mishchenko M I, L D Travis et A A Lacis (2006). Multiple scattering of light by particles. New York, USA, Cambridge university press.

Mishchenko M I, L D Travis et D W Mackowski (1996). "T-Matrix Computations of Light Scattering by Nonspherical Particles: a Review." Journal of Quantitative Spectroscopy and Radiative Transfer **55**(5): 535-575.

Moes J J, M M Ruijken, E Gout, H W Frijlink et M I Ugwoke (2008). "Application of process analytical technology in tablet process development using NIR spectroscopy: Blend uniformity, content uniformity and coating thickness measurements." International journal of pharmaceutics **357**(1): 108-118.

Montet C, M Sentis et F Onofri (2014). Characterization of nanopaticles aggregates by static light scattering. Int. Conf. on Lasers and Interactions with Particles (LIP-2014), Marseille, France.

Mourant J, P Freyer, A Hielscher, A Eick, D Shen et T Johnson (1998). "Mechanisms of light scattering from biological cells relevant to noninvasive optical-tissue diagnostics." Applied Optics **37**(16): 3586-3593.

Mul F F M (2004). Monte-Carlo Simulations of Light Scattering in Turbid Media, Internal report University of Twente.

Mul F F M (2011). "Monte-Carlo simulation of light scattering in turbid media." Internal report, University of twente **Version 3**.

Muller M E (1959). "A comparison of methods for generating normal deviates on digital computers." Journal of the ACM (JACM) **6**(3): 376-383.

Neumann J (1951). Techniques used in connection with random digits. Washington D.C, USA, U.S. Government Printing Office.

Ng T N, W S Wong, M L Chabinyc, S Sambandan et R A Street (2008). "Flexible image sensor array with bulk heterojunction organic photodiode." Applied Physics Letters **92**(21): 213303.

Nieminen T A, V L Y Loke, A B Stilgoe, G Knoner, A M Branczyk, N R Heckenberg et H Rubinsztein-Dunlop (2007). "Optical Tweezers Computational Toolbox." Journal of Optics A **9**(8): 196-203.

Nikolic K, D S S Bello, T Delbruck, S-C Liu et B Roska (2011). "High-sensitivity silicon retina for robotics and prosthetics." SPIE Electronic Imaging & Signal Processing.

Nomura K, H Ohta, A Takagi, T Kamiya, M Hirano et H Hosono (2004). "Room-temperature fabrication of transparent flexible thin-film transistors using amorphous oxide semiconductors." Nature **432**(7016): 488-492.

Nussenzveig H M et H Moysés (1979). "Complex angular momentum theory of the rainbow and the glory." JOSA **69**(8): 1068-1079.

Nussenzweig H M (1992). Diffraction effects in semiclassical scattering. Cambridge, Cambridge University Press.

Onofri F (1999). " Critical Angle Refractometry: for simultaneous measurement of particles in flow size and relative refractive index." Part. and Part. Syst. Charact. **13**(3): 119-127.

Onofri F (2005). Diagnostics optiques des milieux multiphasiques. HDR, Université de Provence, Marseille, France.

Onofri F et S Barbosa (2012). Chapter II: Optical particle characterization. London, Wiley-ISTE.

Onofri F et S Barbosa (2012). Diffusion de la Lumière (Chapitre 1). Paris, Boutier.

Onofri F, S Barbosa, O Touré, M Woźniak et C Grisolia (2013). "Sizing highly-ordered buckyball-shaped aggregates of colloidal nanoparticles by light extinction spectroscopy." Journal of Quantitative Spectroscopy and Radiative Transfer **126**: 160-168.

Onofri F, L Bergougnoux, J L Firpo et J Misguich-Ripaud (1999). "Velocity, size and concentration measurements of optically inhomogeneous cylindrical and spherical particles." Applied Optics **38**(21): 4681-4690.

Onofri F, G Grehan et G Gouesbet (1995). "Electromagnetic Scattering From a Multilayered Sphere Located in an Arbitrary Beam." Applied Optics **34**(30): 7113-7124.

Onofri F, M Krzysiek et J Mroczka (2007). "Critical Angle Refractometry and Sizing for Bubbly Flow Characterization." Opt. Lett **32**(14): 2070-2072.

Onofri F, M Krzysiek, J Mroczka, K-F Ren, S Radev et J-P Bonnet (2009). "Optical characterization of bubbly flows with a near-critical-angle scattering technique." Exp. in Fluids **47**(4-5): 7731-7732.

Onofri F, M A Krzysiek, S Barbosa, V Messager, K-F Ren et J Mroczka (2011). "Near-critical-angle scattering for the characterization of clouds of bubbles: particular effects." Appl. Opt. **50**(30): 5759-5769.

Onofri F, S Radev, M Sentis et S Barbosa (2012). "A physical-optics approximation of the near-critical-angle scattering by spheroidal bubbles." Optics Letters **37**(22): 4780-4782.

Ouedraogo M (2005). Arc-en-Ciel global: développement d'une technique non intrusive de mesure d'indices et de tailles. Stage Ingénieur 2ème année, Université de Provence, Marseille, France.

Pine D J, D A Weitz, P M Chaikin et E Herbolzheimer (1988). "Diffusing wave spectroscopy." Physical Review Letters **60**(12): 1134-1134.

Poicelot P (1957). "Sur le théorème de Babinet au sens de la théorie électromagnétique." Annales des Télécommunications **12**(12): 410-414.

Purcell E M et C R Pennypacker (1973). "Scattering and Absorption of Light by Nonspherical Dielectric Grains." American Astronomical Society **186**: 705-714.

Raković M et G Kattawar (1998). "Theoretical analysis of polarization patterns from incoherent backscattering of light." Applied optics **37**(15): 3333-3338.

Ren K-F, F Onofri, C Rozé et T Girasole (2011). " Vectorial complex ray model and application to two-dimensional scattering of plane wave by a spheroidal particle." Opt. Lett **36**(3): 370-372.

Ren K F, G Grehan et G Gouesbet (1997). "Scattering of a Gaussian Beam by an Infinite Cylinder in GLMT-Framework, Formulation and Numerical Results." Journal of the Optical Society of America A **14**(11): 3014-3025

Roth N, K Anders et A Frohn (1991). "Refractive index measurements for the correction of particle sizing methods." App. Opt. **30**(33): 4960-4965.

Roth N, K Anders et A Frohn (1991). "Simultaneous measurement of temperature and size of droplets in micrometer range." J. of Laser. Appl. **2**(1 ): 37-42.

Roth N, N Anders et A Frohn (1994). "Determination of size, evaporation rate, and freezing of water droplets using light scattering and radiation pressure." Part. Part. Syst. Charat. **11**(3): 207-211.

Sankar S V, K M Ibrahim, D H Buermann, M J Fidrich et W D Bachalo (1993). An integrated Phase Doppler/Rainbow refractometer system for simultaneous measurement of droplet size, velocity, and refractive index. 3rd Int. Cong. on Optical Partical Sizing, Yokohama (JAPAN).

Scheffold F (2002). "Particle sizing with diffusing wave spectroscopy." Journal of dispersion science and technology **23**(5): 591-599.

Schelkunoff S A (1992). "Some Equivalence Theorems of Electromagnetics and Their Application to Radiation Problems." Bell system technical journal **15**(1): 92-112.

Snabre P et A Arhaliass (1998). "Anisotropic scattering of light in random media: incoherent backscattered spotlight." Applied optics **37**(18): 4017-4026.

Ungut A, G Grehan et G Gouesbet (1981). "Comparisons between geometrical optics and Lorenz-Mie theory." Applied Optics **20**(17): 2911-2918.

Wang L H et W R Chen (1997). "Optimal beam size for light delivery to absorption-enhanced tumors buried in biological tissues and effect of multiple beam delivery–A Monte Carlo study." Appl Opt **36**(31): 8286–8291.

Wang X, Y Gang et L V Wang (2002). "Monte Carlo model and single-scattering approximation of the propagation of polarized light in turbid media containing glucose." Applied optics **41**(4): 792-801.

Waterman P C (1965). "Matrix Formulation of Electromagnetic Scattering." Proceedings of the IEEE **53**(8): 805-812.

Watson D (2004) "Diffraction from raindrops (Physics lecture notes)."

Wolf E, G Maret, E Akkermans et R Maynard (1988). "Optical coherent backscattering by random media: an experimental study." Journal de Physique **49**(1): 63-75.

Wozniak M (2012). Characterization of nanoparticle aggregates with light scattering techniques. PHD thesis (in english), Provence university.

Wriedt T (2009). "Light Scattering Theories and Computer Codes." Journal of Quantitative Spectroscopy and Radiative Transfer **110**(11): 833-843.

Wu L, H Yang, B Y X. Li et G Li (2007). " Scattering by large bubbles: Comparisons between geometrical-optics theory and Debye series." J. Quant. Spectros. Radiat. Transfer **108**(1): 54-64.

Xu F, K F Ren, X Cai et J Shen (2006). "Extension of geometrical-optics approximation to on-axis Gaussian beam scattering. II. By a spheroidal particle with end-on incidence." Applied Optics **45**(20): 4990-4999.

Xu R (2002). Particle Characterization: Light Scattering Methods. New York, Kluwert Academic Publisher.

Young M (1971). "Pinhole Optics." Applied optics **10**(12): 2763-2767.

Yuhan Y ( 2012). Diffusion de lalumière par un objet irrégulier et application à l'imagerie des sprays, Thése de l'université de Rouen, Rouen, France.

Zernike F (1934). "Beugungstheorie des schneidenver-fahrens und seiner verbesserten form, der phasenkontrastmethode." Physica **1**(7-12): 689-704.

Zhang X (2011). Modélisation du brouillard durant la campagne ParisFog: approche prédictive et étude de l'effet des hétérogénéités spatiales, Thèse de l'université Joseph Fourier, Grenoble, France.

Zhou J, N V Lafferty, B W Smith et J Burnett (2007). Immersion lithography with numerical apertures above 2.0 using high index optical materials. Optical Microlithography XX, San Jose, CA.

# Résumé

Dans le cadre d'un consortium entre centres de recherche publics et industriels, ce travail de thèse de doctorat s'est attaché à démontrer l'intérêt des détecteurs photo-organiques (OPS) pour la caractérisation des suspensions et écoulements diphasiques. Les principes de plusieurs granulomètres permettant la caractérisation de ces milieux lorsqu'ils sont confinés dans une cuve cylindrique transparente (configuration standard du *Process Analytical Technology*) ont été proposés. Pour évaluer et optimiser les performances de ces systèmes, un code de simulation de type Monte-Carlo a été spécifiquement développé. Ce dernier permet de prendre en compte les nombreux paramètres du problème comme le profil du faisceau laser, les différentes surfaces spéculaires composant le montage, la composition du milieu particulaire (concentration, diamètre moyen, écart-type, matériau,...), la forme et la position des OPS. Les propriétés de diffusion des particules sont traitées à l'aide des théories de Lorenz-Mie et de Debye, de même qu'un modèle hydride prenant en compte les contributions géométriques et physiques. Pour les milieux dilués (diffusion simple), l'analyse repose sur l'inversion des diagrammes de diffusion obtenus sur une large plage angulaire ou au voisinage de singularités optiques remarquables (arc-en-ciel, diffusion critique, diffraction). Pour les milieux denses (diffusion multiple), les pistes étudiées reposent sur l'analyse des caractéristiques de la tache de rétrodiffusion.

**Mots clés** : détecteurs photo-organiques, granulométrie optique, diffusion de la lumière par des particules, méthode de Monte-Carlo, diffusion simple et multiple, inversion, singularités, tache de rétrodiffusion.

# Abstract

As part of a consortium between academic and industry, this PhD work investigates the interest and capabilities of organic photo-sensors (OPS) for the optical characterization of suspensions and two-phase flows. The principle of new optical particle sizing instruments is proposed to characterize particle systems confined in a cylinder glass (standard configuration for *Process Analytical Technologies*). To evaluate and optimize the performance of these systems, a Monte-Carlo model has been specifically developed. This model accounts for the numerous parameters of the system: laser beam profile, mirrors, lenses, sample cell, particle medium properties (concentration, mean & standard deviation, refractive indices), OPS shape and positions, etc. Light scattering by particles is treated either by using Lorenz-Mie theory, Debye, or a hybrid model (that takes into account the geometrical and physical contributions). For diluted media (single scattering), particle size analysis is based on the inversion of scattering diagrams obtained over a wide angular range or near optical singularities (rainbow, critical scattering, diffraction). For dense media (multiple scattering), the solutions foreseen are based on the analysis of the backscattering spotlight characteristics.

**Key words**: organic photo-sensors, optical particle sizing, Monte-Carlo model, light scattering by particles, single and multiple scattering, inversion, optical singularities, backscattering spotlight.